\documentclass[showkeys,aps,showpacs,nofootinbib,floatfix]{revtex4}
\usepackage{bm}
\usepackage{graphicx}
\usepackage{latexsym,color}
\usepackage{hyperref}
\usepackage{amssymb}
\usepackage{natbib}
\usepackage{amsmath,amssymb,amsthm,mathrsfs,amsfonts,dsfont}
\usepackage{color}
\usepackage{hyperref}
\begin{document}
\title{Relativistic thick accretion disks: morphology and evolutionary parameters}
\author{Daniela Pugliese$^{1,\,2}$, Giovanni Montani$^{3,\, 4}$}
\affiliation{\vspace{3mm}
$^1$School of Mathematical Sciences, Queen Mary University of London, Mile End Road, London E1 4NS,  (UK)\\
$^2$Institute of Physics, Faculty of Philosophy \& Science,
  Silesian University in Opava,
 Bezru\v{c}ovo n\'{a}m\v{e}st\'{i} 13, CZ-74601 Opava, Czech Republic
 $^3$ENEA, Unit\`a Tecnica Fusione, ENEA C. R. Frascati, via E. Fermi 45, 00044 Frascati (Roma), Italy
\\
$^4$ Physics Department, ``Sapienza'' University of Rome,\\{ P.le Aldo Moro 5, 00185 (Roma), Italy}\vspace{3mm}
}
\date{\today}
\begin{abstract}
We explore thick accretion disks around rotating attractors. We detail the configurations analysing the fluid angular momentum and finally providing  a characterization of  the disk morphology and  different  possible  topologies.
 Investigating  the  properties of orbiting disks, a classification of attractors, possibly identifiable in terms of their spin-mass ratio, is introduced;  then an attempt to characterize dynamically a series of different disk topologies is discussed,  showing  that some of the  toroidal configuration features  are  determined by the ratio of  the angular momentum of the orbiting matter and the spin mass-ratio of the attractor.
Then we focus   on  ``multi-structured'' disks,  constituted by two o more rings of matter orbiting the same attractor, and we proved that some structures are   constrained in  the dimension of rings, spacing, location and an upper limit  of ring number is provided. Finally, assuming  a polytropic equation
of state we study some specific cases.
\end{abstract}

%
%

\maketitle

\newcommand{\ti}[1]{\mbox{\tiny{#1}}}
\newcommand{\im}{\mathop{\mathrm{Im}}}
\def\be{\begin{equation}}
\def\ee{\end{equation}}
\def\bea{\begin{eqnarray}}
\def\eea{\end{eqnarray}}
\newcommand{\tb}[1]{\textbf{\texttt{#1}}}
\newcommand{\ttb}[1]{\textbf{#1}}
\newcommand{\rtb}[1]{\textcolor[rgb]{1.00,0.00,0.00}{\tb{#1}}}
\newcommand{\btb}[1]{\textcolor[rgb]{0.00,0.00,1.00}{\tb{#1}}}
\newcommand{\il}{~}
\newcommand{\rc}{\rho_{\ti{C}}}
\newcommand{\dd}{\mathcal{D}}
\newcommand{\lie}{\mathcal{L}}
\newcommand{\Mie}{\mathcal{M}}
\newcommand{\Tem}{T^{\rm{em}}}
\section{Introduction}
Accretion disks are one  of the most remarkable environments in  the high energy Astrophysics. In this article  we investigate  thick  accretion disks orbiting a Kerr black hole attractor.  We model the accreting  toroidal matter within  the so called   ``Polish doughnut''  (P-D) hydrodynamic model   introduced  and detailed in a series of works 	
\cite{cc,Pac-Wii,Koz-Jar-Abr:1978:ASTRA:,Abr-Jar-Sik:1978:ASTRA:,Abr-Cal-Nob:1980:ASTRJ2:,Jaroszynski(1980),A1981,astro-ph/0411185,Abramowicz:1996ap,
FisM76,Raine,PuMonBe12}, and then
 developed for many  different attractors  and contexts \cite{2011,Abramowicz:2011xu,Rez-Zan-Fon:2003:ASTRA:,Stuchlik:2012zza,Sla-Stu:2005:CLAQG:,
 astro-ph/0605094,Stu-Sla-Hle:2000:ASTRA:,arXiv:0910.3184,AEA,Adamek:2013dza,Cio-Re,Komiss,Hamersky:2013cza,PuMon13}.  This is a fully general
relativistic model of an opaque and super-Eddington, pressure supported disk,
 based on the Boyer theory of     the equilibrium and rigidity in general relativity \cite{Boy:1965:PCPS:}.
Accretion disks are important structures in the Universe, associated with different  physical phenomena of the high energy sector as Gamma ray bursts (GRBs) or X-ray binaries, they are generally  characterized by well established  geometrical symmetries and constituted by matter and magnetic fields orbiting   an attractor.  The characterization of these objects is  important both to  sketch a model of different phenomena associated to their dynamics  both  for the identification of   the attractor features. Thick accretion disks are usually  associated with very compact objects like  black holes,
and thus they represent   good tracers  for the possible recognition of black hole sources,  giving  rise to physical processes  useful  to catch information of   a possible
black holes presence and their characterization.
Some aspects of  the  rotating attractors for example  are still to be defined as  the presence of some ``magic'' spin-mass ratios emerged from the   Quasi-Periodic Oscillation (QPOs) analysis\cite{Stuchlik:2013esa,KA021,KA02,AKBHRT,SS11,TAKS05,RH05a,RH05,RYMZ,RVW,SM10,
Nagar:2006eu,Abramowicz:2011xu}.
These objects represent a  challenge  for the   current theoretical scenarios where   the jet formation and dynamics,  the  Active Galactic Nuclei (AGN) or
GRBs processes are currently described, in the end accretion disks are  directly involved   in   the equilibrium phases of the attractors.
Some important aspects of  their  structure and morphology are still unclear  needing to be contextualized in the  observational manifestation of the phenomena around compact objects.
The location of the innermost boundary of the disk, the accretion mechanism,  the QPOs and the instability  in general   are  aspects   still very  much debated.
In the Boyer model we are considered here many   features of the disk dynamics and morphology  like the thickness, the stretching  on the equatorial plane and   the location of the disk,  are predominantly managed and determined by the geometric properties of spacetime via an effective potential function  regulating the pressure gradient in  the Euler law.
The effective potential, however, contains  two  essential components: a  geometrical one, related to the properties of the  spacetime background  and  a dynamical one related to the orbiting matter   by means of the fluid angular momentum here  assumed constant along  the disk (see also \cite{Lei:2008ui}).
A third element  adjusting the P-D model  is   the spacetime symmetries envisaged  here by the Killing vectors of the Kerr metric,  and the dynamical symmetries of the fluid configurations
taken in circular (time independent) motion: in other words  a stationary toroidal topology with equatorial plane of symmetry aligned with the equatorial plane of the axially symmetric source.
According to the  ``Boyer's condition'' for  the
analytic theory of equilibrium configurations of   rotating perfect fluids \cite{Boy:1965:PCPS:},   the boundary of any stationary, barotropic, perfect fluid body is determined  by the
equipotential surface,  therefore
surfaces of constant pressure, defined by the gradient of a
scalar function (i.e. the effective potential). This  property holds  if
the relativistic frequency $\Omega$ turns to be function of the fluid angular momentum  $\ell$ only or $\Omega=\Omega(\ell)$
(von Zeipel condition)
\cite{Koz-Jar-Abr:1978:ASTRA:,Jaroszynski(1980),M.A.Abramowicz,Chakrabarti,Chakrabarti0}.
Paczynski realized  that an ad hoc distribution of angular momentum  is a  physically
reasonable assumption, as
during the evolution of dynamic processes, the functional form of the angular
momentum and entropy distribution depends on the initial conditions of the system and on
the details of the dissipative processes. Even if in real situations the fluid  angular momentum would be  determined by  different factors including dissipative processes, the current  models require  assumptions on the viscosity turning again in the adoption of some ad hoc  functions \cite{Balbus2011,Shakura1973}.

From  this theoretical framework three topological classes emerge: the {closed} toroidal configurations, the {open} configurations
and finally  self-crossing surfaces {with a cusp}, which can be either closed or open. The
closed  surfaces  correspond  to stationary equilibrium configurations, the fluid
 filling  any closed surface,  on the other hand the open ones  are important for the modelization of some
dynamical situations as matter funnels or jets.
The crossed surfaces are associated to non-equilibrium situations and, in the case of closed crossed  surfaces, to   disks accreting  onto the black hole.
As  theorized by  Paczy\'nski    from the study of Roche lobe in the accretion disks of the binary
systems, accretion from thick disks is a consequence of the strong gravitational field of the
attractor  realized  by the relativistic
Roche lobe overflow,  neglecting  therefore the role of any dissipative effects like viscosity or resistivity  \cite{Boy:1965:PCPS:,Raine}.  According to Paczy\'nski mechanism
\cite{Abr-Jar-Sik:1978:ASTRA:,Koz-Jar-Abr:1978:ASTRA:,Jaroszynski(1980),A1981}, the
accretion onto the source (black hole) is driven through the vicinity of the cusp (corresponding  to the inner edge of
the disk) in the self crossed configurations  driven by a violation of the hydrostatic equilibrium,
\cite{Koz-Jar-Abr:1978:ASTRA:}. This same mechanism has been proved to be also an
important stabilizing mechanism against the thermal and viscous instabilities locally,
and against the so called Papaloizou and Pringle instability globally \cite{Blaes1987,A1981,Abramowicz:2008bk,Pac-Wii,cc,Koz-Jar-Abr:1978:ASTRA:,
Abr-Jar-Sik:1978:ASTRA:,Jaroszynski(1980),Abr-Cal-Nob:1980:ASTRJ2:,Abramowicz:1996ap,
FisM76,Lei:2008ui,F-D-02,Abramowicz:1997sg}.

In the present article  we focus in particular on  the rotating fluid angular momentum, the location of the maximum and minimum points  of the hydrostatic pressure, the disk center and inner and outer edge of the configurations.  The model, in the regions close to the static limit, is also studied. It will be    convenient to analyze the  accretion disk properties in terms of the  ratio $\bar{\ell}\equiv\ell/ a$ as  an important  parameter  for these  models, we motivate this statement and propose  a comparative analysis for the fluid   in terms of $\ell$ and $\bar{\ell}$.
Then we   detail the morphological and dynamical properties   of Boyer  configurations for different spacetimes.
As a result of this analysis we introduce nine classes of attractors identified by  their spin-mass ratio. Associated to attractor  classes, we consider six  orbital regions related  with the different topological structures of the Boyer fluids.
These classifications open up the possibility of studying thick accretion  disks  by an array or sequence of configurations, elaborated  varying     one of the  two model parameters $\mathbf{p}\equiv(K,\ell)$,  where $K$ is a constant  naturally established  from   the effective potential.
This will be the starting point for the analysis of the second part of this work, where a more general class of configurations, including the P-D tori,  will  be analyzed,  these configurations  turn out in different topologies not arising from  the potential critical points, see also \cite{Abramowicz:2011xu,Raine,BAF2006,Hawley1990,Abramowicz:2004vi}. The array  of configurations will be fitted with a dynamic interpretation useful for  the comparison  with numerical simulations in  more extensive  dynamic models simulating, for example,  the interaction  with some  matter environments. The P-D   analytic model has been in fact  used as  starting condition for numerical studies of black hole accretion, indeed simulations of accretion flows verify
the   agreement with the
model predictions even  in  global
magnetohydrodynamic numerical simulation e.g. \cite{Igumenshchev,Shafee} and \cite{Fragile:2007dk,DeVilliers,Hawley1987,Hawley1990,Hawley1991,Hawley1984,
arXiv:0910.3184,astro-ph/0605094,Stu-Kov:2008:INTJMD:,Raine,Abramowicz:2011xu,Fon03}.
The sequences will be then considered for the  ``multi-structured''  disks, or multiple toroidal surfaces  made by a number of thick rings orbiting    the same attractor. Accretion disks can be structured in two or more rings,  they can be
considered in a variety of models for
planetary  disks
or in the binary systems with non necessary complanar rings,
 or the  Galaxy  rings.
From  a mutual destabilization among  the rings, a  non-equilibrium stage  for  the entire structure could arise,  driven     according to Paczy\'nski mechanism
of violation of the hydrostatic equilibrium for the single  toroidal ring. This destabilization  would be  accompanied by  feeding, or   exchange of
fluid elements, among  the rings.
What is relevant here however, is that the geometric P-D  model  allows to constrain the number of the rings, size and properties of the angular momentum of the fluid,  we can classify different ring  structures and therefore the  ``multi-structured'' torii.  Some of these  configurations are constrained in number of rings  and the angular  momentum of each ring, as well as the ring spacing and dimension being different as they  orbit attractors  of the different  nine   classes.

\medskip

This  article is organized as follows: in Sec.\il(\ref{Sec:model})  we introduce  the thick  accretion Polish doughnut (P-D)  model and  the fluid  effective potential for the  toroidal configurations in a  Kerr spacetime background.
In Sec.\il(\ref{Sec:fisa}) we explore the  properties  of the fluid by means of the effective potential   introducing   and detailing
a classification of nine classes of
 Kerr attractors.
Section (\ref{Sec:barl})  is devoted to the   analysis of  the fluid configuration   with respect to  the angular momentum $\bar{\ell}\equiv\ell/a$.
In Sec.\il(\ref{Sec:Morfology}), we consider a more general class of configurations which  includes as particular case the Boyer surfaces of the P-D accretion disks. In Sec.\il(\ref{Sec:limite.statico}) we investigate  some aspects of the surfaces close to  the static limit, then we present the  different sequences of torus configurations in  Sec.\il(\ref{Subsec:sequence}) and Sec.\il(\ref{Sec:K.L}). General considerations on some  limiting  cases are in Sec.\il(\ref{Subsec:limitcases}). The multiple structured  thick configurations are analyzed in Sec.\il(\ref{Sec:multipleP-D}). Finally the case of the  polytropic equation of state and some aspects of the Boyer disk morphology are explored in Sec.\il(\ref{SeC:poly}). This article ends in  Section\il(\ref{Sec:sum-con}) where some concluding remarks are presented.
\section{Fluid configuration on the Kerr spacetime}\label{Sec:model}
We consider a perfect fluid orbiting  in  the Kerr
spacetime background,  where the   metric tensor can be
written in Boyer-Lindquist (BL)  coordinates
\( \{t,r,\theta ,\phi \}\)
as follows
\begin{equation} \label{alai} ds^2=-dt^2+\frac{\rho^2}{\Delta}dr^2+\rho^2
d\theta^2+(r^2+a^2)\sin^2\theta
d\phi^2+\frac{2M}{\rho^2}r(dt-a\sin^2\theta d\phi)^2\ ,
\end{equation}
here $M$ is a mass parameter and the specific angular momentum is given as $a=J/M$, where $J$ is the
total angular momentum of the gravitational source and  $\rho^2\equiv r^2+a^2\cos\theta^2$, $\Delta\equiv r^2-2 M r+a^2$, in the following it will be also  convenient to introduce  the quantity
$\sigma \equiv\sin\theta$. We will consider  the Kerr black hole (\textbf{BH}) case defined by $a\in ]0,M[ $,  the extreme black hole source $a=M$, and the non-rotating  limiting case $a=0$ of the  Schwarzschild metric.
 The horizons $r_-<r_+$ and the static limit $r_{\epsilon}^+$ are respectively
\bea
r_{\pm}\equiv M\pm\sqrt{M^2-a^2};\quad r_{\epsilon}^{+}\equiv M+\sqrt{M^2- a^2 \cos\theta^2},
\eea
it is $r_+<r_{\epsilon}^+$ on the planes  $\theta\neq0$  and it is $r_{\epsilon}^+=2M$ i on the equatorial plane $\theta=\pi/2$.
In the  {region $r\in]r_+,r_{\epsilon}^{+}$[} ({\em ergoregion}) it is  { $g_{tt}>0$} and $t$-Boyer-Lindquist coordinate becomes spacelike,
this fact implies that a  { static observer} cannot exist inside
the ergoregion.
In this work we investigate toroidal  configurations of a perfect fluids orbiting a Kerr attractor, it will be therefore convenient to consider first the properties of the test particle circular motion.
Since the metric is independent of $\phi$ and $t$, the covariant
components $p_{\phi}$ and $p_{t}$ of a particle four--momentum are
conserved along its  geodesic, or\footnote{We adopt the
geometrical  units $c=1=G$ and  the $(-,+,+,+)$ signature, Latin indices run in $\{0,1,2,3\}$.  The   four-velocity  satisfy $u^a u_a=-1$. The radius $r$ has unit of
mass $[M]$, and the angular momentum  units of $[M]^2$, the velocities  $[u^t]=[u^r]=1$
and $[u^{\varphi}]=[u^{\theta}]=[M]^{-1}$ with $[u^{\varphi}/u^{t}]=[M]^{-1}$ and
$[u_{\varphi}/u_{t}]=[M]$. For the seek of convenience, we always consider the
dimensionless  energy and effective potential $[V_{eff}]=1$ and an angular momentum per
unit of mass $[L]/[M]=[M]$.}
\be\label{Eq:after}
{E} \equiv -g_{ab}\xi_{t}^{a} p^{b},\quad L \equiv
g_{ab}\xi_{\phi}^{a}p^{b}\ ,
\ee
are  constants of motion, where  $\xi_{t}=\partial_{t} $  is
the Killing field representing the stationarity of the Kerr geometry and  $\xi_{\phi}=\partial_{\phi} $
is the
rotational Killing field, the vector $\xi_{t}$ is   spacelike in the ergoregion.
The
 momentum $p^a= \mu u^a$ of the  particle with  mass $\mu$ and four-velocity $u^{a}$
can be normalized so that
$g_{ab}u^{a}u^{b}=-k$, where $k=0,-1,1$ for null, spacelike and timelike
curves, respectively.
In general,  we may interpret $E$, for
timelike geodesics, as representing the total energy of the test particle
 coming from radial infinity, as measured  by  a static observer at infinity, and  $L$ as the angular momentum  of the particle.
Then,
introducing  the scalar quantities \(
\Lambda\equiv u^r,\; \Sigma\equiv u^t, \;  \Phi\equiv u^{\varphi},\;
\Theta\equiv u^{\theta}\, ,\)
%
we can write Eq.\il(\ref{Eq:after}) as
\bea\label{Eq:ELSF}
{E}=-(g_{tt} \Sigma +g_{\phi t} \Phi ),\quad L=(g_{\phi \phi}\text{  }\Phi +g_{\phi t} \Sigma)
,\quad
\Sigma= \frac{\mathcal{E} g_{\phi \phi}+g_{\phi t} L}{g_{\phi t}^2-g_{tt} g_{\phi \phi}},\quad\Phi = \frac{\mathcal{E} g_{\phi t}+g_{tt} L}{-g_{\phi t}^2+g_{tt} g_{\phi \phi}}.
\eea
using Eqs.\il(\ref{Eq:ELSF})  the  { normalization condition}  on the four-velocity
can be solved  for ${E}$  to obtain the two solutions,
\bea\label{Eq:completa}
{E}_{\pm}=\frac{-g_{\phi t} L\pm\sqrt{ \left(g_{\phi t}^2-g_{tt} g_{\phi\phi}\right) \left[L^2+g_{\phi\phi}(g_{\theta \theta}  \Theta ^2+g_{rr}\Lambda ^2 + k)\right]}}{g_{\phi\phi}}.
\eea
 The case of a  circular configuration is defined by   the constraint
$\Lambda=0$, as we assume the motion on the  fixed plane $\sigma=1$ no
motion is  in the $\theta$ angular direction and it is  $\Theta=0$ (the  Kerr metric is  symmetric  under reflection through  the  equatorial hyperplane $\theta=\pi/2$).
Within these assumptions Eq.\il(\ref{Eq:completa}) leads to the definition of the effective potential $V_{eff}(a;L,r)\equiv \left. E_{\pm}/\mu\right|_{\Lambda=0}$, on the equatorial plane.
It represents that value of  the particle energy
at which
the (radial) kinetic energy of the particle vanishes \cite{MTW,chandra42,RuRR,Pu:Kerr}.
The investigation of the test particles circular motion on the equatorial plane  is
then reduced to the study of motion in the effective potential $V_{eff}(a;L,r)$. Furthermore,
Kerr metric (\ref{alai}) is invariant under the application of any two different transformations: $x^a\rightarrow-x^a$
  as one of the coordinates $(t,\phi)$ or the metric parameter $a$, and  the  function $V_{eff}$ is invariant under the mutual transformation of the parameters
$(a,L)\rightarrow(-a,-L)$, thus we  limit our analysis to the case of  positive values of $a$
for corotating  $(L>0)$ and counterrotating   $(L<0)$ orbits.
Circular orbits are therefore described by
\be\label{Eq:Kerrorbit}
\dot{r}=0,\quad V_{eff}={E}/\mu,\quad \partial V_{eff}/\partial r=0.
\ee
Some  notable  radii regulate the particle dynamics, namely the \emph{last circular orbit} for timelike particles  $r_{\gamma}^{\pm}$,  the \emph{last  bounded orbit}  is $r_{b}^{\pm}$, and the \emph{last stable circular orbit} is $r_{lsco}^{\pm}$ with angular momentum and energy  $(E_{\pm}. \mp L_{\pm})$ respectively,  where $(\pm)$ is for  counterrotating or corotating orbits with respect to the attractor. The explicit  expression of these orbits and $(E_{\pm}. \mp L_{\pm})$  is well known in the literature, we  refer  for example to \cite{Pu:Kerr}, they are given in Sec.\il(\ref{Po}).  Timelike  circular orbits  can fill  the spacetime region $r>r_{\gamma}^{\pm}$, stable orbits are in $r>r_{lsco}^{\pm}$ for counterrotating and corotating particles respectively, and  $E_{\pm}(r_b^{\pm})=1.$
It is convenient to  introduce also   the angular frequency  $\Omega$ and the specific angular momentum $\ell$ as follows
\bea\label{Eq:flo-adding}
\Omega \equiv\frac{\Phi }{\Sigma }&=&-\frac{{E} g_{\phi t}+g_{tt} L}{{E} g_{\phi \phi}+g_{\phi t} L}= -\frac{g_{t\phi}+g_{tt} \ell}{g_{\phi\phi}+g_{t\phi} \ell},\quad
\ell\equiv\frac{L}{{E}}=-\frac{\Phi  +g_{\phi t} \Sigma }{g_{tt} \Sigma +g_{\phi t} \Phi } =-\frac{g_{t\phi}+g_{\phi\phi} \Omega }{g_{tt}+g_{t\phi} \Omega },
\eea
we can write now the effective potential $V_{eff}^{\pm}(a;L,r)$  in (\ref{Eq:completa}) in terms of the angular momentum $\ell$ using Eq.\il(\ref{Eq:flo-adding}), as
\be\label{Eq:def-lvl}
u_t^2=V_{eff}^{\pm}(a;\ell,r)^2=\frac{g_{\phi t}^2-g_{tt} g_{\phi \phi}}{g_{\phi \phi}+2 \ell g_{\phi t} +\ell^2g_{tt}}=-\frac{(g_{tt} \Sigma +g_{\phi t} \Phi )^2}{g_{tt} \Sigma ^2+2 g_{\phi t} \Sigma\Phi +g_{\phi\phi}\Phi^2}=\frac{{E}^2 \left(g_{\phi t}^2-g_{tt} g_{\phi\phi}\right)}{{E}^2 g_{\phi\phi}+2 {E} g_{\phi t} L(\ell)+g_{tt}  L(\ell)^2}.
\ee
  or explicitly:
\bea\label{Eq:VltoVL}
V_{eff}(\ell)= \pm\sqrt{\frac{g_{\phi t}^2-g_{tt} g_{\phi \phi}}{g_{\phi \phi}+2 \ell g_{\phi t} +\ell^2g_{tt}}},\quad V_{eff}^{\pm}(L)=\frac{-g_{\phi t} L\pm\sqrt{ \left(g_{\phi t}^2-g_{tt} g_{\phi\phi}\right) \left(L^2+g_{\phi\phi} \right)}}{g_{\phi\phi}}.
\eea
In this work  we consider a one-species particle perfect  fluid (simple fluid),  where
\be\label{E:Tm}
T_{a b}=(\varrho +p) u_{a} u_{b}+\  p g_{a b},
\ee
is the fluid energy momentum tensor,  $\varrho$ and $p$ are  the total energy density and
pressure, respectively, as measured by an observer moving with the fluid. For the
symmetries of the problem, we always assume $\partial_t \mathbf{Q}=0$ and
$\partial_{\varphi} \mathbf{Q}=0$, being $\mathbf{Q}$ a generic spacetime tensor
(we can refer to  this assumption as the condition of  ideal hydrodynamics of
equilibrium).
  The timelike flow vector field  $u^a$  denotes now the fluid
four-velocity.
The motion of the fluid  is described by the \emph{continuity  equation} and the \emph{Euler equation} respectively:
\bea\label{E:1a0}
u^a\nabla_a\varrho+(p+\varrho)\nabla^au_a=0\, ,\quad
(p+\varrho)u^a\nabla_au^c+ \ h^{bc}\nabla_b p=0\, ,
\eea
where $h_{ab}=g_{ab}+ u_a u_b$ \cite{MTW}.
 We investigate in particular  the case of a fluid circular configuration on  the fixed plane $\sigma=1$, defined by the constraint
$u^r=0$,  as for the circular test particle motion no
motion is assumed in the $\theta$ angular direction, which means $u^{\theta}=0$.
We assume moreover a barotropic equation of state $p=p(\rho)$. The  continuity equation 
 %
is  identically satisfied as consequence of the conditions.
From the Euler  equation (\ref{E:1a0}) we obtain
\be\label{Eq:scond-d}
\frac{\partial_{\mu}p}{\varrho+p}=-\frac{\partial}{\partial \mu}W+\frac{\Omega \partial_{\mu}\ell}{1-\Omega \ell},\quad W\equiv\ln V_{eff}(\ell),\quad \ell\equiv \frac{L}{E},\quad V_{eff}(\ell)=u_t
\ee
where $V_{eff}(\ell)$ is given in Eq.\il(\ref{Eq:VltoVL}) and  the function $W$ is Paczynski-Wiita  (P-W) potential. Assuming  the fluid  is   characterized by the  angular momentum  $\ell$  constant (see also \cite{Lei:2008ui}),  we consider  the equation for  $W:\;  \ln(V_{eff})=\rm{c}=\rm{constant}$ or $V_{eff}=K=$constant.
The procedure described in the present article
borrows from the  Boyer theory on the equipressure surfaces applied to a  P-D  torus \cite{Boy:1965:PCPS:}.
  The Boyer surfaces are given by the surfaces of constant pressure  or\footnote{{More generally $\Sigma_{\mathbf{Q}}$ is the  surface $\mathbf{Q}=$constant for any quantity or set of quantities $\mathbf{Q}$.}}  $\Sigma_{i}=$constant for \(i\in(p,\rho, \ell, \Omega) \), \cite{Raine}, where it is indeed $\Omega=\Omega(\ell)$ and $\Sigma_i=\Sigma_{j}$ for \({i, j}\in(p,\rho, \ell, \Omega) \), the toroidal surfaces  are obtained from the equipotential surfaces, critical points of the effective potential  $V_{eff}(\ell) $ \cite{Boy:1965:PCPS:}.

The functions  $V_{eff}(L)$ and  $V_{eff}(\ell)$ in Eqs.\il(\ref{Eq:VltoVL})  are related by the transformation $L=L(\ell)$ or $\ell=\ell(L)$:
\be\label{Eq:defLLLll}
L(\ell)=\sqrt{\ell^2 V_{eff}(\ell)^2}=
\sqrt{\frac{\left(-g_{\phi t}^2+g_{tt} g_{\phi\phi}\right) k \ell^2}{g_{\phi\phi}+\ell (2 g_{\phi t}+g_{tt} \ell)}}
,\quad \frac{\ell_L^{\pm}}{M}\equiv\frac{-2 a L^2\pm\sqrt{L^2 r\Delta/M^2  \left[a^2 (r+2M)+r \left(L^2+r^2\right)\right]}}{a^2 r+(r-2M) \left(L^2+r^2\right)},
\ee
or in term of $K$:
\be\label{Eq:poEq}
\frac{\ell^{\mp}_K}{M}=-\frac{2 a}{r-2M}\mp\sqrt{\frac{ [r^2+(K^2-1)M r] \Delta }{M^2K^2 (r-2M)^2}}, \quad \mbox{where}\quad
\lim_{r\rightarrow\infty}L(\ell)=\sqrt{\ell^2},\quad
\lim_{r\rightarrow\infty}\ell_L^{\pm}(L)=\pm\sqrt{L^2}
\ee
it is worth noting that it is $\partial_{\ell} L(\ell)\neq 0$. The function $V_{eff}(\ell)$ is related to the energy  ${E}$ of the test particle
as it is   $V_{eff}(\ell)^2=L^2/\ell^2= E^2$, see Eqs.\il(\ref{Eq:after},
\ref{Eq:flo-adding},\ref{Eq:def-lvl}).
Test particle  orbits with the (constant) negative energy, are possible in the ergoregion however in the case of the Kerr-\textbf{BH} spacetime no circular orbits   with this  feature are possible, meaning  no solutions of $\partial_rV_{eff}(L)=0$ and $V_{eff}(L)=E<0$.

The  function  $V_{eff}(\ell)\equiv U_t$  in Eqs.\il(\ref{Eq:VltoVL})  is invariant under the mutual transformation for the parameters
$(a,\ell)\rightarrow(-a,-\ell)$, as for the case of test particle circular orbits we can limit our analysis to  positive values of $a>0$,
for corotating  $(\ell>0)$ and counterrotating   $(\ell<0)$ fluids.  More generally  we adopt the notation $(\pm)$  for  counterrotating or corotating matter  respectively.
\section{On  effective potential $V_{eff}(\ell)$: the fluid configurations}\label{Sec:fisa}
\subsection{Forbidden  orbital regions and angular momentum for  the P-D toroidal configurations}\label{Sec:pad}
\begin{figure}
\centering
\begin{tabular}{cc}
\includegraphics[scale=.33]{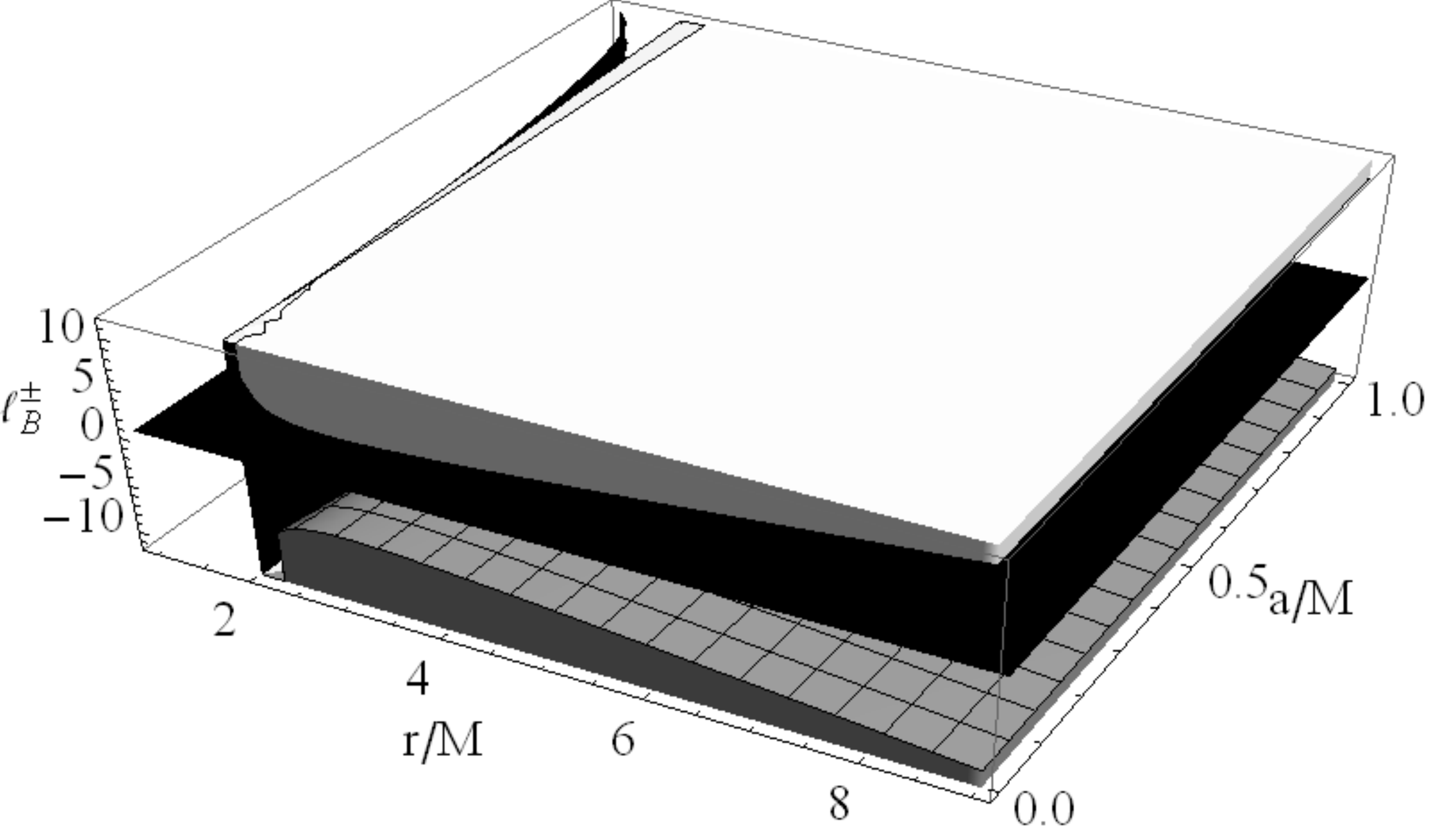}
\includegraphics[scale=.33]{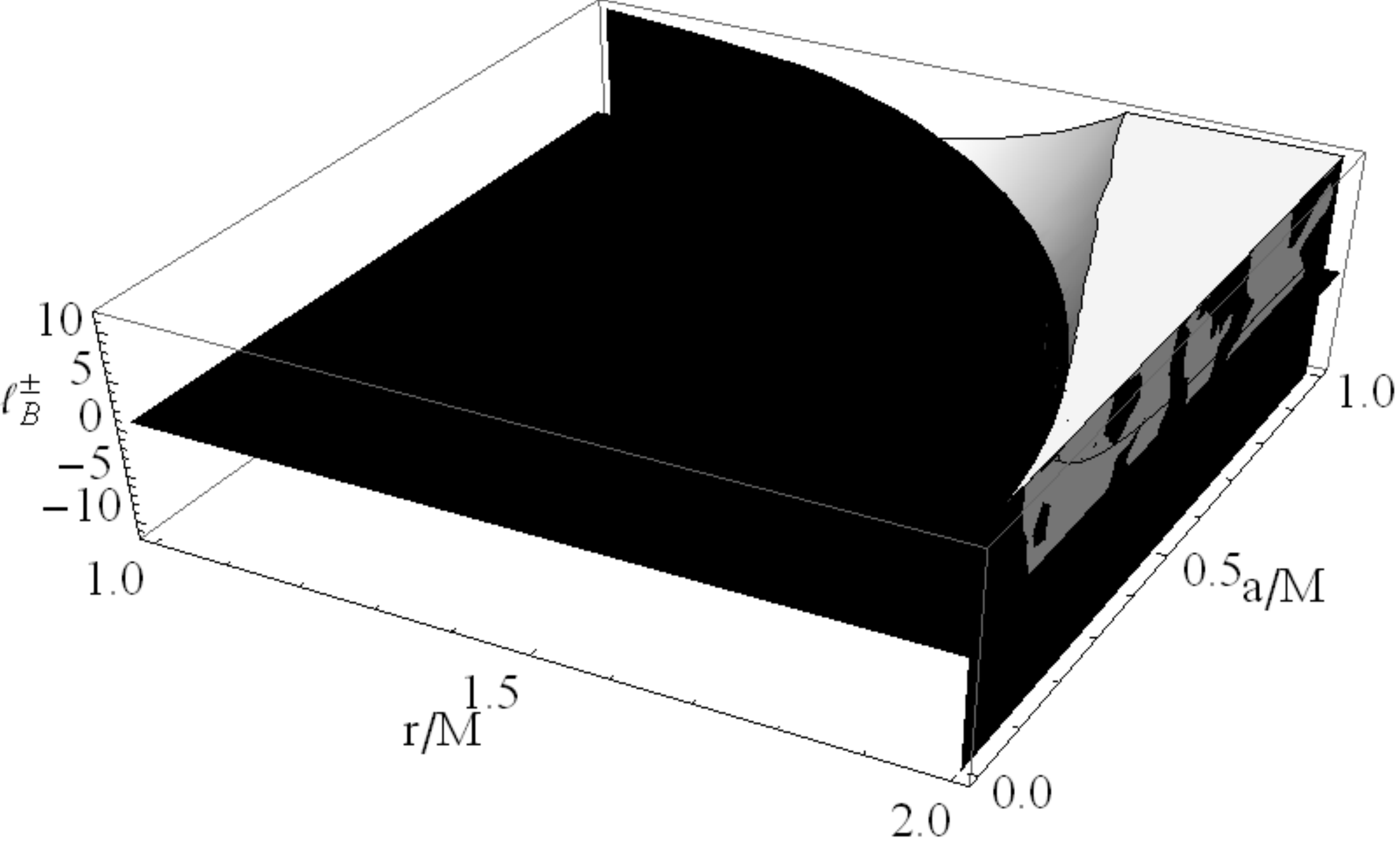}
\end{tabular}
\caption[font={footnotesize,it}]{\footnotesize{Angular momenta $\ell_{\ti{B}}^{\pm}$ as function of  radius $r/M$ and the spin-mass ratio $a/M$ of the spacetime.  Right plot is a zoom in the ergoregion $r\in]r_{\epsilon}^+,r_+[$. TIn the shadded regions the effective potential function $V_{eff}^2$ is not defined. White surface is the plane $\ell_{\ti{B}}^{\pm}=0$, $\ell_{\ti{B}}^+<\ell_{\ti{B}}^-$ is the gray surface. The horizon $r_+$ and the static limit are also plotted (black surfaces). }}\label{Fig:esitotP}
\end{figure}

The effective potential function  $V_{eff}(\ell)$ on the equatorial plane is well defined  (or $V_{eff}(\ell)^2>0$)  in the following cases:
\bea\label{Eq:existe-PO}
\mbox{{Schwarzschild case $a=0$:}} &&  r>r_+ \quad\mbox{and}\quad \ell\in]\ell_{\ti{$\bar{B}$}}^-,\ell_{\ti{$\bar{B}$}}^+[,
\\\nonumber
\mbox{{Kerr case: $a\in]0,M]$}}&&
r\in]r_+,r_{\epsilon}^+[ \quad\mbox{and}\quad\ell\in]-\infty, \ell_{\ti{B}}^-[\cup][\ell_{\ti{B}}^+,\infty[
\\\nonumber
&&
r=r_{\epsilon}^+ \quad\mbox{and}\quad  \frac{\ell}{M}<\frac{2M}{a}+\frac{a}{M},
 \\\label{Eq:existe-PO1}
&& r>r_{\epsilon}^+ \quad\mbox{and}\quad \ell\in]\ell_{\ti{B}}^-,\ell_{\ti{B}}^+[,
\eea
where the following angular momenta are introduced
\bea
&&
{\ell_{\ti{B}}^{\pm}}\equiv\frac{M}{r-2M}\left(-{2 a}\pm{r\sqrt{ \Delta/M^2}}\right)=\left.\ell_{K}^{\pm}\right|_{K=1},\quad \ell_{\ti{$\bar{B}$}}^{\pm}\equiv\left.\ell_{\ti{B}}^{\pm}\right|_{a=0}=
\pm\sqrt{\frac{r^3}{r-2M}},
\eea
the momenta $\ell_{\ti{B}}^{\pm}\neq0$  are not well defined on $r=r_{\epsilon}^+$, but it is
$\ell_{\ti{B}}^{+}>0$ for $r>r_+-\{r_{\epsilon}^+\}$ and  $\ell_{\ti{B}}^{-}>0$ in the ergoregion   Figs.\il(\ref{Fig:esitotP}).
The functions $\ell_{\ti{B}}^{\pm}$ are solutions of  the equation $\ell_{\ti{B}}^{\pm}:\;r_{\ti{B}}^{\pm}(a; \ell)=r$ or explicitly
\be\label{Def:RBB}
r_{\ti{B}}^+\equiv\frac{2 \sqrt{\Delta_{\ell}^+\Delta_{\ell}^-} \cos\left(\frac{1}{3} \arccos\left[-\frac{3 \sqrt{3} \sqrt{M^2\Delta_{\ell}^+\Delta_{\ell}^- }}{(\Delta_{\ell}^+)^2}\right]\right)}{\sqrt{3}},\quad r_{\ti{B}}^-\equiv \frac{2 \sqrt{\Delta_{\ell}^+\Delta_{\ell}^-} \sin\left(\frac{1}{3} \arcsin\left[\frac{3 \sqrt{3} \sqrt{M^2\Delta_{\ell}^+\Delta_{\ell}^-}}{(\Delta_{\ell}^+)^2}\right]\right)}{\sqrt{3}};
\ee
where  $\Delta^{\pm}_{\ell}(a)\equiv\ell\pm a$  and $\left.r_{\ti{B}}^{\pm}\right|_{\ell=\pm a}=0$, { Fig.\il(\ref{Fig:esitotP}).} It is on the other hand:
\be\label{Eq:effper1}
\left.V_{eff}(\ell)\right|_{\ell=a}=\sqrt{\frac{\Delta }{r^2}},\quad
\left.V_{eff}(\ell)\right|_{\ell=-a}=\sqrt{\frac{r \Delta }{8M a^2+r^3}},
\ee
the effective potential at $\ell=\pm a$ is therefore
well defined in all  $r>r_+$. The limiting case of $a=0$  has been addressed in  \cite{PuMonBe12},  and  for the extreme Kerr-\textbf{BH} case it is in particular:
\bea
&&
\mbox{{Extreme Kerr case:}}\quad
a=M,\quad \ell/M=-7 \quad\mbox{and}\quad r>M-\{r=4M\},
\\\nonumber
&&
\ell\in]-7M,2M[\quad\mbox{and}\quad r>M,
\\\nonumber
&&
 \ell/M<-7\cup\ell/M>2 \quad\mbox{and}\quad r \in]r_+,r_{\ti{B}}^-[\cup]r_{\ti{B}}^+,\infty[.
\eea
More generally  for spacetime spins  $a\in]0,M]$ and on any plane $\sigma\in]0,1]$, the fluid effective potential  is well defined in:
\bea\nonumber
 r\in]r_+,r_{\epsilon}^+[\quad\mbox{for}\quad
\ell<\ell_{\sigma}^-,\quad\mbox{at}\quad
r=r_{\epsilon}^+\quad\mbox{for}\quad \ell<\frac{\left(a^2+r^2\right)^2-a^2 \Delta  \sigma ^2}{4 a M r};
\quad\mbox{and in}\quad
r>r_{\epsilon}^+ \quad\mbox{for}\quad  \ell\in]\ell_{\sigma}^-,\ell_{\sigma}^+[,
\eea
with
\bea
\ell_{\sigma}^{\pm}\equiv\frac{2 a Mr \sigma ^2}{2M r-\rho ^2}\pm\sqrt{\frac{\Delta  \rho ^4 \sigma ^2}{\left(\rho ^2-2M r\right)^2}},\quad  \left.\ell_{\sigma}^{\pm}\right|_{\sigma=1}=\ell_{\ti{B}}^{\pm}.
\eea
The conditions above arise from:  $V_{eff}(\ell)^2=U_t^2>0$, as such they regulate some aspects of the hydrodynamics of the  system   but do not define the toroidal P-D topology as  they  are indeed necessary but not sufficient conditions for a P-D accretion  disk could  be formed. To obtain this we should analyse the critical points of the effective potential function $V_{eff}(\ell)$.
However equations \il(\ref{Eq:existe-PO},\ref{Eq:existe-PO1}) and (\ref{Def:RBB}), reveal that the regions  where a fluid configuration can be formed,  as regulated by Eqs.\il(\ref{Eq:scond-d}),  depend   on   $\Delta_{\ell}^{\pm}(a)=\ell\pm a$. Considering that the
configurations can be counterrotating  $(\ell<0)$ and rotating $(\ell>0)$,
as seen from Eqs.\il(\ref{Eq:existe-PO},\ref{Eq:existe-PO1}),
there is a significant difference, in terms of fluid angular momenta, between the regions $r>r_{\epsilon}^+$ and $r\in]r_+, r_{\epsilon}^+[$.
The structure   of these regions  outside the static limit ($r>r_{\epsilon}^+$ on  $\theta=\pi/2$) coincides qualitatively with that of the non rotating case $a=0$ where $r_+=r_{\epsilon}^+$. Indeed, the upper limit $\ell_{lim}$ of the  angular momentum for $r=r_{\epsilon}^+$   diverges as $a$ approaches zero, and  has the limiting values  $\ell_{lim}/M=3$ as $a=M$. The limit  $\ell_{lim}(a)$ is a monotonically decreasing function of  $a/M$,  this implies that  the  range of   angular momentum for corotating fluid approaching the static limit decreases with the black hole spin. Finally the behaviour of the boundaries $\ell_{\ti{B}}^{\pm}$ is very complicated as these are  function of  $(a,r)$, however, in the region $r>r_{\epsilon}^+$ there is a minimum  of the $\Delta_{\ell_{\ti{B}}}=|\ell_{\ti{B}}^+-\ell_{\ti{B}}^-|$ and then   $\Delta_{\ell_{\ti{B}}}$ increases with $a/M$   Fig.\il(\ref{Fig:esitotP}).
\subsection{Analysis of the fluid configurations and pressure-free case}\label{Sec:angularl}
\begin{figure}
\centering
\begin{tabular}{cc}
\includegraphics[scale=.33]{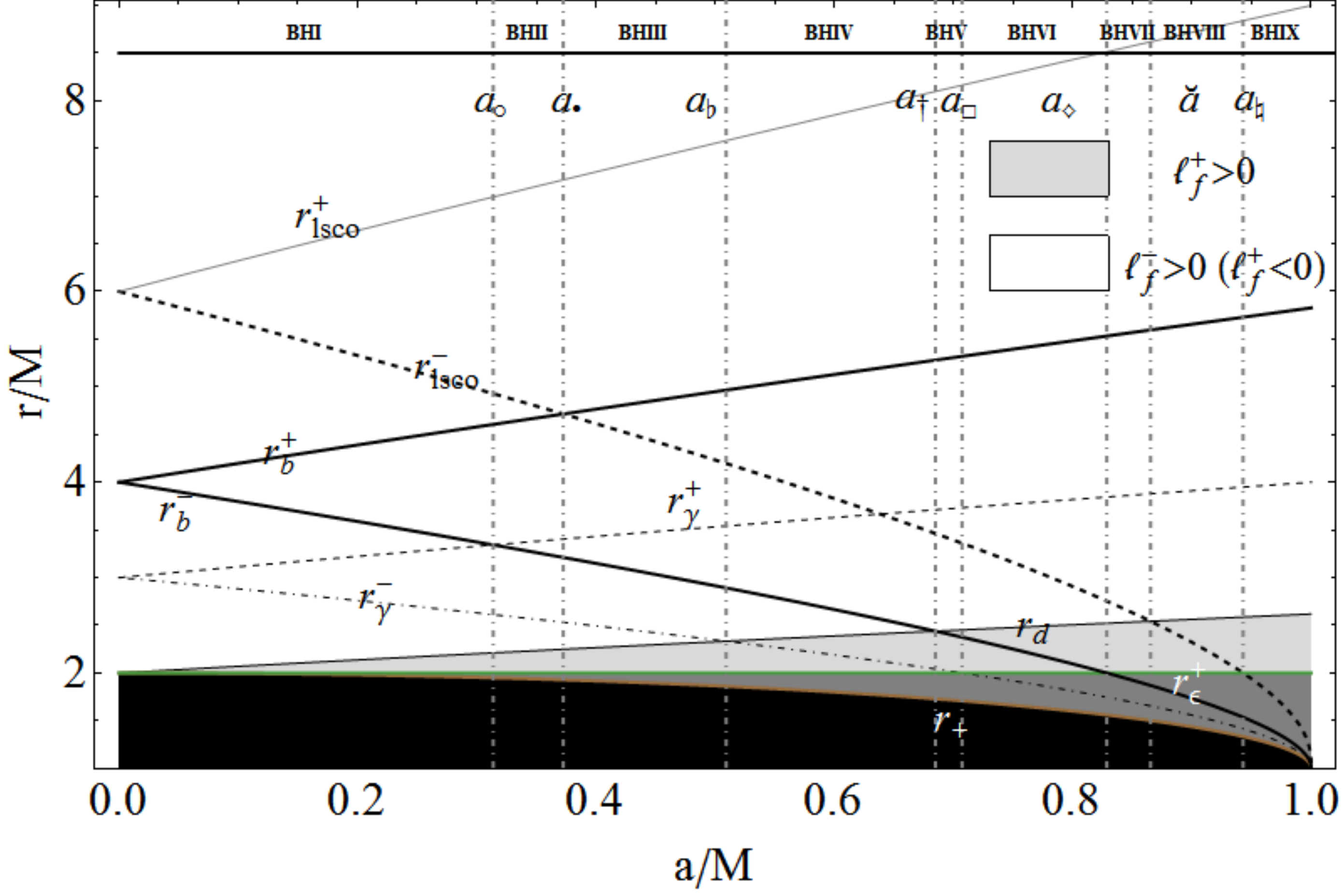}
\includegraphics[scale=.33]{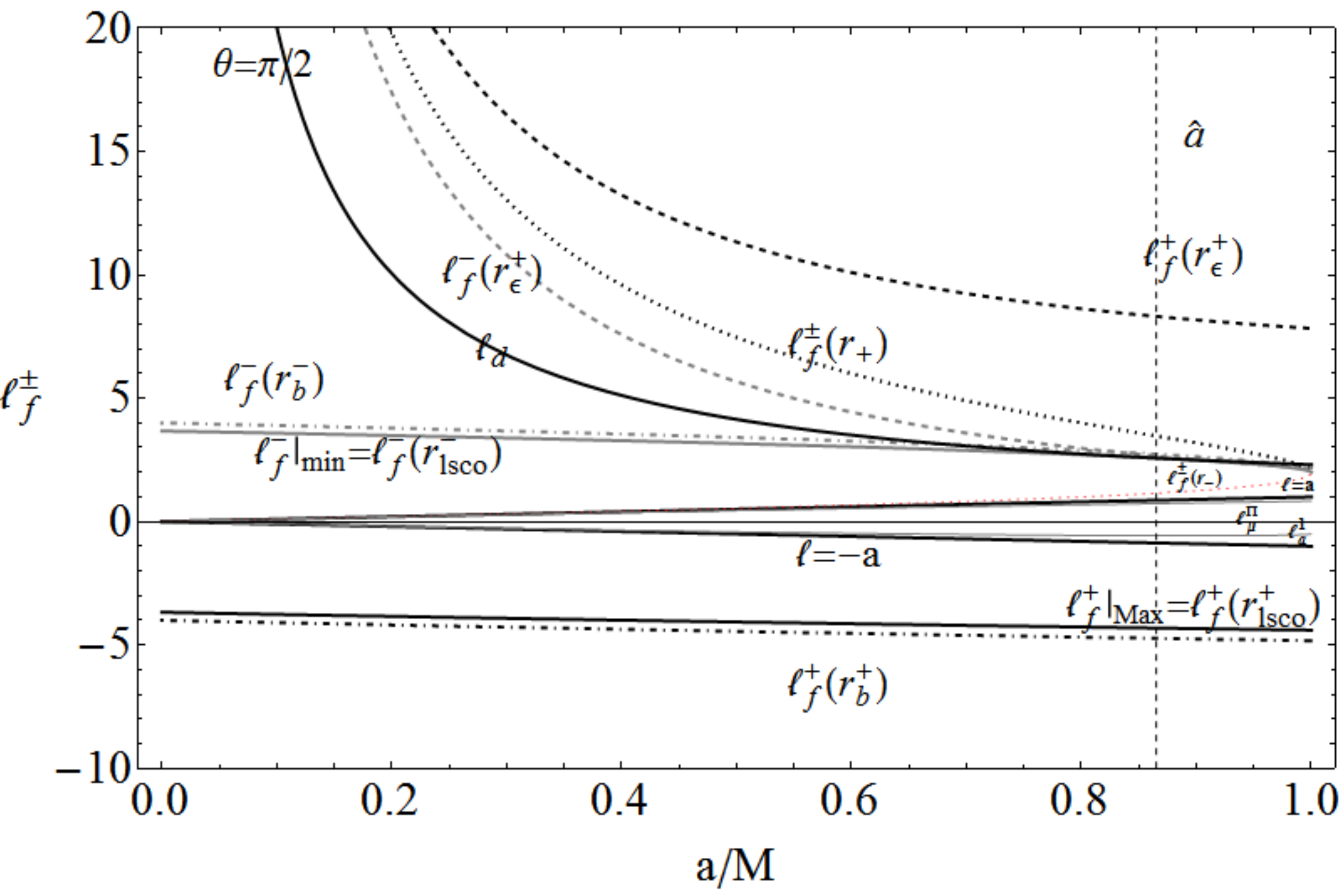}
\end{tabular}
\caption[font={footnotesize,it}]{\footnotesize{Left panel: Radii of the set
$\mathfrak{R}\equiv\{r_d, r_{\epsilon}^+,r_{lsco}^{\pm},r_b^{\pm},r_{\gamma}^{\pm}\}$ as function of black hole spin-mass ratio $a/M$, namely the {last circular orbit} for time-like particles  $r_{\gamma}^{\pm}$,  the {last  bounded orbit}  is $r_{b}^{\pm}$, and the {last stable circular orbit} is $r_{lsco}^{\pm}$,  where $(\pm)$ is for  counterrotating or corotating orbits with respect to the attractor on the equatorial plane $\theta=\pi/2$.  The outer horizon $r_+/M$, and the static limit  $r_{\epsilon}^+=2M$, and $r_d$, are also plotted, the regions of positive fluid angular momentum $\ell_f^{\pm}$, for counterrotating and corotating fluid are shadded.  The gray region is $]r_{\epsilon}^+,r_+[$, the black region $r<r_+$. Vertical lines are the spin in $\mathfrak{A}\equiv\{a_{\circ},a_{\bullet},a_{\flat},
a_{\dag},
a_{\Box},
a_{\diamondsuit},
{\check{a}}
,a_{\natural}\}$ where:  $a_{\circ}\equiv(8 \sqrt{2}-11)M$, $a_{\bullet}\equiv (23-16 \sqrt{2})M$, $a_{\flat}/M\equiv{\sqrt{{7}/{3}}}/{3}$, $a_{\dag}/M\equiv{1}/{2} \left(3 \sqrt{17}-11\right)M$, $a_{\Box}\equiv1/{\sqrt{2}}M$,  $a_{\diamondsuit}\equiv2 \left(\sqrt{2}-1\right)M$, $\check{a}\equiv0.865157M$, $a_{\natural}\equiv2\sqrt{2}/3M$. Nice  classes of black hole spacetimes, $\mathbf{BHI-IX}$, are outlined. Right plot: fluid angular momentum $\ell^{\pm}_{f}$ of the orbits  in units of mass $M$ of $\mathfrak{R}$ as function of $a/M$.}}\label{Fig:radiilf1}
\end{figure}

Many phases of the accretion process in the {P-W} scheme are regulated by the  proprieties of the effective potential function  on the equatorial plane. We  use this property extensively confining our  analysis mostly to a survey on the equatorial plane of symmetry,  projecting each functions  on  $\theta=\pi/2$.
We investigate the orbital regions  where a P-D fluid configuration  exist     studying    the effective potential  critical points on the equatorial plane, in particular we focus on  the relation with the   potential $V_{eff}(L)$  for case of negligible pressure.  The critical points of the  potential $V_{eff}(\ell)$ are related to the critical points of the   potential  $V_{eff}(L)$ for the Keplerian disk by the following relation:
\be\label{Eq:Vder-lL}
\partial_r V_{eff}(\ell)=\partial_r V_{eff}(L)+\partial_L V_{eff}(L)\partial_{r} L(\ell)=0.
\ee
On the other hand, in the regions where   $p=0$  it is  $\Sigma_L\equiv \Sigma_{\ell}$.
 In general one can say: open $\left({O}\right)$ configurations are for $K>1$, closed $\left({C}\right)$  disks for $K\in]0,1[$.
There can be crossed surfaces  $\left(O_{x}, C_{x}\right)$ for each classes as the effective potential has a maximum, that is a minimum of the hydrostatic  pressure, the maximum of pressure on the other hand (minimum of $V_{eff}(\ell)$) are associated to the center  of the disk.
Note that
$\partial_L V_{eff}(L)=0$ at $r>r_{\epsilon}^+$ for every $a\in]0,1]$ and
$\bar{\bar{L}}/M\equiv-2 \sqrt{{a^2}/{(r-2M)r}}$ ($r=M+M\sqrt{{(4 a^2+L^2)/(L^2)}}$), where  $L$ must be negative and it is $\bar{\bar{L}}\neq -L_+$.
The fluid angular momentum, solutions of Eq.\il(\ref{Eq:Vder-lL}) are
\bea\label{Eq:deflfmp}
\ell_{f}^{\pm}:\; \partial_r V_{eff}(\ell_{f}^{\pm})=0,\quad \frac{\ell_{f}^{\pm}}{M}\equiv \frac{a^3+a r (3 r-4M)\pm\sqrt{r^3 \Delta^2/M}}{a^2M-(r-2M)^2 r},\quad\mbox{at}\quad \sigma=1,
\eea
in the {Schwarzschild} and the {extreme } \textbf{BH} cases it is respectively
\bea
\left.\ell_f^{\pm}\right|_{a=0}=\pm\frac{Mr^4}{\sqrt{(r-2M)^2 Mr^5}},
\quad
\left.\ell_f^{\pm}\right|_{a=M}=\frac{M^3\pm M^3\sqrt{r^3/M^3}-rM^2 \left(3\pm\sqrt{r^3/M^3}\right)}{r(r-3M) +M^2}.
\eea
where
$\ell_{f}^->0$, $\forall r>r_+-\{r_{d}\}$, and $\ell_{f}^+>\ell_{f}^->0 (\ell_{f}^+<0)$ {for} $r\in]r_+,r_{d}[ (]r_{d},\infty[)$, moreover  it is
\be
r_d\equiv
 \frac{8M}{3} \cos\left[\frac{1}{6} \arccos\left(\frac{27 a^2}{16M^2}-1\right)\right]^2,
\quad
\ell_f^{\pm}(r_{\epsilon}^+)=\frac{4M^2}{a}+a\pm2 M \sqrt{2},\quad \lim_{r\rightarrow r_\pm}\ell_f^{+}=\lim_{r\rightarrow r_\pm}\ell_f^{-}=\frac{2M r_{\pm}}{a},
\ee
with
\[
r_{d}(\check{a})=r_{lsco}^-(\check{a}) \approx2.54257M,\quad\breve{a}\equiv\frac{1}{3} \sqrt{\frac{1}{2} \left(249-41 \sqrt{33}\right)}M\approx0.865157M,
\]
the solution $r_{d}(a,M):a^2M-(r-2M)^2 r=0$ from Eq.\il(\ref{Eq:deflfmp}) is plotted in Fig.\il(\ref{Fig:radiilf1}) with the $r_{\gamma}^{\pm}$ and $r^{\pm}_{lsco}$.
We  consider then the  set of orbits
$\mathfrak{R}\equiv\{r_d, r_{\epsilon}^+,r_{lsco}^{\pm},r_b^{\pm},r_{\gamma}^{\pm}\}$, and  the spin set $\mathfrak{A}\equiv\{a_{\circ},a_{\bullet},a_{\flat},
a_{\dag},
a_{\Box},
a_{\diamondsuit},
{\check{a}}
,a_{\natural}\}$ associated with  the orbits in $\mathfrak{R}$
 each spin is shown in Fig.\il(\ref{Fig:radiilf1}) as the cross between pairs  of  orbits $r_i\in\mathfrak{R}$.

Furthermore,
we can express  conditions in Eqs.\il(\ref{Eq:existe-PO},\ref{Eq:existe-PO1}) in terms of the fluid angular momentum and the limiting photon orbits $r_{\gamma}^{\pm}$ as follows:
\bea
&&
\mbox{{Kerr case $a\in]0,M[$:}}
\quad
\\\nonumber
&&
\ell<\ell_{\gamma}^+\cup \ell\in]\ell_{\gamma}^-,\ell_f^-(r_+)[\cup \ell>\ell_f^-(r_+) \quad\mbox{and}\quad r\in]r_+,r_{\ti{B}}^-[\cup]r_{\ti{B}}^+,\infty[,
\\\nonumber
&&
\ell\in]\ell_{\gamma}^+,\ell_{\gamma}^-[\quad\mbox{and} \quad  r>r_+,
\quad
\ell=\ell_f^-(r_+)\quad\mbox{and}\quad  r>r_{\ti{B}}^+,
\\\label{Eq:pol-prop}
&&
  \ell=\{\ell_{\gamma}^+,\ell_{\gamma}^-\} \quad\mbox{and}\quad r>r_+-\{r_{\ti{B}}^+\},
\eea
see Fig.\il(\ref{Fig:esitotP}) where
\bea
\ell_{\gamma}^{\pm}\equiv\ell_f^{\pm}(r_{\gamma}^{\pm})=-\frac{a}{M}\mp 6 \cos\left[\frac{\arccos\left(\pm\frac{a}{M}\right)}{3}\right],
\eea
more generally we introduce the notation $\mathbf{Q}_{b}^{\pm}\equiv \mathbf{Q}(r_{b}^{\pm})$ and also $\mathbf{Q}_{lsco}^{\pm}\equiv \mathbf{Q}(r_{lsco}^{\pm})$, for any quantity $\mathbf{Q}$ as well.
The angular momentum of  the test  particles orbits  and the fluid angular momentum in the P-D configurations are related  as follows:
 $\ell_f^+=\frac{-L_+}{E_+}$ and $\ell_f^-=\frac{L_-}{E_-}$,
where  $\ell_L^{\pm}(L_\pm)=\ell_{f}^{\mp}$ and $L\left(l_f^{\pm }\right)=L_{\pm }$ with the definitions Eqs.\il(\ref{Eq:defLLLll}), and
 $l_f^{\pm }:\; \left.\partial_rL(\ell)\right|_{\ell_f^{\pm}}=0$, that is the fluid angular momentum  under hydrostatic pressure are the critical points of the angular momentum for test particle motions $L(\ell)$ respect to the radial coordinate, precisely taking into consideration the  existence conditions Eqs.\il(\ref{Eq:pol-prop}) the critical points are
\bea
\mbox{for} \;a\in[0,a_{\flat}]:&&\mbox{in}\;r\in]r_{\gamma}^-, r_{\gamma}^+]\mbox{with}\ \ell_f^-,\quad\mbox{and}\quad r>r_{\gamma}^+,\;\mbox{with}\quad \ell_f^{\pm},\quad a_{\flat}\equiv\frac{1}{3}\sqrt{\frac{7}{3}}M\approx0.509175M,\\
\mbox{for}\;a\in]a_{\flat},M]:&&\mbox{in}\;r\in]r_{\gamma}^-,r_d[,\;\mbox{with}\;\ell_f^-\;\mbox{and}\;
r=r_d\;l=\ell_d,\; \mbox{in}\; r\in]r_d, r_{\gamma}^+]\;\mbox{with}\;\ell_f^-
,\quad
r>r_{\gamma}^+\;\mbox{with}\;\ell_f^{\pm}.
\eea
\subsubsection{On the fluid angular momentum}\label{Sec:contro-l}
\begin{figure}
\centering
\begin{tabular}{cc}
\includegraphics[scale=.33]{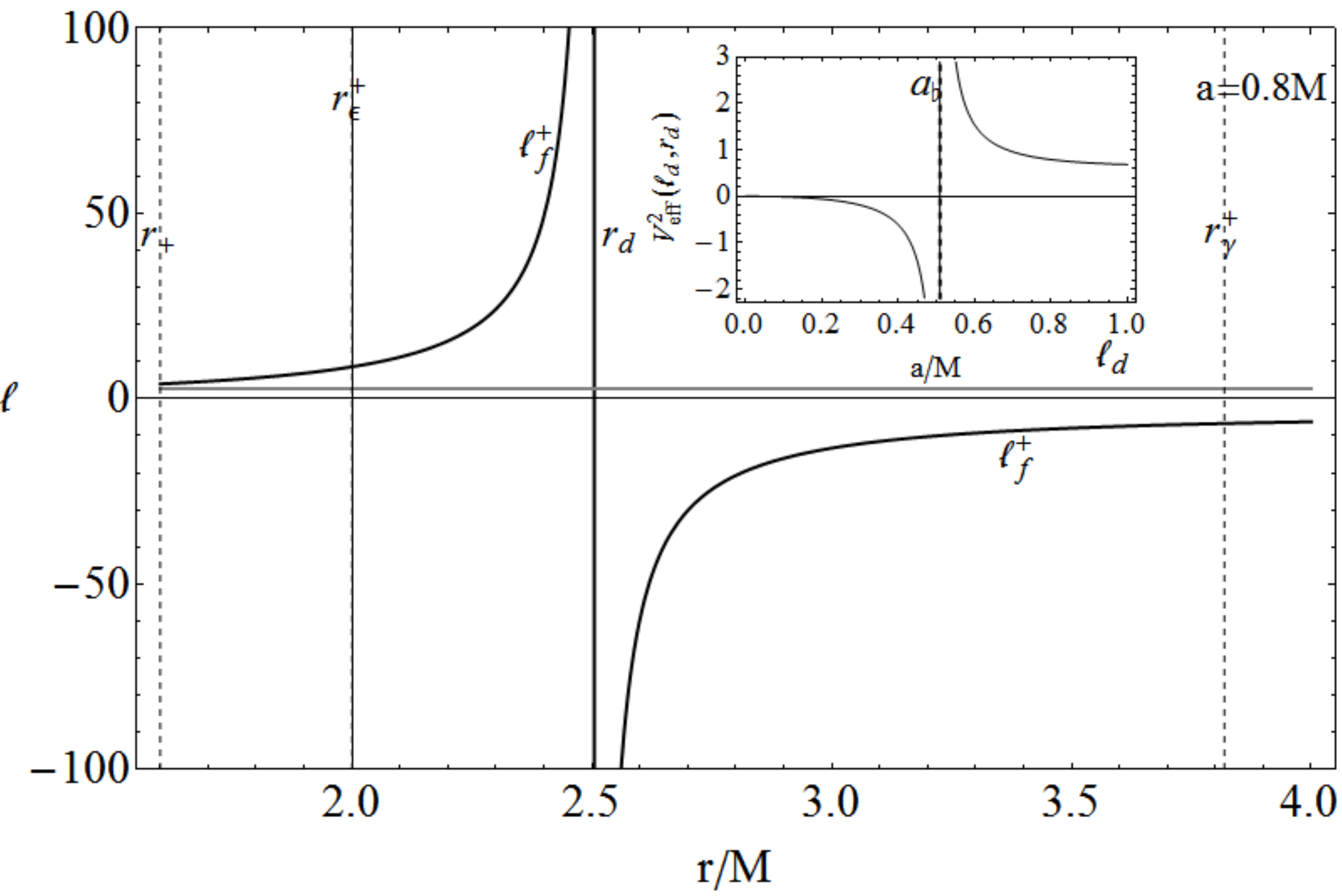}
\includegraphics[scale=.33]{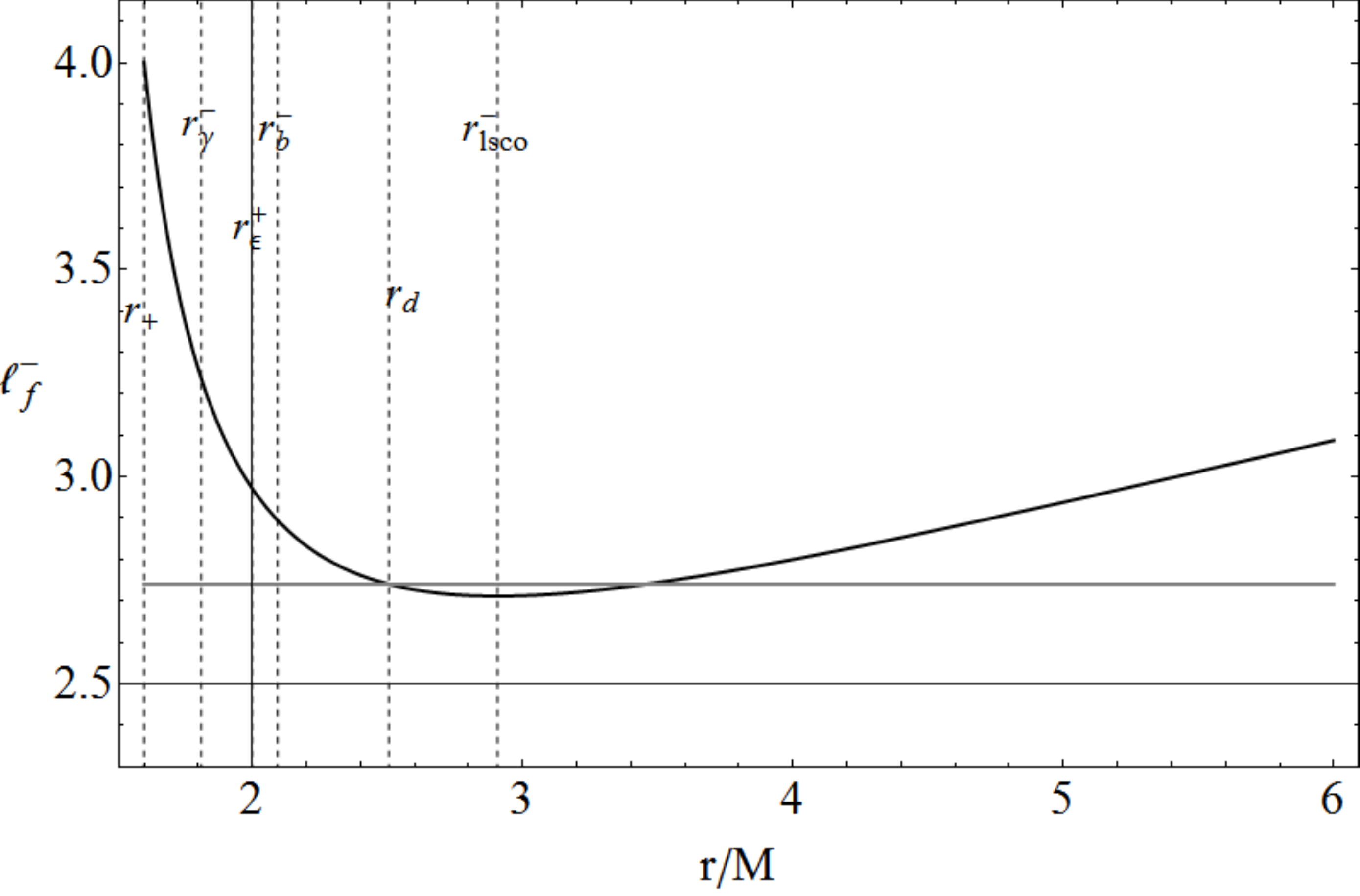}
\end{tabular}
\caption[font={footnotesize,it}]{\footnotesize{ {Left} panel: spacetime spin $a=8/10 M\in\mathbf{BHVI}$, fluid angular momentum in units of mass $M$, $\ell_{f}^+$ (black curve), $\ell_d=2.74035M$ (gray curve) as functions of $r>r_+$, the horizon $r_+=1.6M$ and the static limit $r_{\epsilon}^+=2M$ are plotted (dashed lines). The inset plot is the effective potential  $V^2_{eff}(\ell_d,r_d)$ function of $a/M$. The energy function $V^2_{eff}(\ell_d,r_d)$ is not well defined in $a=a_{\flat}\approx0.51M$  and it is negative  in  $a<a_{\flat}$.  Right panel: spacetime spin $a=8/10 M\in\mathbf{BHVI}$, fluid angular momentum $\ell_f^-$ for corotating matter as function of $r/M$. Dashed lines are the last circular orbit $r_{\gamma}^-$, the static limit $r_{\epsilon}^+$, and the last bounded orbit $r_b^-$, the radius $r_d$, and the last stable circular orbit $r_{lsco}^-$.
The curve $\ell_f^->0$ spacetime spin $a=8/10 M\in\mathbf{BHVI}$ has a minimum point  in $r_{lsco}^-$.
 }}
\label{Fig:LimitPlottransition}
\end{figure}

\textbf{Some notes on the counterrotating configurations}
The fluid momentum associated with the counterrotating configurations is $\ell_f^+$: counterrotating disks can be formed only at $r>r_{\gamma}^+>r_d>r_{\epsilon}^+$, where $\ell_f^+$ sets the critical points of the fluid effective potential as function of the radius $r$. The case  $\ell=0$ is not a  value for $\ell_f^{\pm}$ and it is not a critical point for the effective potential $V_{eff}(\ell)$ (for $\sigma\in[0,1]$), we could say there is no continuum sequence of configurations, parameterized by $\ell$, from the corotating  $(\ell>0)$ to the counterrotating $(\ell<0)$ fluids.
On the orbit $r=r_d$, $\ell_f^{\pm}$ are not well defined,
however
the pressure has still a critical point, the associated angular momentum is:
\bea
l_d=\left.\frac{a^4+2 a^2 (r-2M) r+r^4}{2 a \left[a^2+r (3 r-4M)\right]}\right|_{r_d}>0,\quad a\neq a_{\flat},
\eea
where $\ell_f^-$  approaches $\ell_d$  in the limit $r\approx r_d$, Fig.\il(\ref{Fig:LimitPlottransition}).
The function $\ell_d>0$ can be seen as a ``transition'' fluid angular momentum and  $r_d:\; \ell_f^+\lessgtr0$ for $r\gtrless r_d$  as a  ``transition'' orbit
Fig.\il(\ref{Fig:LimitPlottransition}): the critical points at $(r_d, \ell_d)$  exist only in the spacetimes  $a\in]a_{\flat},M]$, where $V_{eff}(r_d,\ell_d)=1$ as $a$ increases  and  it goes  to infinity as $a=a_{\flat}$, precisely it is
   $K>1$ in $a\in]a_{\flat},a_{\dag}[$,  $K=1$ for $a= a_{\dag}$, $K\in]0,1[$ in the   spacetimes $a\in]a_{\dag},M]$, where $a_{\dag}\equiv{1}/{2} \left(3 \sqrt{17}-11\right)M:\; r_b^-(a_{\dag})=r_d(a_{\dag})$.
 The effective potential is well defined at $\ell<0$ in the ergoregion (at any plane) but has no critical points for $\ell<0$.
Then, it is
$\ell_f^+<0$
when
$L_{+}\lessgtr 0$  and $E_+\lessgtr0$, however
the positive  solutions  $\ell_f^+$ can exist for
$r\in]r_+,r_d[$  but  both  $L_+$ and $ E_+$ are well (real) defined when   $r>r_{\gamma}^+$, moreover in the \textbf{BH}-case there are no circular orbits  with $E<0$ and $(L_+, E_+)$ are positive and  defined for $r>r_{\gamma}^+$.
In $r\in] r_+, r_d[$
the couple  $(E_-,L_-)$ is well defined and it is
$\ell_f^-=L_-/E_-$, then it is $ r_d(a_{\flat})=r_{\gamma}^-(a_{\flat})$, at $r<r_{\gamma}^-$   no circular orbits with $L_-$ are possible,   it is $r_d( \check{a} )=r_{lsco}^- (\check{a})$.
This analysis  however
does not    specify the  configuration topology, in order to do this
we need to study   Eq.\il(\ref{Eq:Vder-lL}) by taking  a second derivation with respect to $r$ and to consider the values of the parameter  $K\in[0,\infty[$
to establish the possible existence of a cusp and  to fix the   $({O},{C})$ classes.

\textbf{{Orbital regions of extreme fluid angular momentum}}
We investigate  the  critical points of the  fluid angular  momentum  i.e. the solutions of $\left.\partial_r\ell_{f}^{\pm}\right|_{\theta=\pi/2}=0$.  These  orbits are related to the  orbits of maximum and minimum   particle angular momentum regulating   the  case of  the Keplerian disks or  null pressure $p$ configuration, see Figs.\il(\ref{Fig:radiilf1})-right.

\textbf{{On the counterrotating fluid angular momentum}}
  The angular momentum  for  counterrotating fluids  $\ell^+_f<0$ increases with the   the orbital distance from the center  i.e. $\partial_r\ell_{f}^{+}>0$, in $r\in ]r_+,r_{\epsilon}^+]$.
Critical points of  the fluid angular momentum on the equatorial plane  exist in
 $r\in[6M,9M]$, on $r=r^+_{lsco}$, line of saddle points for   the effective potential, as $a/M$ varies in $[0,1]$ see Figs.\il(\ref{Fig:radiilf1})-left, from the case
at $a=0$ in $r_{lsco}^+=6M$ to $a=M$, $r_{lsco}^+=9M$.
The counterrotating angular momentum increases with the radius in $]r_{+},r_{lsco}^+[-\{r_d\}$ that is  in $r\in]2M,6M[$,
 while the angular momentum decreases always with the radius (on the equatorial plane) far enough  from the attractor i.e. $r>9M$ and in general
in $r>r_{lsco}^+$.
However there are no turning points or minimum for the specific fluid momentum but there is a {maximum} point at  $\left.\ell_f^+\right|_{\ti{Max}}$ as it is $\partial_r^2\ell_f^+<0$  on the orbit $r_{lsco}^+$.
The maximum value $\ell^+_f(r_{lsco}^+)$ is plotted in Fig.\il(\ref{Fig:radiilf1}) for $a\in[0,M]$.

\begin{figure}
\centering
\begin{tabular}{lcr}
\includegraphics[scale=.3]{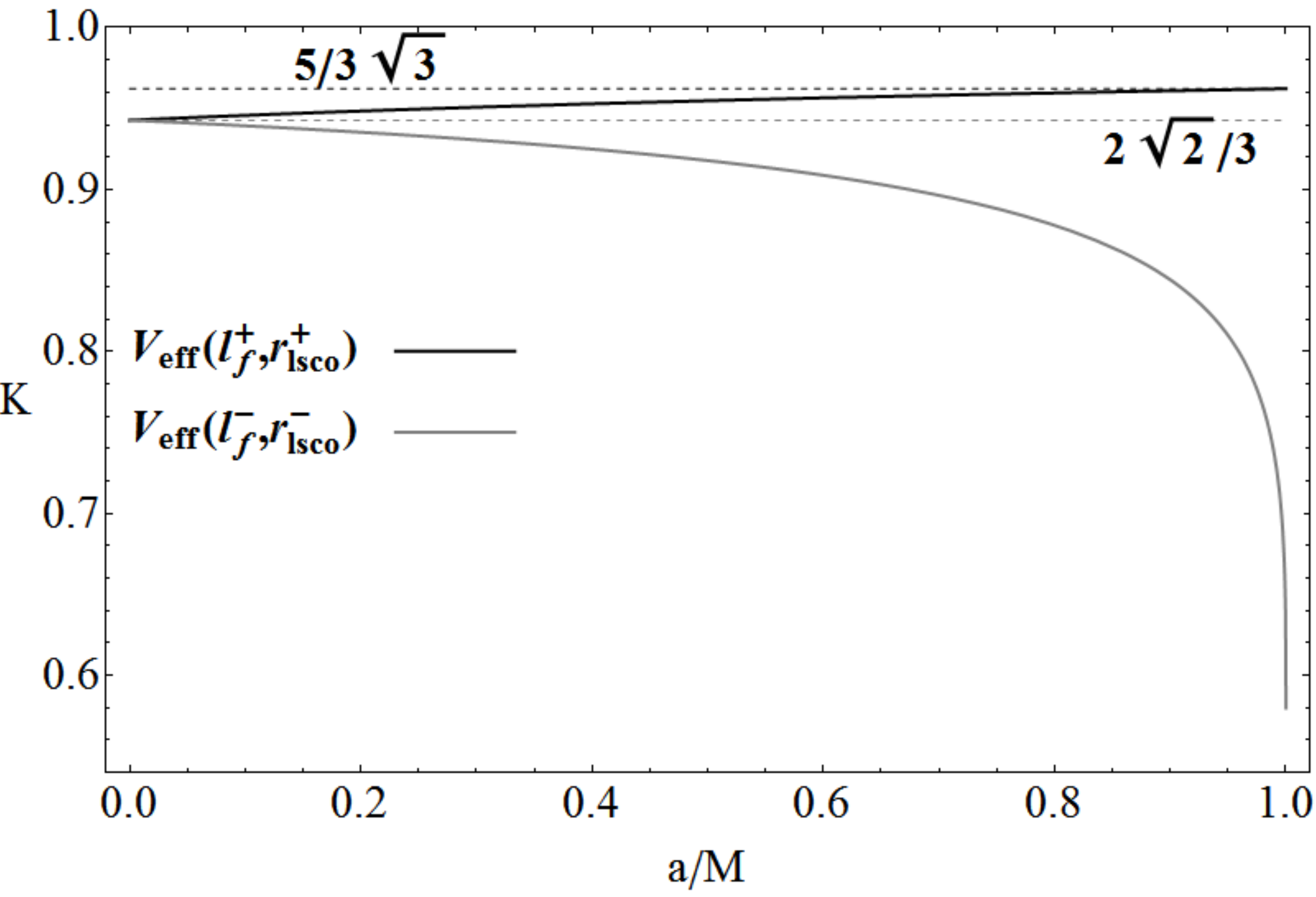}
\includegraphics[scale=.3]{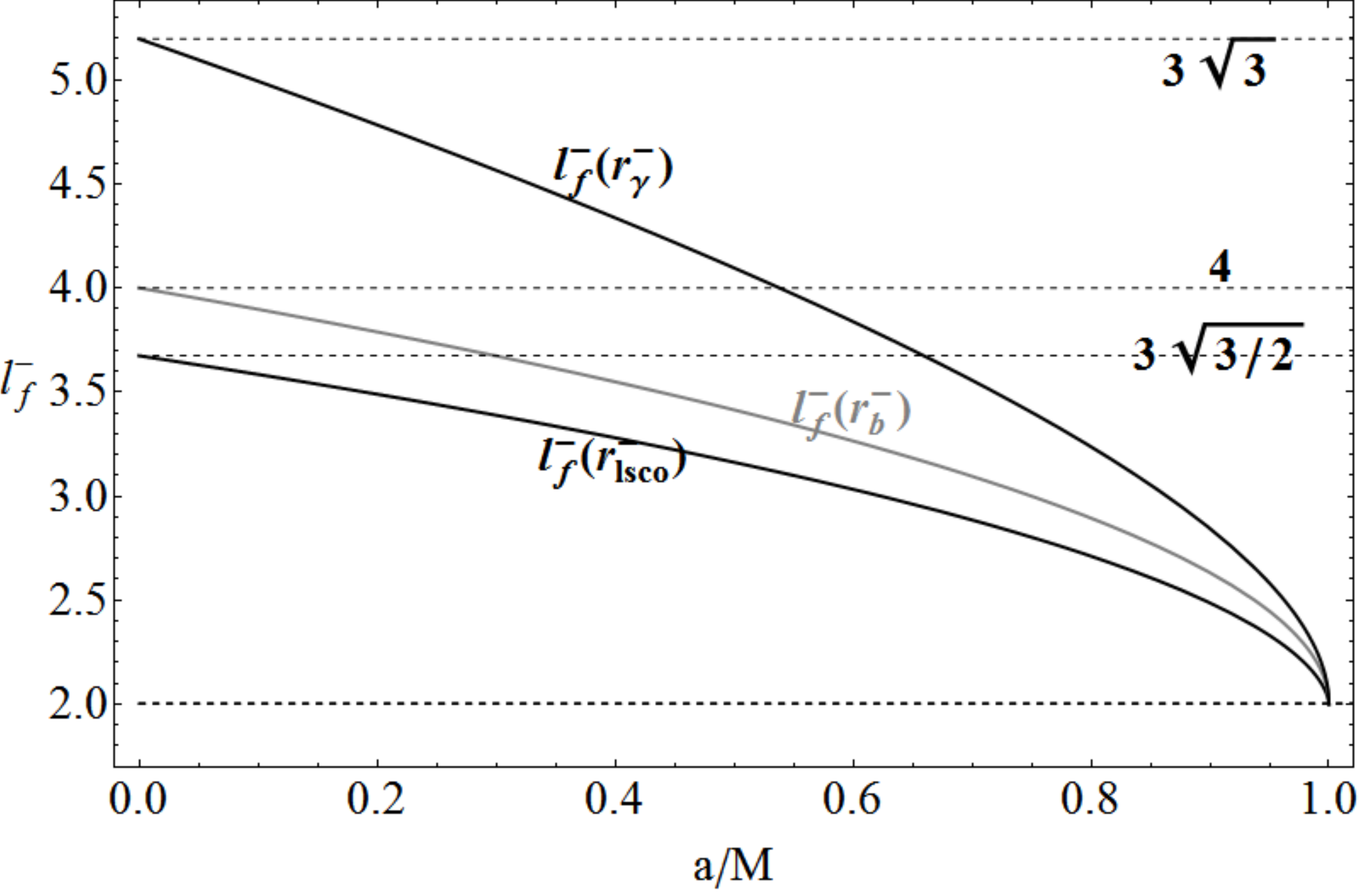}\\
\includegraphics[scale=.3]{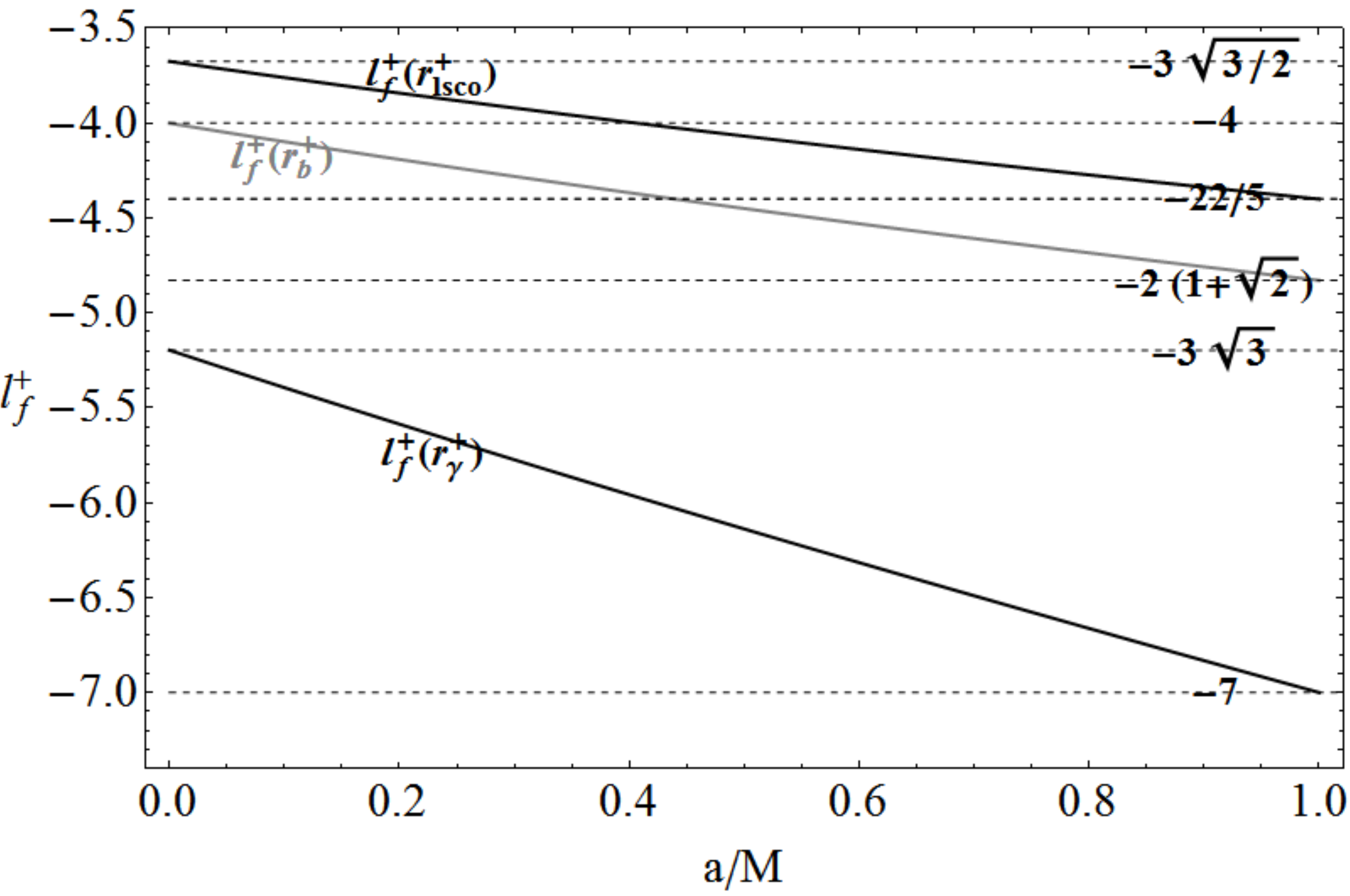}
\end{tabular}
\caption[font={footnotesize,it}]{\footnotesize{Left top panel: plot of the effective potential   $V_{eff}(\ell_{f}^{\pm},r_{lsco}^{\pm})$ as function of $a/M$. At $a=M$ the  $r_{lsco}^{-}$ reaches the horizon. Right top  and bottom panels    angular momentum, in units of mass $M$, $\ell_f^{-}$ $(\ell_f^{+})$ at $\{r_{lsco}^-,r_b^-, r_{\gamma}^-\}$, $(r_{lsco}^+,r_b^+, r_{\gamma}^+)$ for different spacetimes.  It is $\ell_f^{\mp}(r_{\gamma}^{\mp})\gtrless\ell_f^{\mp}(r_{b}^{-})
\gtrless\ell_f^{\mp}(r_{lsco}^{\mp})$.
In terms of the fluid angular momentum only ${C}_{x}$ configurations are at $K<1$ and $\ell\in]\ell_b^+,\ell_{lsco}^+[$   and $\ell\in]\ell_{lsco}^-,\ell_b^-[$,
 ${C}$ disks are for $K\in]K_{lsco}^{\pm},1[$,  $\ell<\ell_{lsco}^+<0\cup\ell>\ell_{lsco}^->0$
and $  r>r_{lsco}^{\pm}$,
  $ {O}_{x}$ disks  at $K>1$ $\ell\in ]\ell_{\gamma}^+,\ell_b^+[\cup]\ell_b^-,
  \ell_{\gamma}^-[$, see text.}}\label{Fig:tema}
\end{figure}

\textbf{{The corotating fluid angular momentum }}
The critical points of the angular momentum  $\ell_{f}^{-}$
 are in the range  $r\in]M,6M]$,
in particular for $a=0$ in  $r=6M$ and
at $a\in]0,\check{a}[$ at $r=r_{lsco}^-$.
It is $\partial_r\ell_{f}^->0$ in  $r>6M$, for any attractor with $a/M\in[0,1]$, but  in the case of a
 Schwarzschild geometry  ($a=0$) increases in $r>6M$
and $a\in]0,\check{a}]$ in $r>r_{lsco}^-$
and for sources $a\in]\check{a},M]
$ in $r>r_{lsco}^- -\{r_d\}$.
There are {minimum} points  at $r_{lsco}^-$.
 The angular momenta $\ell_{lsco}^{\pm}$  decrease with $a/M$, see  Figs.\il(\ref{Fig:radiilf1}).
It is worth  noting that there is a region of minimum points in the ergoregion as $a\in]a_{\natural},M]$
where at  $a_{\natural}\equiv2\sqrt{2}/3 M$  it is $r_{lsco}^-(a_{\natural})=r_{\epsilon}^+$. Nevertheless  it is  always $K<1$ on these peculiar orbits, that means the possibility for  $C$ configuration to  form, see Fig.\il(\ref{Fig:tema}).
\subsection{Classes of Kerr attractors  and disks configurations}\label{Sec:s-pf}
In this section we  investigate the existence of the  thick configurations  with angular momentum $\ell=\ell_f^{\pm}$ considering  separately the  following three  cases: \textbf{(I)}$K>1$, \textbf{(II)} $K=1$, \textbf{(III)} $K\in]0,1[$,
where opened $\left({O}\right)$ configurations are for $K>1$, closed $\left({C}\right)$ for $K\in]0,1[$.
For each classes, $(O, C)$ there can be crossed  surfaces or $(O_x, C_x)$,  where the hydrostatic pressure is minimum  and the   accretion can occur.
P-D configurations can exist in
$r>r_{\gamma}^{\pm}$ with  $|\ell|< |\ell_{f}(r_{\gamma}^{\pm})|$ for counterrotating or corotating matter respectively.  We detail the situation as follows.
\textbf{(I) for $K>1$}    there are only {maximum}   points for the effective potential on $r=r_f^{\pm}$ solutions
\(r_f^{\pm}:\; a^2 M(\Delta_{\ell}^-)^2-4 M^3 (\Delta_{\ell}^-)^2 r+2M (\Delta_{\ell}^- +\ell) (\Delta_{\ell}^-)  r^2-\ell^2 r^3+Mr^4=0\)
correspondingly  there are  $ {O}_{x}$-configurations  only with:
\(\ell\in ]\ell_{\gamma}^+,\ell_b^+[\cup]\ell_b^-,\ell_{\gamma}^-
[\).
These configurations fill
 the orbital regions  $r\in]r_{\gamma}^{\pm},r_{b}^{\pm}[$ closest to the attractor,  these can be  accretion points.
\textbf{(II) for   $K=1$}  there are only maximum points  of $V_{eff}(\ell)$, thus ${O}_{x}$-configurations, on $r_b^{\pm}$ with $\ell_f(r_{b}^{\pm})$ respectively.
\textbf{(III) for $K\in]0,1[$} the fluid is characterized by maximum, minimum  or saddle points for the effective potential.
 For $K\in]K_{lsco}^{\pm},1[$
the  minimum points, centers of  ${C}$-configurations, are  in $r>r_{lsco}^{\pm}$   with $\ell<\ell_{lsco}^+<0\cup\ell>\ell_{lsco}^->0$.
The {maximum points} of the effective potential with $K<1$ (minima of the pressure as function of the orbital distance from the source)  are associated to the closed crossed, ${C}_{x}$ surfaces with  the angular momentum  $\ell\in]\ell_b^+,\ell_{lsco}^+[$   and $\ell\in]\ell_{lsco}^-,\ell_b^-[$, the radius  of these orbits, location of the accretion points, are at
 $r_f^{\pm}\in]r_b^{\pm}r_{lsco}^{\pm}[$.
As the $C$-disk is formed, it can thicken  approaching the source,  leading to  the $C_x$-accretion   morphology,  and eventually it could rise in  the unstable modes with open branches as  $O_x$, thus one could say that  the disk follows, during its evolution  the sequence of configurations  $\mathfrak{B}=[C, C_x, O_x]$. One can argue  that the entire evolution of a P-D disk, from the  formation   to   the accretion and  finally to the open configuration (or reversing with an excretion process see for example \cite{2011,astro-ph/0605094,Stu-Sla-Hle:2000:ASTRA:,arXiv:0910.3184}) could  occur  for example as one of  the following sequences of P-D configurations: $\mathfrak{B}=[{C},{C}_{x},{O}_{x},{O}]$ (or   even   $\mathfrak{B}=[{O} ,{O}_{x},{C}_{x}, {C}]-[{C}_{x},{C}]$).  The configurations in the sequence $\mathfrak{B}$ are all Boyer surfaces, correspondent to the pressure critical  points, however the  dynamical evolution   between each configurations  of $\mathfrak{B}$, is due to a (correlated) change  of the model parameters $\mathbf{p}\equiv (K,\ell)$,  the transition configurations between one Boyer surface to another  of the sequence  $\mathfrak{B}$ represents  a transition  in the dynamical evolution of the disks interacting with the external environment.
This evolution can be clearly followed through  the shifting  of  the effective potential and therefore the $K$-parameter, this aspect will be investigated in details in Sec.\il(\ref{Sec:s-pf}). As   it is $r_{\gamma}^->r_{\gamma}^+$, corotating and counterrotating thick disks  can be in, or share part of  the same orbital region.
Here we  construct  a  set of corotating and  counterrotating configurations   keeping one parameter of the couple  $\mathbf{p}\equiv(K,\ell)$ constant and   changing the other one in the set of values  for the formation of a P-D surface.
We consider then the  set of  orbits
$\mathfrak{R}\equiv\{r_d, r_{\epsilon}^+,r_{lsco}^{\pm},r_b^{\pm},r_{\gamma}^{\pm}\}$:
correspondingly the regions between subsequent orbits of $\mathfrak{R}$  identify a slot of the sequence   $\mathfrak{B}$.
The properties of the P-D configurations
 turn out to be so distinctive that they allow the introduction of a complete classification of
Kerr spacetimes according to the  spin set $\mathfrak{A}\equiv\{a_{\circ},a_{\bullet},a_{\flat},
a_{\dag},
a_{\Box},
a_{\diamondsuit},
{\check{a}}
,a_{\natural}\}$ as shown in Fig.\il(\ref{Fig:radiilf1}), each class is characterized by different physical features of the thick disks that could be of
particular relevance in observational astrophysics.
We locate nine classes of Kerr attractors, named  accordingly   $\mathbf{BHI-BHIX}$  and  identifiably  in terms of their  spin mass ratio class  in the ranges  with boundaries  in  $\mathfrak{A}$ as illustrated  in Fig.\il(\ref{Fig:radiilf1}).
 Each class has different characteristics in the arrangement of the Boyer surfaces  and therefore in their possible  evolution between one configuration to another.
As a consequence of this   the  analysis of the toroidal configurations could be a  possible way to recognize a source from the analysis of  the toroidal configuration and in particular the fluid  angular momentum.  In the  following we  define each class  of attractors  discussing  their  main properties in terms of the  characteristic sequence $\mathfrak{B}$.
\begin{description}
\item[The \textbf{BHI}  class: \({a\in[0, a_{\circ}]}\)] It is  $a_{\circ}\equiv(8 \sqrt{2}-11)M$ where  \(r_b^-(a_{\circ})=r_{\gamma}^-(a_{\circ})\), this class of metrics  includes the special case of the  Schwarzschild geometry.
In general  in the \textbf{BHI}  spacetimes  the orbits $r_i\in\mathfrak{R}$ can  be ordered as follows:
    $r_{lsco}^+>r_{lsco}^->r_b^+>r_{b}^->r_{\gamma}^+>r_{\gamma}^->r_d>r_{\epsilon}^+>r_+$.
The set of possible configurations is
$\mathfrak{B}_{I}\!=\![C^{\pm}, (C^- C^+_x), C_x^{\pm},(C_x^{-} O_x^+), O_x^{\pm}, O_x^-]$, each of the six slot of $\mathfrak{B}_{I}$ is associated to one of the  six regions $[r_i,r_j[$ with boundaries  $r_i\in\mathfrak{R}$ where the hydrostatic pressure has critical points, and for \({a\in[0, a_{\circ}]}\) it includes the radii  in $r>r_{\gamma}^-$.
With $()^{\pm}$ we denote the disks of counterrotating or corotating fluids respectively, each slot   of the six-dimension arrow $\mathfrak{B}_{I}$ is intended for  a radial region where a couple or a single kind of configuration can be located starting from  the region farthest from the source thus  the farthest  $C$ morphology to the closest  $O_x$ one. Since each slot of $\mathfrak{B}_{I}$ identifies one of the regions in Fig.\il(\ref{Fig:radiilf1}) determined by $\mathfrak{R}$, then the  configuration $C_x$ are to be understood crossing two regions, the center of the configuration placed  in the outer region, and the  cross point in the inner one. The couple of different configurations in $\mathfrak{B}_{I}$  means a multiple configuration located in one orbital region, for example in $r>r_{lsco}^+$, they may both closed  $C^{\pm}$ of  corotating  and counterrotating fluids but the spacing and  the exact number of  permitted disks, has to be further discussed and it will be addressed  in Sec.\il(\ref{Sec:multipleP-D}).
The more we move towards the interior, that is, in regions closer to the  attractor, and more unstable configurations appear leading to a  transition from $C$ to $C_x$ up to $O_x$-configurations.
\item[The \textbf{BHII} class:{ \( a\in ]a_{\circ},a_{\bullet}] \)} ]  It is $a_{\bullet}\equiv (23-16 \sqrt{2})M$  where  \( r_b^+(a_{\bullet})=r_{lsco}^-(a_{\bullet})\). In these spacetimes it is $r_{b}^->r_{\gamma}^+$, correspondingly $\mathfrak{B}_{II}\!=\![C^{\pm},( C^- C^+_x), C_x^{\pm},(C_x^{-} O_x^+), C_x^-, O_x^-]$.  Corotating surfaces are therefore allowed in a larger  number of  orbital regions respect  to \textbf{BHI} geometries.
    Moreover there is a  simultaneous decreasing of number of   regions where both corotanting and counterrotating are possible, for example in this case as a consequence of the  permutation between the couple $(r_b^-, r_{\gamma}^+)$
the double $O_x^{\pm}$ in \textbf{BHI} class is reduced to the $C_x^-$ configuration in  \textbf{BHII}.
\item[The \textbf{BHIII}  class: $ {a\in]a_{\bullet},a_{\flat}] }$]  It is $a_{\flat}\equiv{\sqrt{{7}/{3}}}/{3}M$ where  \(r_{\gamma}^-(a_{\flat})=r_d(a_{\flat})\). Respect to  the \textbf{BHII} geometries  the  \textbf{BHIII} spacetimes   are characterized  by a further permutation of  the couple $(r_{lsco}^-,r_b^+)$ and   it is $\mathfrak{B}_{III}\!=\![C^{\pm},(C^- C^+_x),( C^- O^+_x), (C^-_x, O_x^+), C_x^-, O_x^-]$.
   There is a reduction in the number of  regions where $C_x^+$ are permitted, but the    orbital extension  in which these solutions are  admitted increases with the spin.
The difference with  \textbf{BHII} spacetimes is the third slot of the sequences $ (\mathfrak{B}_{II},\mathfrak{B}_{III})$, in  $\mathfrak{B}_{III}$ filled   with a corotating closed and open crossed counterrotating  disks.
    \item[The \textbf{BHIV}  class: {\( a\in]a_{\flat},a_{\dag}]\)} ] It is $a_{\dag}/M\equiv{1}/{2} \left(3 \sqrt{17}-11\right)M$   where
  \(r_{b}^-(a_{\dag})=r_d(a_{\dag})\). Respect to the class \textbf{BHIII} there is a further permutation  between  $(r_d, r_{\gamma}^-)$.
It is convenient to introduce  the  spin:
$a_{\ddag}\equiv0.638284738M$ splitting the set $\mathbf{BHIV}$ as $\mathbf{BHIV}=\mathbf{BHIVa}\cup\mathbf{BHIVb}$.
For  \textbf{BHIVa} spacetimes it is $a<a_{\ddag}$  and
 $\mathfrak{B}^{<}_{IV}\!=\![C^{\pm},( C^- C^+_x),( C^- O^+_x),(C_x^- O^+_x), C^-_x, O_x^- ]\equiv\mathfrak{B}_{III}$, and   the orbital region of the last slot with $ O_x^- $ configurations is  crossed by $r_d$.
For \textbf{BHIVb} spacetimes it is  $a>a_{\ddag}$,  there is a permutation $(r_\gamma^+, r_{lsco}^-)$ and  as a consequence  it is
$\mathfrak{B}^{>}_{IV}\!=\![{C}^{\pm},( C^- C^+_{x}),( C^- O^+_{x}),C^-, C^-_x, O_x^-]$.
%
\item[The \textbf{BHV}  class: {\( a\in]a_{\dag},a_{\Box}]\)}] It is   $a_{\Box}\equiv1/{\sqrt{2}}M$ where \(r_{\gamma}^-(a_{\Box})=r_{\epsilon}^+(a_{\Box})\)
 and   $\mathfrak{B}_{V}\!=\mathfrak{B}^{>}_{IV}$.   However respect to the  \textbf{BHIV} geometries,  in the spacetimes belonging  to  the \textbf{BHV}  class there is the  further  permutation  in the couple $(r_d, r_b^-)$, as of consequence of this the last  slot of  $\mathfrak{B}_{V}$, filled with $O_x^-$,  and a part of $C^-_x$-region are  now entirely contained in $r<r_d$.
\item[The \textbf{BHVI}  class:{ \(a\in]a_{\Box},a_{\diamondsuit}]\)} ] It is $a_{\diamondsuit}\equiv2 \left(\sqrt{2}-1\right)M$ where  \(r_b^-(a_{\diamondsuit} )=r_{\epsilon}^+(a_{\diamondsuit} )\), the sequence of configurations is  $\mathfrak{B}^{>}_{VI}=\mathfrak{B}^{>}_{IV}$. For the spacetimes with spin $a>a_{\Box}$ there is a  further  permutation in the arrow $\mathfrak{R}$ involving   the static limit $r_{\epsilon}^+$, for \textbf{BHVI} spacetime    in particular  there is a permutation in the couple   $(r_{\gamma}^-,r_{\epsilon}^+)$: as a consequence of this the last slot of    $\mathfrak{B}^{>}_{VI}$ filled with  $O_x^-$ configurations, crosses the static limit and  it is partially contained in $r<r_{\epsilon}^+$.
%
%
\item[The \textbf{BHVII}  class:{ $a\in]a_{\diamondsuit},{\check{a}}]$} ] It is  $\check{a}\equiv0.865157M$ where    $ r_{lsco}^-({\check{a}}) =r_d({\check{a}})$.  The sequence of configurations is  $\mathfrak{B}_{VII}=\mathfrak{B}^{>}_{IV}$ and a  permutation is in $(r_d, r_{\epsilon}^+)$, the region filled with the   $O_x$ disks and a part of $C_x^-$ configurations  are  contained in the ergoregion.
\item[The \textbf{BHVIII}  class:{ $a\in]{\check{a}},a_{\natural}]$} ]  It is  $a_{\natural}\equiv2\sqrt{2}/3M$ where \(r_{lsco}^-(a_{\natural})=r_{\epsilon}^+(a_{\natural})\).  The sequence of configurations is  $\mathfrak{B}_{VII}=\mathfrak{B}^{>}_{IV}$. For these sources a  further  permutation occurs  in the couple $(r_d, r_{lsco}^-)$, it follows that, respect to \textbf{BHVII} sources,  a part of $C^-$ region is at $r<r_d$.
%
\item[The \textbf{BHIX}  class: {\( a\in]a_{\natural},M]\) }] The extreme black hole case is a special case of \textbf{BHIX} spacetimes\footnote{We note here that the spacetimes with  limiting spins $a/M\in\{0.9953,0.99979\}$ for the
Aschenbach effect \cite{Aschenbach,Stuchlik:2004wk}   belong to the class  \textbf{BHIX}}.
A further permutation respect to the situation in \textbf{BHVIII}  geometries,  is in the couple $(r_{lsco}^-,r_{\epsilon}^+)$ leading  to the $C_x^-$ configuration entirely contained, and the  $C^-$ configurations  partially  contained in the region $r<r_{\epsilon}^+$.
\end{description}
The  limiting spacetimes with spin $ a_i \in \mathfrak{A} $ require further investigation.
In Secs.\il(\ref{Sec:KU1ana}--\ref{Sec:KM1ana}) we  detail  and comment the nine classes of attractors and the P-D torii,
while the  multiple configurations in  the sequences $\mathfrak{B}$ will be discussed in Sec.\il(\ref{Sec:multipleP-D}) and the P-D configurations   near the static limit  or in the region $r<r_{\epsilon}^+$, possible in the
\textbf{BHVI-IX} spacetimes,
  are  considered      in details in Sec.\il(\ref{Sec:limite.statico}).
%
\subsubsection{Configurations   at $K=1$.}\label{Sec:KU1ana}
We consider the critical points of the effective potential  for  $K=1$, solutions of
\(
V_{eff}(\ell)=K=1\)  and  \(\partial_r V_{eff}(\ell)=0
\).
No {minimum} and {saddle} points are possible i.e. $\partial_r^2 V_{eff}(\ell)\ngeq0$. Maximum points are located  on the orbits $r_{\kappa}^-$:
\bea\label{Eq:firsr-rver}
r_{\kappa}^{\pm}\equiv\frac{1}{4M} \left(\ell^2\pm \sqrt{-16\left(\frac{1}{4} l (4M+l)-aM\right)\left(\frac{1}{4} l(4M-l)-aM\right)}\right),
\\
a\in[0,M[,\quad \ell=\{\ell_b^+,\ell_b^-\},\quad
a=M,\quad r=(3+2 \sqrt{2})M\quad \ell=\ell_b^+.
\eea
On the other hand one can equally say orbits are in
 \bea\label{Eq:con-nap-spac}
 &&
a\in[0,M[-\{a_{\dag}\},\quad r=r_b^{\pm}\quad\mbox{and}\quad \ell=\ell_f^{\pm},
\quad\mbox{and}\quad
a=\{a_{\dag}, a=M\}\quad\mbox{with}\quad\ell=\ell_f^{+}\quad\mbox{in}\quad r=r_b^+
\\\nonumber
&& a_{\dag}\equiv\frac{1}{2} \left(3 \sqrt{17}-11\right)M\approx0.684658M,\quad r_{b}^{\pm}\equiv2M\pm a+2\sqrt{M} \sqrt{M\pm a},
 \eea
it is  $a_{\dag}: r_{b}^-=r_d$, and  $r_{b}^{\pm}$ are the
 marginally bounded orbits for test particles: critical points of $r_b^{\pm}:\; \partial_rV_{eff}(r_b^{\pm},\mp L_{\pm})=0,\;   \partial_r^{2}V_{eff}(r_b^{\pm},\mp L_{\pm})<0, \;  V_{eff}(r_b^{\pm},\mp L_{\pm})=1$.
There are   open ${O}_{x}$ configurations. The surface $r_{\kappa}^-(a,\ell)$ includes the radii $r_{b}^{\pm}(a)$, where the cusps are located  i.e. $\exists\ell(a): \; r_{\kappa}^-(a,\ell(a))=r_{b}^{\pm}(a)$, indeed it is:
\(r_{\kappa}^-(\ell_{b}^{\pm})=r_b^{\pm}\).
Then there are  two cusps correspondent  to two different angular momenta for $a\neq0$ and  $a\neq M$, in the  particular case of non-rotating  attractor   $a=0$ it is   $r_b^{\pm}=4M$ and  $\ell=\pm 4M$. In the spacetimes \textbf{BHVII-IX} these peculiar configurations of  corotating fluids can fill the region $r<r_{\epsilon}^+$, see Fig.\il(\ref{Fig:radiilf1}).
\subsubsection{Configurations at $K\in]0,1[$.}\label{Sec:Km1ana}
We analyse the case $K\in]0,1[$
 to find the maximum points of the effective potential ($\partial^2_r V_{eff}(\ell)<0$) associated to the cross points of the configurations  ${C}_{{x}}$, and minimum  points ($\partial^2_r V_{eff}(\ell)>0$) centers of the closed ${C}$-configurations.
We will consider also the critical points $r_{f}:\  \partial_r V_{eff}(\ell)=0$, $ \partial_r^2 V_{eff}(\ell)=0$, saddle point and limiting orbits of instability.
For the  non rotating case of the {Schwarzschild metric  ($a=0$)} minimum points are at $r>6M$ with $\ell=\pm\sqrt{\frac{Mr^3}{(r-2M)^2}}$. The maximum points are  located in the lower orbital region $r\in]4M,6M[$. A saddle  is located at $r=4M$, an extensive analysis of the non rotating background  can be found in \cite{PuMonBe12}.

{For a Kerr attractor  ($a\in]0,1[$) we summarize the situation as follows:}

\textbf{Minimum points}
Minimum points of the effective potential $V_{eff}(\ell)$ are located in $r>r_{lsco}^{\pm}$.
They are  maximum points of the hydrostatic pressure and locate the disk center, and then with the instability points  they  contribute to determine  the disk extension on the equatorial plane and the location of the inner boundary.
However
to analyze properly the  {minimum} points we should consider two classes of sources, defined by the spins $\check{a}$:
The first set  is the  \textbf{BHI-VII} classes where { $a\in]0,\check{a}]$}, and  $\check{a}:\; r_{lsco}^-=r_d$; according with the analysis for the case of null pressure minimum points are located in  $r\in]r_{lsco}^-,r_{lsco}^+]$ with  $\ell=\ell_f^->0$, for  corotating disk, and $r>r_{lsco}^+$ with  $\ell=\ell_f^{\pm}$ where  corotating and counterrotating disks are possible.
Finally the second  set is for  $a/M\in]\check{a},M[$-\textbf{BHVIII-IX}. Minimum points are located in  $r\in]r_{lsco}^-,r_d[$ with  $\ell=\ell_f^->0$ for  corotating disks, $(r=r_d,\ell=\ell_d)$, at $r\in]r_d,r_{lsco}^+]$ with $\ell=\ell_f^-$, and  $r>r_{lsco}^+$ with $\ell=\ell_f^{\pm}$.
Counterrotating closed  disks (minimum points) are possible only at $r>r_{lsco}^+$.
The angular momentum $\ell_f^-$ has a bounded orbital extension and penetrates the ergoregion at $a=a_{\natural}$ after that it extended theoretically up to the horizon.
For \textbf{BHVIII} sources with  $a\in]\check{a},a_{\natural}[$ the region with $\ell_f^-$ has  the static limit as the upper bound,
the region with $\ell_f^-$-tori within and out the ergoregion is then  possible only for sufficiently large spin.
The extension of the   orbital region increases with  the spin, up to the maximal extension for  spin close to $a=M$.
 The transition orbit   for the configurations here analyzed  is instead   $r_{lsco}^+$ that defines the lower boundary for the counterrotating fluids. However there are no critical points   with $\ell=0$ for  $K\in]0,1[$.
This analysis can be then restated in terms of the  orbits of minimum points, namely the orbits $r(a;\ell)$ where the  minimum points are for $\ell<\ell_{lsco}^+<0\cup\ell>\ell_{lsco}^->0$ at  $r=r_f^{\pm}$  and in the non rotating case it is $\ell_{lsco}^+=-\ell_{lsco}^-=-3 \sqrt{{3}/{2}}M.$
The solutions are $r_f^{\pm}:\; \ell_f^{\pm}=\ell$.
From the analysis in Sec.\il(\ref{Sec:contro-l})  it follows that  the angular momentum $\ell_f^+<0$  decreases  with $r>r_{lsco}^+$.

The {saddle}, orbits $r_{f}$, are  on the boundary $r_{lsco}^{\pm}$ where  $\ell=(\ell_{lsco}^+,\ell_{lsco}^-)$.

\textbf{Maximum points}
The {maximum points} of the effective potential with $K<1$ (minima of the pressure as function of the orbital distance from the source) are associated to the closed crossed ${C}_{x}$ surfaces located in the regions  $r\in]r_b^{\pm},r_{lsco}^{\pm}[$. We consider the situations as follows:

\textbf{1} for  $a\in]0,a_{\bullet}[$-\textbf{BHI-II}-classes  orbital regions are $r\in]r_b^-, r_b^+]$ with $\ell=\ell_f^-$,  and $r\in]r_b^+,r_{lsco}^-[$ with $\ell=\ell_f^{\pm}$ and finally $r\in[r_{lsco}^-,r_{lsco}^+[$  where $\ell=\ell_f^+$.
In the spacetime with $a=a_{\bullet}$ it is in particular $r\in]r_b^-,r_b^+[$ with $\ell_f^-$ and
$r\in]r_b^+,r_{lsco}^+)$ with $\ell_f^{+}$.

\textbf{2}  for $a\in]a_{\bullet}, a_{\dag}]$-\textbf{BHIII-IV} classes  it is $r\in ]r_b^-, r_{lsco}^-[ $ with $\ell_f^{\pm}$ and
$r\in]r_b^+,r_{lsco}^+[$ with $\ell_f^{+}$.
As for the case $p = 0$,   configurations at $\ell_f^+$ are    counterrotating. However these orbits are in $r>r_{\epsilon}^+$. No torus configurations with  $\ell=0$ are allowed.

\textbf{3} for $a\in]a_{\dag},\check{a}[$-\textbf{BHV-VII}-classes   solutions are as follows
$r\in]r_b^-,r_d[$ with $\ell_f^+$, $r\in]r_d,r_{lsco}^-[$ with  $\ell_f^-$ and
$r\in]r_b^+,r_{lsco}^+[$  with $\ell_f^+$.
For the spacetime  $a=\check{a}$ it is $r\in]r_b^-,r_d[\cup]r_b^+,r_{lsco}^+[$  with $\ell_f^+$.
In the  spacetime with spin $a_{\lozenge}$  these  orbits cross the  static limit.

\textbf{4 } for $a\in]\check{a},M[$-\textbf{BHVIII-IX}-classes,
maximum points  are in $r\in]r_b^-,r_{lsco}^-[$  with$\ell_f^-$ and $]r_b^+,r_{lsco}^+[$  with angular momentum  $\ell_f^+$.
Interestingly  for sufficiently high spin the first set of corotating orbits, located at $]r_b^-,r_{lsco}^-[$ and $\ell_f^->0$, are unstable and partly included in the ergoregion
 and there are indeed  regions with both  corotating and counterrotating matter.

\textbf{Extreme BH case ($a=M$):}
In the {extreme BH-case} {minimum points}  are in
in $r\in]({M}/{2}), \left(3+\sqrt{5}\right)M[$ at $\ell=\ell_f^-$, in $r=\frac{1}{2} \left(3+\sqrt{5}\right)M $ at  $\ell=\frac{1}{4} \left(7+\sqrt{5}\right)M$, and  $r/M\in] \left(3+\sqrt{5}\right)/2,9]$ with $\ell=\ell_f^-$, and finally  $r>9M$ with angular momentum $\ell=\ell_f^{\pm}$, i.e. $\ell<-22/5\cup\ell>2$ .
{Maximum points} are in  $r/M\in]5.82843,9[ $ with $\ell_f^-$
that is $\ell/M\in]2 \left(-1-\sqrt{2}\right),-{22}/{5}[$.
On the other hand a saddle point is located at $r=9M$ with $\ell/M=-22/5$.

\textbf{Range of the fluid angular momentum}
 {Maximum points} of the effective potential are characterized by angular momentum
 $|\ell/M|\in]3 \sqrt{{3}/{2}},4[$ for $a=0$, and for Kerr spacetime  $a\in]0,M]$ it is $\ell\in]\ell_b^+,\ell_{lsco}^+[$   and $\ell\in]\ell_{lsco}^-,\ell_b^-[$.

At $r=r_b^+$, we need to introduce the spin
$a_{\bullet}:\;r_b^+=r_{lsco}^-$: critical points are at $\ell_f^-$, maximum points are in the spacetimes $a\in]0,a_{\bullet}[$-\textbf{BHI-II}, minimum points are  on $a\in]a_{\bullet},M]$-\textbf{BHIII-IX}
and then a {saddle} point is on $a_{\bullet}$
$\ell= 3.31081M $ and  $r = 4.71573M$.

\textbf{Maximal extension of  the $K$-parameter}
We introduce the notations: $K_{lsco}^{\pm}\equiv V_{eff}(\ell_f^{\pm},r_{lsco}^{\pm})$ and  $K_{b}^{\pm}\equiv V_{eff}(\ell_f^{\pm},r_{b}^{\pm})$,  see Figs.\il(\ref{Fig:tema});
the $K$ parameter lies  in a  bounded range $[K_{min},K_{Max}]\subset[0,+\infty[$.
it is $K_{Max}=V_{eff}(\ell_f^{\pm},r_b^{\pm})=1$ and
$K_{min}=V_{eff}(\ell_f^{\pm},r_{lsco}^{\pm})$.
The  boundary values    depend on the spacetime spin,
and they are  different for counterrotating and corotating orbits: the $K^-_{lsco}$  ( for corotating fluids) decreases as the  spacetime spin-mass ratio increases,
on the contrary $K^+_{lsco}$ (for counterrotating fluids) decreases with $a/M$.
\subsubsection{On the pressure critical points at $K>1$}\label{Sec:KM1ana}
Finally  we consider the  configurations defined  by the conditions
$
V_{eff}(\ell)=K>1$, $ \partial_r V_{eff}(\ell)=0$,
this case
is   associated with the open configurations generally interpreted as hydrodynamics jets.
There are no {minimum} or {flexes}, but there are {maximum} points on $r=r_f^{\pm}$, more precisely in a Kerr spacetime
$a\in]0,1]$, maximum points are for:
$\ell\in ]\ell_{\gamma}^+,\ell_b^+[\cup]\ell_b^-,\ell_{\gamma}^-
[\subset ]-7,-4[ \cup ]2,3\sqrt{3}[$,
The orbits  $r_f^{\pm}$   cross  the radius $r_{b}^{\pm}$
at $r=r_b^-$ for $\ell=l_f^+$ in the spacetimes  $a/M\in]0,a_{\circ}[$-\textbf{BHI}, where $a_{\circ}:\;r_b^-=r_{\gamma}^+$.
Or in terms of the fluid momentum $\ell_f^{\pm}$:
 {maximum points} for counterrotating fluid $(\ell_f^{+})$ are in the class   \textbf{BHI} in the range
$r\in]r_{\gamma}^+,r_b^+[$ and in the classes  $a\in]a_{\circ},M]$-\textbf{BHII-III}  on $r\in]r_{\gamma}^+,r_b^+[$.
For corotating fluids  $(\ell_f^{-})$ maximum points are in the \textbf{BHI-III} spacetimes where  $a\in[0, a_{\flat}]$  in the range   $r\in]r_{\gamma},r_b^-[$,  and in the \textbf{BHIV} class with spin $a\in]a_{\flat},a_{\dag}[$ in the range  $r\in] r_{\gamma}^-,r_{\gamma}^+[$ and $r\in]r_{\gamma}^+,r_b^-[$, and finally  for \textbf{BHV-IX} attractors where   $a\in[a_{\dag},M[$ in $r\in]r_{\gamma}^-,r_b^-[$.
For completeness we note here that the surfaces $K=V_{eff}=0$, are not critical points of the potential.
\subsection{Analysis of the fluid configurations for the angular momentum $\bar{\ell}=\ell/a$}\label{Sec:barl}
The orbital regions and the angular momentum where  the effective potential $V_{eff}(\ell)$  is well defined  depend on the difference   $\Delta^{\pm}_{\ell}(a)\equiv\ell\pm a$ as in Eqs.\il(\ref{Def:RBB}),  and one could introduce the rationalized dimensionless  quantities $(r_{\ti{B}}^{\pm}/a,\ell/a)$.
More generally,  many   properties of the fluid effective potential are determined  by  the rationalized dimensionless quantity   $\bar{\ell}\equiv\ell/(a\sigma^2)$ where:
\bea
V_{eff}(\bar{\ell})=
\sqrt{\frac{\Delta  \rho^2}{r^4+a^4 \left(\bar{\ell}^2-1\right) \left(\sigma ^2-1\right)-a^2 r \left[\bar{\ell}^2 (r-2M)-2 r+4 M \bar{\ell} \sigma +(r-2M) \sigma ^2\right]}},
\eea
and it is
\bea\label{Eq:effper2}
\ell=-a\sigma^2\quad V_{eff}(\ell)=\sqrt{\frac{\Delta  \rho ^2}{\rho ^4+8M a^2 r \sigma ^2}},\quad\ell=a\sigma^2\quad V_{eff}(\ell)=\sqrt{\frac{\Delta }{\rho ^2}},
\eea
see also Eq.\il(\ref{Eq:effper1}). In the limiting case of the non-rotating  Schwarzschild solution one could use   the rationalized angular momentum $\bar{\ell}=\ell/\sigma$  to take into  account,  by means of the term $\sigma$,   of  the  spherical symmetry   of the spacetime \cite{PuMonBe12}, while for a rotating geometry we can consider $\bar{\ell}=\ell/\sigma^2$ only in the extreme-Kerr case $a=M$.
In this section we study  the P-D fluid configurations in terms on the  rationalized  parameter $\bar{\ell}\equiv\ell/a$.
We show
the existence of a   limit on the maximum ratio $\bar{\ell}$ for  the   P-D model: in  some cases the condition for the  existence  of these configurations is determined by the ratio $\bar{\ell}$ only.
There are {no} critical points for  $\bar{\ell}\in[-1,1]$ i.e.
P-D configurations must have angular momentum whose magnitude is greater  of the spacetime spin mass ratio,
moreover also the momentum-spin  ratio $\bar{L}\equiv L/a$, in the case of zero hydrostatic pressure or the  Keplerian disk, is bounded to circular orbits with $|\bar{L}|>\mu$, see also discussion in\footnote{For simplicity we use here all dimensionless quantities, we introduce the rotational version of the Killing vectors $\xi_t$ and $\xi_{\phi}$ i.e. the
canonical vector fields
$\tilde{V}\equiv(r^2+a^2)\partial_t +a\partial_{\phi}$
 and $\tilde{W}\equiv\partial_{\phi}+a \sigma^2 \partial_t$ then the contraction
 the geodesic four-velocity with $\tilde{W}$ leads to the (non-conserved) quantity
$L-E a \sigma^2$,
 function of the conserved quantities $(E,L)$, the spacetime parameter  $a$ and the polar coordinate $\theta$; on the equatorial plane it then reduces on $ L-E a$.
When we consider the principal null congruence
$
\gamma_{\pm}\equiv\pm\partial_r+\Delta^{-1} \tilde{V}$
the angular momentum $L=a \sigma^2$ that is $\bar{\ell}=1$ (and $E=+1$, in proper unit), every principal null geodesic is then characterized by $\bar{\ell}=1$, on the horizon it is
$L=E=0$.} \cite{ONeill95,chandra42}.
However  as the conditions $L(\ell)>a$ and  $\ell\leq a$  cannot be  fulfilled together,
or $\ell(L)\in]-a,0[$ and $L>a$, as it is $\ell\equiv L/E$ (this quantity being related to  the apparent impact parameter,
of the light radii), then we could consider separately the orbits at $E/\mu>1$ (unbounded orbits) and $E/\mu\in(0,1)$ (bounded orbits),
so that  it will be convenient in the  following to consider    three    classes of configurations.
\textbf{(I) Configurations with $K\in]0,1[$}
The maximum extension of
rationalized angular momentum  at $r>r_{\epsilon}^+$, for the minimum points (torus centers) in the case of corotating  fluids,
is
$\bar{\ell}>5-2 \sqrt{2} $.
There are {no} fluid configurations
 in the range $\bar{\ell}\in[1,2]$.
Regarding the counterrotating configurations with minimum points (closed counterrotating torii), the  angular momentum of the fluid  is $\bar {\ell} < -22/5 $ or  $\ell<-3 \sqrt{3/2}M$.
For the saddle points  of the fluid effective potential
at $r>r_{\epsilon}^+$ it is $\bar{\ell}>5/2$ or
$\ell/M\in]{5 \sqrt{2}}/{3}, 3 \sqrt{{3}/{2}}]$,
and for  the counterrotating configurations it is
$\ell/M\in[-(22/5) ,-3 \sqrt{3/2}]$,
and $\bar{\ell}\leq -{22}/{5}$.
Finally for the maximum points of the potential
the maximum extension of the rationalized angular momentum  outside the static limit   is $\bar{\ell}>5/2$ where  $\ell/M\in]{5 \sqrt{2}}/{3},4[$.
Regarding the counterrotating  configurations with maximum points, the  angular momentum of the fluid must be    $\bar {\ell}<-22/5 $ and $\ell/M\in] -2 \left(1+\sqrt{2}\right),-3 \sqrt{{3}/{2}}[$.

\textbf{(II) Configurations with $K>1$}
For  $\ell>a$  only maximum points exist at $a/M\in[0,1[$
with $\ell/M \in]2,3 \sqrt{3}[$,  and $\bar{\ell}>2$.
At $r>r_{\epsilon}^+$, it is $\bar{\ell}>1$  then $\bar{\ell}>2+\sqrt{2}$  and  $\ell/M\in]2\sqrt{2},3 \sqrt{2}[$.
Counterrotating unstable configurations are characterized by   $\ell/M\in]-7,-4[$ and  $\bar{\ell}<2 (-1- \sqrt{2})$.

\textbf{(III) Configurations with $K=1$}
For $K=1$ there are only maximum points.
As $\ell>a$
 maximum points  are  for $ a\in[0,M[$ with  $\ell=\ell_b^-$ and $r=r_{\kappa}^-$,   where the angular momentum is limited in the range  $\ell/M\in]2,4]$ and   $\bar{\ell}>2$.
For  $\ell<-a$
there are only maximum points for $a\in[0,M]$  (note here is included the extreme Kerr case) and $\ell=\ell_b^+$ with $r=r_{\kappa}^-$,
 and $\ell/M\in[-2 (1+\sqrt{2}),-4 ]$ and $\bar{\ell} \leq -2(1+\sqrt{2})$,  out of the ergoregion.
In conclusion corotating configurations with $K=1$ can exist only if  $\bar{\ell}>2$ that is the fluid angular momentum doubles the black hole spin mass ratio, in the ergoregion these configurations can be formed only when  $\bar{\ell}\in]2 ,2+ \sqrt{2}[$.
Counterrotating configurations can  exist  with the upper limit on the ratio
$\bar{\ell} \leq -2(1+\sqrt{2})$.
\section{Morphology of the Boyer surfaces and variation in the model parameters}\label{Sec:Morfology}
In this section we study a more general class of matter configurations  which includes the  P-D torus   as a special case.
It is then convenient to analyze the zeros of the function $\mathcal{V}_{eff}(\mathbf{p})\equiv V_{eff}(\ell)-K$ on the equatorial plane  i.e. we set  $\sigma=1$ and $y=0$ as  in the cartesian coordinates: $x=r\cos\theta$, $y=r\sin \theta $\cite{PuMonBe12}.
Solutions    on  $\theta=\pi/2$, are at  $y_i=y_i(a;\mathbf{p})$ with $i\in\{1,2,3\}$ and $K\neq1$ where:
\bea\nonumber
 && y_1=\frac{2M\left[\alpha_{\lambda}(K^2-1)  \cos\left(\frac{1}{3} \arccos\beta_{\lambda} \right)-1 \right]}{3 (K^2-1)},
 \quad
y_2=-\frac{2M\left[1+ \alpha_{\lambda} (K^2-1) \sin\left[\frac{1}{6} (\pi +2 \arccos\beta_{\lambda} )\right]\right]}{3 (K^2-1)},
  \\\nonumber
&&y_3=-\frac{2M\left[1+ \alpha_{\lambda} (K^2-1) \cos\left[\frac{1}{3} (\pi +\arccos\beta_{\lambda})\right]\right]}{3 (K^2-1)},
 \quad\alpha_{\lambda}=\sqrt{\frac{4M^2+3 (K^2-1) \left[K^2\Delta_{\ell}^-\Delta_{\ell}^+ +a^2\right]}{M^2(K^2-1)^2}},
 \\
 &&\label{Eq:5raidiff}\beta_{\lambda} =-\frac{9 (1-K^2) \left[K^2\Delta_{\ell}^-\Delta_{\ell}^++a^2\right]-8M^2-27 K^2 (K^2-1)^2 (-\Delta_{\ell}^-)^2}{\alpha_{\lambda}^3M^2(1-K^2)^3},
  \eea
  $y_i$ are regulated by the difference $(K^2-1)$ and $\Delta_{\ell}^{\pm}$ introduced in Eq.\il(\ref{Def:RBB}), the particular  solutions at $K=1$ will be discussed  in Sec.\il(\ref{SeC:K1}).
In general
there are three real solutions $y_i>0$ with $y_2<y_3<y_1$ , then
the crossing points with the equatorial plane are represented  for example with $y_{ijk}$ for the set $\{y_i,\; y_j,\;y_k\}$.
We introduce   here  the quantities:
\be\label{Eq:uni-b}
\ell_{\mu}^\Pi\equiv-2M+2 \sqrt{M}\sqrt{M+a}<a,\; a\neq0,\quad \ell_a^1\equiv-a+a \sin\left(\frac{\arcsin \frac{a}{M}}{3}\right)<0,
\ee
where in $]0,1[$ it is
$-a<\ell_a^1<\ell_{\mu}^\Pi
<a
<\ell^-_f(r_-)<\ell_{lsco}^-$ moreover it is
$\ell_f^{\pm}(r_+)\in]\ell_f^- (r_{\epsilon}^+),\ell_f^+ (r_{\epsilon}^+)[$,
see also Fig.\il(\ref{Fig:radiilf1})
and the  quantities
  $K_i:\; i\in\{1,...,4\}$ are $V_{eff}(a;\ell,r_{f}^{\pm})$ where $r_{f}^{\pm}(a;\ell):\; \ell_f^{\pm}(a,r)=\ell$.
The zeros of $\mathcal{V}_{eff}(\mathbf{p})$ ($\Sigma_{{V}_{eff}}$ surfaces) include but do not reduce to  the critical values of the fluid effective potential, therefore we explore here the more general set of solutions\footnote{We point out that the (Boyer) solutions  analyzed in  Secs.\il(\ref{Sec:model},\ref{Sec:fisa}) are associated to the critical points of the fluid effective potential at constant angular momentum as the closed $\mathbf{C}$ topology is centered in $r_{min}$ and/or crossed in  $r_{Max}$, here we focus on a on more general configurations defined as surfaces  $\mathcal{V}_{eff}=K$ which obviously include surfaces centered  (so that with an inner  and outer edges).}  providing different classifications according to  the couple of parameters $\mathbf{p}=(K,\ell)$. Considering the surface cross with the plane of symmetry we set different regions of variations for $\mathbf{p}$  which  include the P-D sector where the Boyer surfaces are defined. We  provide a surface classification   by fixing one of the model parameters and let the other change.
As discussed in Sec.\il(\ref{Sec:s-pf}) the different configurations  may represent different stages of time-evolution of orbiting matter, describing individual moments of the    evolution of one single  fluid configuration   in accretion.
However,   the  Boyer  theory considered  here  is able to model a (dynamical) stationary but not evolutive situation.
As a consequence of this, varying the couple $\mathbf{p}$, we can find  a sequence  of equilibrium configurations, each of them labelled by the fixed couple $\mathbf{p}$, not connected to each others within the theory by any dynamical law  which could bind chronologically  the different surfaces.  Thus, considering the six-dimensionally (time independent) array $\mathfrak{B}$ introduced in Sec.\il(\ref{Sec:s-pf}), we could properly consider  here a set of nine six-dimensional  matrices $\mathfrak{B}_{\mathbf{p}}$ on the surface $\Sigma_{\mathbf{p}}\equiv\Sigma_{K}\otimes\Sigma_{\ell}$ (or $\Sigma_{K}\otimes\Sigma_{\bar{\ell}}$), each for the nine \textbf{BH}-class of spacetimes, of elements  defined by the  fixed index $\mathbf{p}$. Then we could consider the ``projection''  $\mathfrak{B}_{\mathbf{p}_j}\equiv\mathfrak{B}_{\mathbf{p}}/\Sigma_{\mathbf{p}_i}$ of $\mathfrak{B}_{\mathbf{p}}$  on the constant surface $\Sigma_{\mathbf{p}_i}$, i.e. the  sequence (array or column) of elements on  the  constant $\mathbf{p}_i$ surfaces, and pointing    $\mathbf{p}_j $ as a chronological parameter, meaning that  we assume it  to follow  an   evolutionary model and providing  a sequence of evolutionary phases of one single configuration labelled by $\mathbf{p}_i$. Therefore  we relate the ordered sequences of  equilibrium configurations (or a part of this sequence) $\mathfrak{B}_{\mathbf{p_j}}$ to the history of a single disk (at fixed $\mathbf{p}_i$), independently by the dynamic law, that cannot be induced from the model itself, furthermore some  stages of formation and thickening of the disk   to be dynamically interpreted  need to be described by theories that include the interaction of the disk with source of matter  from which to accrete,  a  material  embedding of the disk that the   hydrodynamical model  here considered does not provide. $\mathfrak{B}_{\mathbf{p_j}}$, is thus a  (non necessary six dimensional) sequence of elements that figure different morphological phases of the $\mathbf{p}_i$-disk, each slot of $\mathfrak{B}_{\mathbf{p_j}}$ stands for an evolutive stage of the configuration (in this sense $\mathfrak{B}_{\mathbf{p_j}}$ could be considered time-dependent by means of its dependence from the $\mathbf{p_j}$ parameter). It is clear that the  real  evolution  can even  occur along a diagonal or any other  sequence of  elements of $\mathfrak{B}_{\mathbf{p}}$. However accretion is usually   modeled in terms of  angular momentum transport  inside the matter \cite{Abramowicz:2011xu}, thus we expect a more  realistic choice for an evolutive or a ``chronological'' parameter would be the fluid  momentum $\ell$. It should  be noted here  then $\ell$=constant  for any P-D solution (this is  a model assumption, see  \cite{Lei:2008ui}) and  $\ell$=constant on $\Sigma_t$. The results we provide could be easily rearranged according to a known dynamical law or  by comparison with  numerical simulations considering  the matrix elements  following  a different order.  So that we actually propose   here the comparison of this scheme  with an  evolutionary dynamical theory   (see in \cite{Igumenshchev,Shafee,Fragile:2007dk,DeVilliers,Hawley1987,Hawley1990,
Hawley1991,Hawley1984,arXiv:0910.3184,astro-ph/0605094,Stu-Kov:2008:INTJMD:,
Raine,Abramowicz:2011xu}.)
   where   it is shown how GRMHD-simulation  fits  with the hydrodynamical thick models:
we should recognize the matrix elements  and identify then a proper exact chronological order.
We analyze the sequences of  models at $\ell$ fixed (sequences $\mathfrak{B}_{K}$) for $K>1$  in (\ref{Subsec:KMlf})
and $K<1$   for the
 corotating  $(a\ell>0)$ fluid   configurations in (\ref{subsub:corot.mo}) and  the
counterrotating  $(a\ell<0)$  ones in (\ref{subsub:contro.mo}). The regions where the P-D configurations emerge  are highlighted in the lists below.
Despite the dependence of the effective potential from the \textbf{BH} spin,  the structure of the  classes is mostly independent from $a/M$ but,  following  also the discussion in Sec.\il(\ref{Sec:barl}), we consider the two cases  $|\ell|<a$ and  $|\ell|>a$, thus  we will reconsider the solutions in terms of the rationalized fluid momentum $\bar{\ell}=\ell/a$. In Sec.\il(\ref{Subsec:sequence}) we analyse a sequence of torus shapes in evolution considering the
{fluid configurations (belonging to sequences $\mathfrak{B}_{\bar{\ell}}$) at $\bar{\ell}>1$} (\ref{Subsun:barellM1}) and  $\bar{\ell}<1$ (\ref{Subsubsec:lbarm1}).
Sec.\il(\ref{Sec:limite.statico})  addresses some aspects of surfaces close to the static limit and clarifies certain model features in  the regions close to the ergoregion.
In
Sec.\il(\ref{Subsec:limitcases}) some particular cases are studied:
the case
{$K=1$} in (\ref{SeC:K1}),
 $\bar{\ell}=\pm1$  in (\ref{subsubsec:lpm1}),
  $\ell=0$ in (\ref{subsubsec:l0}) and
{the Schwarzschild case} is considered in (\ref{subsubsec:Schw}).
Finally   Sec.\il(\ref{Sec:multipleP-D})  discusses the existence of possible contemporaneous multiple P-D configurations, or intertwined and
ringed  P-D tori  (loops of disks),   Sec.\il(\ref{SeC:poly}) outlines some general considerations on the model morphology   for different   attractors and different values of the couple $\mathbf{p}$,   the case of  polytropic equation of  state for the orbiting fluid is also considered.

\subsection{Some notes on the surfaces close to  the static limit}\label{Sec:limite.statico}
In this section we focus on the P-D configuration  close to the static limit,  introduced  in   Sec.\il(\ref{Sec:fisa}).
\begin{figure}
\centering
\begin{tabular}{cc}
\includegraphics[scale=.31]{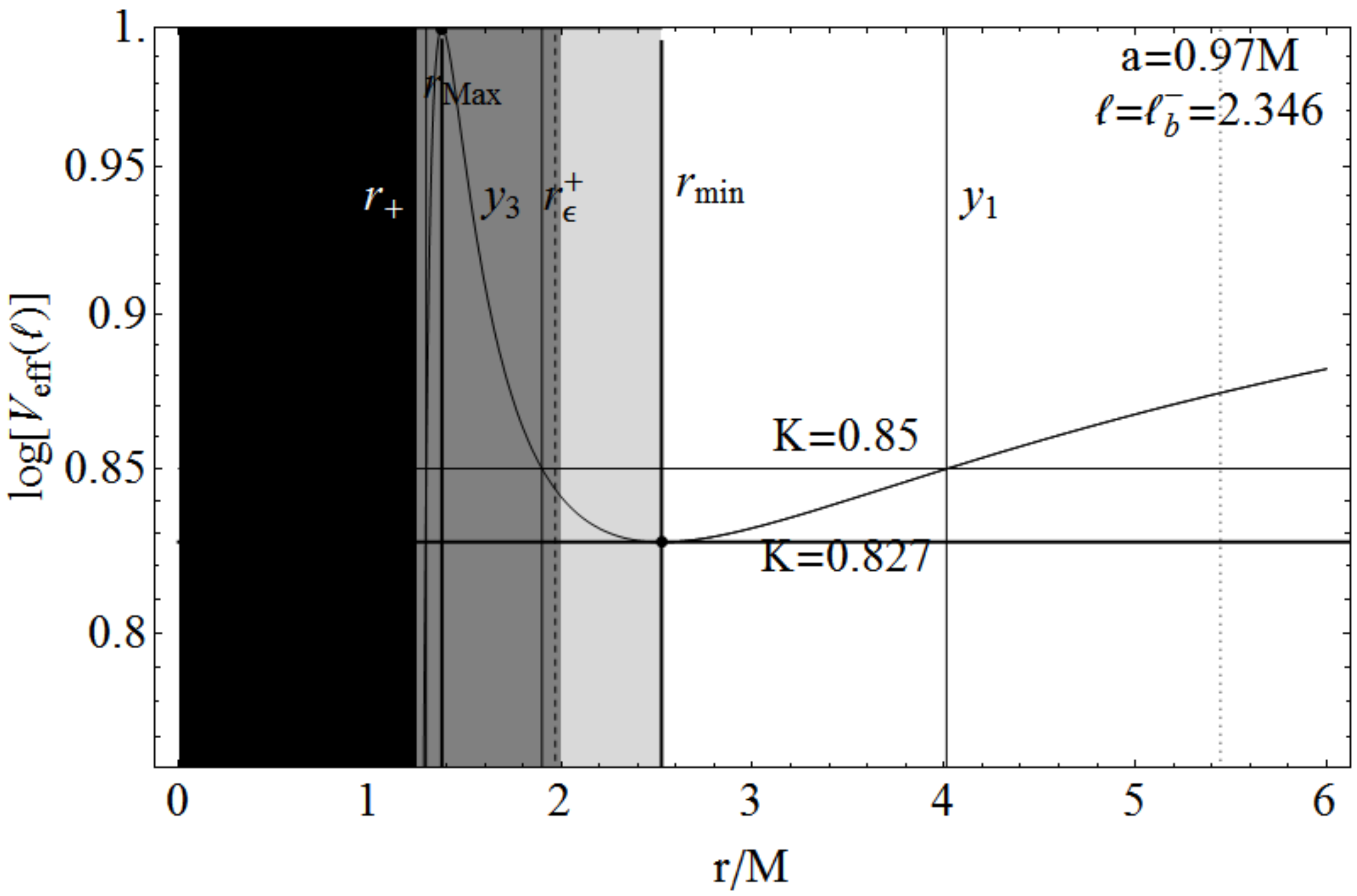}
\includegraphics[scale=.31]{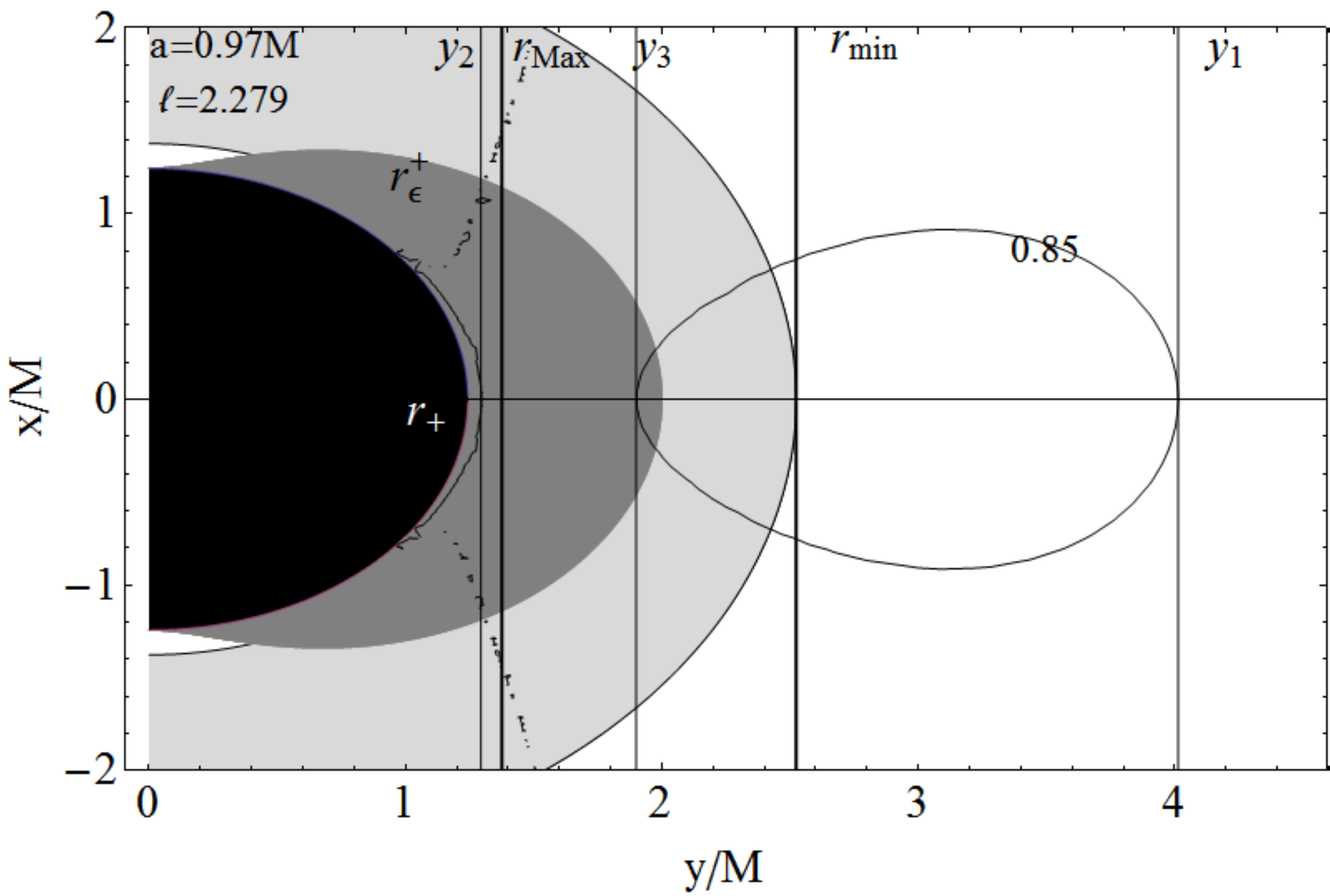}
\end{tabular}
\caption[font={footnotesize,it}]{\footnotesize{Left panel: logarithm plot of the effective potential $V_{eff}(\ell)$ as function of $r/M$. Right plot: the Boyer surface in the plane $(x,y)$ for a spacetime spin-mass ratio $a/M=0.97\in\mathbf{BHIX}$ and fluid angular momentum $\ell=2.279$ in units of mass $M$.
Black regions are $r<r_+$ where $r_+$ is the  black hole outer  horizon, gray region is $]r_+,r_{\epsilon}^+]$ where $r_{\epsilon}^+$ is the static limit, light-gray region is $]r_{\epsilon}^+,r_{min}]$, the critical points of $V_{eff}(\ell)$ are signed in the plots: $r_{min}$ is a minimum point of $V_{eff}(\ell)$  and $r_{Max}$ a maximum point.
The surfaces and the points at $K=V_{eff}(\ell)=$constant are black  lines, the point $y_1$ (left plot) corresponds to the outer boundary of the disk section (right plot), $y_3$ to the inner one, and $y_2$ to the outer boundary of the innermost surface, $r_{min}$ sets the diks center.}}\label{Fig:PSAX01u}
\end{figure}
\begin{figure}
\centering
\begin{tabular}{cc}
\includegraphics[scale=.293]{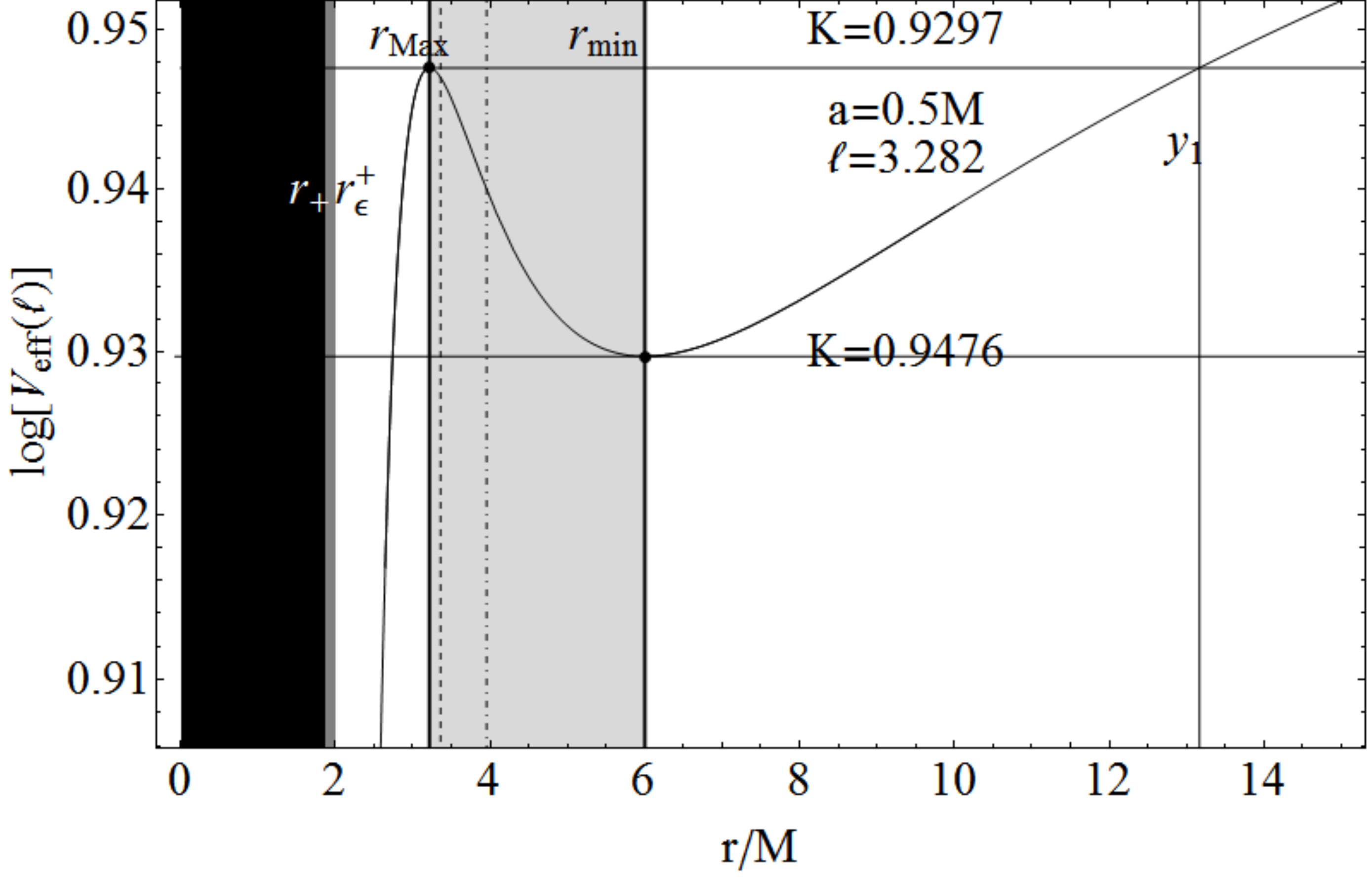}
\includegraphics[scale=.293]{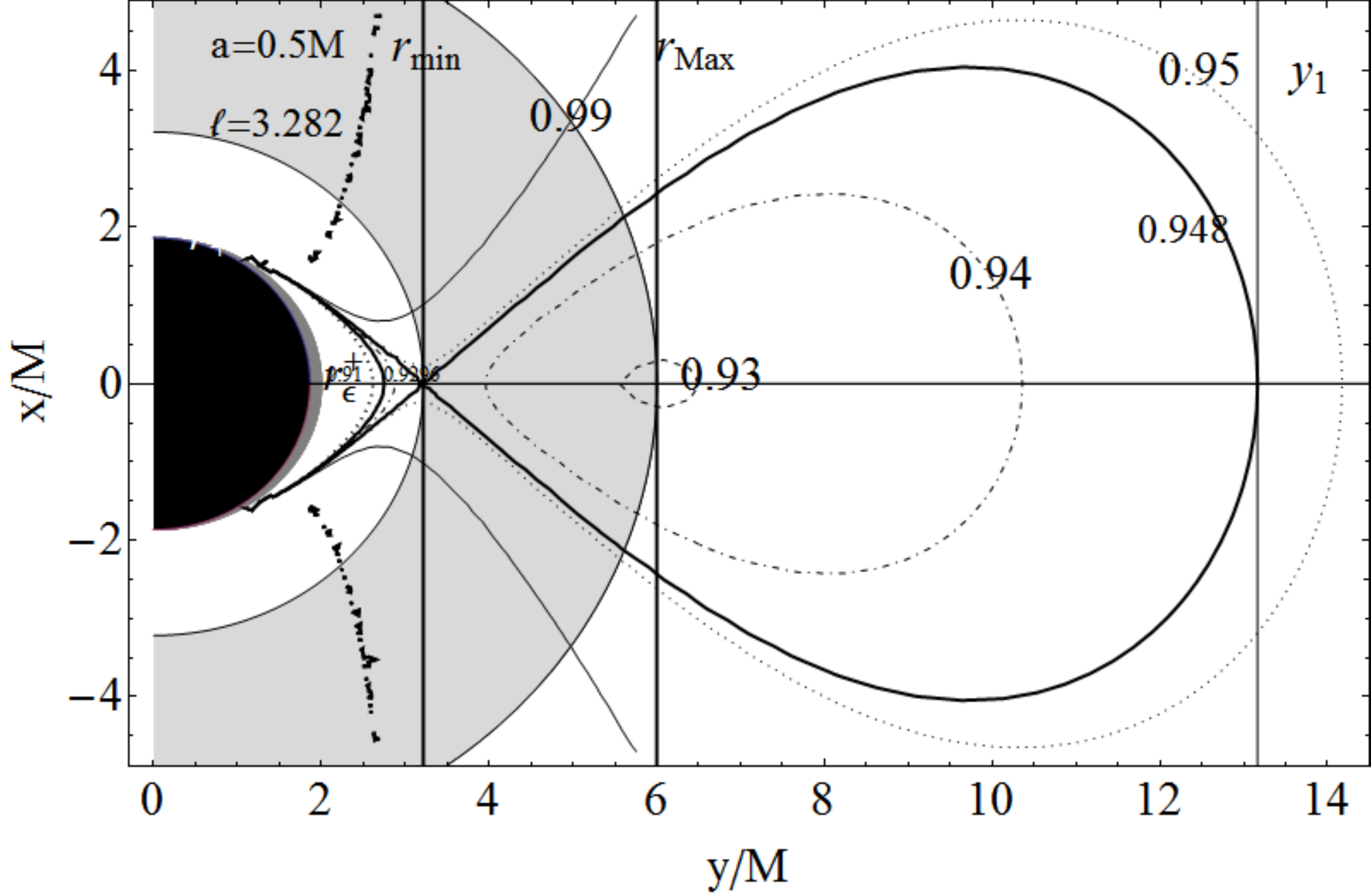}
\\
\includegraphics[scale=.293]{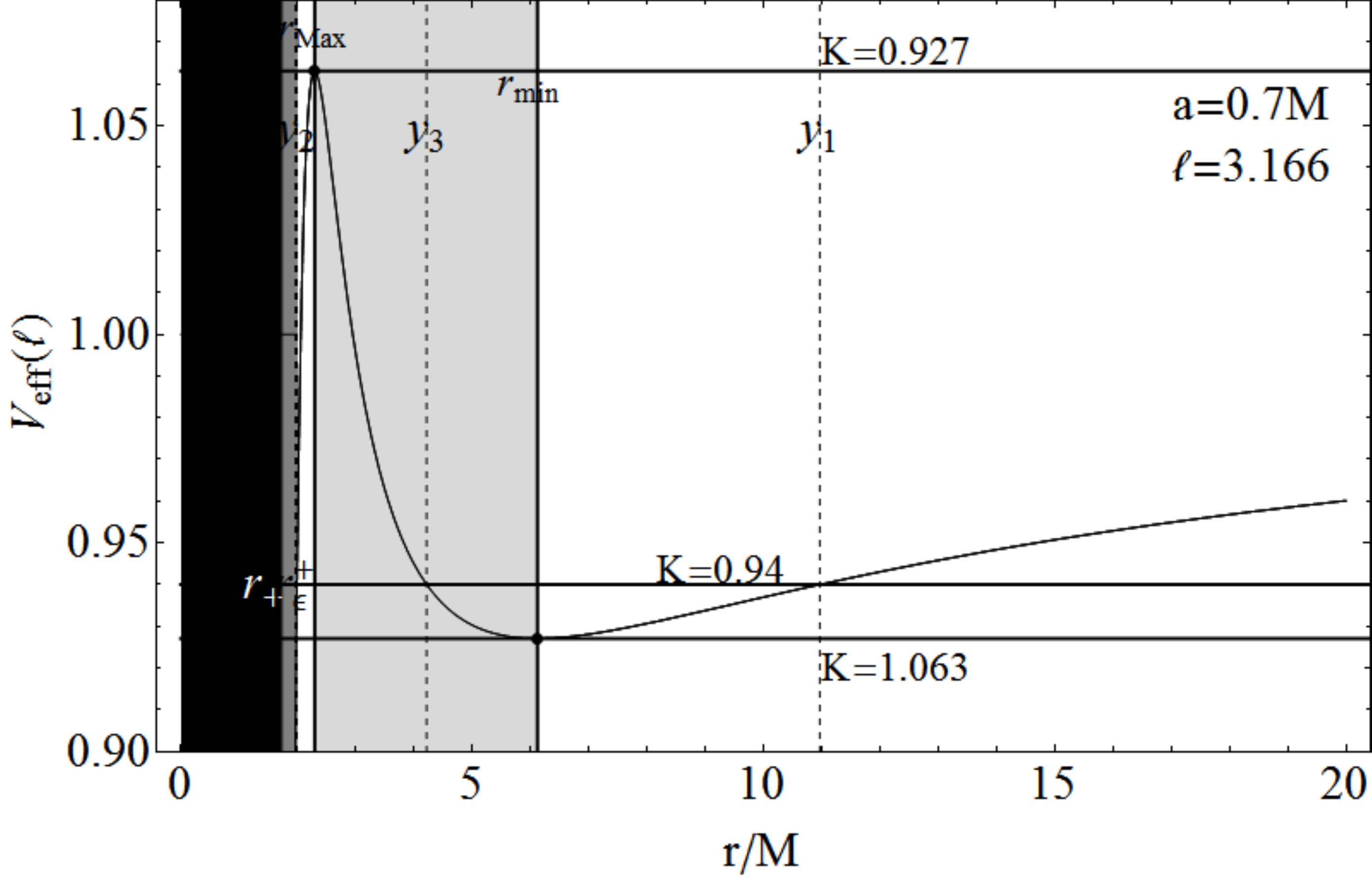}
\includegraphics[scale=.293]{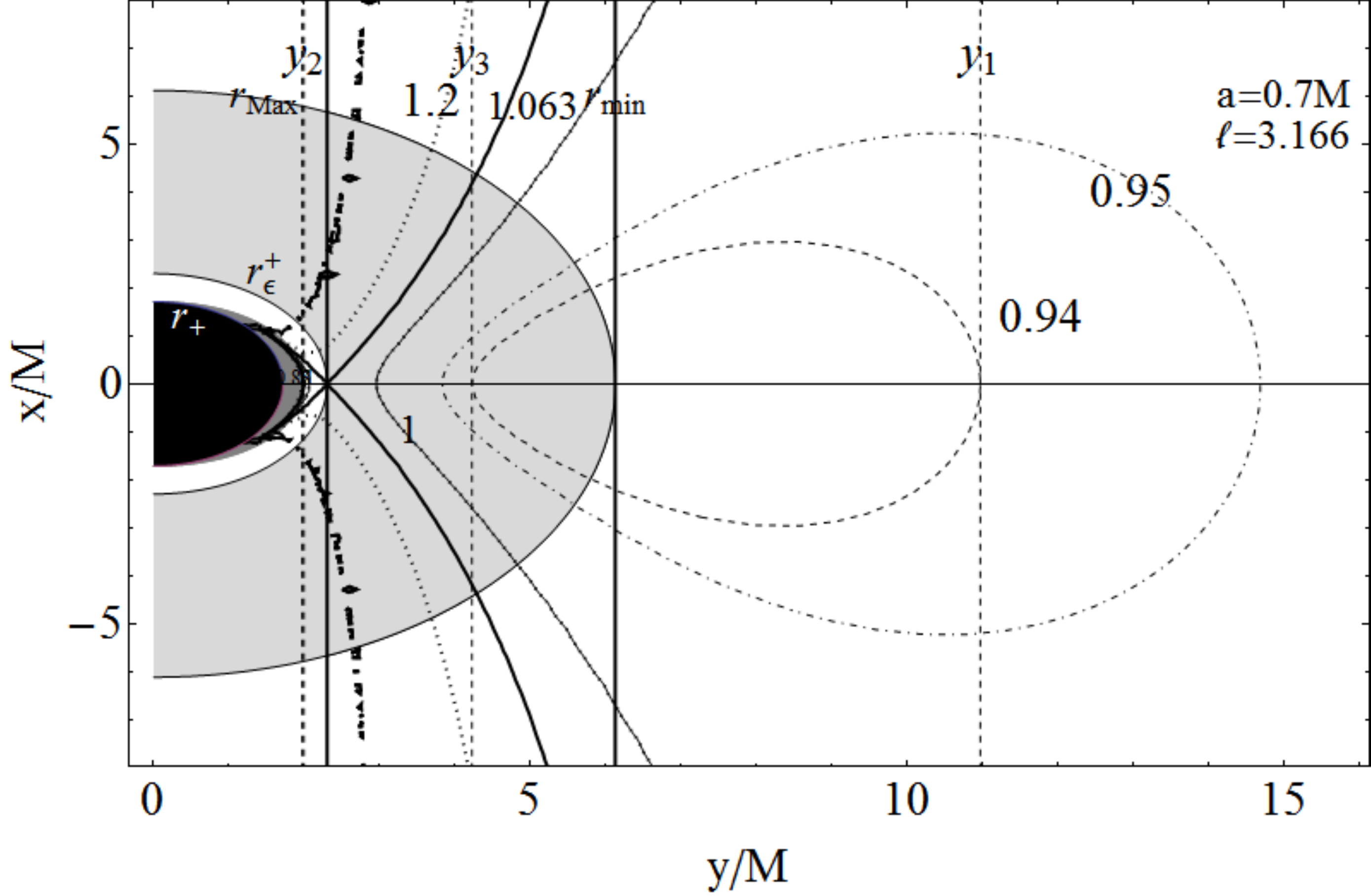}
\\
\includegraphics[scale=.293]{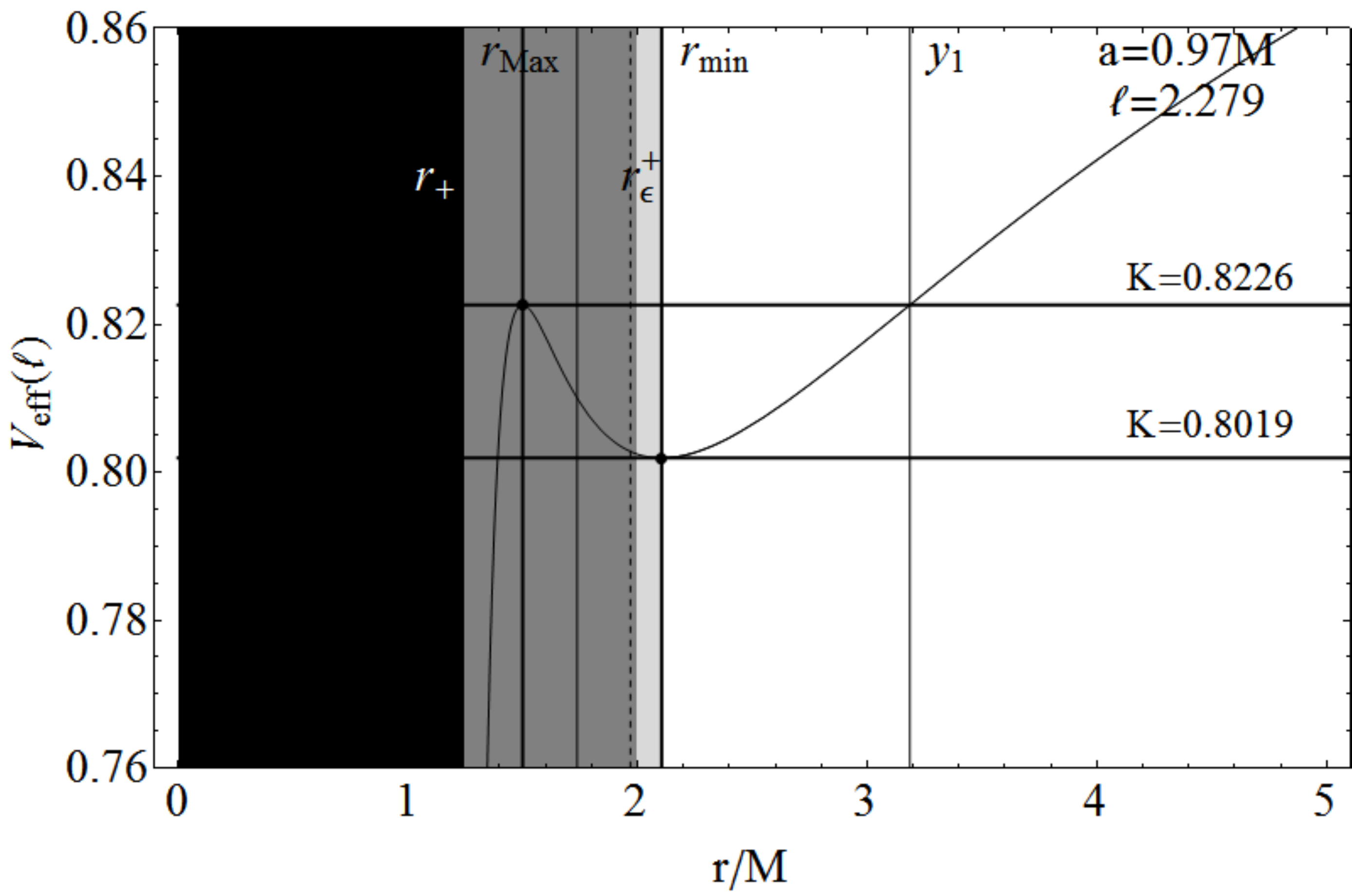}
\includegraphics[scale=.293]{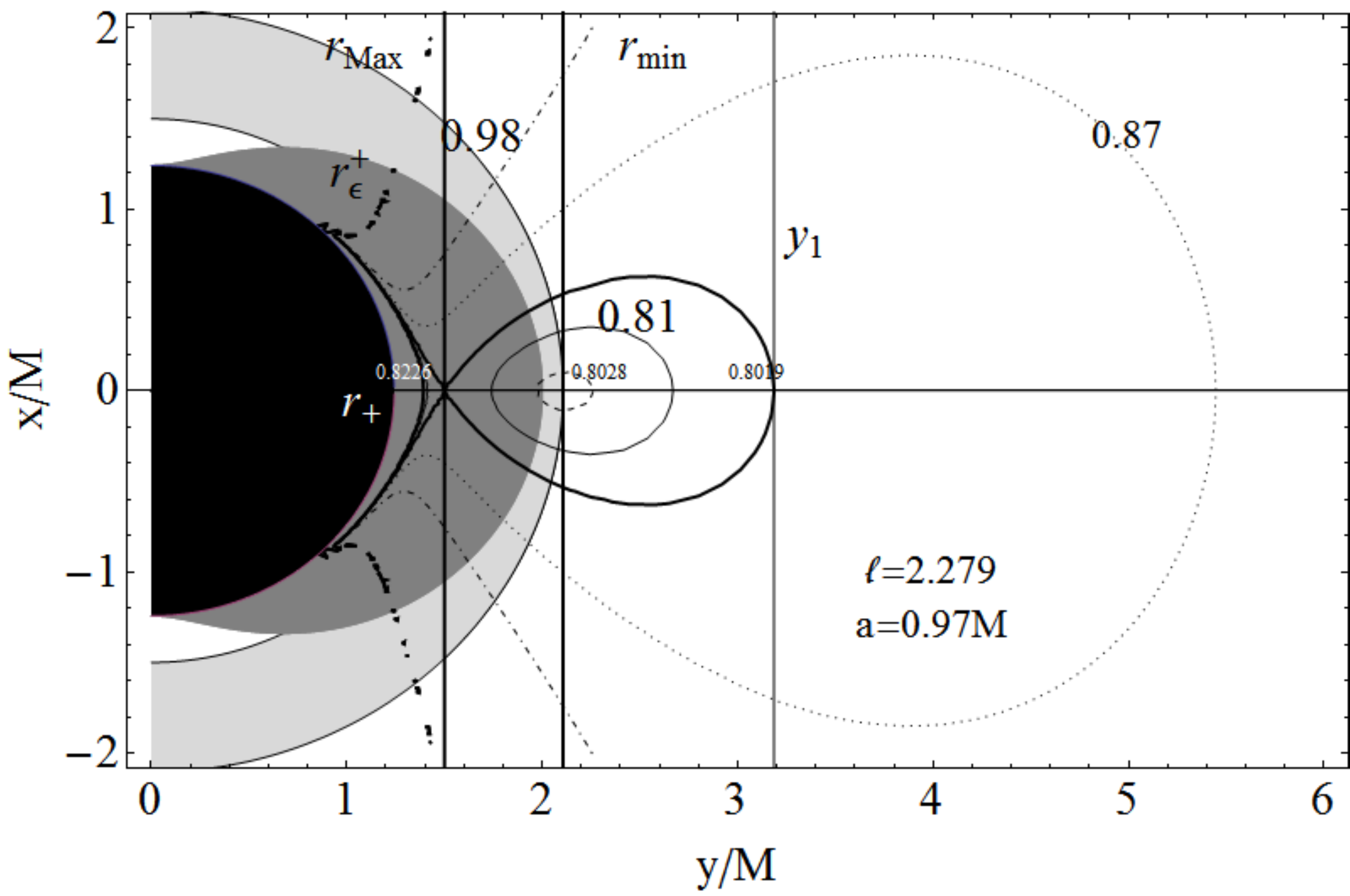}
\\
\includegraphics[scale=.293]{PSAX01u}
\includegraphics[scale=.293]{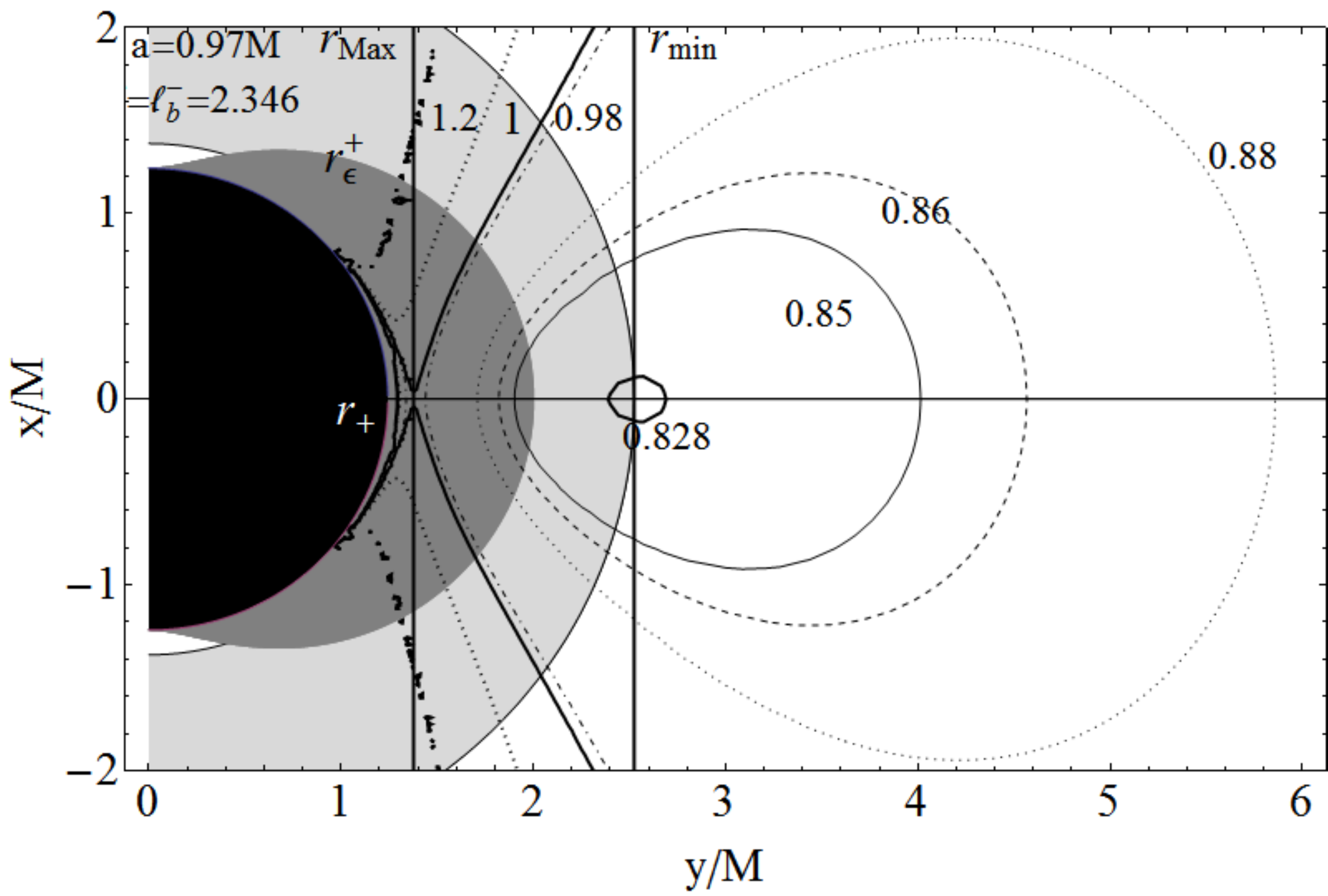}
\end{tabular}
\caption[font={footnotesize,it}]{\footnotesize{Left panels: logarithm plot of the effective potential $V_{eff}(\ell)$ as function of $r/M$. The angular momentum $\ell$ is in units of mass $M$. Right panels: the Boyer surface in the plane $(x,y)$  (sequences $\mathfrak{B}_{K}\equiv\mathfrak{B}_{\mathbf{p}}/\Sigma_{\ell}$  of P-D tori loops).
Black regions are $r<r_+$ where $r_+$ is the outer black hole outer  horizon, gray region is $]r_+,r_{\epsilon}^+]$ where $r_{\epsilon}^+$ is the static limit, light-gray region is $]r_{\epsilon}^+,r_{\min}]$, the critical points of $V_{eff}(\ell)$ are signed in the plots: $r_{min}$ is a minimum point of $V_{eff}(\ell)$  and $r_{Max}$ a maximum point.
The surfaces and the points at $K=V_{eff}(\ell)=$constant are black  lines, the point $y_1$ (left plot) corresponds to the outer boundary of the disk section, $y_3$ to the inner one, and $y_2$ to the outer boundary of the innermost surface, $r_{min}$ sets the disks center. First line for spacetime at $a=0.5M \in \mathbf{BHIII}$,
second line for $a=0.7M \in \mathbf{BHV}$ and third  and fourth  lines
$a=0.97M\in \mathbf{BHIX}$.
Each contours in the right plots is at $K=$constant.
 It is $\ell=\ell_f^+(r_{Max})$, where $r_{Max}$ is the maximum point of the effective potential $K_{Max}$. The first and third row of plots is at  $K_{Max}<1$.
Fourth line is
$K_{Max}=1$, second line
$K_{Max}>1$.
The radii $r: V_{eff}(a,\ell)=K$, within the  critical points are shown in the plot of the  effective potential.
 }}\label{Fig:cPSAX01g}
\end{figure}

Figures\il(\ref{Fig:PSAX01u},\ref{Fig:cPSAX01g}) show  the Boyer surfaces crossing   $r_{\epsilon}^+$   and eventually, the penetration of the disk surface in the region $r<r_{\epsilon}^+$.
We  consider the orbital region $\Delta_{crit}\equiv[r_{Max}, r_{min}]$, whose boundaries correspond to the  maximum and minimum points of the effective potential respectively (the minimum and maximum of the hydrostatic pressure).
The innermost boundary of the P-D configuration,  $y_3:\; y_2<y_3<y_1$, must be   $y_3\in\Delta_{crit}$
the outer one is $y_1>r_{min}$,
the center of the disk is located on $r_{min}$,
the cross of the surfaces $C_{x}$  is at $r_{Max}$ and there are closed crossed surfaces  if  $K_{Max}<1$. Then at fixed $\ell$ the closed disk,  as a point in $r_{min}$ (ring of particles at $p=0$), it grows  (with $K$) to fill the entire region up to $r_{Max}$ where the accretion occurs:
there are closed surfaces only if the constant  parameter $K\in[K_{min},K_{Max}]$ where $K_{min}\equiv V_{eff}(\ell, r_{min})$ and $K_{Max}\equiv V_{eff}(\ell, r_{Max})$,
 light-gray regions of Figs.\il(\ref{Fig:PSAX01u},\ref{Fig:cPSAX01g}),
such that there exist at last a solution at $K= V_{eff}(\ell, r)$  for the parameter couple  $\mathbf{p}\equiv(K,\ell)$ fixed.  A further  matter configuration  (with solution $y_2$) closest to the black hole is  at $r<r_{max}$.
However this scheme   foreseen also areas in the space of $\mathbf{p}$-parameter, in \textbf{BHIX} spacetimes, entirely   contained in the ergoregion i.e.
$r_{min}<r_{\epsilon}^+$ and $r_+<y_3<r_{min}<y_1<r_{\epsilon}^+$, this implies  the existence of at last a right neighborhood $I^+_{\ti{$r_{min}$}}\subset\Sigma_{\epsilon}^+$ of the minimum radius   (where $\Sigma_{\epsilon}^+\equiv]r_+,r_{\epsilon}^+[$) with $\partial_rV_{eff}>0$ and $\partial_r^2V_{eff}>0$, or  $y_1 \in I^+_{\ti{$r_{min}$}}:\; D(y_1,y_3)=0,\; D(y_1,r_{min})>0$ where $D(r_1,r_2)\equiv V_{eff}(r_1)-V_{eff}(r_2)$. It can be shown that  if $D(y_1,r_{min})>0$ exists for some fixed $\mathbf{p}$, it remains  small  requiring a fine-tuning (on $10^{-3}$ for  $K$). In the  ergoregion the  Killing vector $\xi_t^{a}$ becomes spacelike, but still the associated constant of motion (now it can be $E<0$) is well defined (see also discussion in \cite{Pugliese:2014ela}).
\begin{figure}
\centering
\begin{tabular}{cc}
\includegraphics[scale=.3]{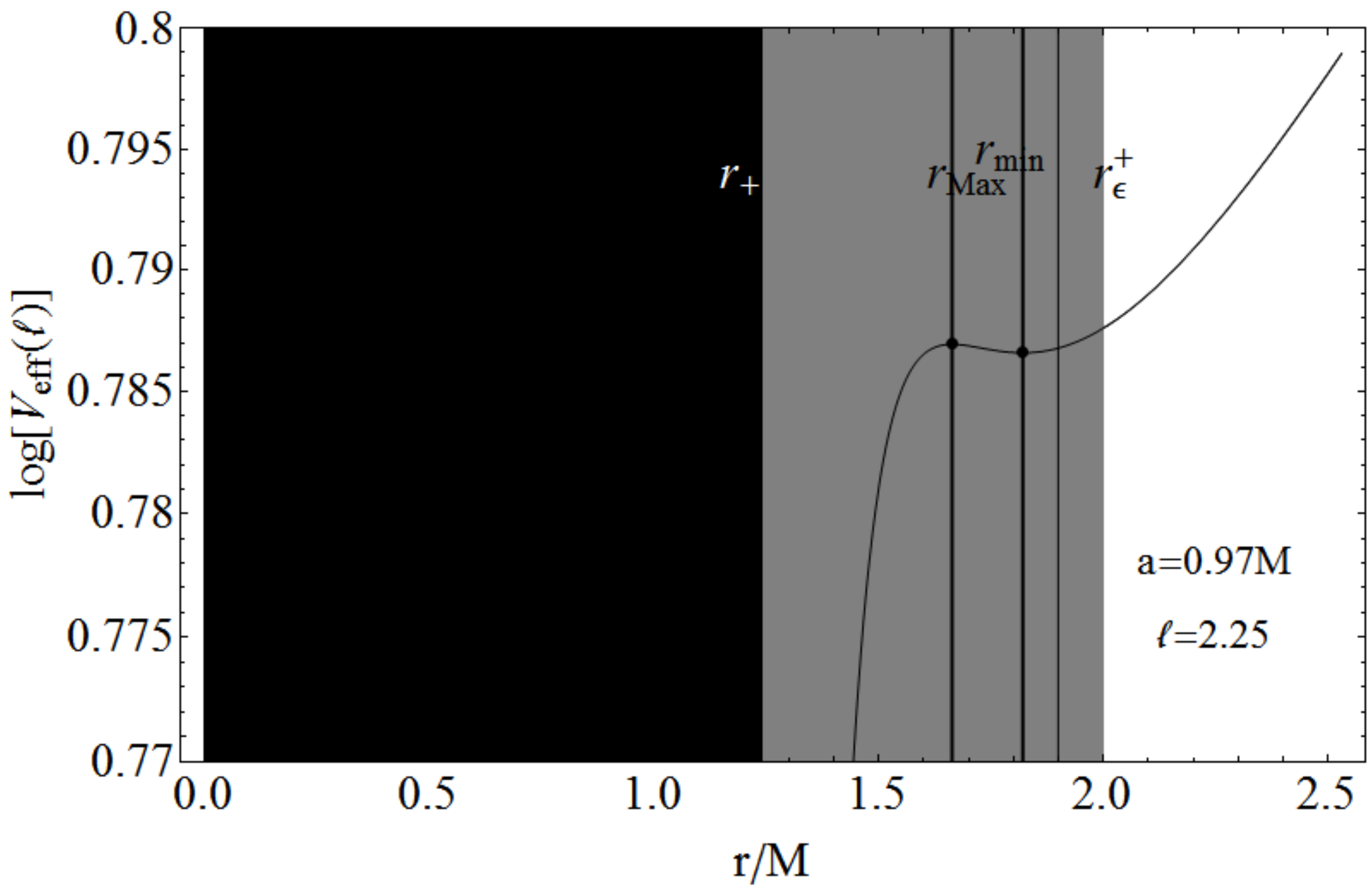}
\includegraphics[scale=.3]{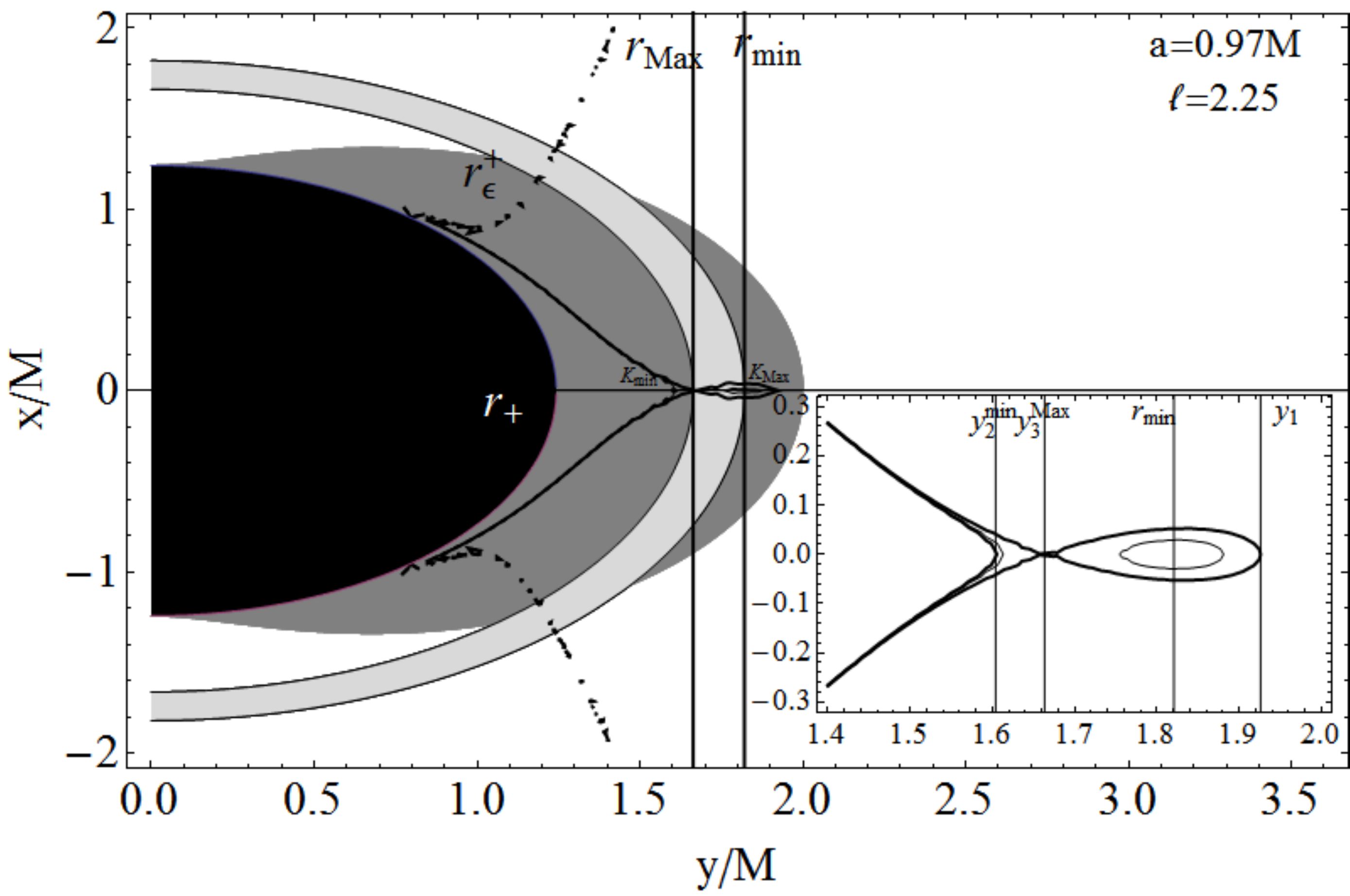}
\end{tabular}
\caption[font={footnotesize,it}]{\footnotesize{Left panel: effective potential $V_{eff}(\ell)$
versus the orbital radius $r/M$ at $a=0.97M$ (\textbf{BHIX} spacetime) and orbital fluid angular momentum $\ell=2.25$ in units of mass $M$.
Right  panel: Boyer surfaces at fixed $K$  (sequences $\mathfrak{B}_{K}\equiv\mathfrak{B}_{\mathbf{p}}/\Sigma_{\ell}$  of P-D tori loops).
Black region is $r<r_{+}$ ($r_+$ being the outer horizon),  the ergoregion  (dark gray) is  $r\in]r_+,r_{\epsilon}^+[$ where $r_{\epsilon}^+$ is the static limit.
The critical points of the effective potential are both in the ergoregion, the maximum, located in
$r_{Max}=1.66346M$ and minimum $r_{min}=1.82079M$ are signed by thick black lines.
Closed configuration are centered in $r_{min}$,  the crossing points are at $r_{Max}$.
 In the right panel are   configurations entirely contained in the ergoregion.  Light gray region is the $[r_{Max},r_{min}]$ the contours of the  Boyer surfaces associated to the $K_{Max}\equiv V_{eff}(r_{Max})=0.786937$ and $K_{min}\equiv V_{eff}(r_{min})=0.786588$ (crossed closed $C_x$) are in black thick lines. Inside panel, the surface  at $K=0.7865\in]K_{mnin},K_{Max}[$ is  closed and contained  in the ergoregion. The surfaces boundaries $y_1=1.92614M$, $y^{Max}_3=1.66346M$ (configuration of maximum $K$) and $y^{min}_2=1.60392M$ (configuration of minimum $K$) are signed.
}}\label{Fig:PreattyM}
\end{figure}
Fig.\il(\ref{Fig:PreattyM}) shows the  limiting case of a closed Boyer surface entirely contained in the ergoregion: both the critical points of the effective potential are included in $\Sigma_{\epsilon}^+$.
A cross can occur for closed surfaces in the \textbf{BHVI-IX} spacetimes
, that is where $r_{\gamma}^-<r_{\epsilon}^+$,
 since the inner boundary satisfies  $y_2<r_{Max}<r_{\gamma}^-$ there could be  funnels of matter crossing the static limit  from the accretion point.
The ergoregion is filled with  ``orbits'' at $K=1$ only in the \textbf{BHVII-III-IX} spacetimes:
$
a\in]a_{\diamondsuit},M[$, $\ell=\ell_b^-$, $a_{\diamondsuit}\equiv2 \left(\sqrt{2}-1\right)M\approx0.828427M$,
and at $a_{\diamondsuit}$ it is  $r_b^-=r_{\epsilon}^+$ where  $\ell\approx2.82843M$.
And at $r\in]r_+,r_{\epsilon}^+[$ it is $\ell/M\in] 2,2 \sqrt{2}[$  and   $\bar{\ell}\in]2,2+ \sqrt{2}[$. 
The critical points at $K\in]0,1[$ are both minima and maxima.
For the minimum points inside the ergoregion the situation is as follows: for the \textbf{BHIX} attractors ( $ a/M\in] a_{\natural},M]$) minima are  in $r\in]r_{lsco}^-,r_{\epsilon}^+[$, with  angular momentum $\ell\in]\ell_{lsco}^-,\ell_f^-(r_{\epsilon}^+)[$.
Configurations  with $\bar{\ell}>1$ are inside the ergoregion
in  \textbf{BHIX}, and  that is  {all} the configuration inside the ergoregion are $\ell>a$.
The maximum extension of
normalized angular momentum in $r
\in]r_{lsco}^-, r_{\epsilon}^+[$  is
$\bar{\ell}\in]2, 5/2[$
but  $ \ell/M\in]2,(5
\sqrt{2})/3[$.
 Inside the ergoregion {saddle points} are located in the \textbf{BHIX} spacetimes  with $r_{lsco}^-$ with $\ell_f^-$, that is $\ell_{lsco}^-$ and $r_f^{\pm}(\ell_{lsco}^-)$. A {seddle point} exist on the static limit at $a=a_{\natural}$ where $\ell={5 \sqrt{2}}/{3}M$. In terms of the rationalized angular momentum it is $\bar{\ell}\in ] 2, 5/2[$ and  $\ell/M\in] 2,(5 \sqrt{2}/3[$.
{Maximum points}  in  the  ergoregion are located in
$a\in]a_{\diamondsuit}, a_{\natural}]$ with   $\ell=\ell_f^-$, for \textbf{BHIX} sources maxima exist in the region $\in]r_{lsco}^-, r_{\epsilon}^+[$.
A maximum is located on the static limit for \textbf{BHVII-VIII} spacetimes where
$a\in]a_{\diamondsuit}, a_{\natural}[$ and $\ell=\ell_f^-(r_{\epsilon}^+)$.
Or
maximum points, located on the orbits $r_f^{\pm}(\ell,a)$ are in the \textbf{BHVII-VIII} spacetimes with
$\ell\in]\ell_f^-(r_{\epsilon}^+),\ell_b^-[$
and for the  \textbf{BHIX}-class  with $\ell\in]\ell_{lsco}^-, \ell_b^-$.
Then it is
$\bar{\ell}\in]2, 2+\sqrt{2}[$  and  $\ell/M\in] {5 \sqrt{2}}/{3},2 \sqrt{2}[$.
Configurations $K>1$ in the ergoregion are for
$\bar{\ell}\in]2, 5[$ and $\ell/M\in] 2, {5}/{\sqrt {2}}[$.
On the static limit we should consider the spacetime with  $a_{\Box}\equiv1/\sqrt{2}M$ where $r_{\gamma}^-=r_{\epsilon}^+$  with $\ell=\ell_f^-(r_{\epsilon}^+)$. For  \textbf{BHVI} spacetimes, characterized by
$
a\in]a_{\Box},a_{\diamond}[$,  maxima are with
$\ell\in]\ell_f^-(r_{\epsilon}^+),\ell_{\gamma}^-[$ in \textbf{BHVI},
then
 in \textbf{BHVII-III}  with $\ell\in]\ell_b^-,\ell_{\gamma}^-[$, while at  $a=a_{\natural}$  with $\ell\in]\ell_b^-,\ell_f^-(r_{\epsilon}^+)[$, finally  \textbf{BHIX} with $\ell\in]\ell_b^-,\ell_{\gamma}^-[$.
Further consideration on dynamics in this region can be found
in Sec.\il(\ref{subsubsec:l0}) where the matter configuration at  $\ell=0$ ($L=0$) is considered.
\subsection{A sequence of torus shapes in evolution}\label{Subsec:sequence}
In this section we explore the sequences $\mathfrak{B}_{	 \bar{\ell}}\equiv\mathfrak{B}_{\mathbf{p}}/\Sigma_{K}$ fixing   $K\in]0,1[$, and considering  different values of  the rationalized angular momentum  $\bar{\ell}\equiv \ell/a$, thus we distinguish the corotating fluids  with  $\bar{\ell}>1$  in Sec.\il(\ref{Subsun:barellM1}) and in Sec.\il(\ref{Subsubsec:lbarm1}) the configurations
$\bar{\ell}<1$, which include negative values for the fluid angular momentum or counterrotating  configurations.   The extreme case of ``steady'' fluid  respect to the central object, in other words $\bar{\ell}=1$, or the counterrotating  case  $\bar{\ell}=-1$ will be considered in   Sec.\il(\ref{Sec:barl}).
Then it is  convenient to  introduce  the angular momenta $\ell_K^{i} (a;K)\;i\in\{1,2,3\}$ such that $\ell_K^i: V_{eff}=K,\; \partial_r V_{eff}=0$. One can solve the first equation to get $\ell_K^{\pm}(a; r,K)$,  the second equation gives the solutions $r_i$ as in Eq.\il(\ref{Eq:5raidiff}) that is used in   $\ell_K^{\pm}(a; r_i, K)$ as in Eq.\il(\ref{Eq:poEq}). 
It is  $\ell_K^{\pm}(a; r_i, K_{lsco}^{\pm})=\ell_{lsco}^{\pm}$   and  $\ell_K^{\pm}(a; r_i, K_{b}^{\pm})=\ell_{b}^{\pm}$.
\subsubsection{Fluid configurations at $\bar{\ell}>1$}\label{Subsun:barellM1}
We investigate   the  corotating configurations at  $\bar{\ell}>1$,   we consider  the sequences $\mathfrak{B}^{>}_{\bar{\ell}}\equiv\left.\mathfrak{B}_{\mathbf{p}}\right|_{\bar{\ell}>1}/\Sigma_{K}$ the critical points for the hydrostatic pressure in this case are analysed in  Sec.\il(\ref{Sec:barl}):
three main regions,\textbf{Region I}--\textbf{Region III}, for the $K$ parameter can be recognized and different phases for the angular momentum parameter namely:
\begin{description}
\item[
\textbf{Region I}:]
$K\in]0,K_{lsco}^-]$; \textbf{1.} $(\ell\in]a,\ell_{K}^2[,y_{123})$, \textbf{2.}
$(\ell_{K}^2, y_{13})$, \textbf{3.}
$(\ell>\ell_{K}^3,y_1)$.
\begin{figure}
\centering
\begin{tabular}{cc}
\includegraphics[scale=.3]{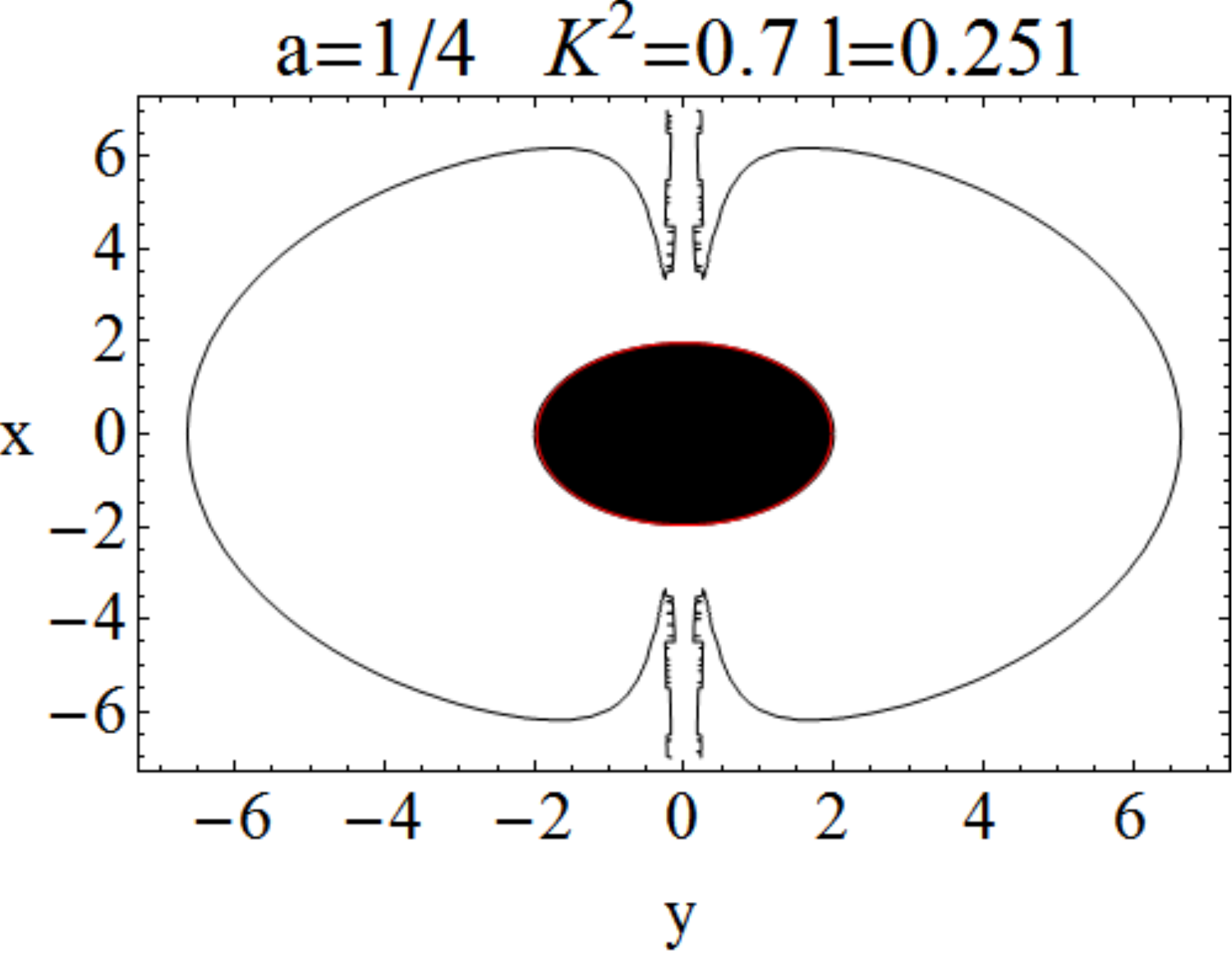}
\includegraphics[scale=.3]{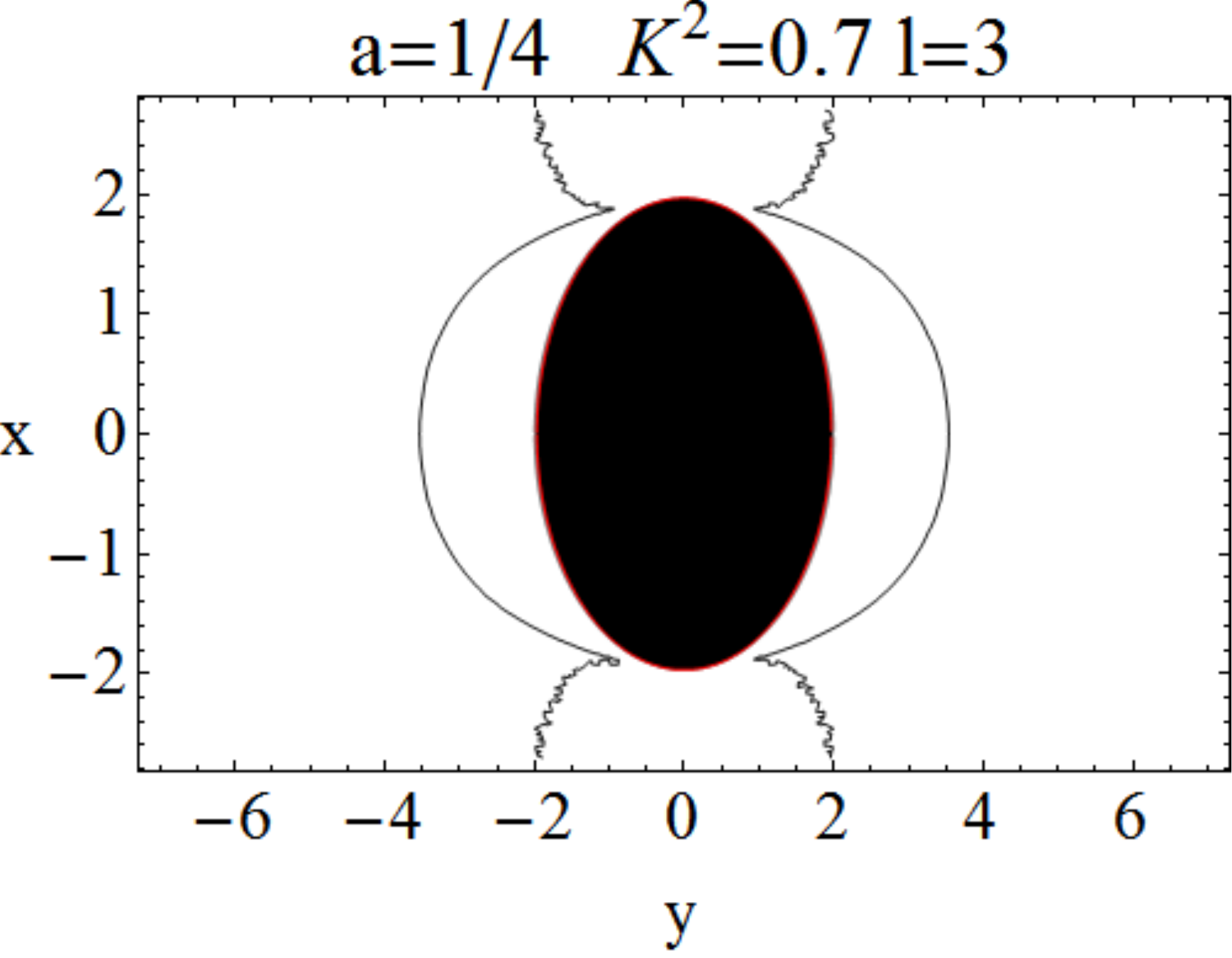}
\includegraphics[scale=.3]{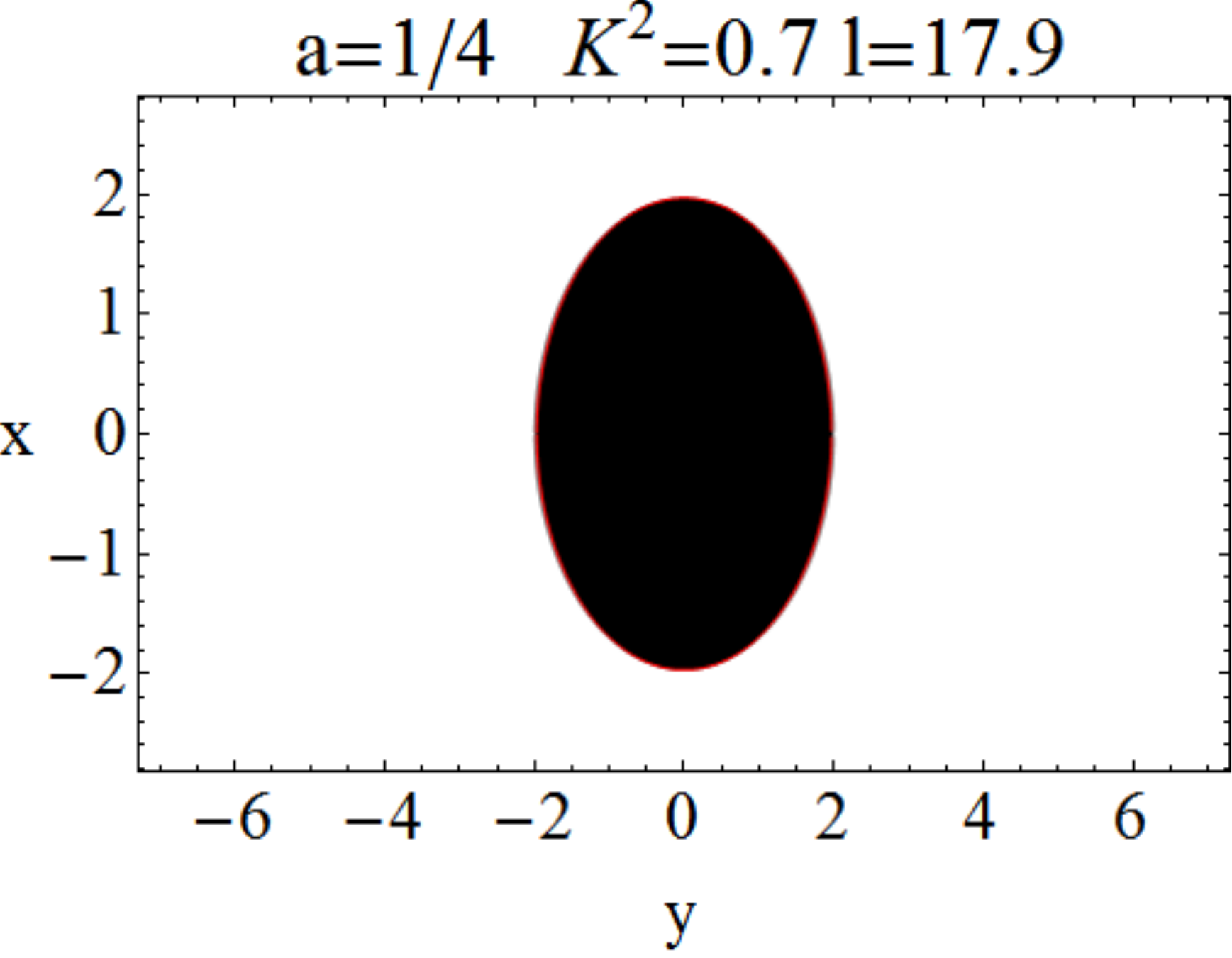}
\end{tabular}
\caption[font={footnotesize,it}]{\footnotesize{Configurations at $K<1$, $\ell>a$. Region $K\in]0,K_{lsco}^-]$.  The spacetime spin is $a=0.25M\in\mathbf{BHI}$, the outer horizon is at $r_+=1.96825M$ the static limit $r_{\epsilon}^+=2M$, where $K^2=0.7$  (sequences $\mathfrak{B}_{\ell}\equiv\mathfrak{B}_{\mathbf{p}}/\Sigma_{K}$), it is $\ell_{{K}}^i=\{0.233022, 0.270991\}$, and  $\ell_f(r_+)=15.746$,
 $
\ell_{\epsilon}^+=13.4216$,
$\ell_f(r_d)=8.22857$,
$\ell_\gamma^-=4.67535$,
$\ell_b^-=3.73205$,
$\ell_{lsco}^-=3.43993$,
 in units of mass $M$.
}}\label{Fig:TSStairCase}
\end{figure}
This region does not include closed  (Boyer) surfaces
or ${C}$-configurations, that is the effective potential has not minimum points. As shown in  Figs.\il(\ref{Fig:TSStairCase}), the orbiting matter rotates around the attractor
with a very clear evolution:
increasing  the angular momentum    the configuration approaches
 the source,   the torus becoming  thinner (see also discussion in \cite{PuMonBe12} for the case $a=0$), the solutions around
the rotation axis  spread on the  equatorial plane, we note also  that at the axis of rotation there is a singularity due to the adopted frame. In some cases the surfaces cross  the equatorial plane   very close to the region $[r_{\epsilon}^+,r_+]$.
Then we introduce  a new morphological  type, fat torii,  denoted as
  ${B}$-configurations, often  with opened funnels, see also  for a general discussion of the different torii \cite{Abramowicz:2011xu,
Raine,
BAF2006,
Hawley1990,
Abramowicz:2004vi}. These surfaces could be associated to the innermost configurations surrounding  the  black hole, always present with the closed $C$ configurations and  correspondent to  the solution $y_2$ leading to the accretions  at the instability point where $y_2=y_3$, matching the outer $C$ disk in a  $C_x$ morphology.
\item[\underline{\textbf{Region II}}:] $K\in]K_{lsco}^-,K_{lsco}^+[$:
\textbf{1.}
$(\ell\in]a,\ell_{K}^2[,y_{123})$, \textbf{2.} $(\ell=\ell_{K}^2, y_{13})$, \textbf{3.} $(\ell\in]\ell_{K}^2,\ell_{K}^3[,y_1)$, \textbf{4.}
$(\ell=\ell_{K}^3,y_{13})$, \textbf{5.}   $(\ell\in]\ell_{K}^3,\ell_{K}^1[,y_{23})$, \textbf{6.}
$(\ell=\ell_{K}^1,  y_2)$,  \textbf{7.} $ \ell>\ell_{K}^1$
see Fig.\il(\ref{Fig:TSMaxime}). In this case there are  closed surfaces and  ${B}$-configurations  associated with lower angular momentum.
With increasing angular momentum a pattern similar  to \textbf{Region I}  appears,   starting with the  ${C}_{x}$ configurations, it decreases in thickness, separates in the two Boyer lobes and then it disappears, leaving an open, not crossed configuration that is one could consider the sequence
    $\mathfrak{B}^{>}_{\bar{\ell}}=[B, {C}_{x}, {C}, {O}]$. In this region we considered both the limiting values $K_{lsco}^{\pm}$ even if we analyze the corotating matter only. The evolutive order in  $\mathfrak{B}^{>}_{\bar{\ell}}$ should  follow the decreasing  angular momentum $\bar{\ell}$ to figure  properly an accretion process  onto the black  hole, from $C$-topology to the $B$ one, starting from  a former   opened $O$ one.
\begin{figure}
\centering
\begin{tabular}{cc}
\includegraphics[scale=.23]{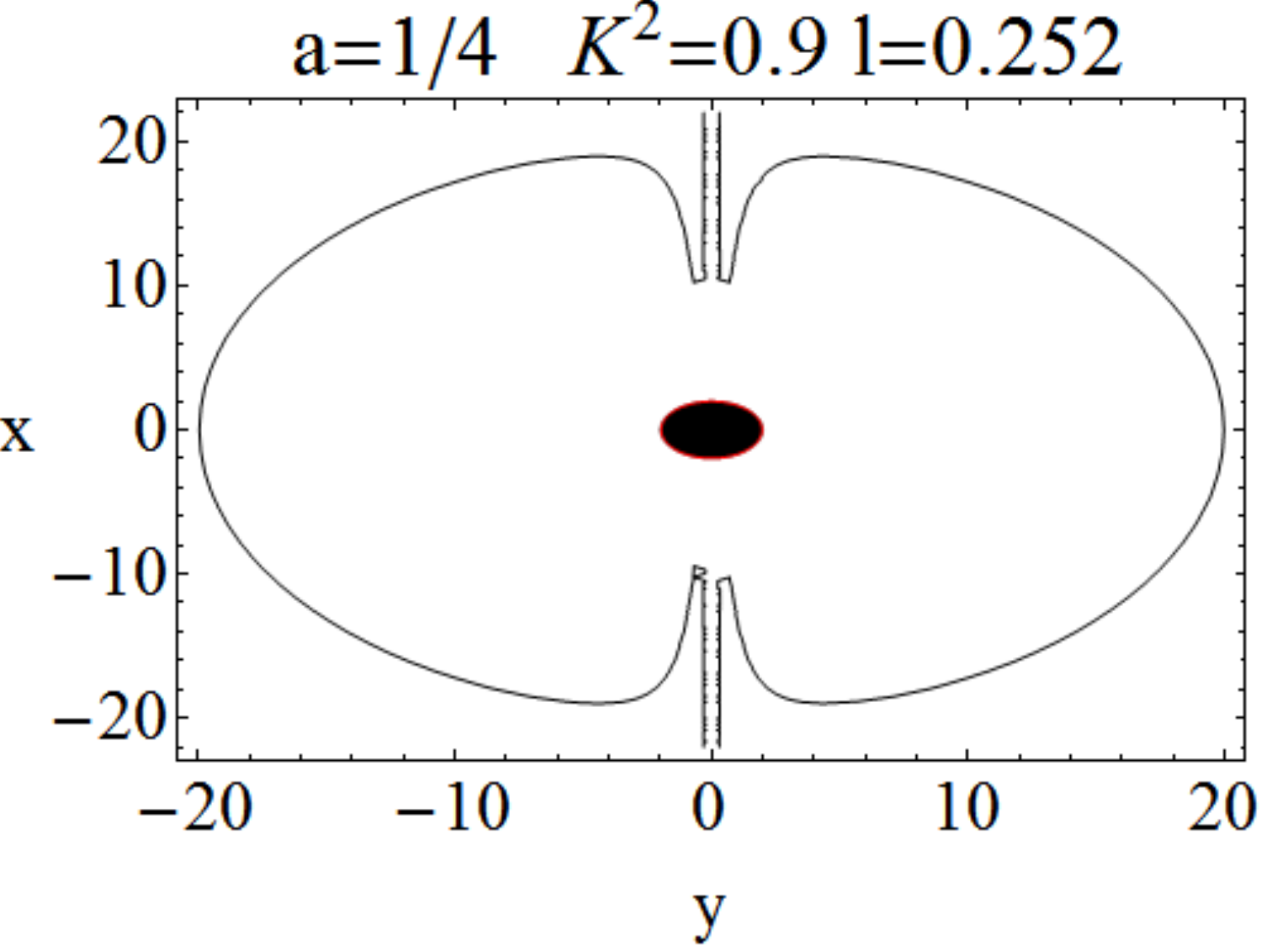}
\includegraphics[scale=.123]{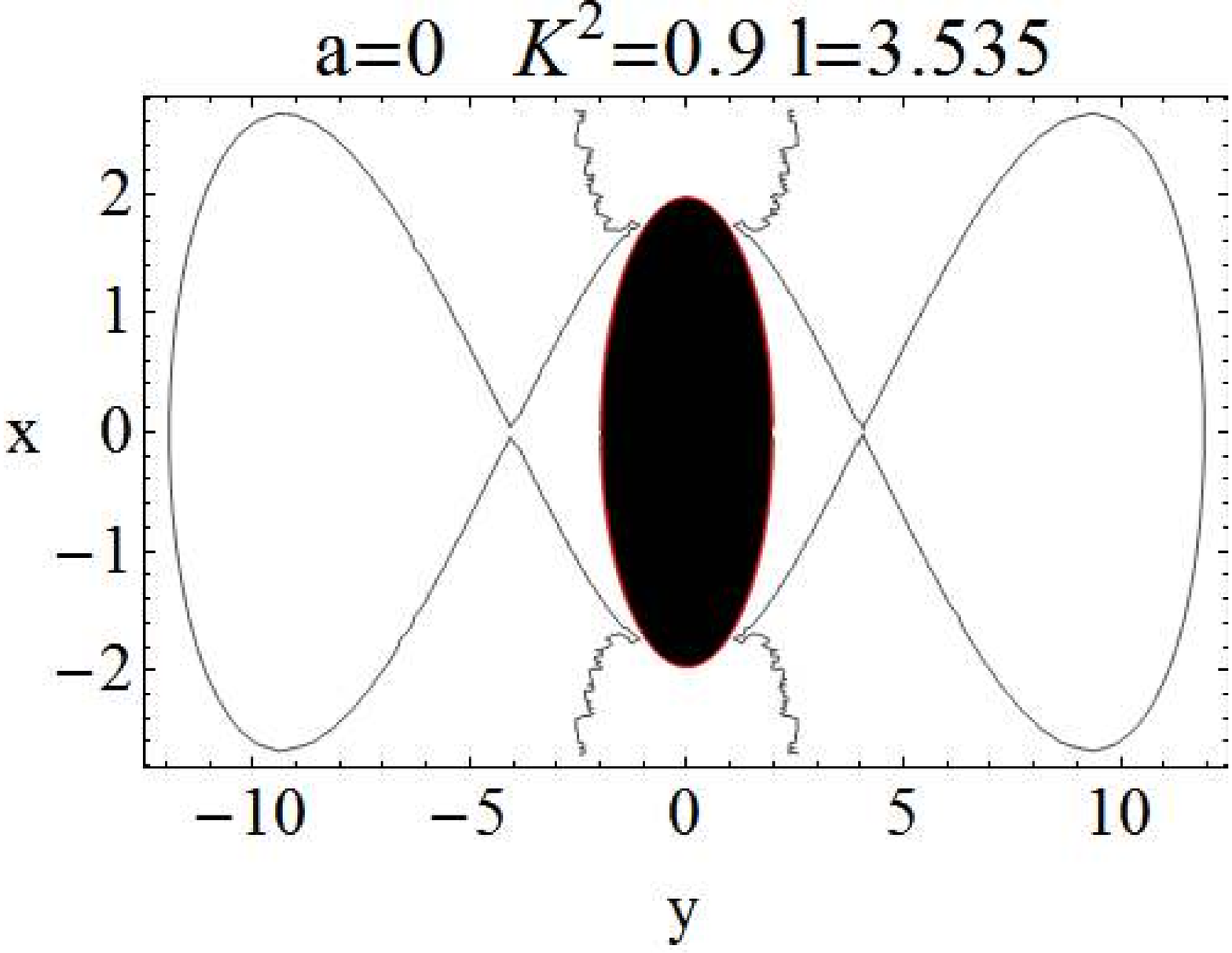}
\includegraphics[scale=.123]{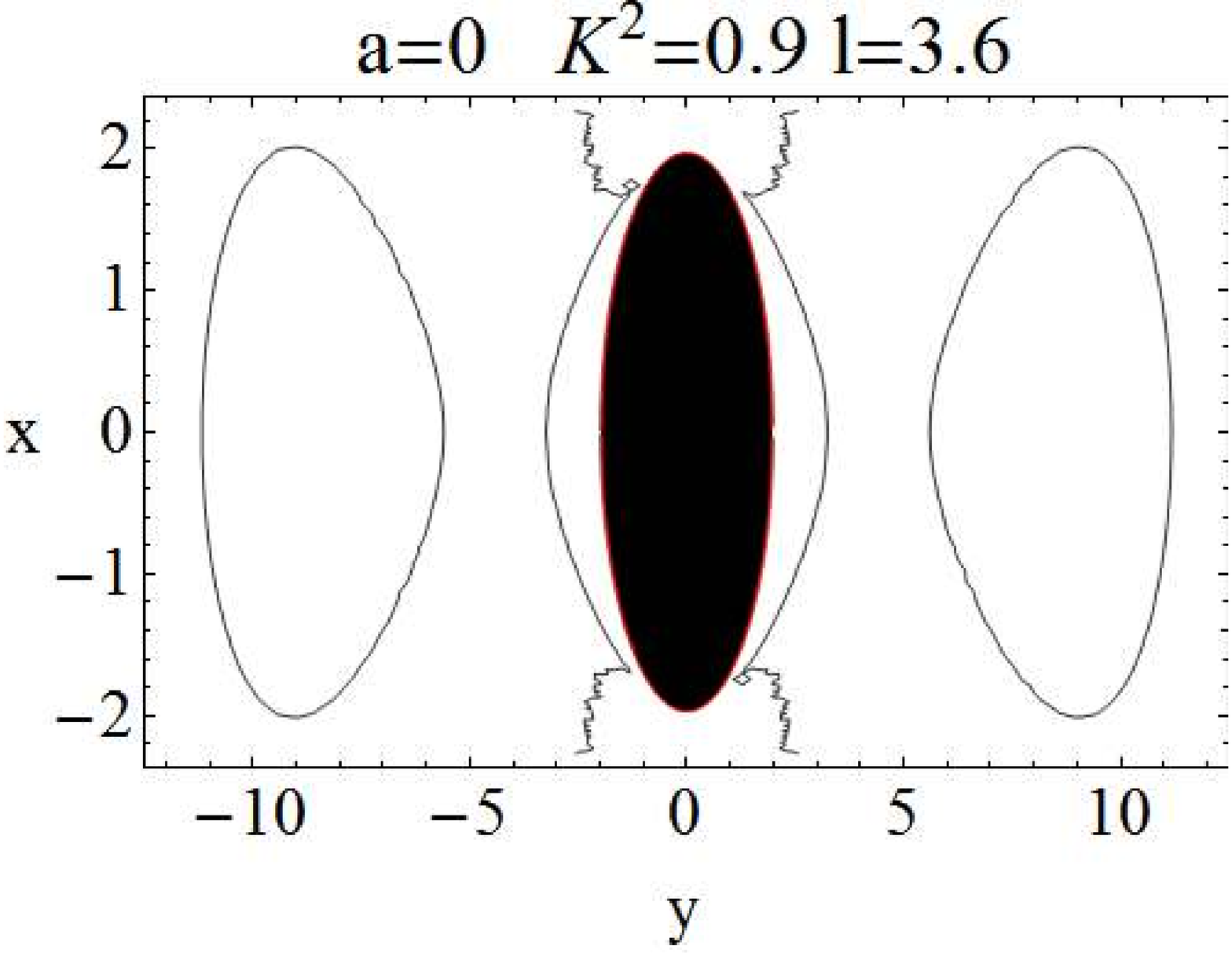}
\includegraphics[scale=.123]{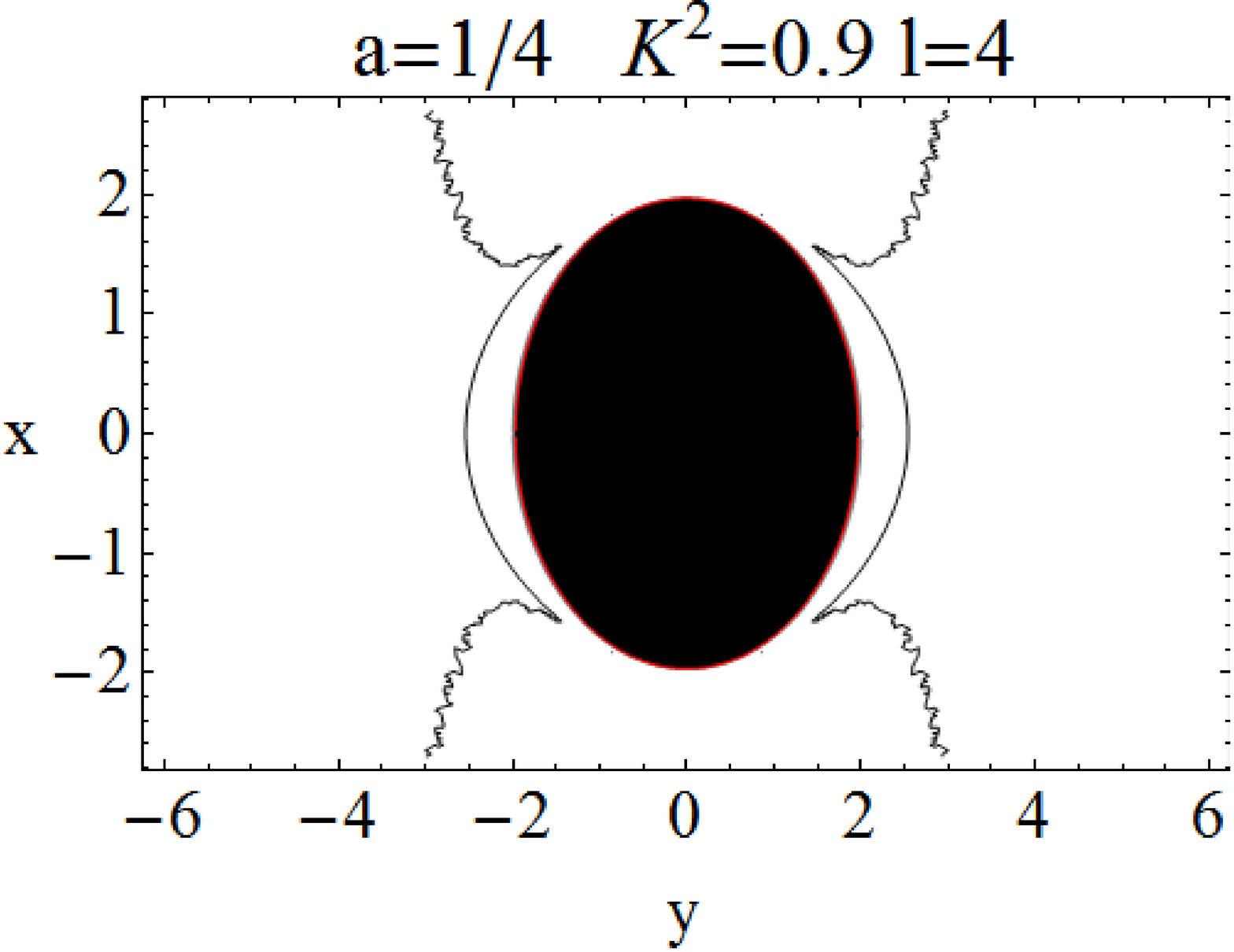}
\end{tabular}
\caption[font={footnotesize,it}]{\footnotesize{Configurations at $K<1$ and  $\ell>a$. Region $K\in]K_{lsco}^-,K_{lsco}^+[$.  The spacetime spin is $a=0.25M\in\mathbf{BHI}$, the outer horizon is at $r_+=1.96825M$ the static limit $r_{\epsilon}^+=2M$,  (sequences $\mathfrak{B}_{\ell}\equiv\mathfrak{B}_{\mathbf{p}}/\Sigma_{K}$)  where $K^2=0.9$, it is $\ell_{K}^i=\{ 0.235225,0.268791,3.53586, 3.69391\}$
and $\ell_f(r_+)=15.746$, $
\ell_{\epsilon}^+=13.4216$, $\ell_f(r_d)=8.22857$,
$\ell_\gamma^-=4.67535$,
$\ell_b^-=3.73205$,
$\ell_{lsco}^-=3.43993$. The fluid angular momentum is in units of mass $M$.}}\label{Fig:TSMaxime}
\end{figure}
\item[\underline{\textbf{Region III}}]:
$K\in[K_{lsco}^+,1[$:
\textbf{1.} $(\ell\in]a,\ell_{K}^1[, y_{23})$, \textbf{2.}
$(\ell_{K}^1, y_3)$, \textbf{3.}
$(\ell\in]\ell_K^4,\ell_K^5[,  {y_1})$, \textbf{4.}
$(\ell_K^5,y_{13})$, \textbf{5.}
$(\ell\in]\ell_K^5,\ell_K^6[, {y_{123}})$, \textbf{6.}
$(\ell_K^5,{y_{12}})$, \textbf{7.}
$(\ell>\ell_K^6, {y_1})$.
This is an articulated region.   The surfaces approach the attractor  increasing the angular momentum.  Figs.\il(\ref{Fig:TSPrediction})
 show a sequence of shapes   analogue to \textbf{Region II}: with increasing orbital angular momentum
the basic sequence of surfaces is:
$	\mathfrak{B}^{>}_{\bar{\ell}}=[{B},{C}_{x}, {C}, {O}, {O}_{x}, {O}]$. The last  open configuration of $	\mathfrak{B}^{>}_{\bar{\ell}}$ disappears into the black hole increasing $\ell>\ell_k^6$, . In this case the evolutive sequence $	 \mathfrak{B}^{>}_{\bar{\ell}}$  is quite articulated  and to figure  a disk  evolution towards the accretion  we should consider decreasing values of the rationalized angular momentum, neglecting then the starting sequence of opened configurations.
\begin{figure}
\centering
\begin{tabular}{cc}
\includegraphics[scale=.123]{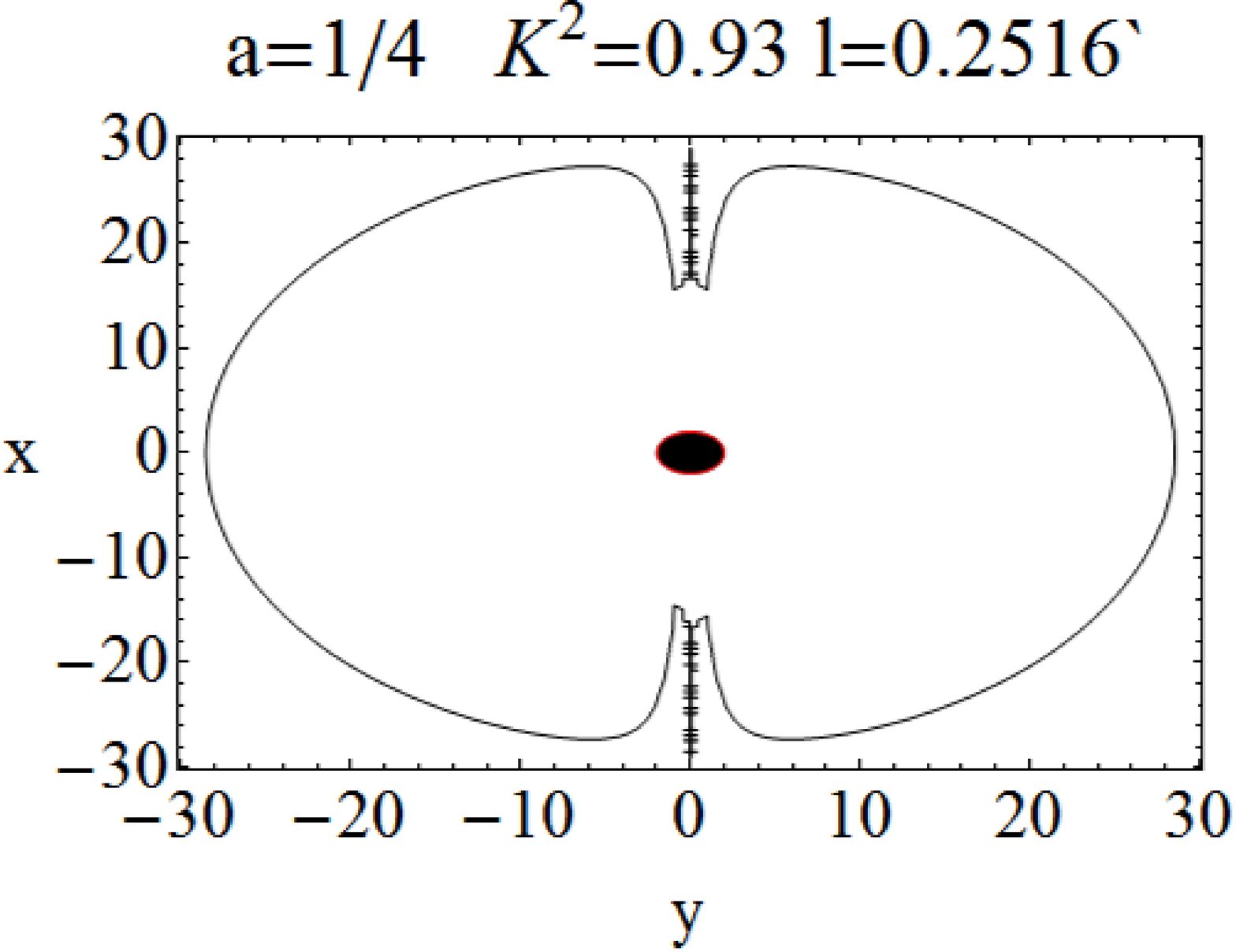}
\includegraphics[scale=.123]{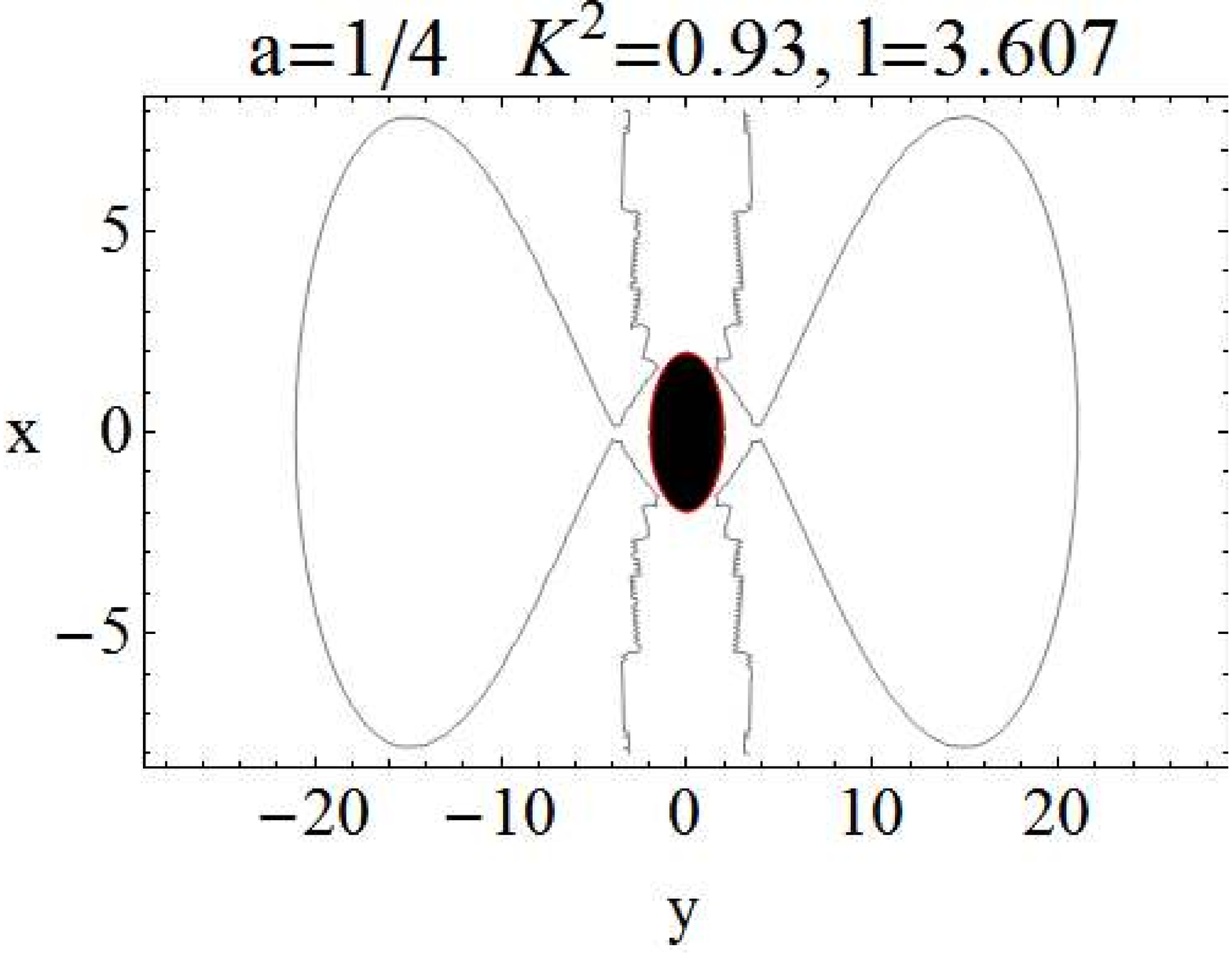}
\includegraphics[scale=.123]{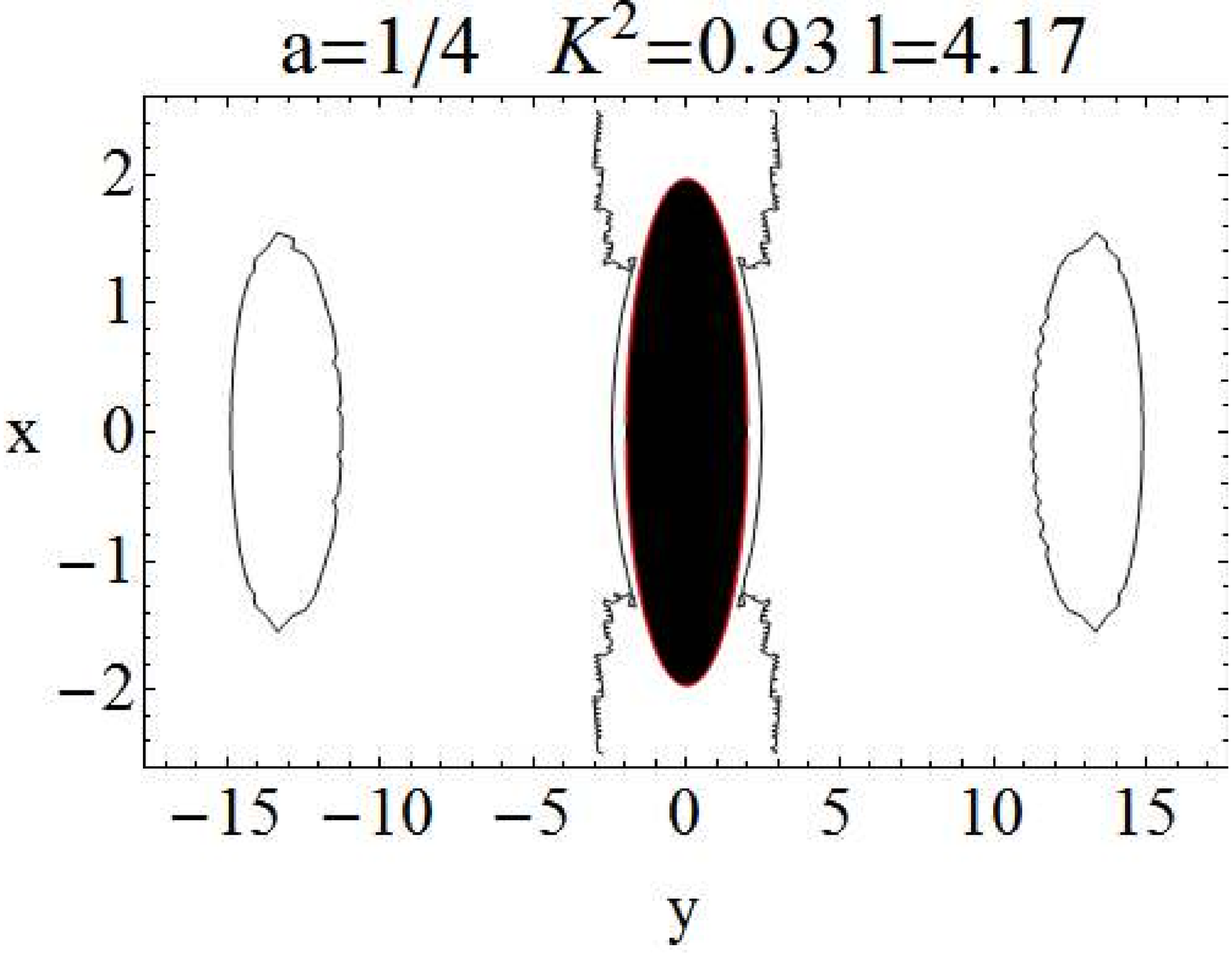}
\includegraphics[scale=.123]{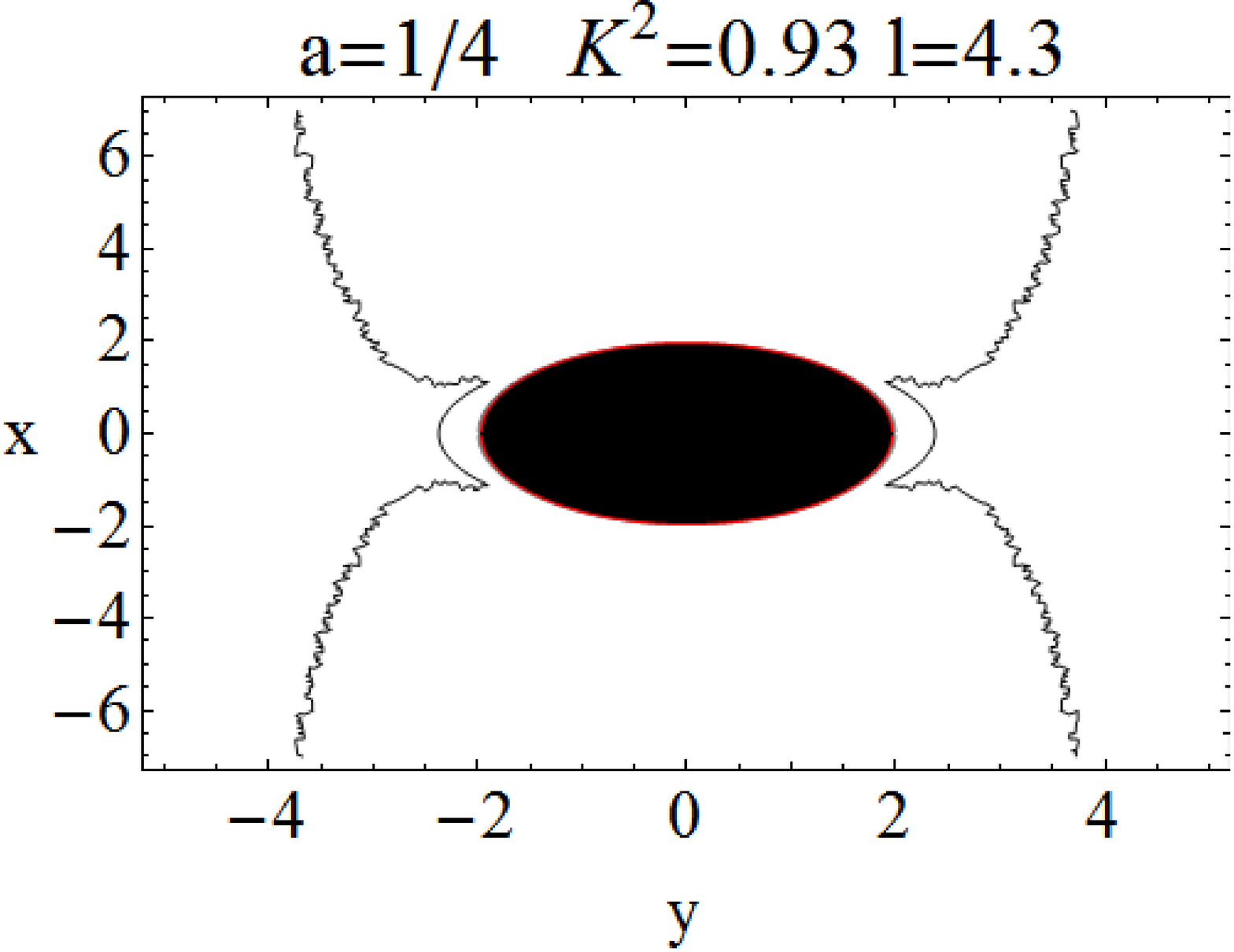}
\includegraphics[scale=.123]{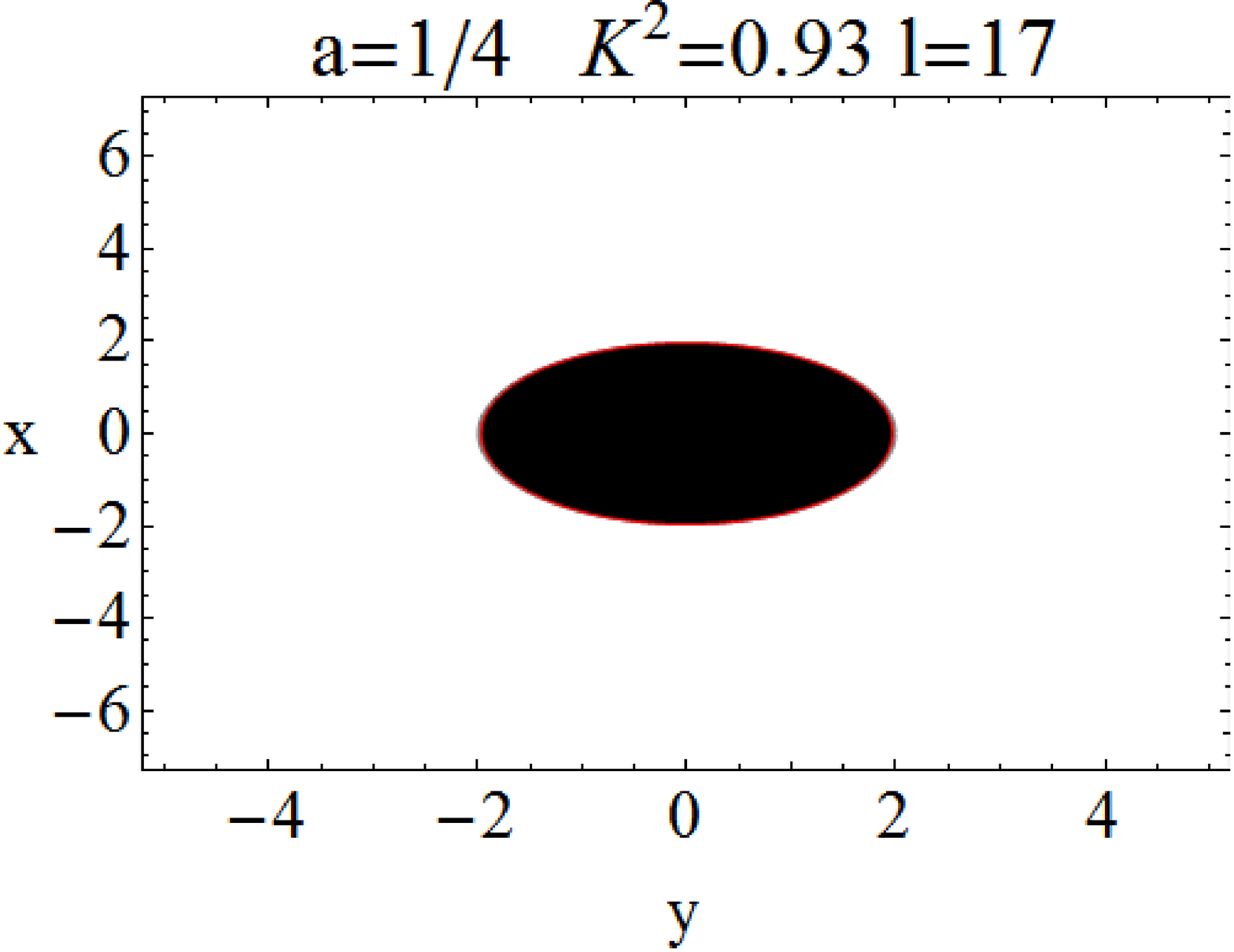}
\end{tabular}
\caption[font={footnotesize,it}]{\footnotesize{Configurations at  $K<1$, $\ell>a$ in \textbf{Region III}  $K\in[K_{lsco}^+,1[$  (sequences $\mathfrak{B}_{\ell}\equiv\mathfrak{B}_{\mathbf{p}}/\Sigma_{K}$), see Sec.\il(\ref{Subsun:barellM1}).  The spacetime spin is $a=0.25M\in\mathbf{BHI}$, the outer horizon is at $r_+=1.96825M$ the static limit $r_{\epsilon}^+=2M$,  it is $K^2=0.93$ and $\ell_{K}^i\in\{ -4.29142, -4.01885,0.235492,
    0.268524,3.60729,
    4.19897\}$, and  $\ell_f(r_{+})^{\pm}=15.746$,
$\ell_{\epsilon}^+=13.4216$,
$\ell_f(r_d)=8.22857$,
$\ell_\gamma^-=4.67535$,
$\ell_{lsco}^-=3.43993$,
$\ell_b^-=3.73205$.
and
$\ell_\gamma^+=-5.68011$,
$\ell_b^+=-4.23607$
$\ell_{lsco}^+=-3.88089$.}}\label{Fig:TSPrediction}
\end{figure}
 \end{description}
This  analysis overlooks a small region of  $K$-parameter very close to the zero that  would require a fine-tuning of the configuration parameters.
 Figure (\ref{Fig:tema})-left shows \textbf{Region II} and \textbf{Region III} on the plane $(K, a/M)$. A discussion on the maximum extension of the $K$-parameter  has been addressed in Sec.\il(\ref{Sec:Km1ana}).
 \subsubsection{Fluid configuration at $\bar{\ell}<1$}\label{Subsubsec:lbarm1}
In this section we focus on the  corotating and  counterrotating configurations at $\bar{\ell}<1$ and the sequence $\mathfrak{B}^{<}_{\bar{\ell}}\equiv\left.\mathfrak{B}_{\mathbf{p}}\right|_{\bar{\ell}<1}/\Sigma_{K}$
: the situation is much more detailed as we approach  the limits $\bar{\ell}\lessapprox0
$ and $\bar{\ell}\gtrapprox
0$.  In this case we can  identify three regions for the $K$-parameter and different phases for the evolution of the $\ell$-parameter. As shown in Sec.\il(\ref{Sec:barl}), the critical points for the pressure can be only at $\bar{\ell}<-1$, therefore only counterrotating P-D configurations are allowed with $\bar{\ell}<1$.  The only limiting value for the $K$-parameter is $K_{lsco}^+(a)=V_{eff}(a;\ell_{lsco}^+)$  associated to counterrotating orbits only.
\begin{description}
\item[\textbf{Region I}:]
 $K\in]0,K_{lsco}^+[$; \textbf{1.} $(\ell<\ell_K^1\,{y_1})$
  \textbf{2.} $(\ell_K^1, y_{13})$,
 \textbf{3.}  $(\ell\in]\ell_K^1,a[\;{y_{123}})$. This case is illustrated in Fig.\il(\ref{Fig:TSBHouse}).
 There are both corotating and  counterrotating   ${B}$-configurations: decreasing   the magnitude of the  orbital angular momentum   the configuration stretches along the axis  on the equatorial plane. In Tables.\il(\ref{Fig:closed-Kmin-lneg}) it is show the set of  counterrotating and  corotating fluids respect  a change in  $K$ and $\ell$. We note that the boundary of this region is determined by  $K_{lsco}^+$  only for both $\ell>0$ and $\ell<0$, however only counterrotating configurations at $\bar{\ell}<1$ can give rise to P-D tori, as discussed in Sec.\il(\ref{Sec:Km1ana}) and Sec.\il(\ref{Sec:barl}).
 \begin{figure}
\centering
\begin{tabular}{cc}
\includegraphics[scale=.13]{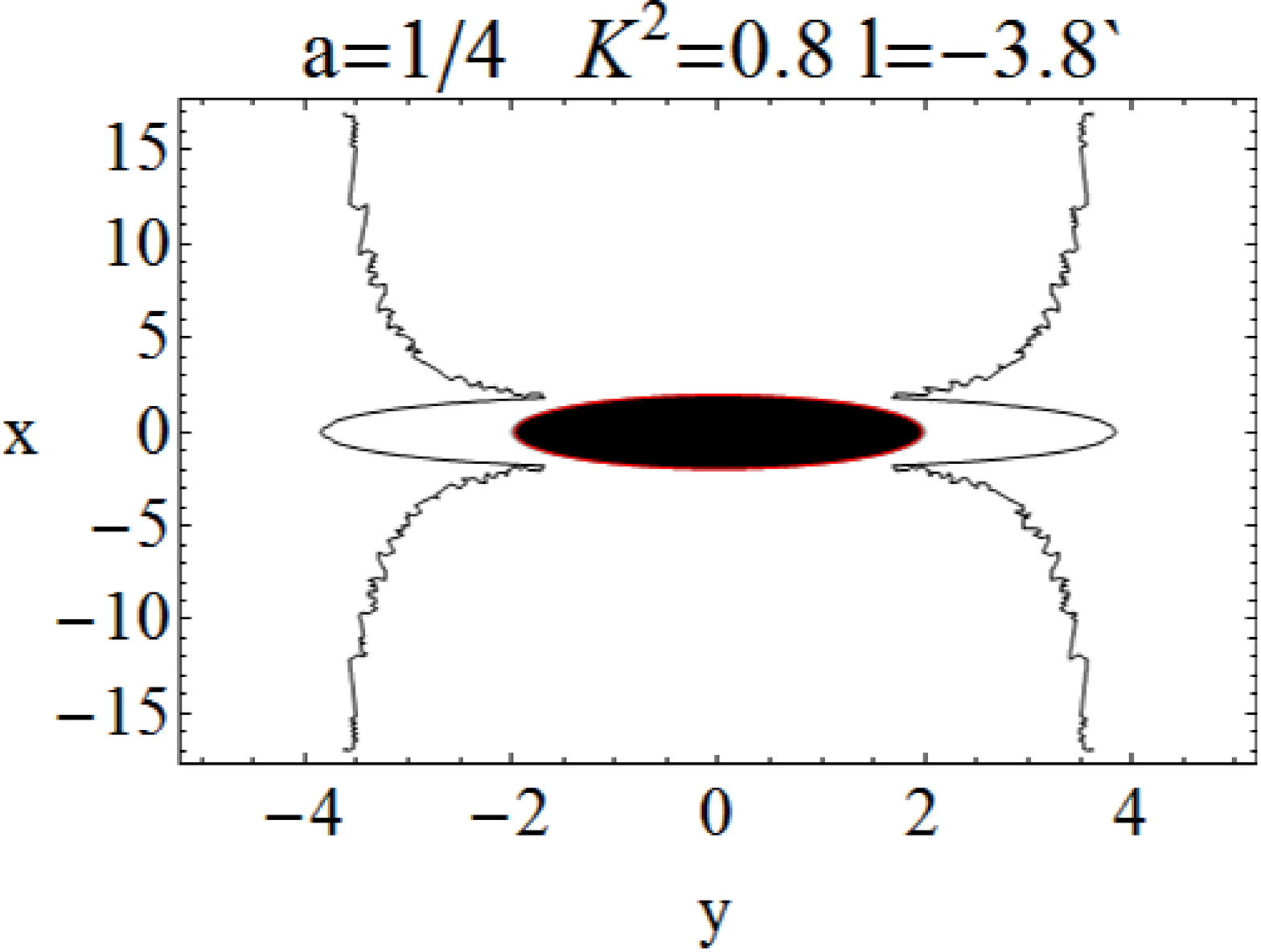}
\includegraphics[scale=.13]{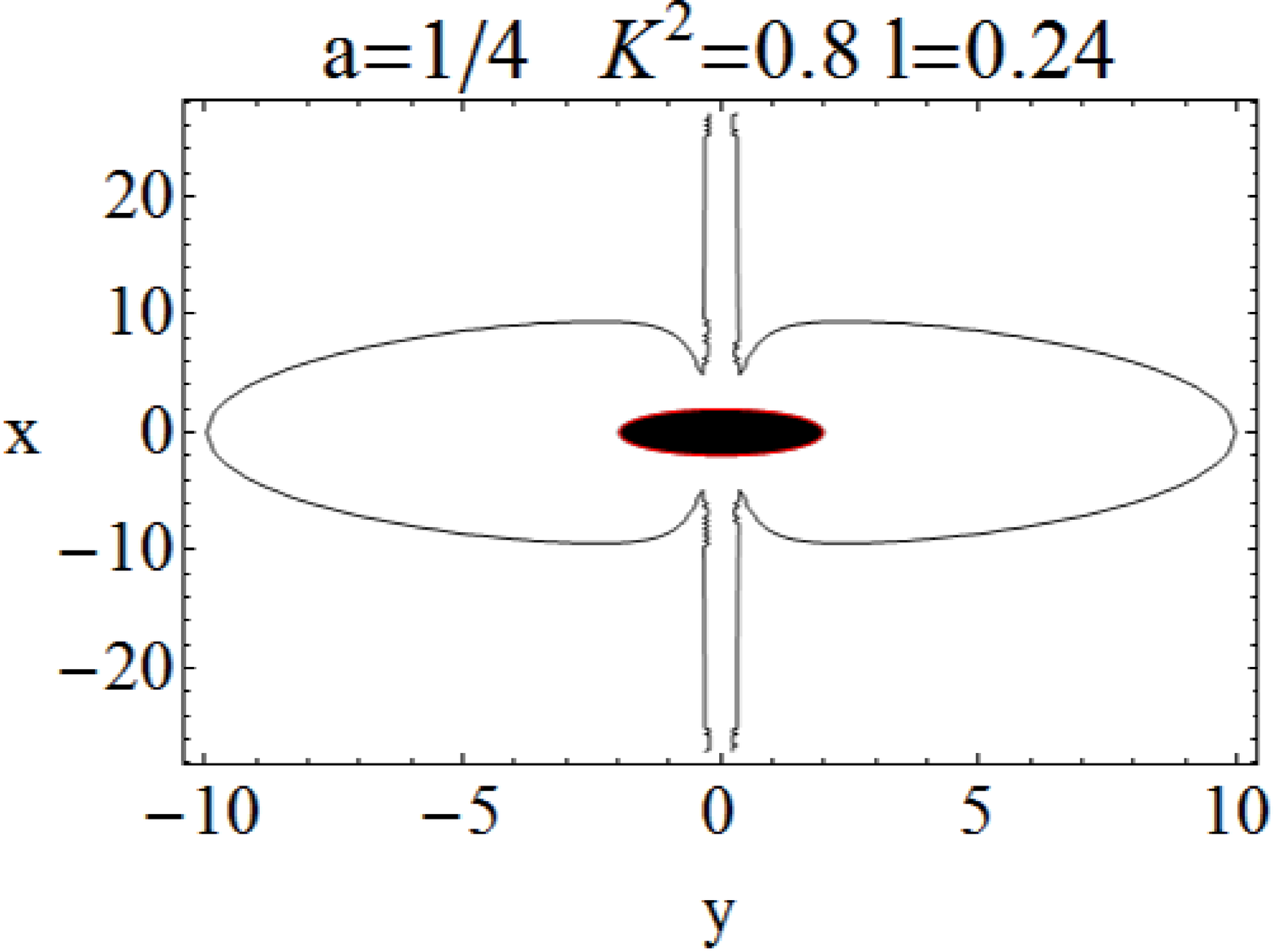}
\end{tabular}
\caption[font={footnotesize,it}]{\footnotesize{Configurations at  $K<1$, $\ell<a$ in  \textbf{Region I} $K\in]0, K_{lsco}^+[$ see Sec.\il(\ref{Subsubsec:lbarm1}).  The spacetime spin is $a=0.25M\in\mathbf{BHI}$,  (sequences $\mathfrak{B}_{\ell}\equiv\mathfrak{B}_{\mathbf{p}}/\Sigma_{K}$) where $K=0.81$, the outer horizon is at $r_+=1.96825M$ the static limit $r_{\epsilon}^+=2M$, $\ell_{K}^i=\{0.234227,0.269787\}$, in units of mass $M$.}}\label{Fig:TSBHouse}
\end{figure}
 \item[\underline{\textbf{Region II}}:]
 $K_{lsco}^+$: \textbf{1.}  $(\ell<\ell_K^3, {y_1})$,
  \textbf{2.} $(\ell_K^3\;y_{13})$,
\textbf{3.} ($\ell\in]\ell_K^3,a[\;{y_{123}})$. This is a limiting case and, accordingly  to the analysis in Fig.\il(\ref
{Fig:tema}) it corresponds to an unstable orbit located in $r_{lsco}^+$.
 \item[\underline{\textbf{Region III}}:] $K\in]K_{lsco}^+,1[$, \textbf{1.} $(\ell<\ell_K^1\; {y_1})$
 \textbf{2.} $(\ell_K^1\;{y_{12}})$,
\textbf{3.} $(\ell\in]\ell_K^1,\ell_K^2[\;{y_{123}})$
\textbf{4.} $(\ell_K^2,{ y_{13}})$
\textbf{5.} $(\ell\in]\ell_K^2,\ell_K^3[
{y_1})$ \textbf{6.} $
(\ell_K^3\; {y_{12}})$ \textbf{7.}
$(\ell\in]\ell_K^3,a[\; {y_{123}})$.
This case is illustrated in  Figs.\il(\ref{Fig:TSModena}) and it includes the Boyer surfaces:
decreasing   the angular momentum magnitude up to zero there are ${B}$ configurations then a closed ${C}$ configuration appears, and only after this stage,  in contrast with the corotating case in \textbf{Region III} of Sec.\il(\ref{Subsun:barellM1}), a  closed-crossed ${C}_{x}$ surface appears,  it is then  followed by a second  ${B}$-configuration with  matter aligned to the axes. With increasing  angular momentum the   fluid   stretches on the
equatorial plane and finally tends to thicken up to the ${B}$-configuration, with thickness  close to the unity, then one can say the set of surfaces is $\mathfrak{B}^{<}_{\bar{\ell}}=[B, C, C_x, B]$.
\begin{figure}
\centering
\begin{tabular}{cc}
\includegraphics[scale=.123]{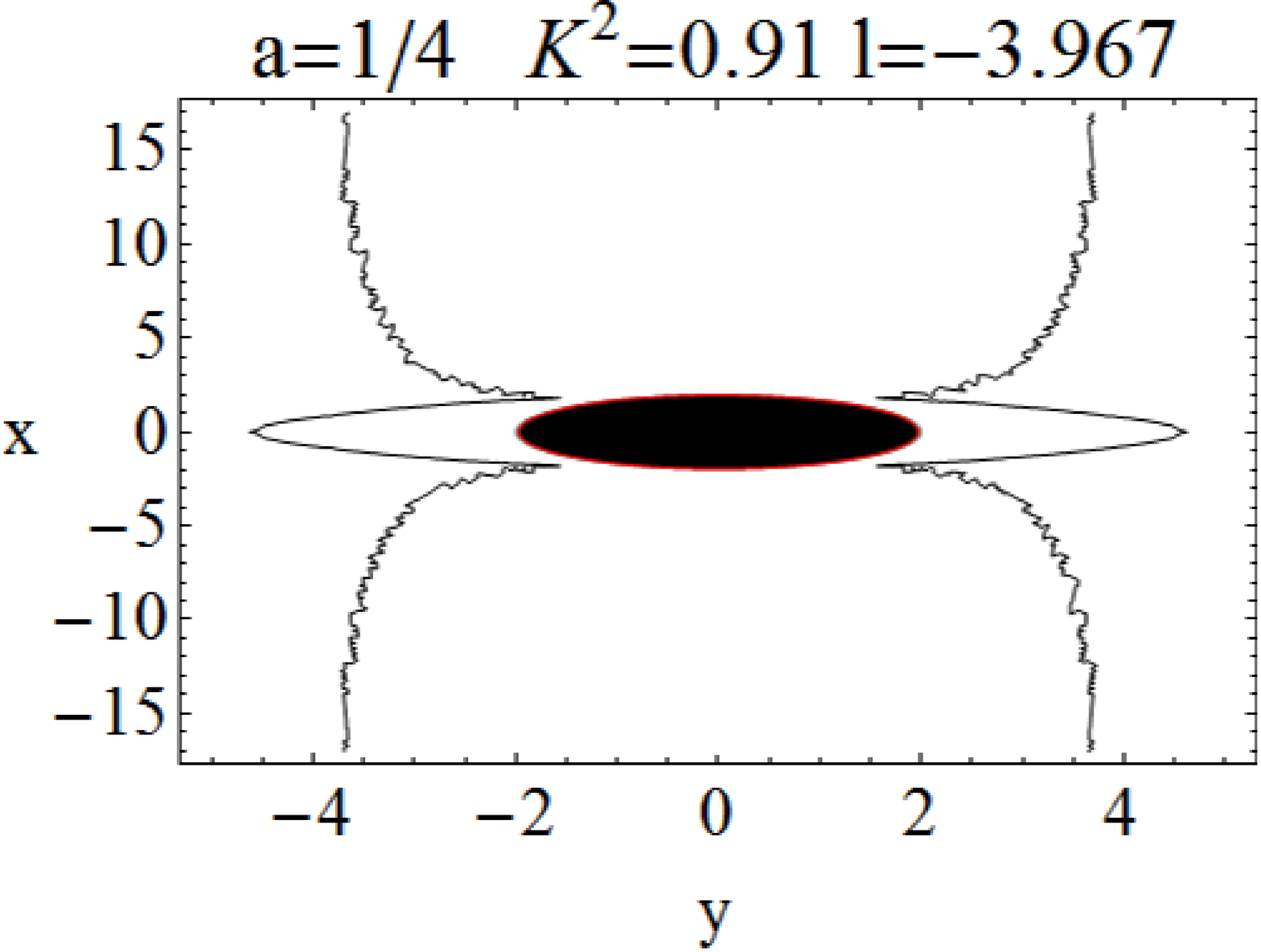}
\includegraphics[scale=.123]{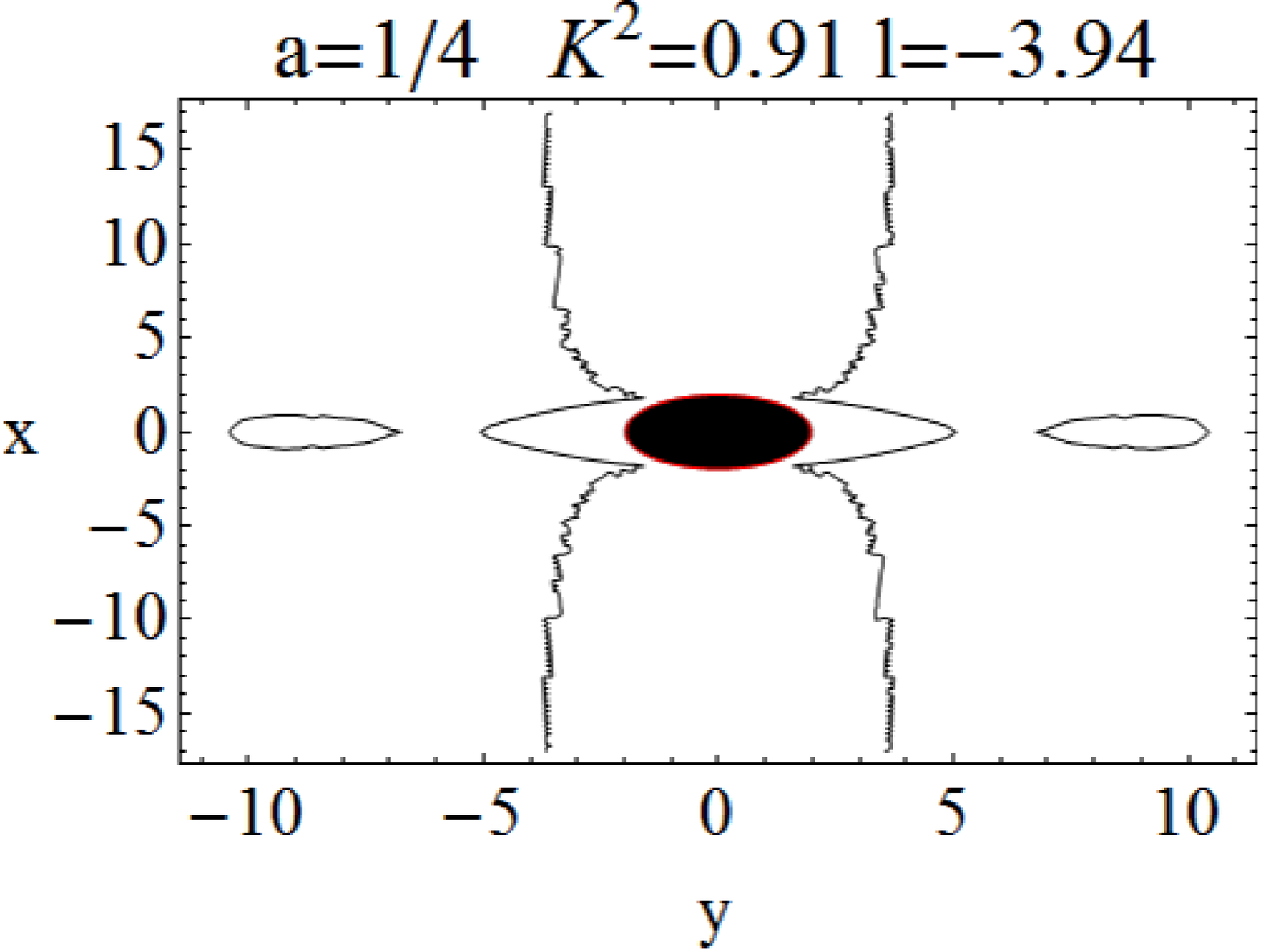}
\includegraphics[scale=.123]{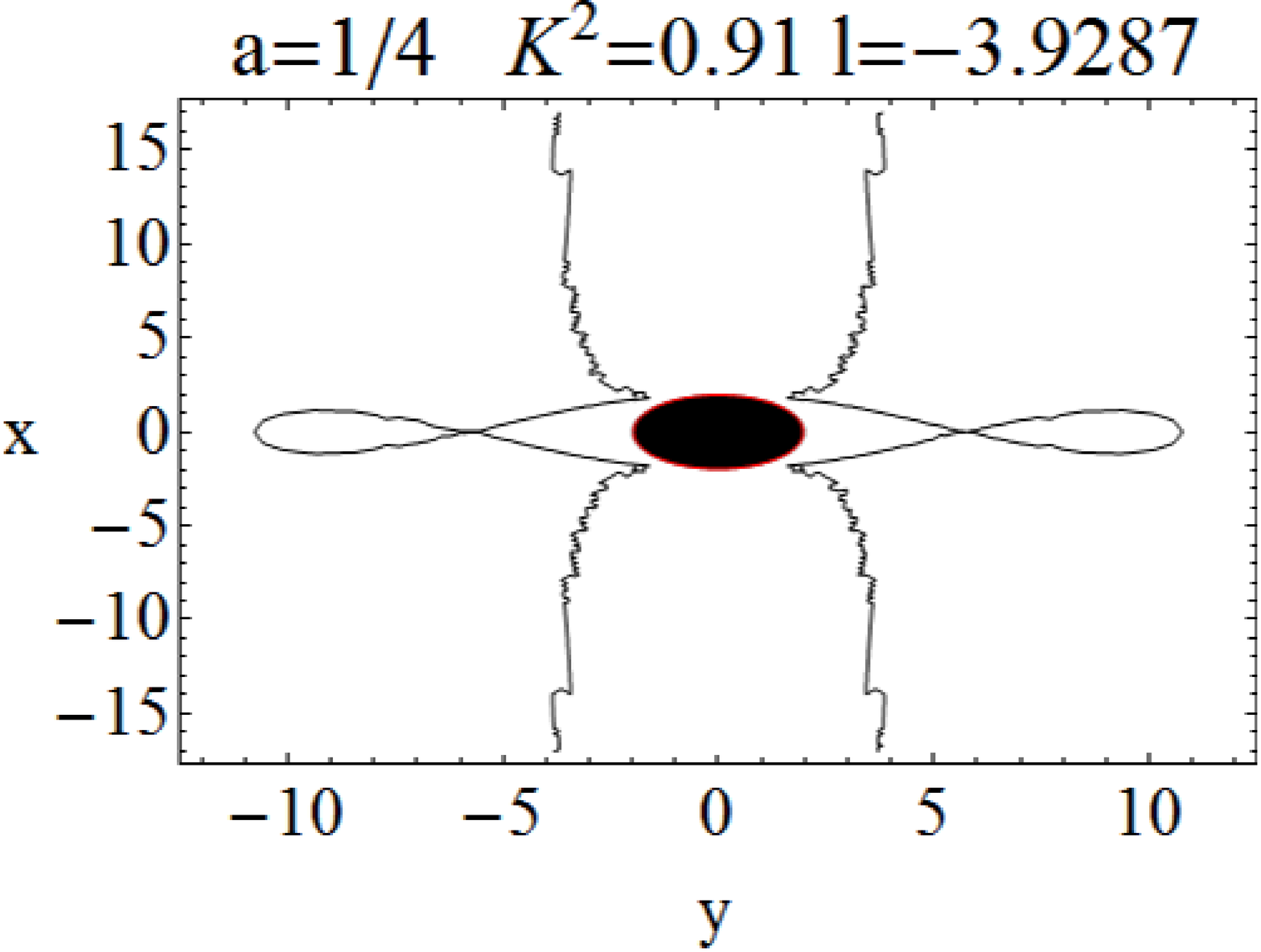}
\includegraphics[scale=.123]{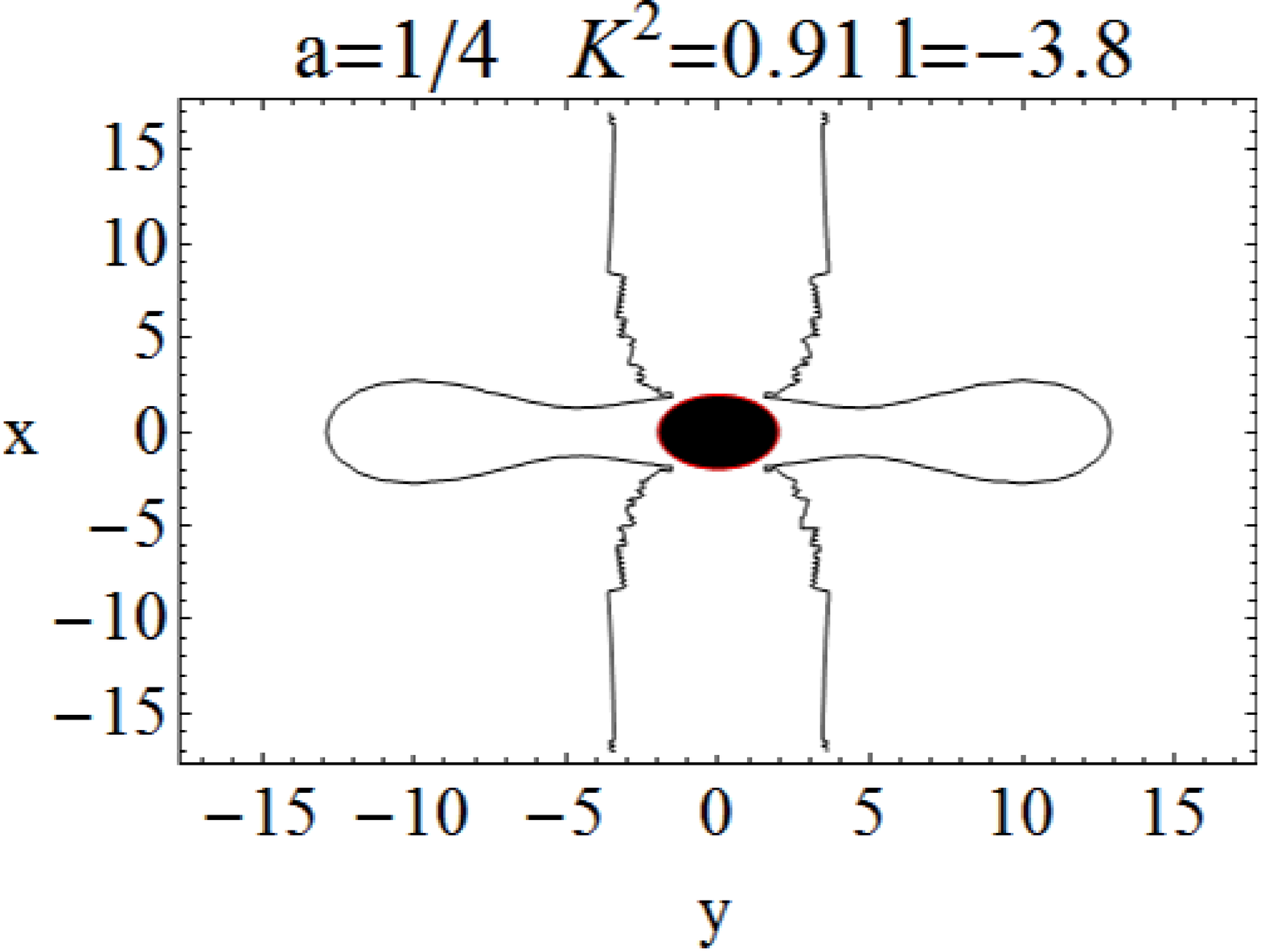}
\includegraphics[scale=.123]{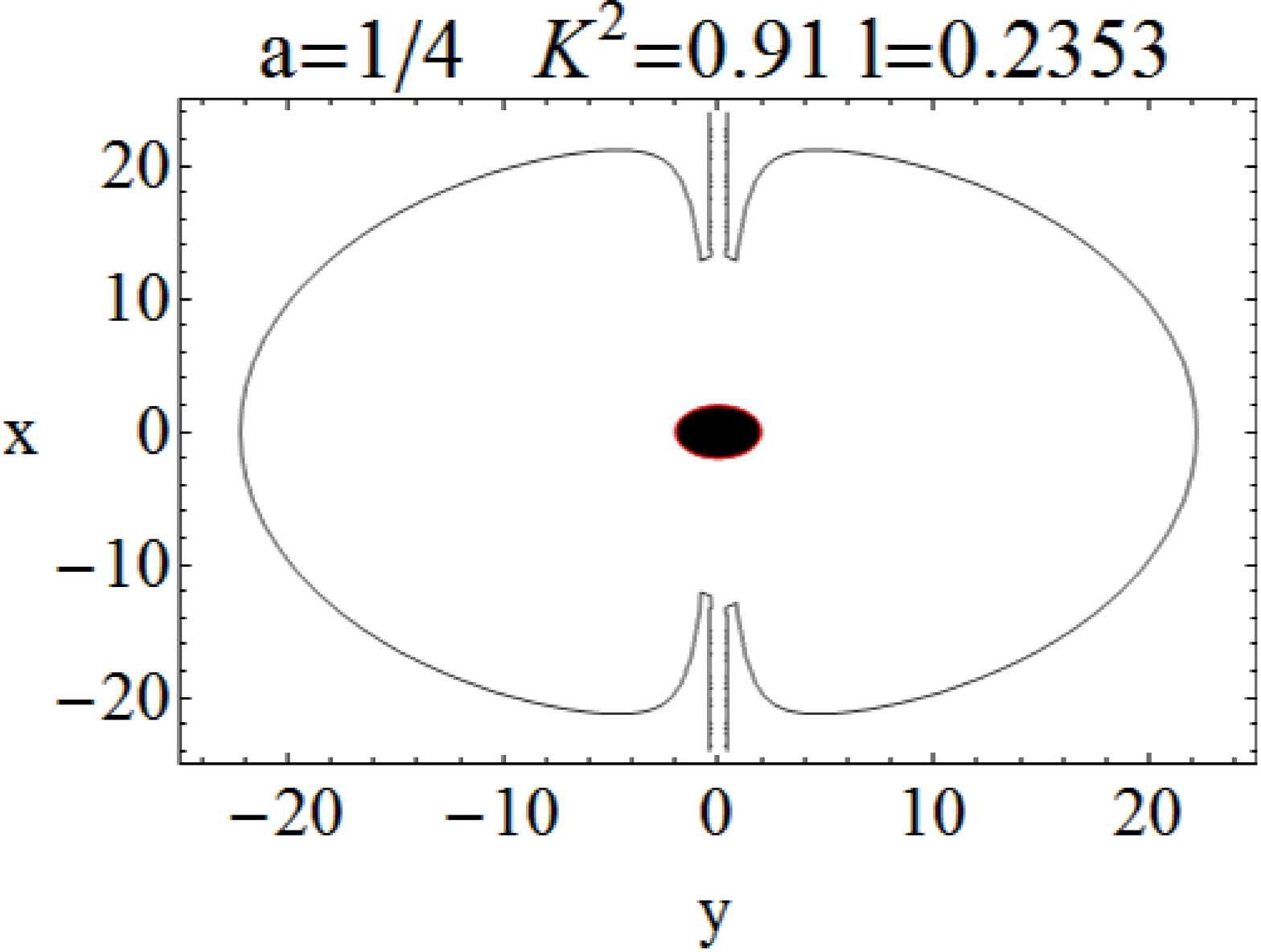}
\end{tabular}
\caption[font={footnotesize,it}]{\footnotesize{Configurations at $K<1$, $\ell<a$. in  \textbf{Region III} $K\in]K^+_{lsco},1[$ see Sec.\il(\ref{Subsubsec:lbarm1}).  The spacetime spin is $a=0.25M\in\mathbf{BHI}$, the outer horizon is at $r_+=1.96825M$ the static limit $r_{\epsilon}^+=2M$, (sequences $\mathfrak{B}_{\ell}\equiv\mathfrak{B}_{\mathbf{p}}/\Sigma_{K}$) where $K^2=0.91$, it is $\ell_K^i=({ -3.96369}, { -3.92872}, {0.235316}, {
   0.2687}, { 3.56152}, { 3.82688})$. The angular momentum is in units of mass $M$.}}\label{Fig:TSModena}
\end{figure}
 \end{description}
\subsection{The model evolution for different $K$ at $\ell$ fixed}\label{Sec:K.L}
Here we analyse  the sequence $\mathfrak{B}_{K}\equiv\mathfrak{B}_{\mathbf{p}}/\Sigma_{\ell}$.
It is convenient to address the discussion  separating the two subcases of  the corotating  $\ell>0$ configurations  in Sec.\il(\ref{subsub:corot.mo}) and  counterrotating $\ell<0$   configurations  in Sec.\il(\ref{subsub:contro.mo}). In Sec.\il(\ref{Subsec:KMlf}) we explore the open configuration $K>1$.
\subsubsection{Corotating  $(a\ell>0)$ fluid   configurations at $K<1$}\label{subsub:corot.mo}
We focus on the corotating orbits $\ell>0$ at $K<1$. The situation is very articulated, it is shown schematically in Table\il(\ref{Fig:closed-Kmin-lneg})-right
 and discussed in six macroregions, defined by fixed $\ell$ and varying
 the  $K$-parameter, each region of angular momentum values include different subregions for the variation of the $K$-parameter.
\begin{description}
\item[\textbf{Region I}] We consider three sub regions of $\ell$-parameters namely:
 $\mathbf{Ia}$ for
$\ell\in]0,\ell_{\mu}^\Pi[$: \textbf{1.} $(0,y_{23})$, \textbf{2.}
$(K\in]0,K_{2}[,y_{123})$. \textbf{3.} $(K_2,y_{13})$, \textbf{4.}
$(K\in]K_2,1[,y_1)$.  $\mathbf{Ib}$ for
$\ell\in[\ell_{\mu}^\Pi,a[$: \textbf{1.} $(0\,y_{23})$, \textbf{2.} $(K\in]0,1[,y_{123})$.
 $\mathbf{Ic}$ for
$\ell=a$: \textbf{1.} $(K\in[0,1[, y_{23})$. This region corresponds to  $\bar{\ell}<a$  for corotating matter  plotted in Figs.\il(\ref{Fig:TSBHouse},
\ref{Fig:TSModena}) and detailed in Sec.\il(\ref{Subsubsec:lbarm1}), the last subregion $\ell=a$, is a limiting case and it will be investigated further in Sec.\il(\ref{subsubsec:lpm1}). As discussed in Sec.\il(\ref{Sec:barl}) there are no P-D tori but in general $B$-surfaces can be formed, there is  one solution  of the equation $\mathcal{V}_{eff}=0$  corresponding  to the  exterior boundary of the disk.
The boundary $\ell_\mu^\Pi <a$ is defined and discussed in Eq.\il(\ref{Eq:uni-b}) and illustrated in
 Fig.\il(\ref{Fig:radiilf1}).
\item[
\textbf{Region II}] This case can be analysed in four ranges of variation of $\ell$.
 $\mathbf{IIa}$  for
$\ell\in]a,\ell_f^{\pm}(r_-)[$ it is \textbf{1.} $(0,y_{23})$, \textbf{2.}
$(K\in]0,1[,y_{123})$,
 $\mathbf{IIb}$  for
$\ell=\ell_f^{\pm}(r_-)$ it is \textbf{1.} $(0,y=3)$, \textbf{2.} $(K\in]0,1[,y_{23})$.
 $\mathbf{IIc}$ for
$\ell\in]\ell_f^{\pm}(r_-),-\ell^{\Pi}_{\mu}]$ it is
 \textbf{1.} $(0,y_{23})$, \textbf{2.} $(K\in]0,1[,y_{123})$.
 $\mathbf{IId}$  for
$\ell\in]-\ell^{\Pi}_{\mu},\ell_{lsco}^-]$ it is  \textbf{1.} $(0, y_{23})$, \textbf{2.} $(K\in]0,K_2[
,y_{123})$, \textbf{3.}
$(K_2,y_{13})$, \textbf{4.} $(K\in]K_2,1[,y_{1})$. The location of the angular momenta as function of $a/M$ is illustrated  in Fig.\il(\ref{Fig:radiilf1}) however it is $\ell_f^{\pm}(r_-)<\ell_{lsco}^-$. No closed Boyer surfaces are possible.
\item[
\underline{\textbf{Region III}}]. The configurations are as follows:
$\mathbf{IIIa}$ for
$\ell\in]\ell_{lsco}^-,\ell_b^-[$ it is then
\textbf{1.} $(0,y_{23})$, \textbf{2.}
$(K\in]0,K_2[,y_{123})$, \textbf{3.}
$(K_2,y_{13})$, \textbf{4.}$(K\in]K_2,K_1[\cup K_3,y_{13})$,
\textbf{5.} $(K\in]K_3,K_4[,y_{123})$,
\textbf{6.} $(K_4,y_{13})$, \textbf{7.}
$(K\in]K_4,1[,y_{1})$.  This region of  fluid angular momentum allows the formation of  P-D configurations, see also Fig.\il(\ref{Fig:tema})
the corresponding class for in terms of the rationalized  angular momentum is \textbf{{Region II}} and \textbf{{Region III}} in Sec.\il(\ref{Subsun:barellM1}). The maximum of the effective potential with $K<1$is associated to the closed crossed ${C}_{x}$ surfaces, when $\ell\in]\ell_{lsco}^-,\ell_b^-[$, as described  in Sec.\il(\ref{Sec:s-pf})
and Eq.\il(\ref{Fig:radiilf1}).
\item[\underline{\textbf{Region IV}}] There is only one set of values for the fluid angular momentum to be considered.
$\mathbf{IVa}$ for
$\ell\in[\ell_b^-,\ell_{\gamma}^-]$ it is
\textbf{1.} $(0,y_{23})$, \textbf{2.}
$(K\in]0,K_2[,y_{123})$, \textbf{3.}
$(K_2,y_{13})$, \textbf{4.} $K\in]K_2,K_3],y_{1})$, \textbf{5.}$(K_3,y_{12})$, \textbf{6.}
$(K\in]K_3,1[,y_{123})$. When $K>1$   there are {maximum}   points only  i.e. $ {O}_{x}$-configurations at $r=r_f^{\pm}$ for
$]\ell_b^-,\ell_{\gamma}^-$; for  $K\in]K_{lsco}^{\pm},1[$
the  minimum points, ${C}$ are for $\ell>\ell_{lsco}^->0$
and $r>r_{lsco}^{\pm}$, see Table\il(\ref{Fig:closed-Kmin-lneg})-right.
\item[
\underline{\textbf{Region V}}] We distinguish two ranges of angular momentum.
$\mathbf{Va}$: for
$\ell\in]\ell_{\gamma}^-,\ell_f^{\pm}(r_+)[$ the situation is as follows
\textbf{1.} $(0,y_{23})$, \textbf{2.} $(K\in]0,K_2[,
,y_{123})$, \textbf{3.} $(K_2,y_{13})$, \textbf{4.}
$(K\in]K_2,K_4[,y_3)$, \textbf{5.}
$(K_4,y_{12})$, \textbf{6.} $(K\in]K_4,1[,y_{123})$,
$\mathbf{Vb}$ for $\ell=\ell_f^{\pm}(r_+)$:
\textbf{1.}$(0,y_2)$,  \textbf{2.}
$(K\in]0,K_3[, y_{123})$, \textbf{3.}
$(K_3,y_{1})$, \textbf{4.}
$(K_4,y_2)$, \textbf{5.} $(K\in]K_4,1[,y_{123})$,
\item[
\underline{\textbf{Region VI}}]
We consider the following set $\mathbf{VIa}$ for
$\ell>\ell_f^{\pm}(r_+)$ it is
\textbf{1.}$(0,y_{23})$, \textbf{2.}$(K\in]0,K_3[,
y_{123})$, \textbf{3.}$(K_3,y_{12})$, \textbf{4.}
$(K\in ]K_3,K_4[,y_{1})$, \textbf{5.}
$(K_4,y_{12})$, \textbf{6.} $(K\in]K_4,1[,y_{123})$ see also  Fig.\il(\ref{Fig:tema}).
\end{description}
The union of  \textbf{Region III-VI} corresponds to $\ell>\ell_{lsco}^-$. Table\il(\ref{Fig:closed-Kmin-lneg})-right summarizes this situation.
\subsubsection{Counterrotating  $(a\ell<0)$  fluid configurations at $K<1$}\label{subsub:contro.mo}
\begin{figure}[t]
\includegraphics[width=.481\textwidth]{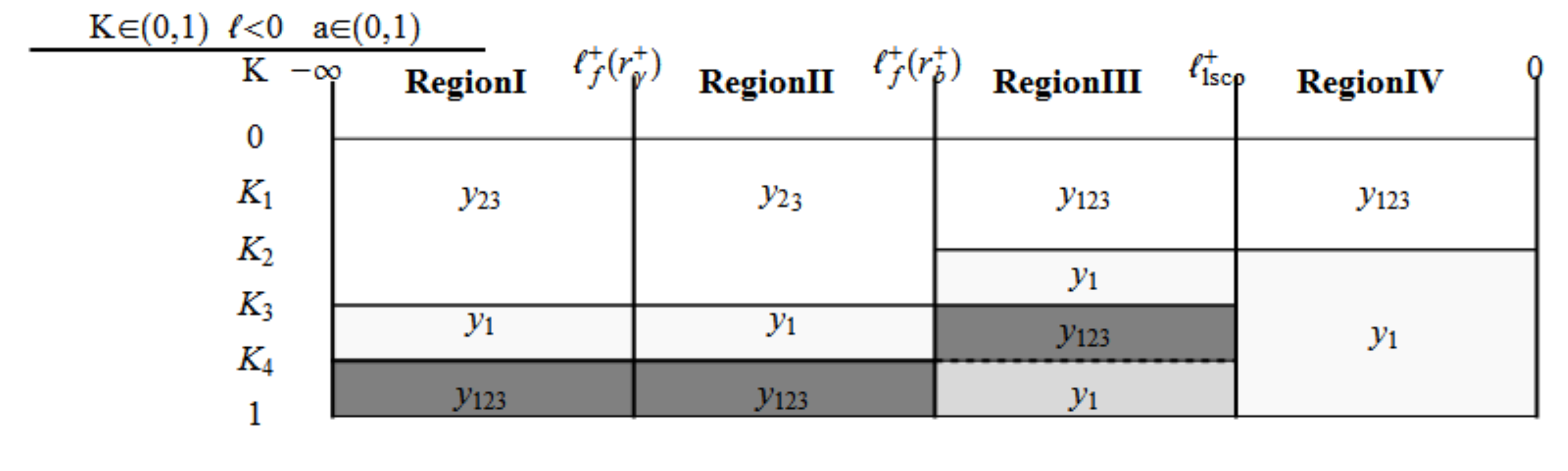}
\includegraphics[width=.51\textwidth]{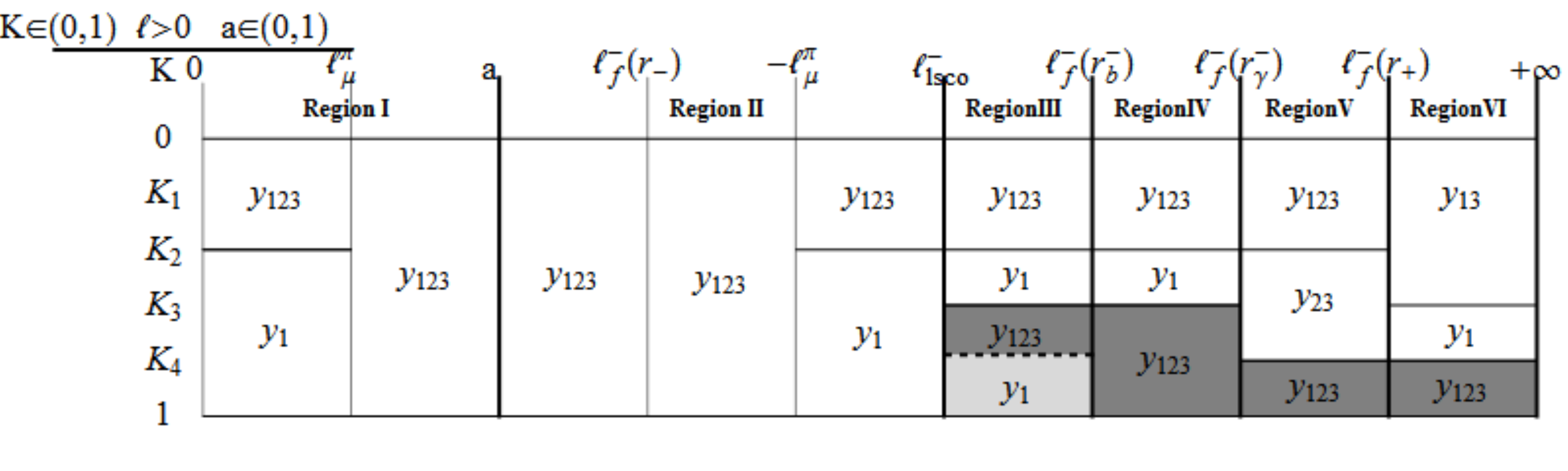}
\caption{Configurations of the fluid angular momentum $\ell=\ell(a)<0$, left panel, and  $\ell=\ell(a)>0$ right panel.  Existence regions of the zeros $y_i$ of $\mathcal{V}_{eff}(\mathbf{p})\equiv V_{eff}(\ell)-K$ for fixed ranges of angular momentum, and  the potential-parameter $K_i=K_i(a,\ell)\in]0,1[$. In the dark gray-shaded  regions closed $C$ Boyer disks exist.}
\label{Fig:closed-Kmin-lneg}
\end{figure}

We focus  on the case $\ell<0$ (counterrotating fluid configurations). We summarize the situation in Table\il(\ref{Fig:closed-Kmin-lneg}). This case $a\ell<0$   is much less articulated then for the corotating fluids and here we can distinguish four regions of  angular momentum:
\begin{description}
\item[\underline{\textbf{Region I}}] for
$\ell<\ell_{\gamma}^+$: \textbf{1.}
$(K\in]0,K_1[, {y_{23}})$, \textbf{2.} $( K_1, y_{13})$, \textbf{3.}
$(K\in]K_3,K_4[,{y_1})$, \textbf{4.} $(K_4,{y_{12}})$, \textbf{5.} $(K\in]K_4,1[,y_{123})$. For angular momentum in this range there are no critical points for the effective potential and no Boyer surfaces.
\item[\underline{\textbf{Region II}}] for
$\ell\in[\ell_{\gamma}^+,\ell_b^+]$: \textbf{1.} $(K\in]0, K_2[,y_{123})$, \textbf{2.} $(K_2,y_{13})$, \textbf{3.} $(K\in]K_2, K_3[,y_1)$, \textbf{4.} $(K_3, y_{12})$, \textbf{5.} $(K\in ]K_3,1[,y_{123})$. The effective potential admits critical  unstable and unbounded orbits, Fig.\il(\ref{Fig:tema}). There are only open $O_x$ surfaces.
\item[\underline{\textbf{Region III}}] for
    $\ell\in]\ell_b^+,\ell_{lsco}^+[$:  \textbf{1.}
$(K\in]0,K_2[,y_{123})$, \textbf{2.} $(K_2,y_{13})$, \textbf{3.} $(K\in ]K_2,K_3[,y_1)$, \textbf{4.} $(K_3, y_{12})$, \textbf{5.} $(K\in]K_3, K_4[, y_{123})$, \textbf{6.} $(K_4, y_{13})$, \textbf{7.}  $(K_4\in]K_4,1[y_1)$.
Critical points are in the  \textbf{Regions \text{I-II-III}}, there are only closed  or closed crossed surfaces: closed crossed ${C}_{x}$ surfaces, where $\ell\in]\ell_b^+,\ell_{lsco}^+[$ while $C$ configurations  are for $\ell<\ell_{lsco}^+<0$, as $K\in]K_{lsco}^{\pm},1[$.
 \item[{\textbf{Region IV}}] for
 $\ell\geq\ell_{lsco}^+$: \textbf{1.} $(K\in]0,K_2[,y_{123})$, \textbf{2.} $(K_2\,y_{13})$, \textbf{3.}
$(K\in]K_2, 1[, y_1)$. There are no closed  $C$-configurations see Table\il(\ref{Fig:closed-Kmin-lneg}) and Figs.\il(\ref{Fig:tema}).
\end{description}
 Tables\il(\ref{Fig:closed-Kmin-lneg})     together  show  the  maximum   and minimum extension of the $\mathbf{p}$ parameter for the existence of a P-D configuration, it is clear    the gap for $\ell\in]\ell_{lsco}^+,\ell_{lsco}^-[$,  where no P-D configurations are possible, and  the presence of the limiting values
$K_4$ and  $K_3$, for the set of $K$-parameter. It is important to note that this analysis does not take into account the attractor spin explicitly but through the angular momentum  $\ell^i_K(a; K)$ or the parameter $K_i(a;\ell)$.
Moreover, there is no evidence of a clear evolutive set $\mathfrak{B}_{K}$ for both the  $a\ell<0$ or  $a\ell>0$ cases. This would confirm  that  a better choice for a  dynamical  parameter could be the fluid angular momentum $\ell$ or $\bar{\ell}$. However we will address more deeply this point in Sec.\il(\ref{subsubsec:Schw}) where, considering a non rotating attractor, we detail the possible  $\mathfrak{B}_{K}$ sequences and we give also some general considerations comparing the  $\mathfrak{B}_{K}$  and $\mathfrak{B}_{\ell}$ sequences.
\begin{figure}[t]
\includegraphics[width=.2\textwidth]{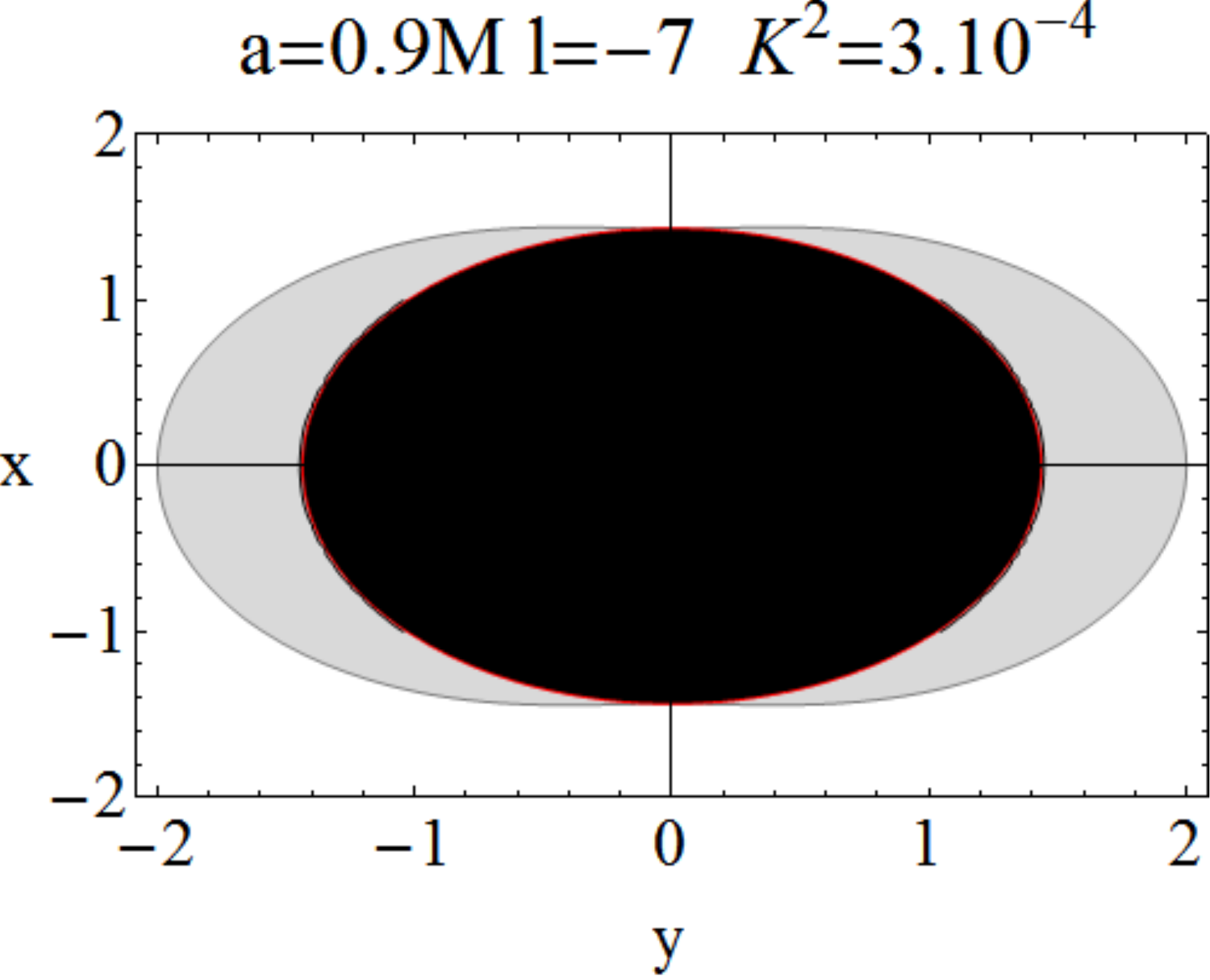}
\includegraphics[width=.2\textwidth]{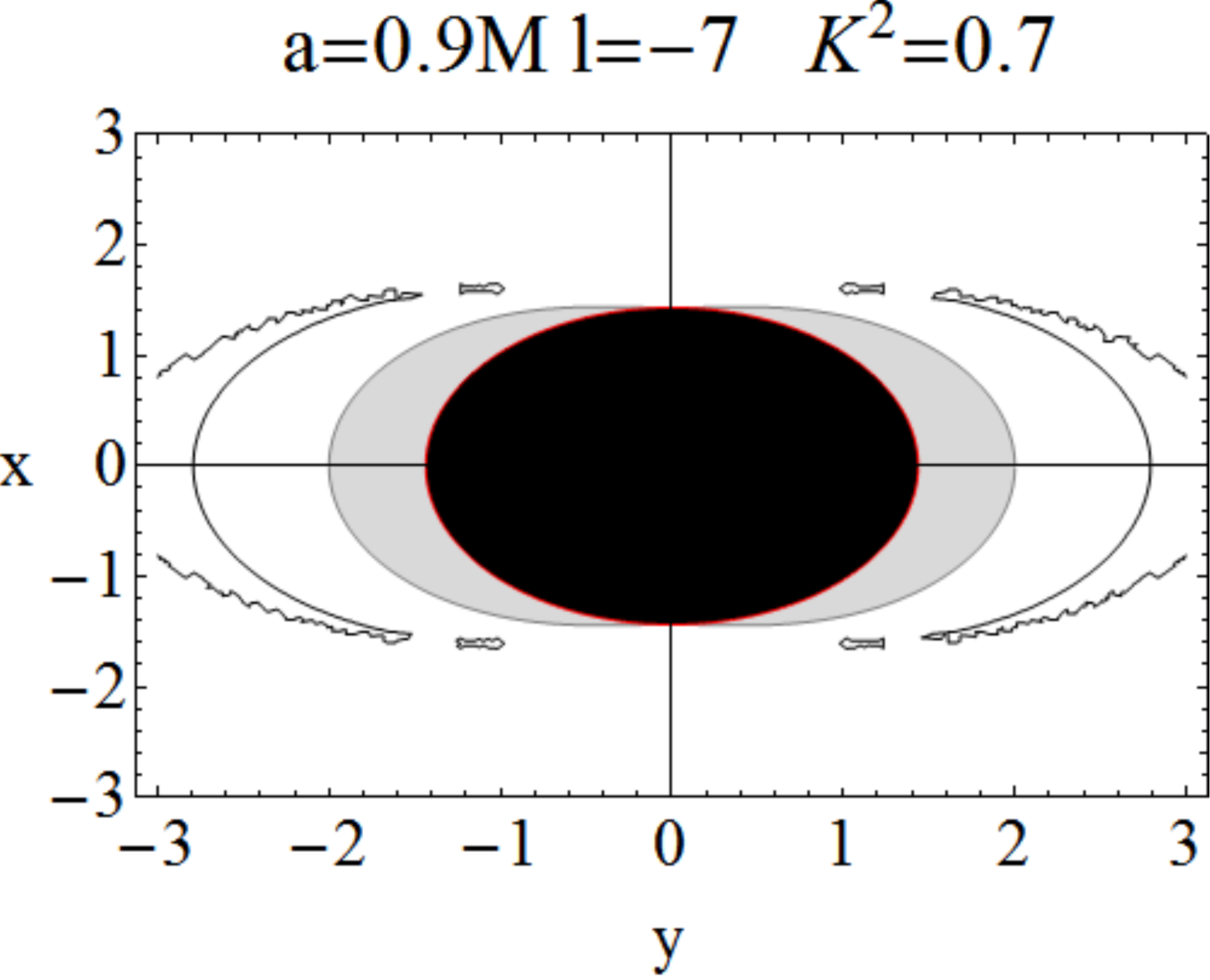}
\includegraphics[width=.2\textwidth]{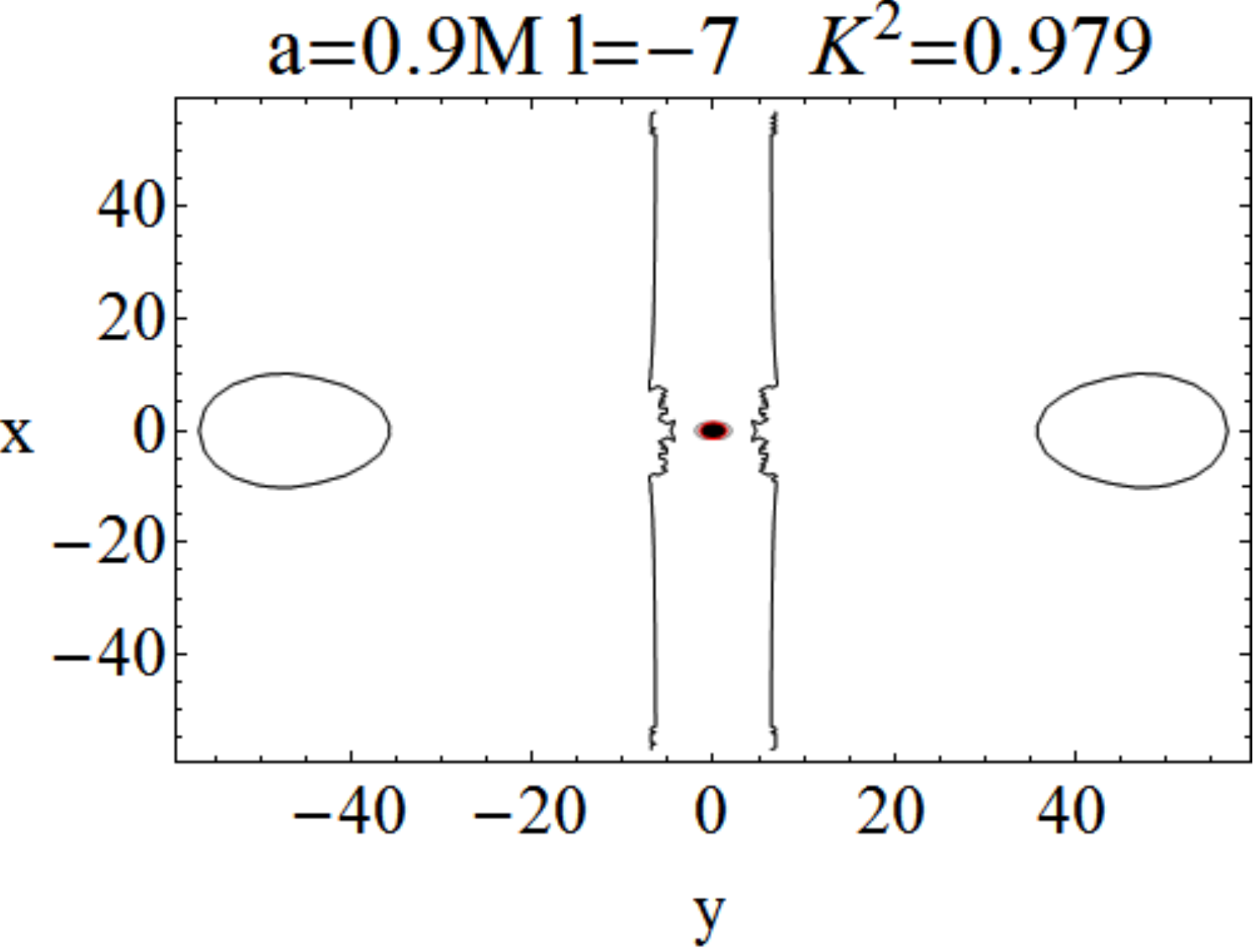}
\\
\includegraphics[width=.2\textwidth]{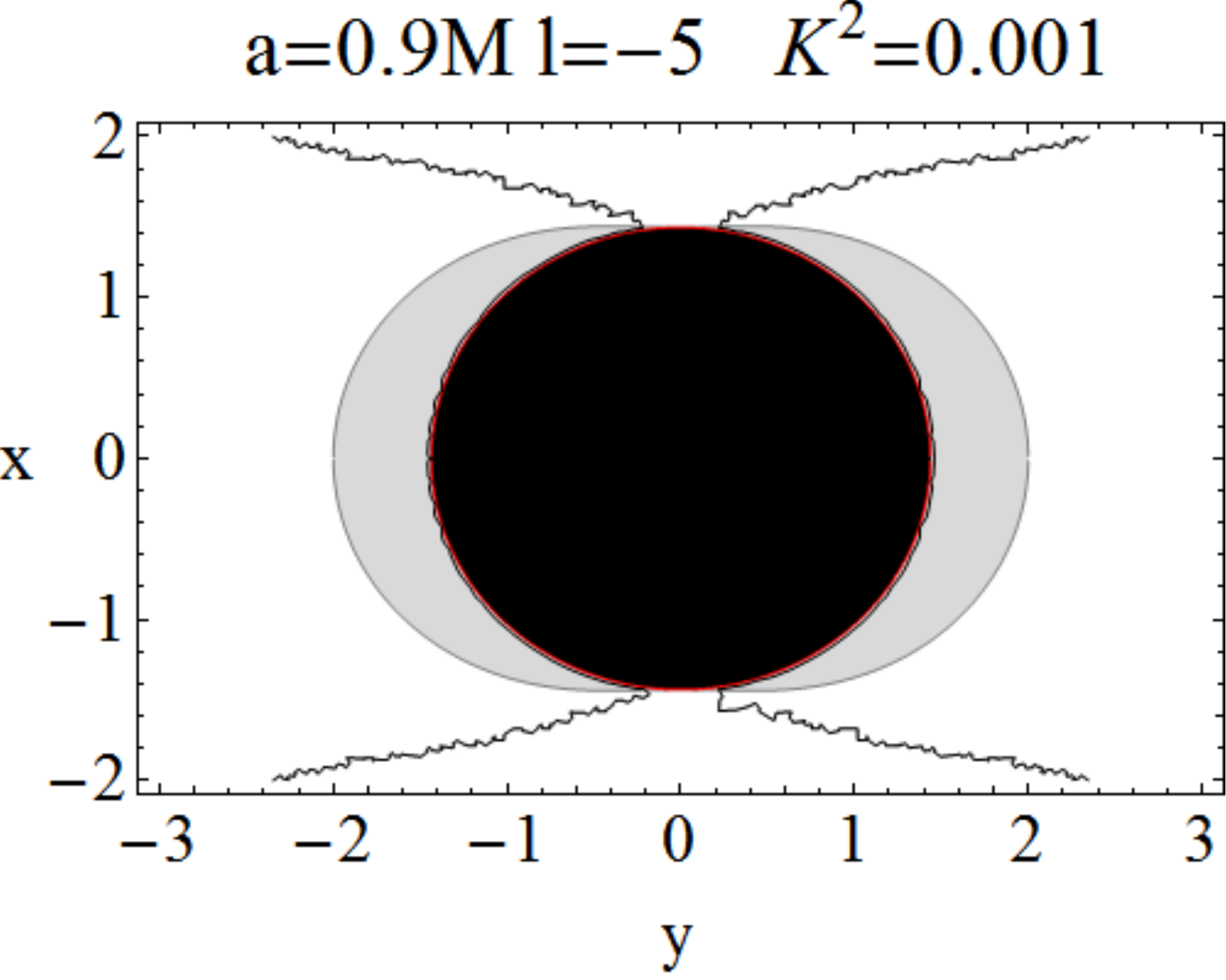}
\includegraphics[width=.2\textwidth]{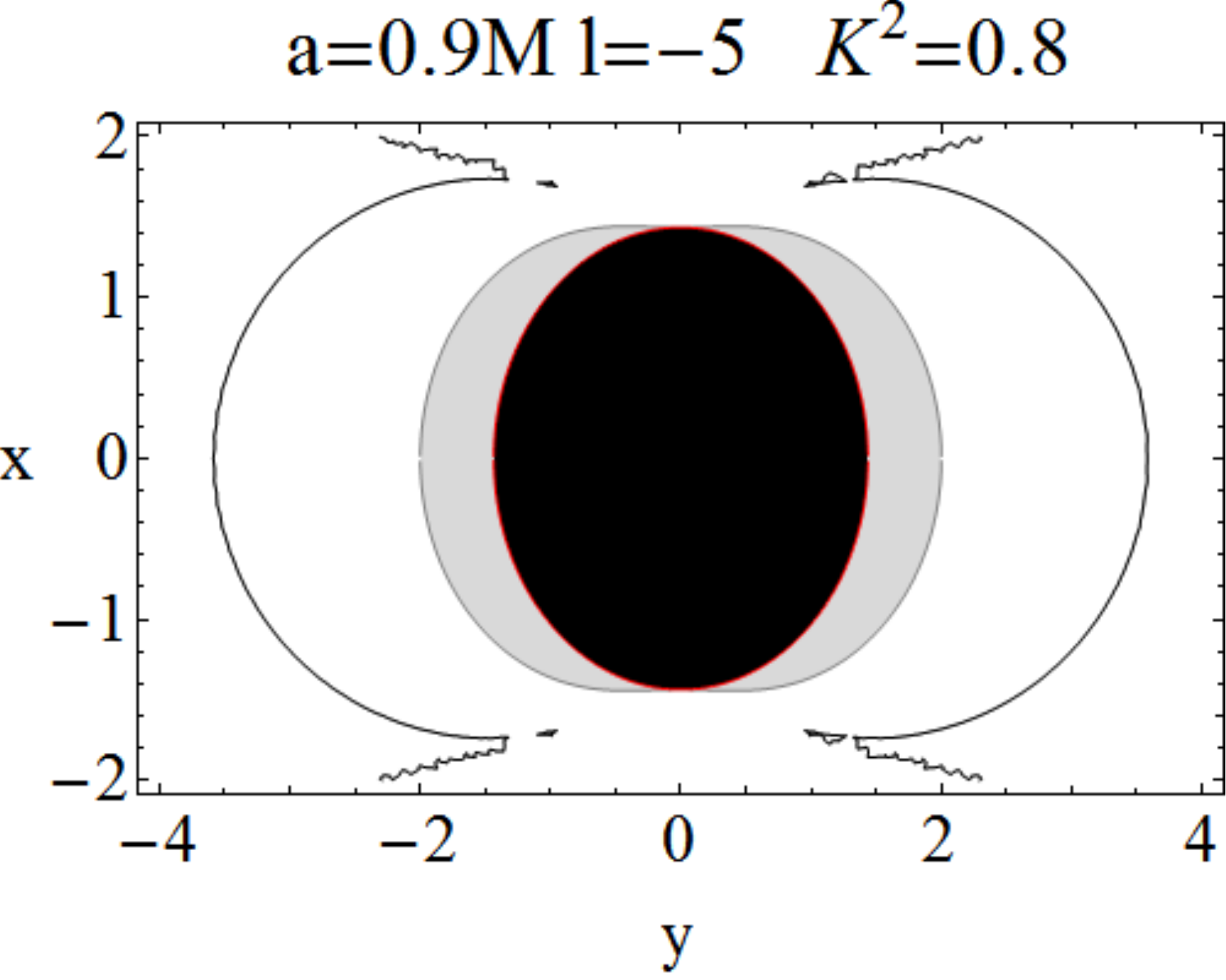}
\includegraphics[width=.2\textwidth]{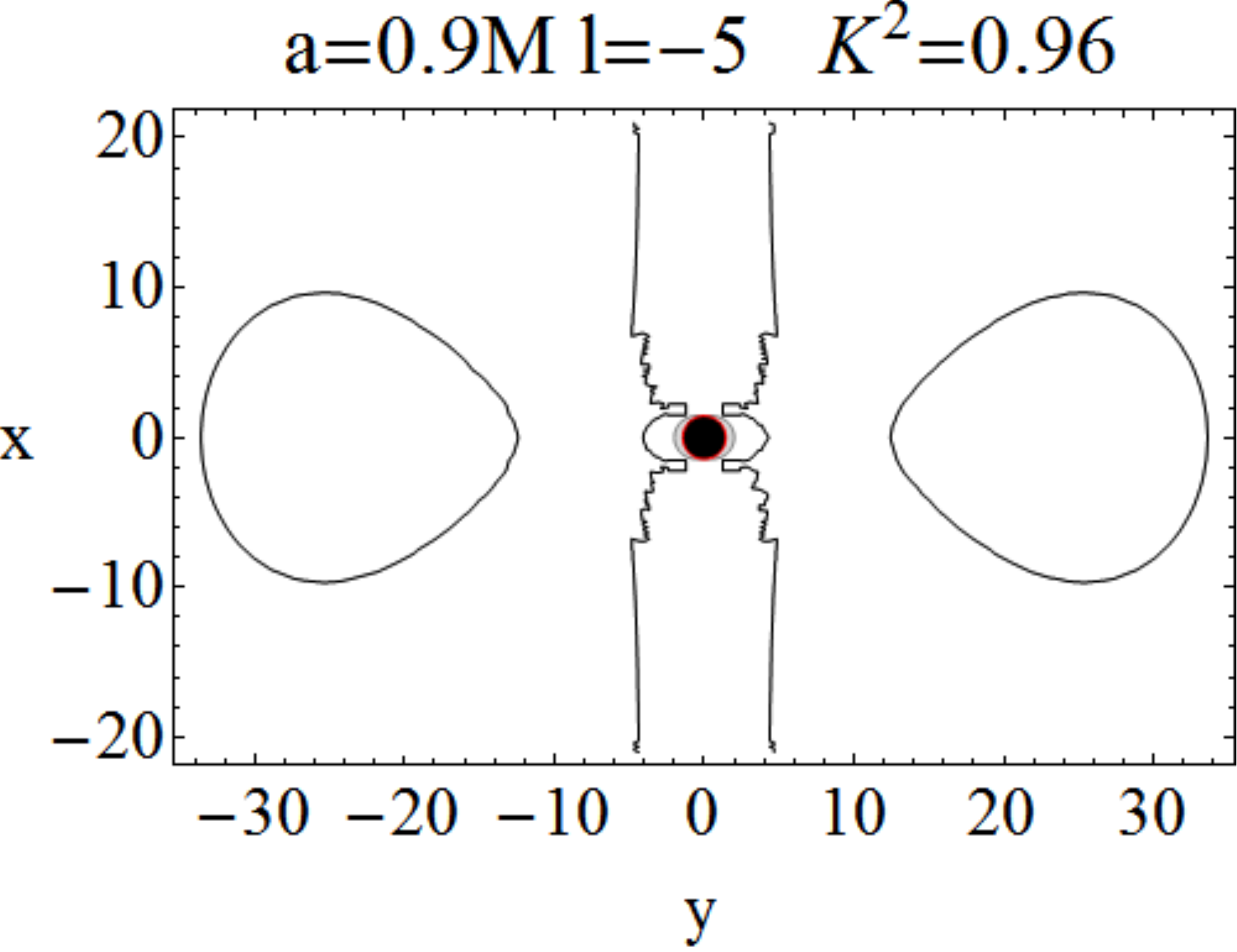}
\\
\includegraphics[width=.2\textwidth]{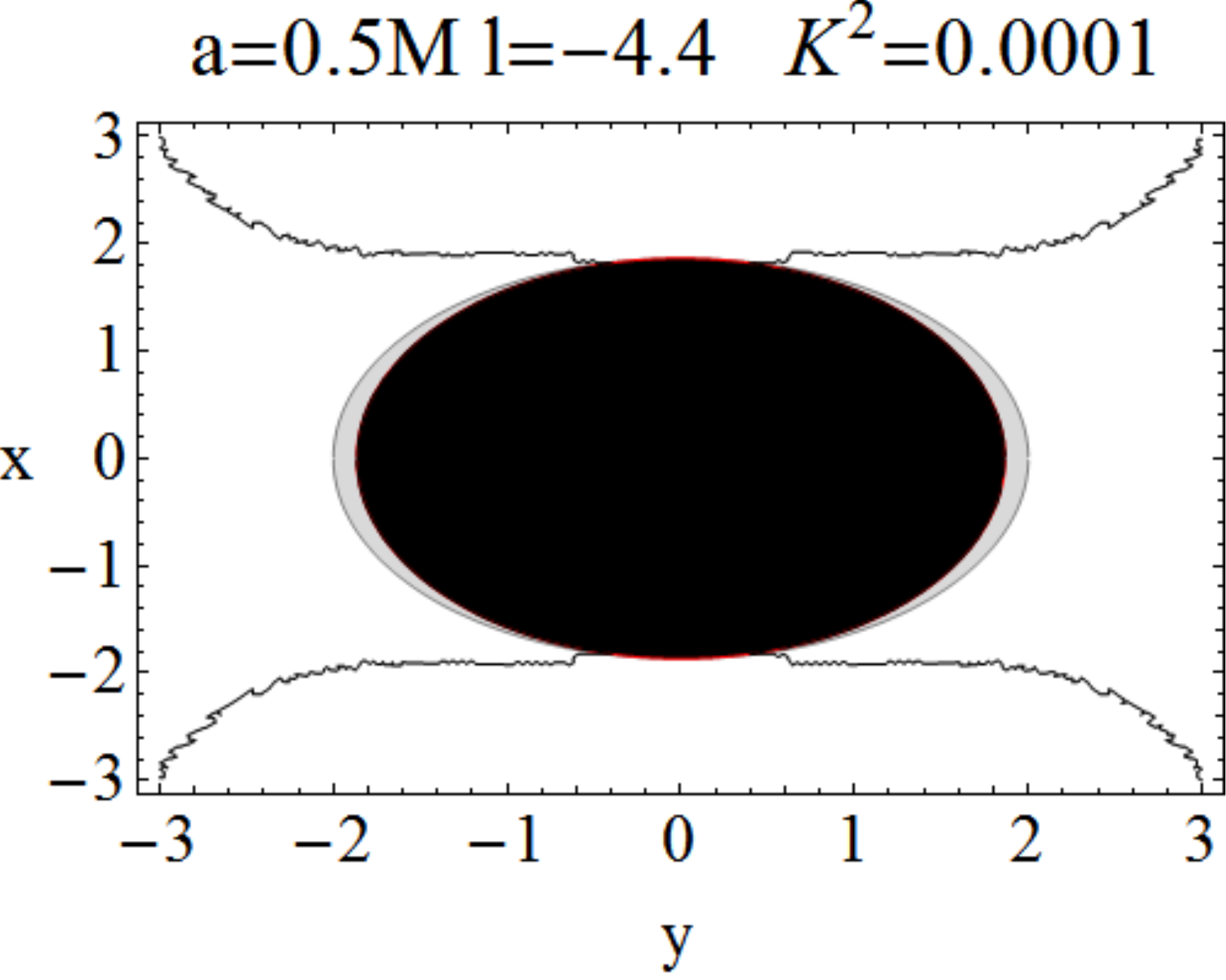}
\includegraphics[width=.2\textwidth]{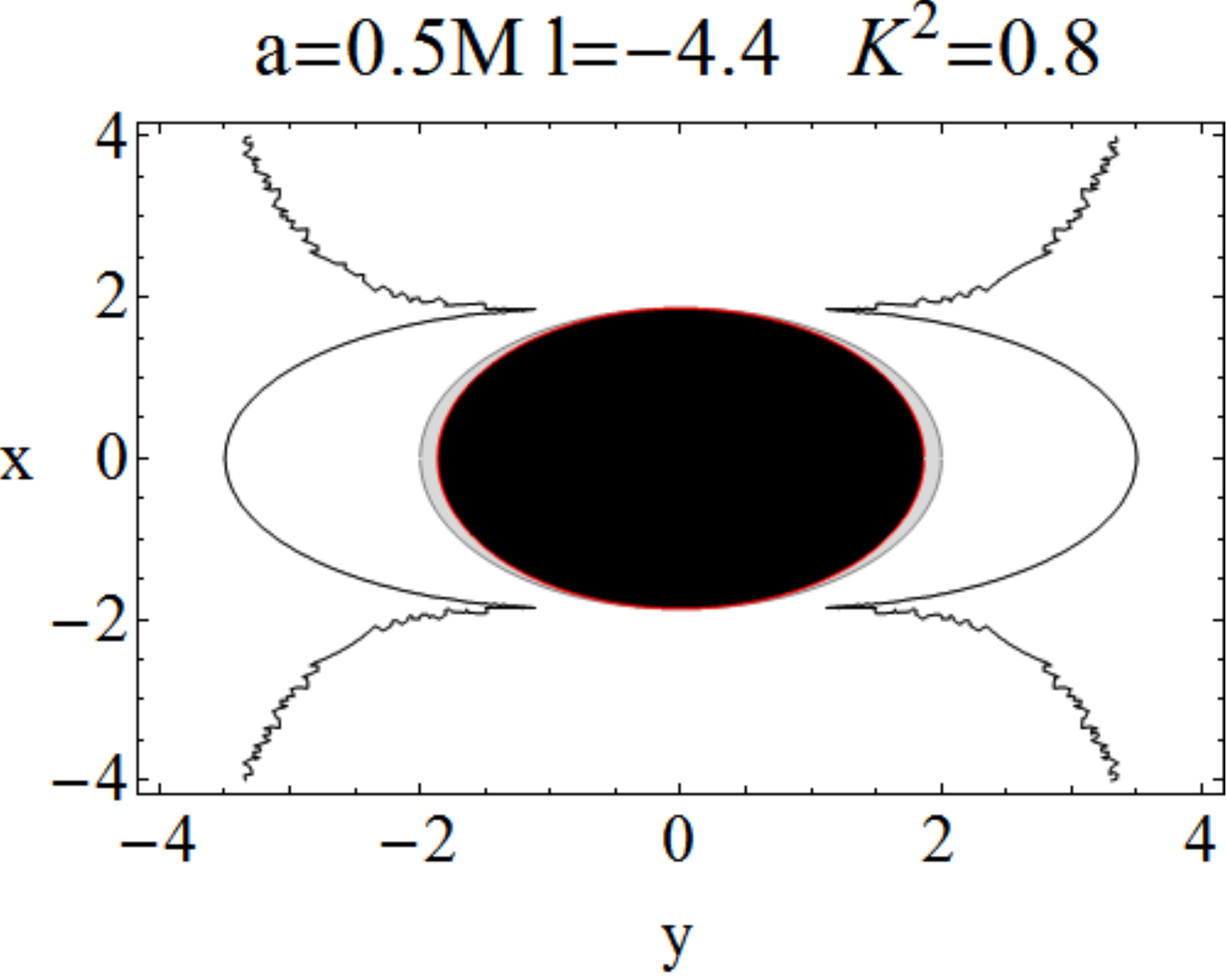}
\includegraphics[width=.2\textwidth]{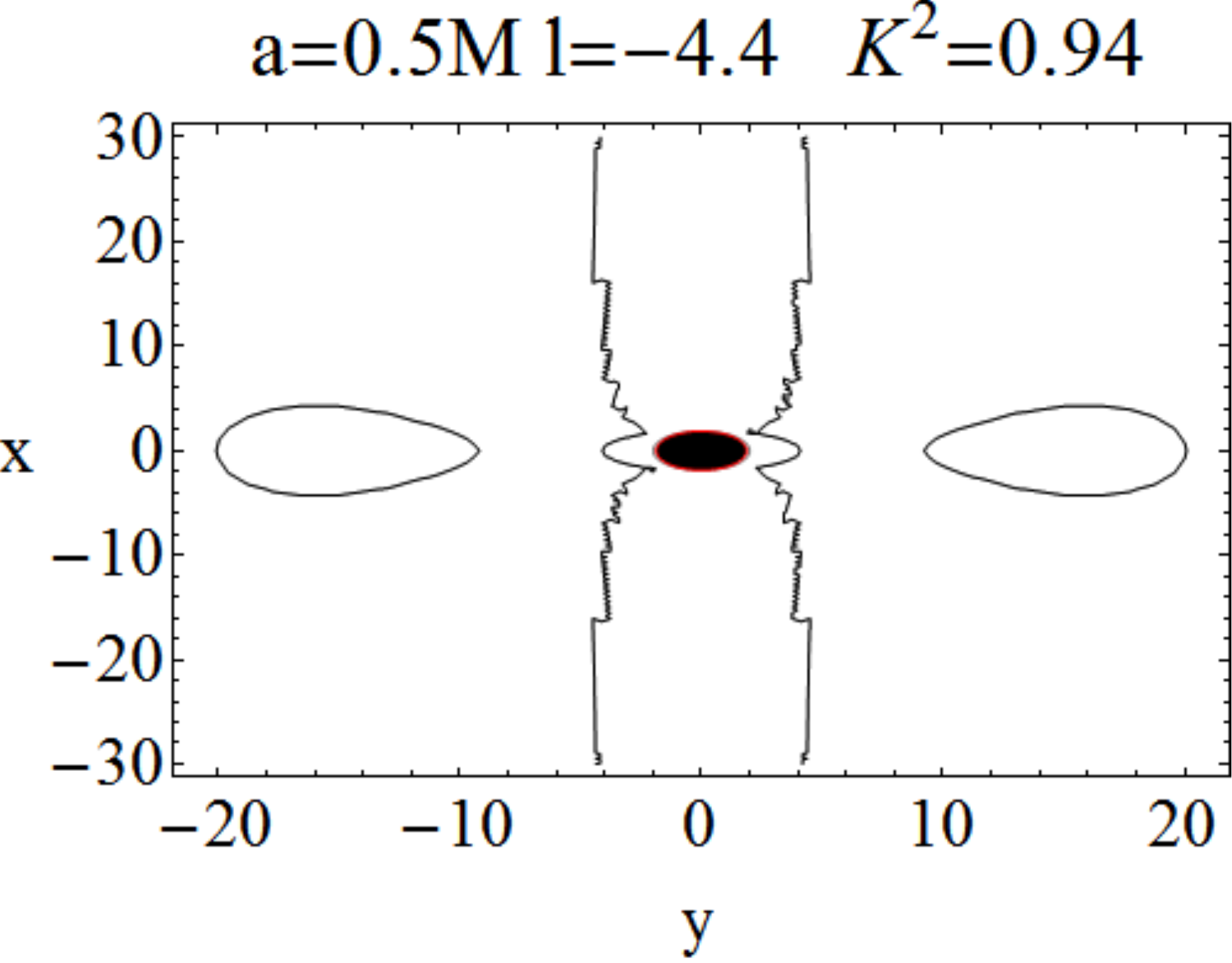}
\includegraphics[width=.2\textwidth]{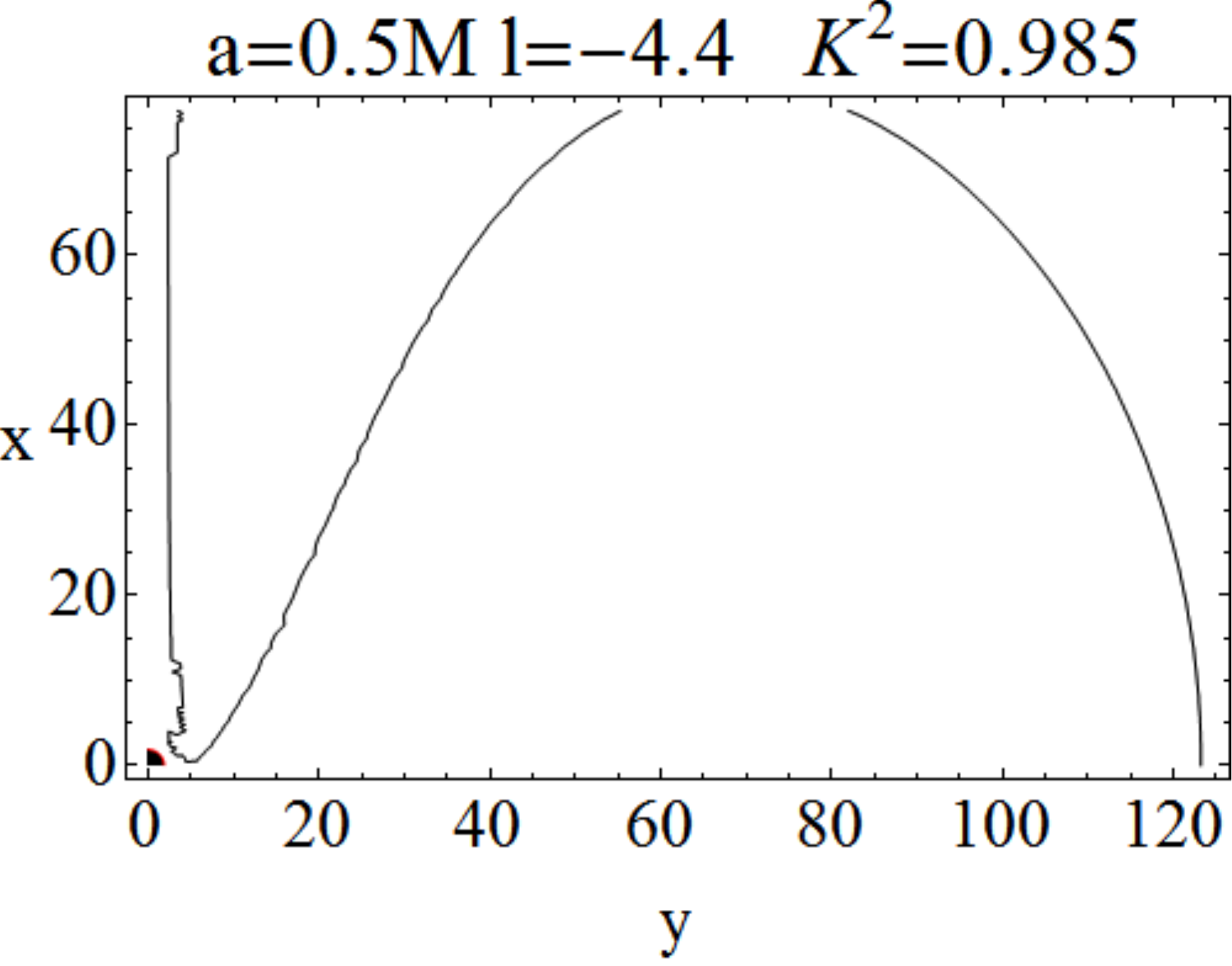}
\\
\includegraphics[width=.2\textwidth]{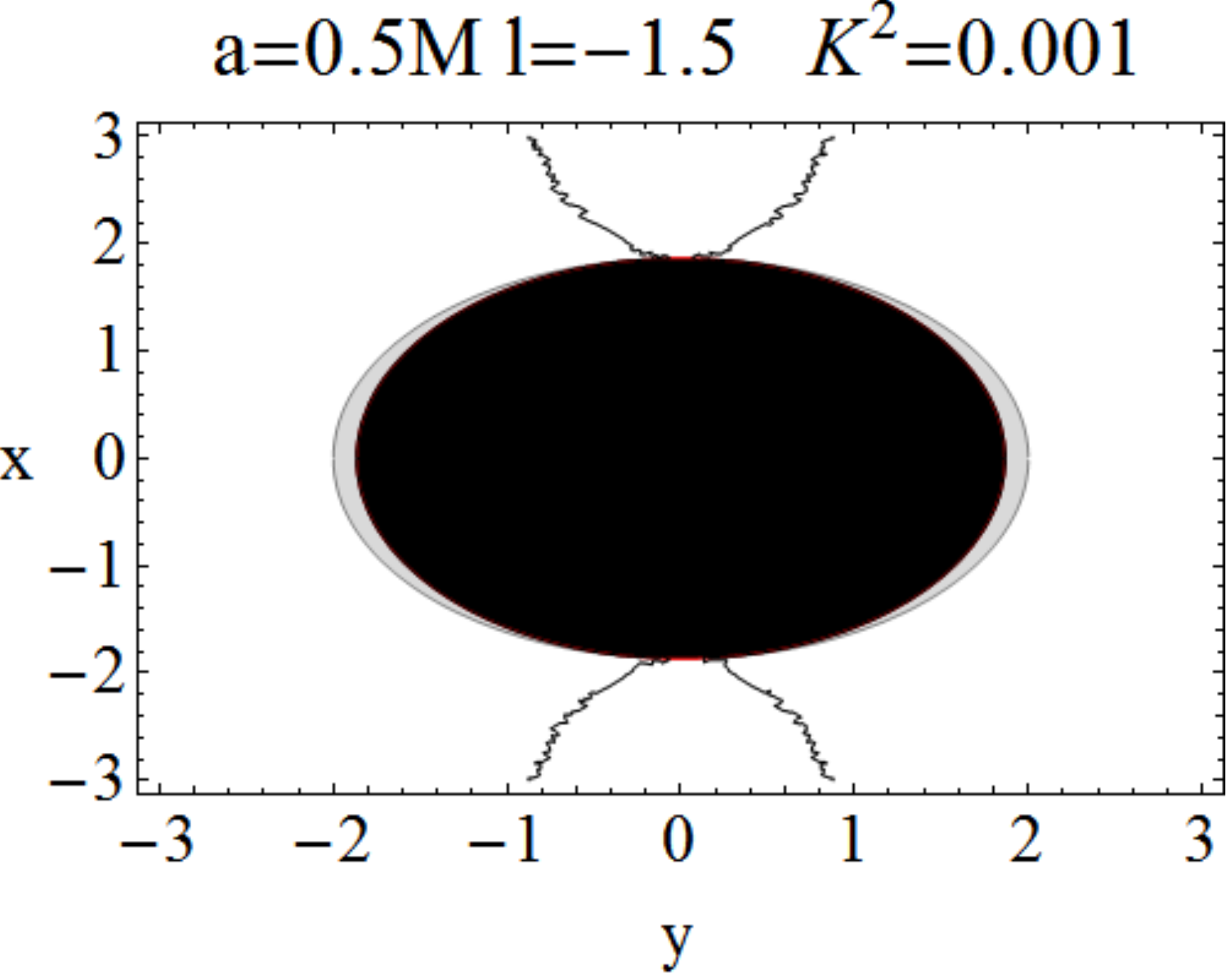}
\includegraphics[width=.2\textwidth]{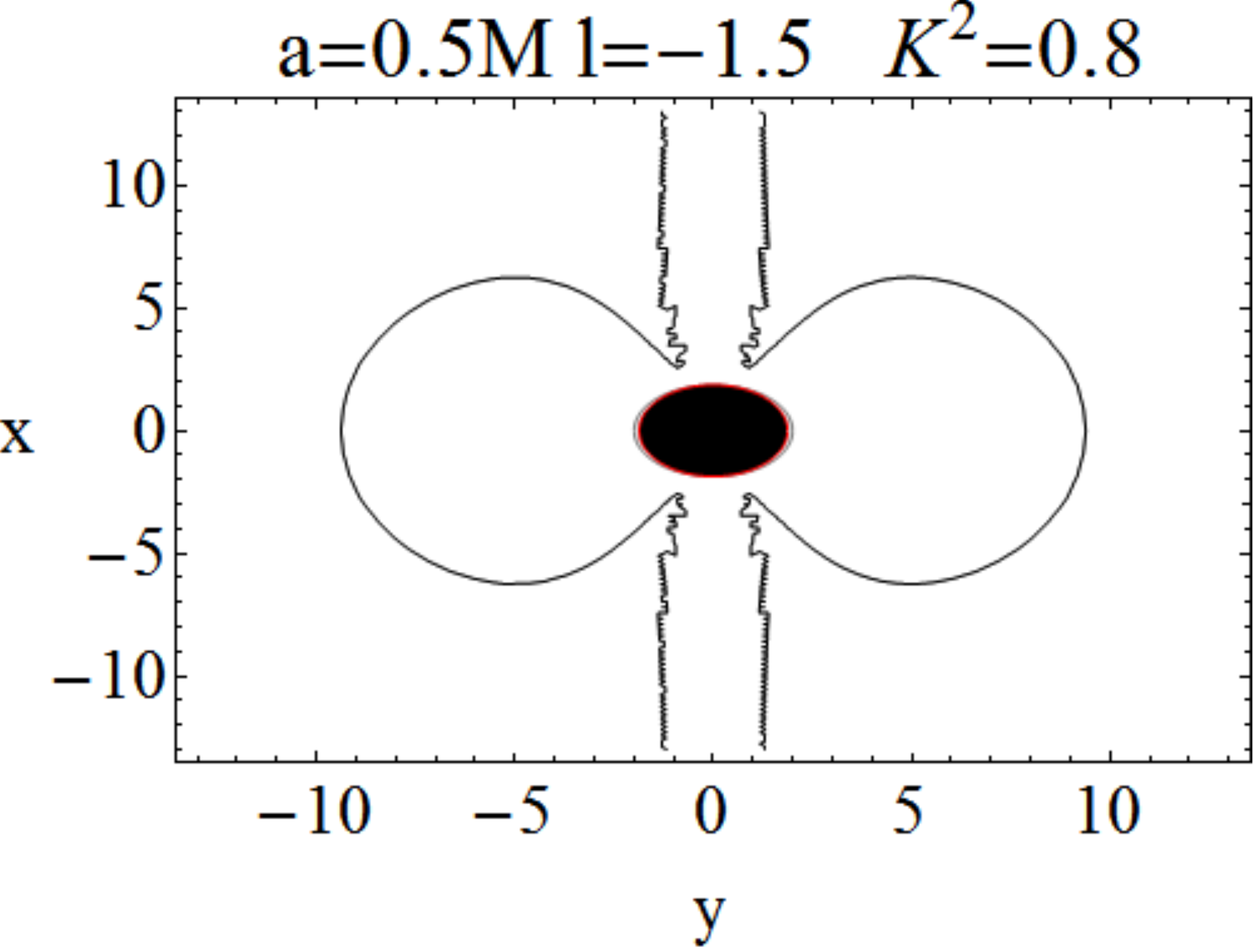}
\includegraphics[width=.2\textwidth]{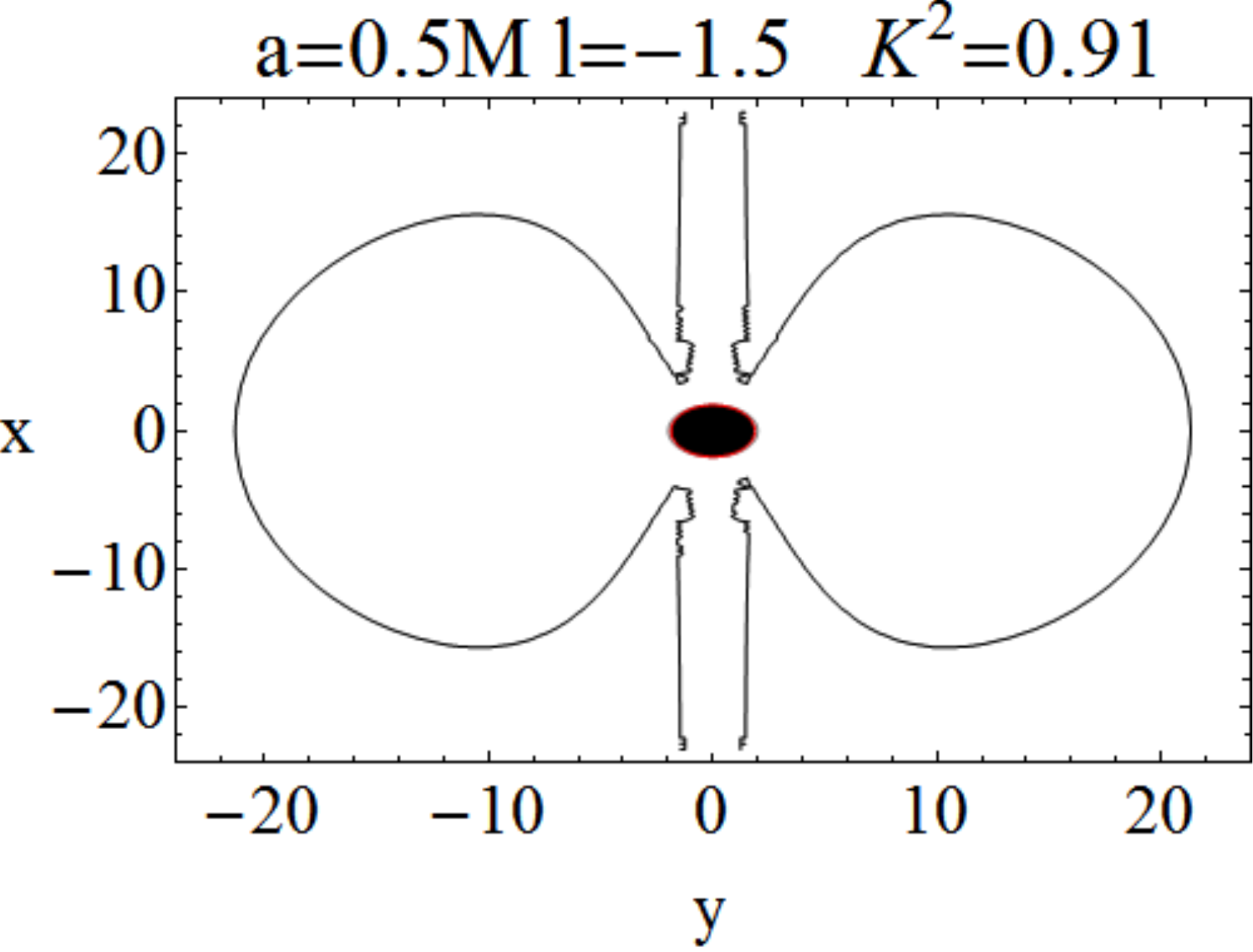}
\caption{Closed Boyer surfaces (sequences $\mathfrak{B}_{K}\equiv\mathfrak{B}_{\mathbf{p}}/\Sigma_{\ell}$) for fixed ranges of the fluid angular momentum $\ell=\ell(a)<0$ in units of mass $M$ and the potential-parameter $K_i=K_i(a,\ell)\in]0,1[$ in different spacetimes. It is:
$a=9/10M\in\mathbf{BHVIII}$  and $\ell_{\gamma}^+= -6.83232$ and $(K_i)^2\in\{0.977871,0.000828153\}$
Second row: $\ell_f^+(r_{b}^+)=  -4.75681$ and $(K_i)^2\in\{0.00146791,0.951018\}$. Third row: $a/M=5/10\in\mathbf{BHIII}$, $\ell_f^+(r_{b}^+)= -4.44949$ and $\ell_f^+(r_{lsco}^+)=-4.06784$ $(K_i)^2\in\{0.000172721,0.932708,0.984353\}$, Bottom row: $a/M=5/10\in\mathbf{BHIII}$, $(K_i)^2\in\{0.00102636, 0.900794\}$}\label{Fig:Anamb-ly}
\end{figure}
\begin{figure}[t]
\includegraphics[width=.2\textwidth]{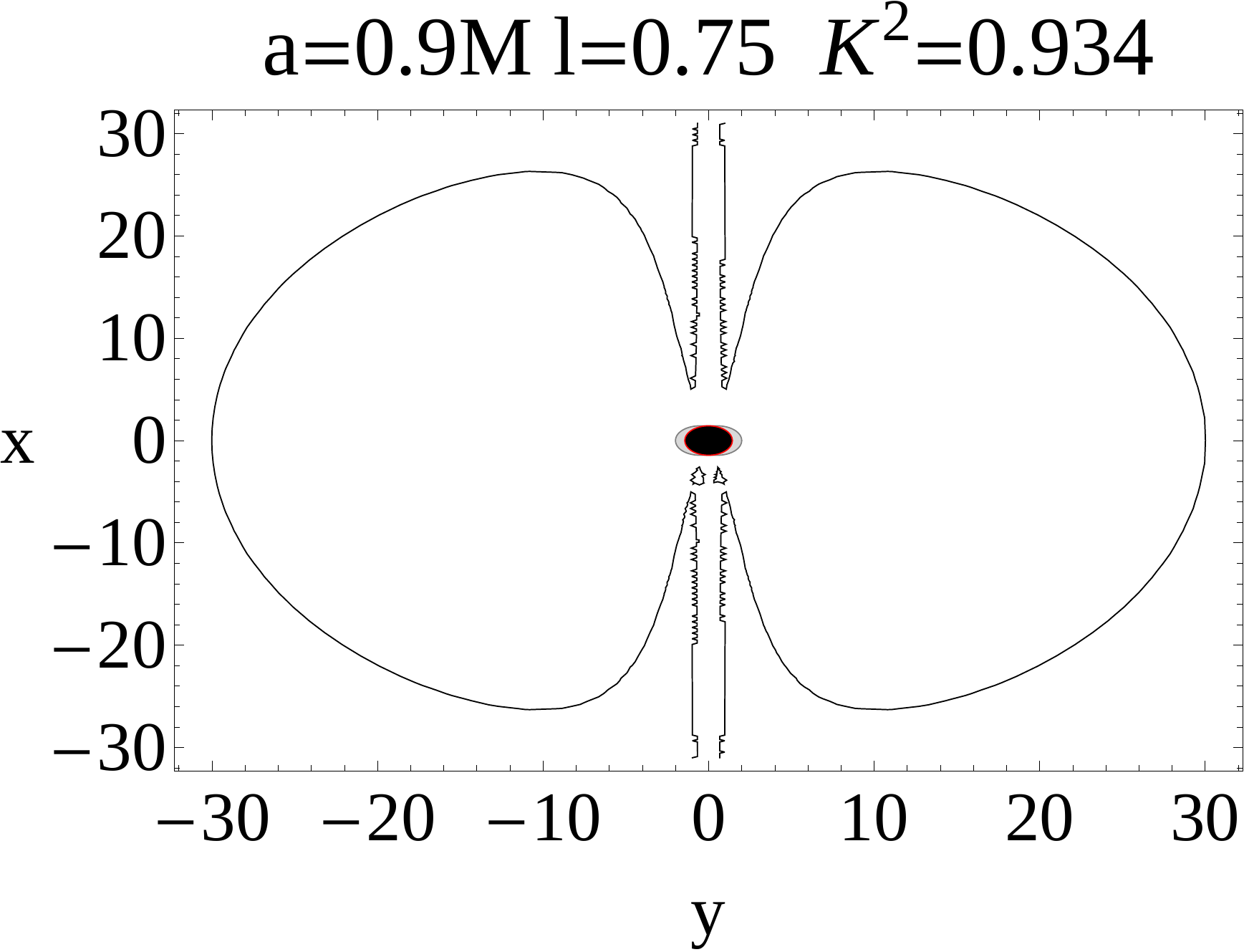}
\includegraphics[width=.2\textwidth]{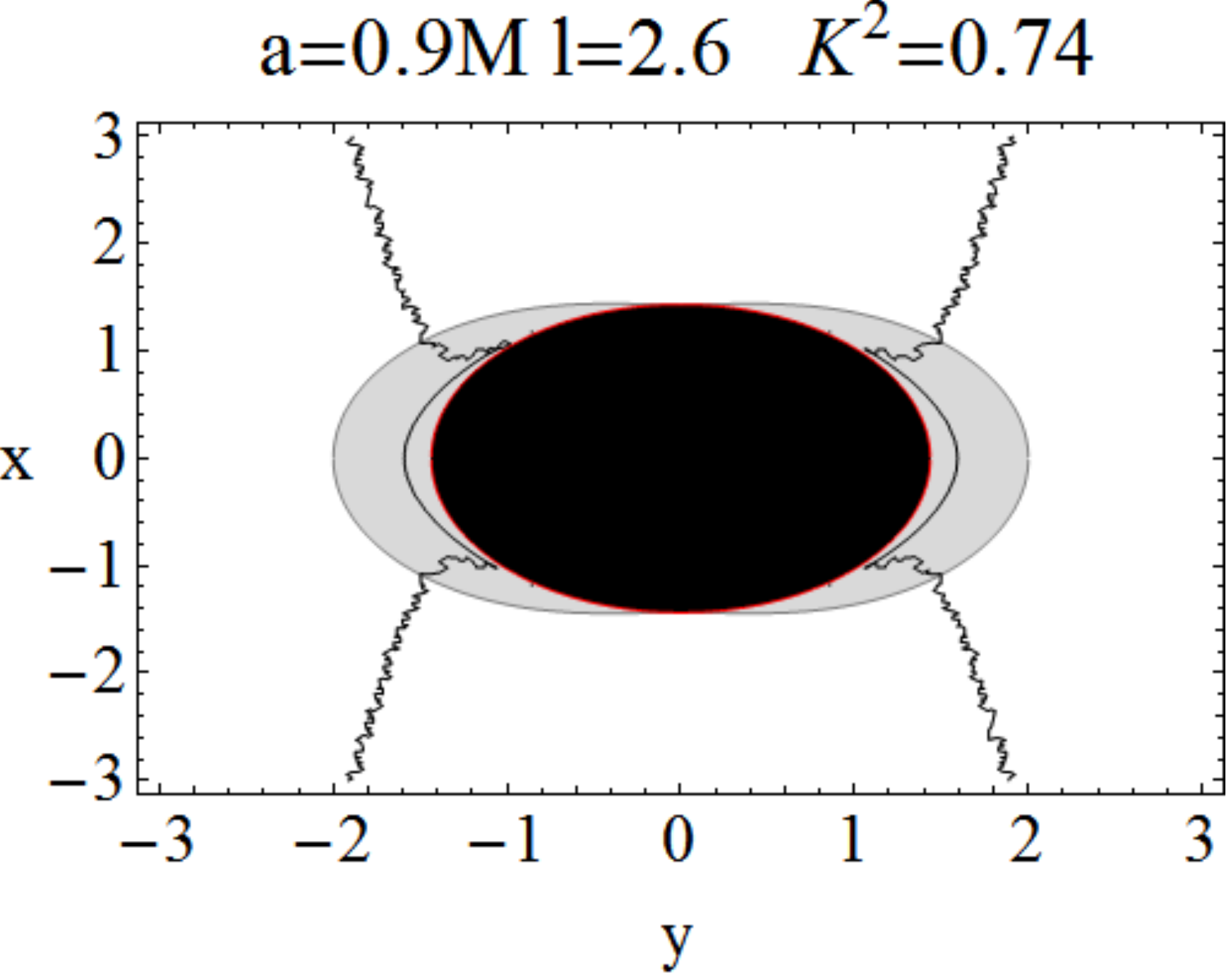}
\includegraphics[width=.2\textwidth]{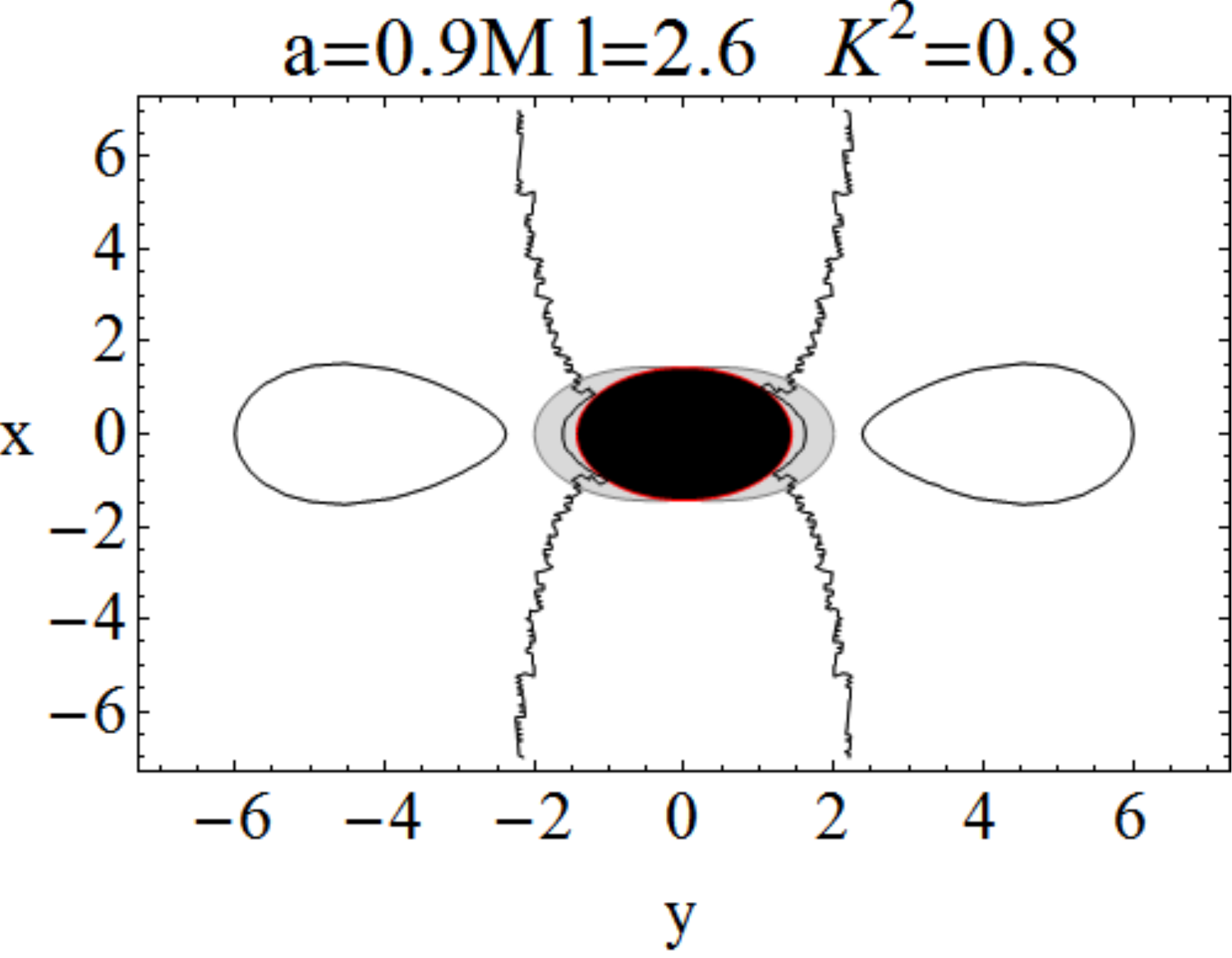}
\includegraphics[width=.2\textwidth]{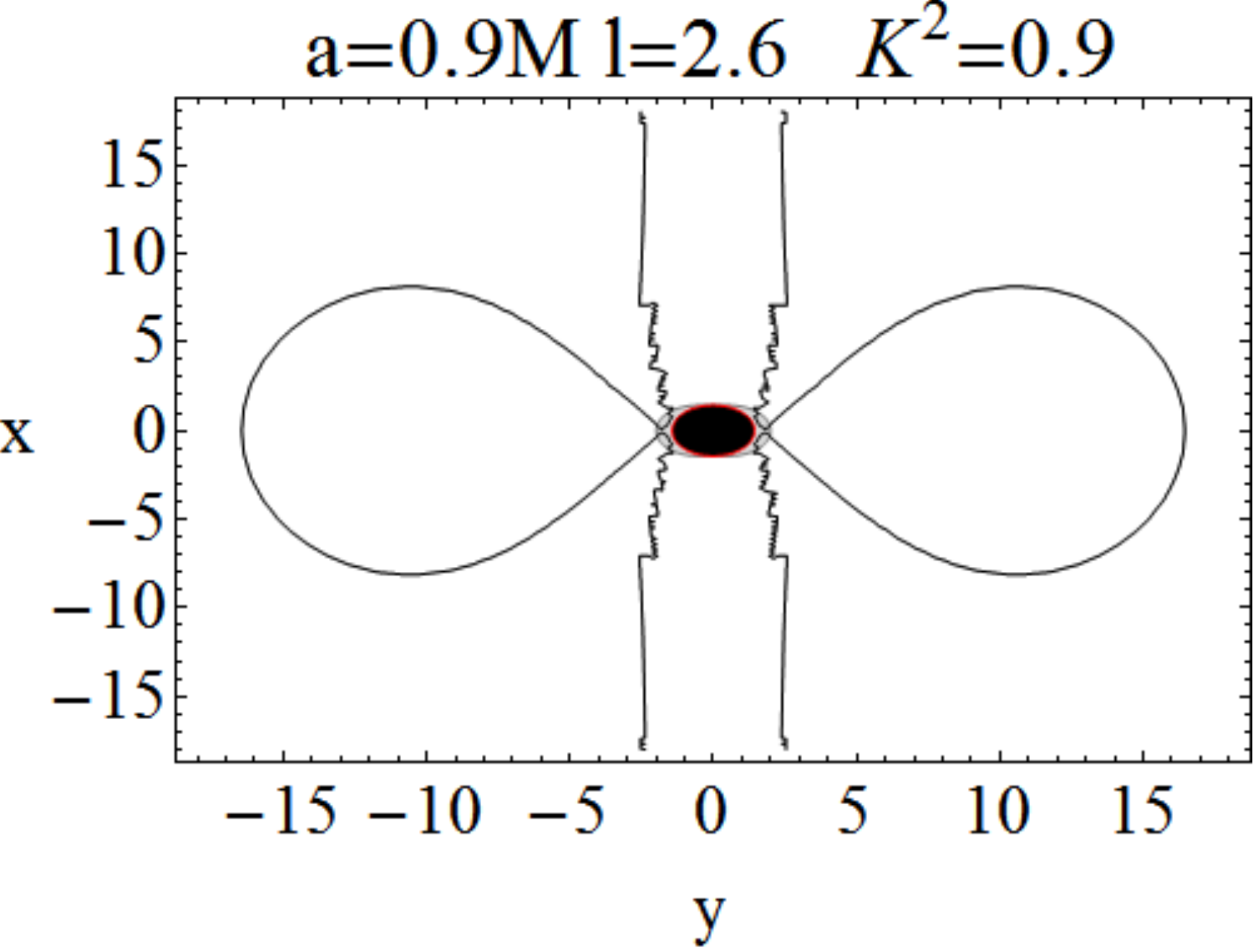}
\\
\includegraphics[width=.2\textwidth]{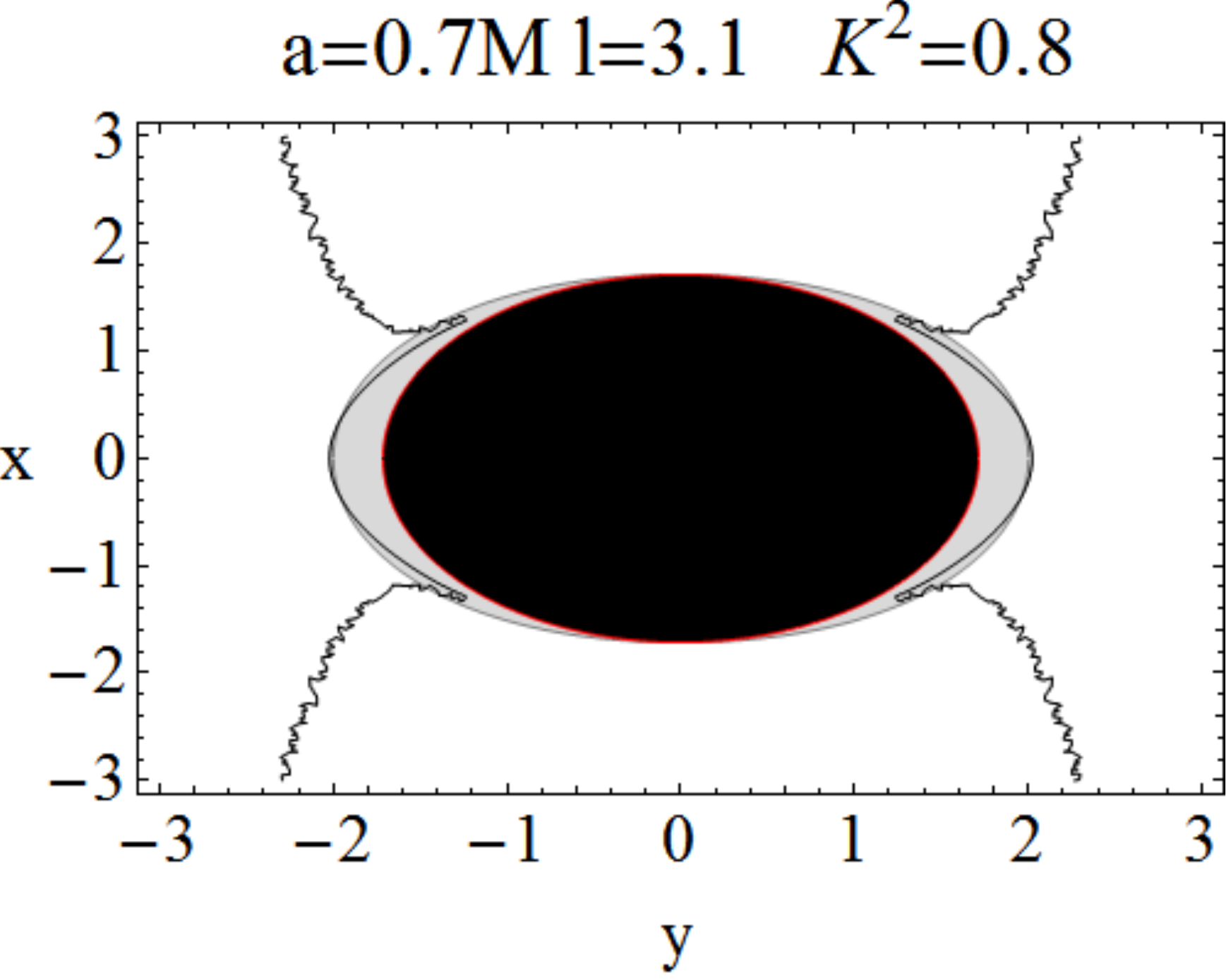}
\includegraphics[width=.2\textwidth]{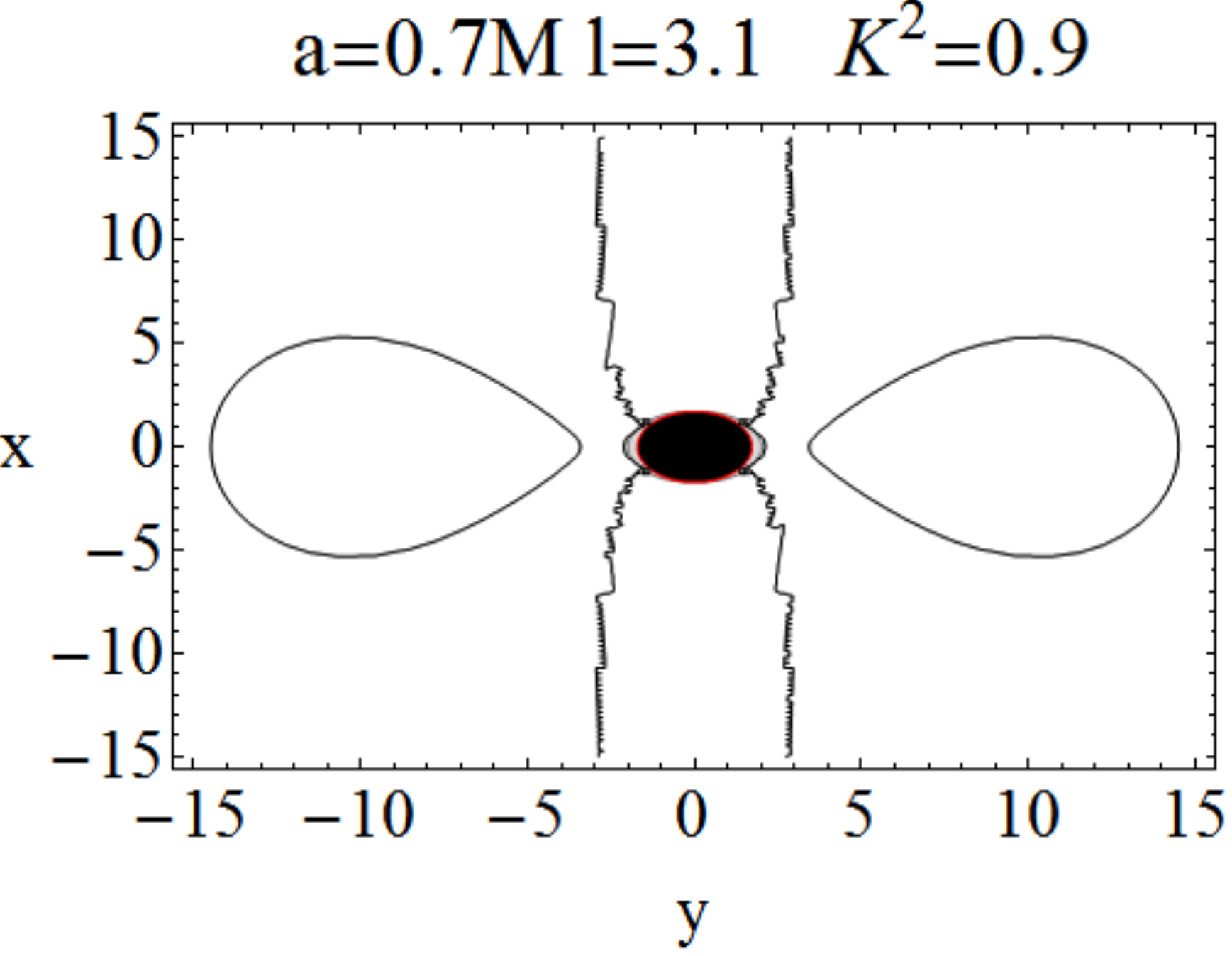}
\includegraphics[width=.2\textwidth]{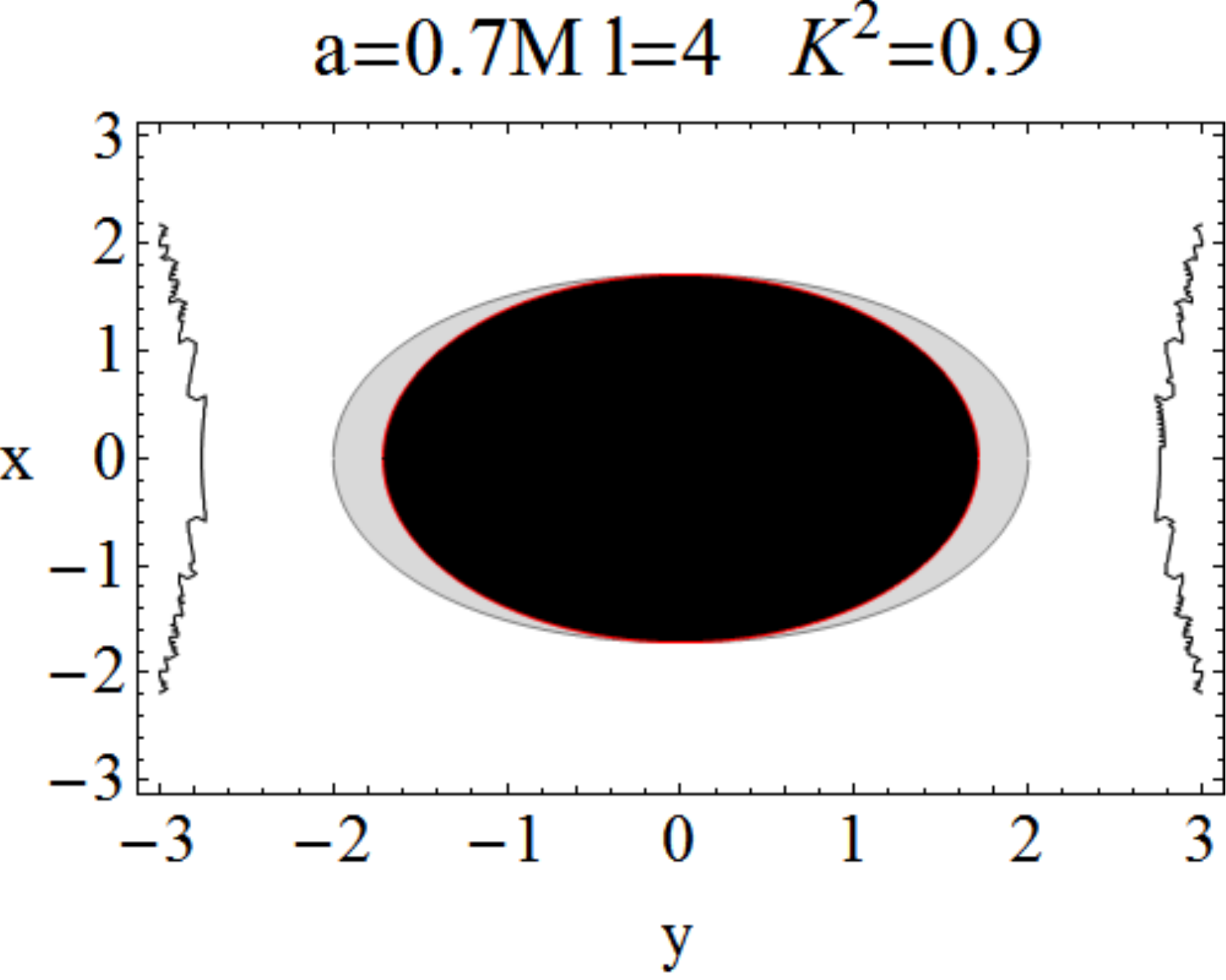}
\includegraphics[width=.2\textwidth]{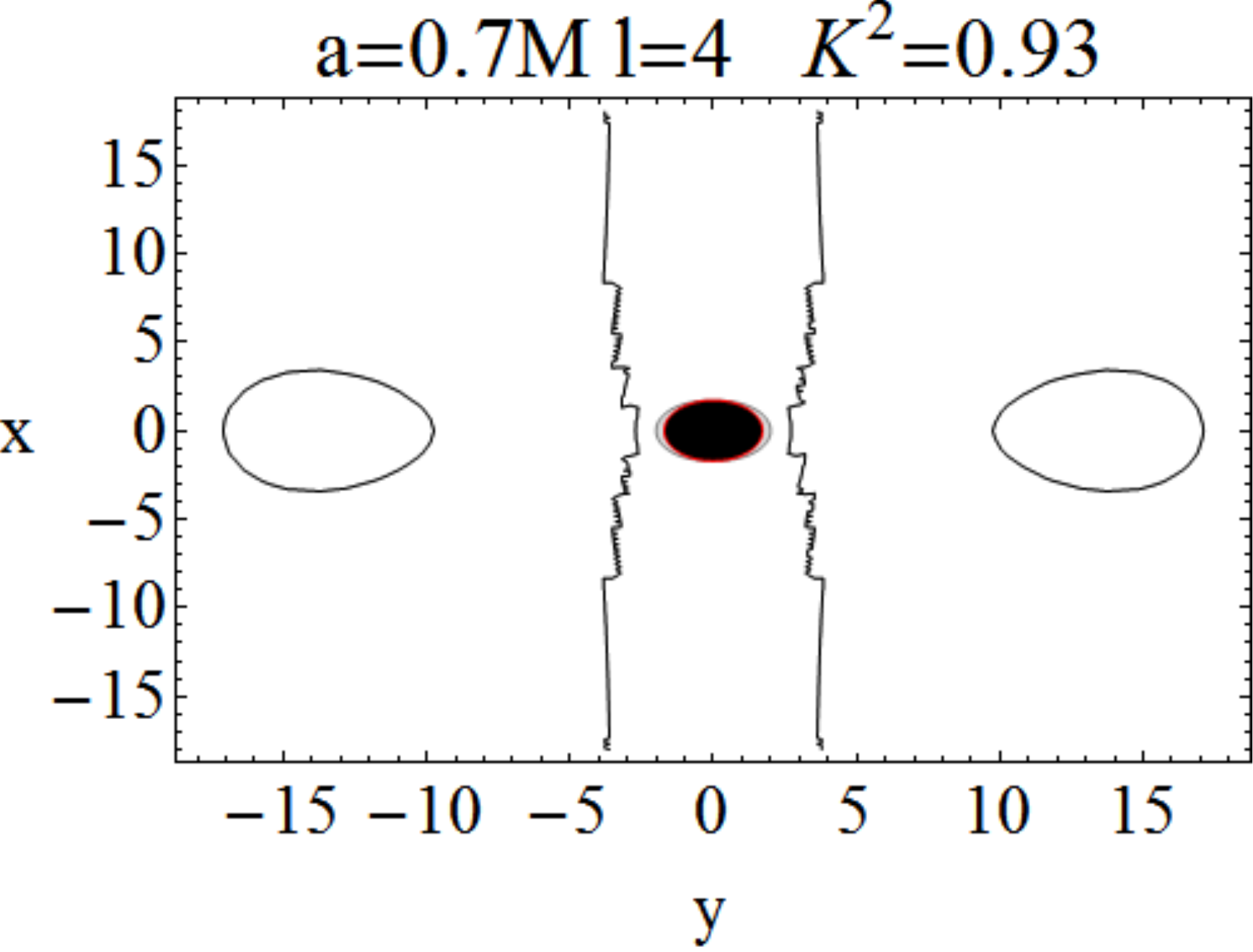}
\\
\includegraphics[width=.2\textwidth]{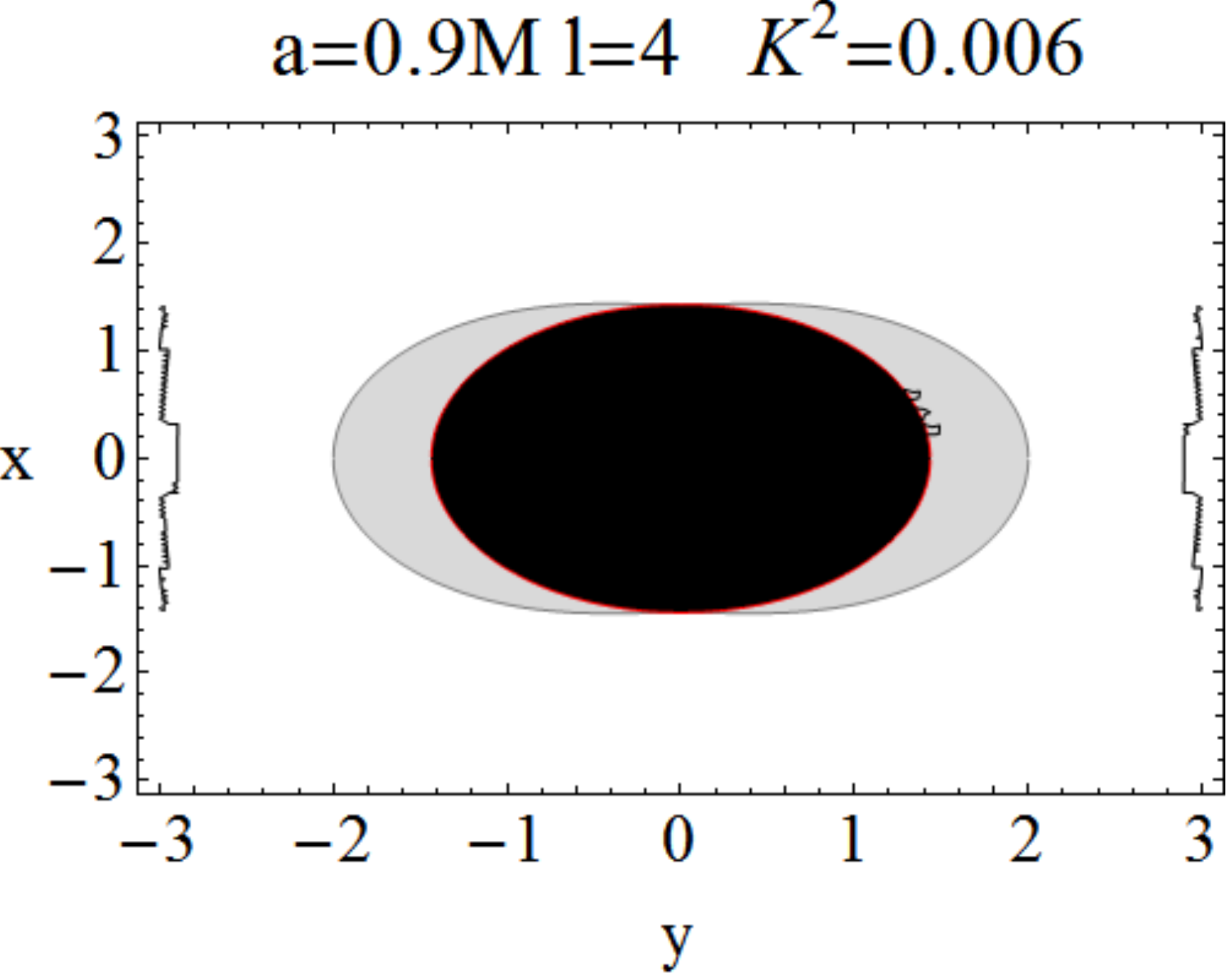}
\includegraphics[width=.2\textwidth]{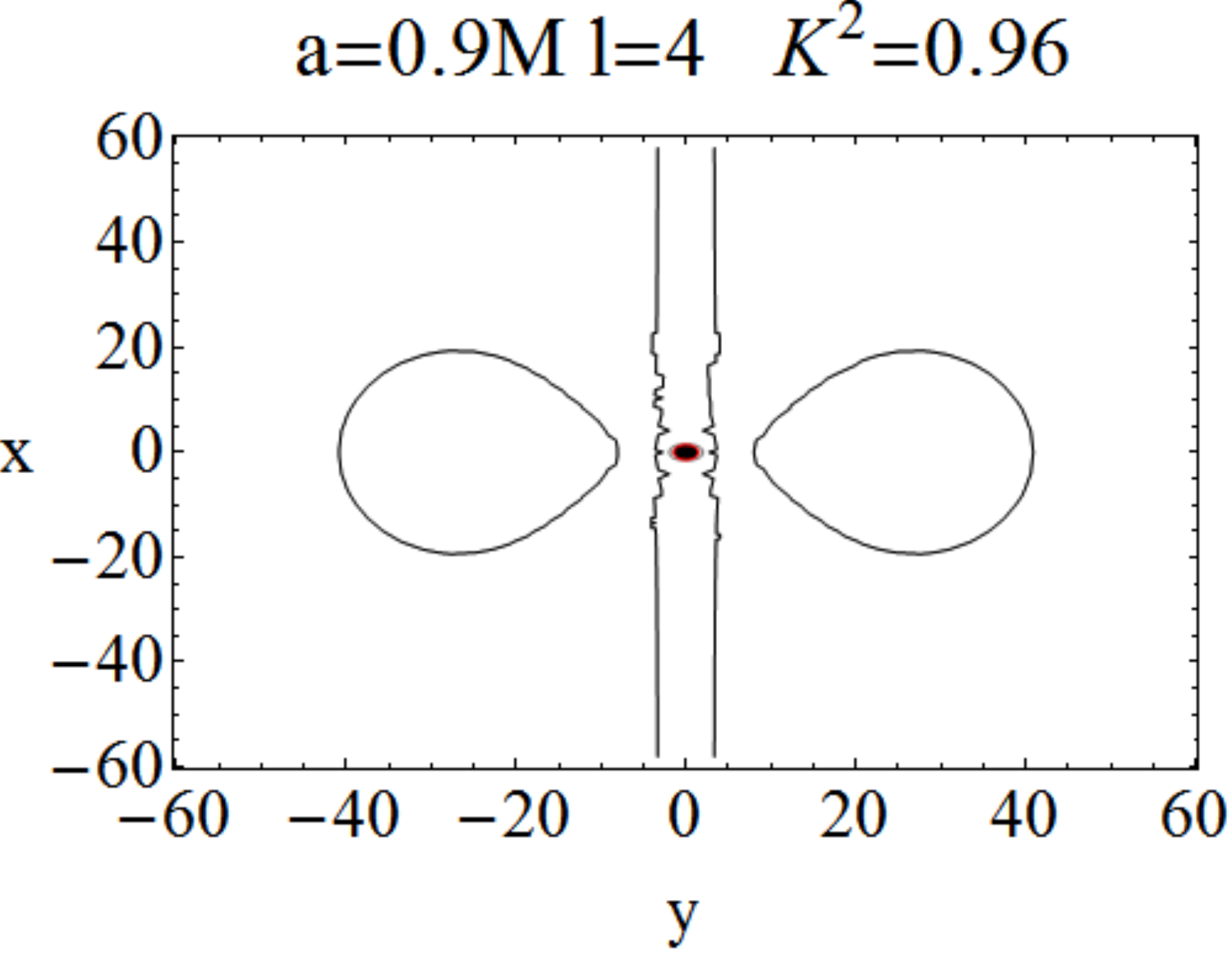}
\includegraphics[width=.2\textwidth]{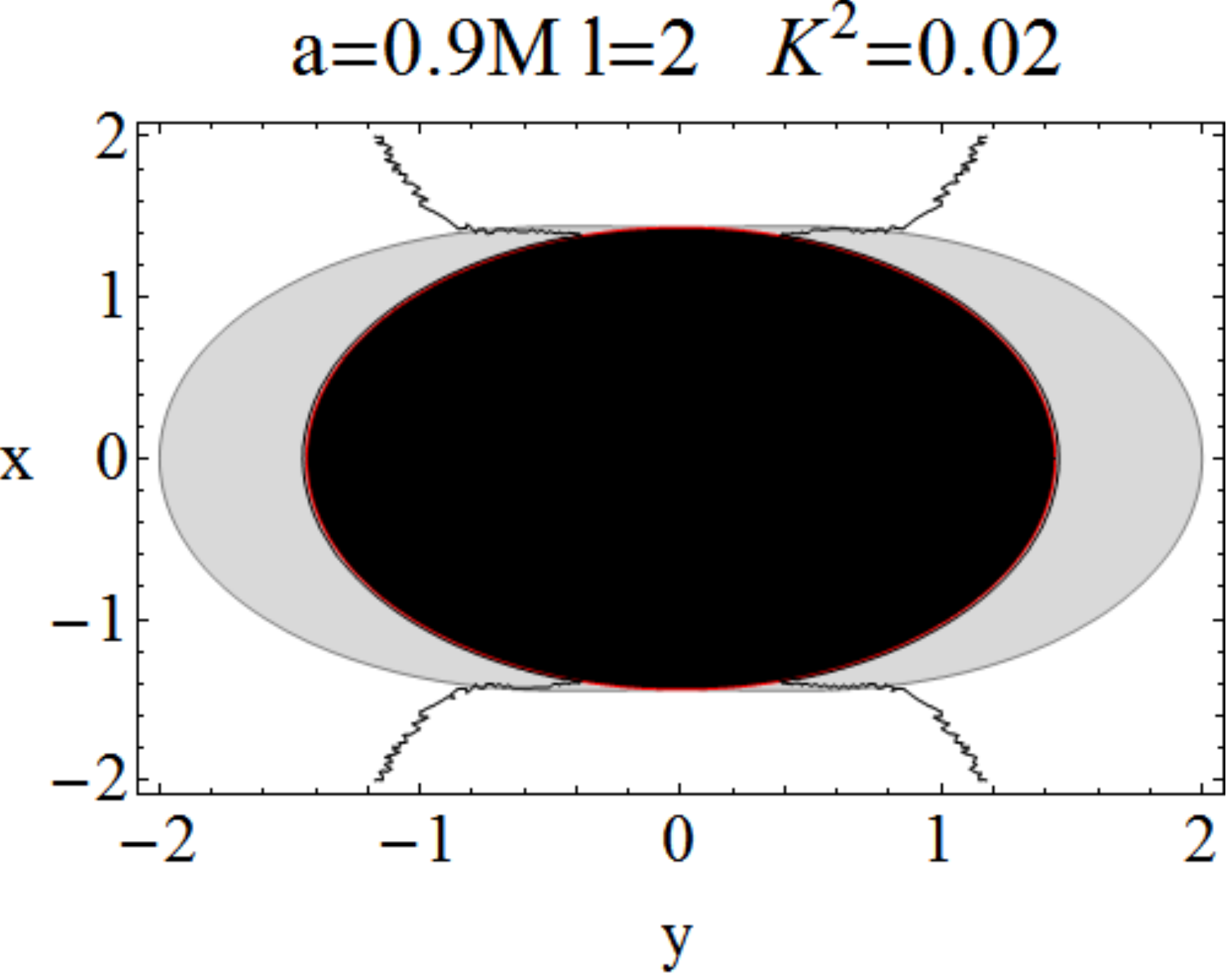}
\includegraphics[width=.2\textwidth]{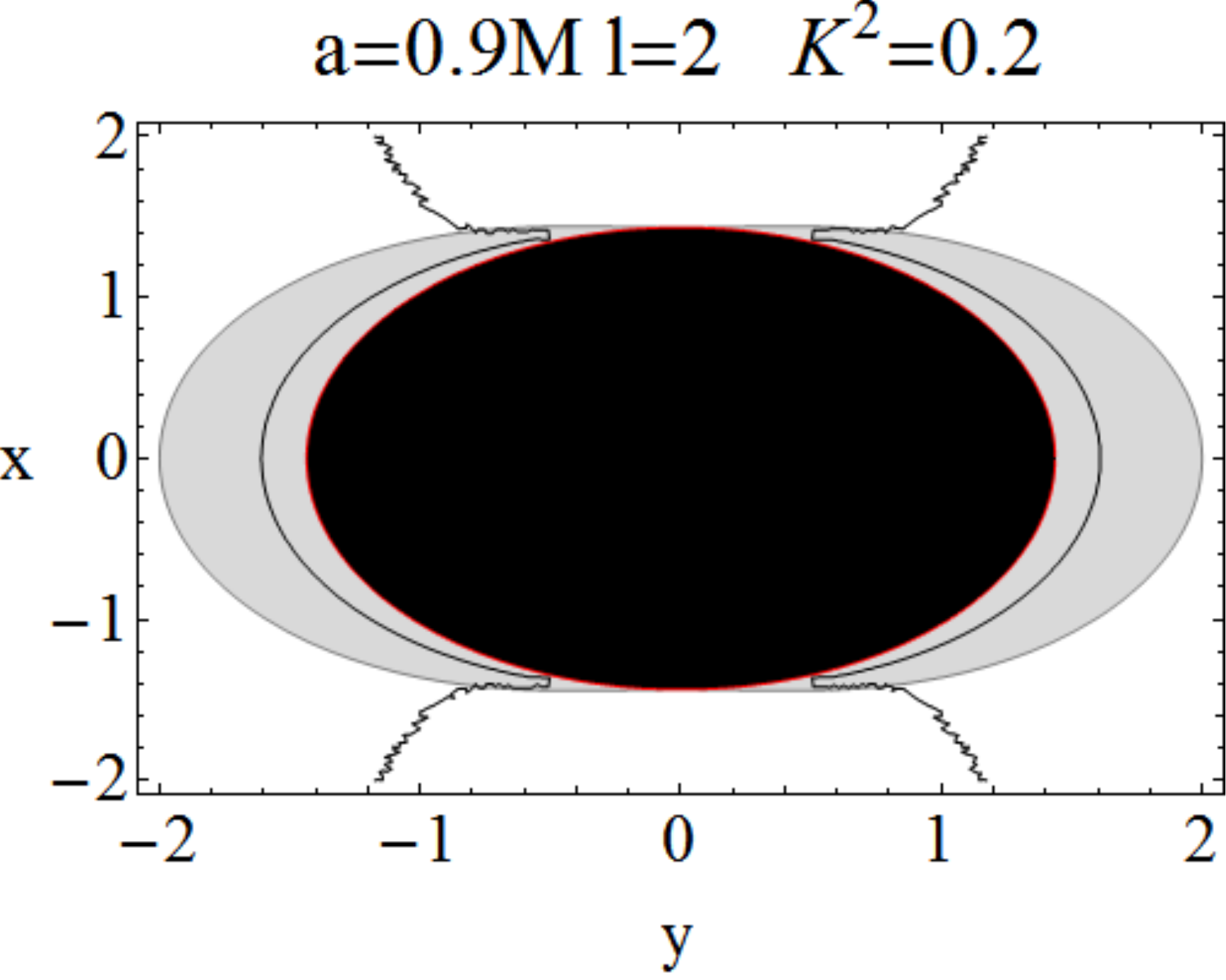}
\caption{Closed Boyer configurations  for fixed ranges of the fluid angular momentum $\ell=\ell(a)>0$ in units of mass $M$ and the  parameter $K_i=K_i(a,\ell)\in]0,1[$ for spacetime
$a=9/10M\in\mathbf{BHVIII}$  and
$a=7/10M\in\mathbf{BHV}$.
}\label{Fig:Sei-f}
\end{figure}
\subsubsection{The evolution of the models for $K>1$ at $\ell$ fixed}\label{Subsec:KMlf}
In this section we consider the  the sequences $\mathfrak{B}_{K}\equiv\mathfrak{B}_{\mathbf{p}}/\Sigma_{\ell}$ for
 $K>1$.  There are no  closed configurations, and in general critical points are in $r\in]r_{\gamma}^{\pm},r_b^{\pm}[$ for corotating  and counterrotating matter respectively, see Sec.\il(\ref{Sec:KM1ana}). At $K>1$ only maximum of the effective potential, or minimum of the pressure, are possible. These surfaces, however, could shape jets   crossing  the equatorial plane in one or more points.

\textbf{Corotanting fluids ($\ell>0$)}
\begin{description}
\item[\textbf{Region I}]
for $\ell\in]
 -\ell_{\mu}^\Pi,a[$ it is: \textbf{1.} $(K\in]1,K_2[, y_{23})$, or \textbf{2.}
 $(K=K_2, {y_2})$ and for
  $\ell=a$: ir is $ (K>1, y_3)$.
This region corresponds to the solutions  with $\bar{\ell}<1$, where critical points cannot exist.
\item[\textbf{Region II}]
For  $\ell \in ]a,\ell_f^{\pm}(r_-)[$ it is $ (K>1, y_{23})$.
For the particular values   $\ell=\ell_f^{\pm}(r_-):$ it is   $(K>1,y_2)$.
In the range   $\ell\in]\ell_f^{\pm}(r_-),\ell_a^1]:  \quad (K>1,y_{23})$,
and finally for  $\ell\in]\ell_a^1,-\ell^{\Pi}_{\mu}[$ it is ,
 $\mathbf{1.}\quad (K\in]1,K_2[, y_{23}), \mathbf{2.} \quad (K_2, {y_2})$. A limit case for configurations of \textbf{Region II}   is the  non rotating background of the Schwarzschild geometry, some configurations at $a=0$ are for example in  Fig.\il(\ref{Fig:TSToti}). However the introduction of a spin for the attractor does not change qualitatively this structure for  the  \textbf{Region II}.
\item[\textbf{\underline{Region III}}]
For fluids with
 $\ell\in]\ell_b^-, \ell_{\gamma}^-[$, solutions are for $\mathbf{1.}\quad  (K\in]1,K_4[\;y_{23})$, and $\mathbf{2.} \quad (K_4,{y_2})$.  In this region  P-D configurations of the type $O_x$  are possible, see also Fig.\il(\ref{Fig:tema}).
 \item[ \textbf{Region IV}]
For corotating fluids with
  $\ell \in[\ell_{\gamma}^- ,\ell_f^{\pm}(r_+)[$, it is  \textbf{1. } $(K>1,y_{23})$,
while in the limit case
   $\ell_f^{\pm}(r_+)$, it is  $(K>1, y_3)$.
\item[\textbf{Region V}]
In this region we consider only fluid configurations with
 $\ell>\ell_f^{\pm}(r_+)$, where solutions are for  $(K>1,y_{23})$.
 \end{description}

\textbf{Counterrotating fluids $\ell<0$}
Here we focus on the case of counterrotating fluids and the sequences   $\mathfrak{B}_{K}$, two regions of values of $\ell$  need to be considered (see also $\mathfrak{B}_{\bar{\ell}}$ sequences considered in Sec.\il(\ref{Subsubsec:lbarm1})).
\begin{description}
\item[\textbf{Region I}:]
$\ell \leq
    \ell_{\gamma}^+$,  \textbf{1.} $(K > 1, y_{23})$
  \item[\textbf{\underline{Region II}}:]
    $\ell\in]\ell_{\gamma}^+,\ell_b^+[$:   \textbf{1.} $ (K\in]1, K_4[,y_{23})$ \textbf{2.} $ (K_4, y_2)$.  In this region  open crossed, $O_x$   are possible.
    \end{description}
The case $a\neq0$ is not qualitatively
    different from the  situation for  a Schwarzschild  attractor, however a careful analysis should take into account the greater or lesser collimation of the open surfaces with respect to the rationalized angular momentum $ \bar{\ell}$,  and in  an extended P-D model  in GRMHD  the    influence of the magnetic field for  the corotating and  counterrotating configurations should be taken into account as well,
 for open solutions could play an important role in the jets analysis even   where there is also a magnetic contribution \cite{Meier,Abramowicz:2011xu}.
\subsection{Some general considerations on the  limiting  cases}\label{Subsec:limitcases}
\subsubsection{Fluid configurations in the   Schwarzschild spacetime}\label{subsubsec:Schw}
We focus now  on the case of non rotating attractors. The P-D models in the   Schwarzschild spacetime have  been extensively analyzed   for example in   \cite{PuMonBe12}, here we  reproduce the analysis in Sec.\il(\ref{Sec:K.L}) for the limiting case $a=0$.
It is convenient to  introduce the  angular momentum $\ell_{\mathfrak{K}}^{\pm}(K)$
\bea
\frac{\ell_{\mathfrak{K}}^{\pm}}{M}\equiv\frac{\sqrt{27+\frac{1}{1-K^2}-\frac{8}{K^2}
\pm\sqrt{\frac{(9K^2-8)^3}{(K^2-1)^2 K^2}}}}{\sqrt{2}}
\eea
 and the ${K}_i(\ell)$ functions or
 \bea
{K}_a\equiv\sqrt{\frac{\ell^2-36M^2-2 \beta_1 \sin\left(\frac{1}{3}
\arcsin\alpha_1\right)}{3 \ell^2-81M^2}},
\quad
{K}_b\equiv\sqrt{\frac{\ell^2-36M^2+2 \beta_1 \cos\left(\frac{1}{3}
\arccos\alpha_1\right)}{3 \ell^2-81M^2}},
\eea
and \( K_a>K_b\) in $\ell>3\sqrt{3/2}M$  and  \( K_a=K_b\) in $\ell=3\sqrt{3/2} M$ with
\bea
\alpha_1&\equiv&\frac{\left(3^9 \,8 \right)M^8-\left(108 \sqrt{2}\right)^2\ell^2M^6+\left(6
\sqrt{51}\right)^2 {\ell^4}M^4-72 {\ell^6} M^2+{\ell^8}}{\ell^2 \beta_1^3},
\quad
\beta_1\equiv(\ell^2-27M^2) \sqrt{\frac{72M^4+(\ell^2-24M^2)^2}{(\ell^2-27M^2)^2}},
\eea
introduced and studied in \cite{PuMonBe12} they are special  cases of  $\ell_K^i(a, K)$ and $K_i(a,\ell)$ for $a=0$.
As it is  $a=0$ there is no need to distinguish between corotating and  counterrotating fluid matter and we can summarize this special case as follows:

\textbf{Fixed orbital angular momentum $\ell$}
We consider the range $K\in]0,1[$ and the evolutive sequences $\mathfrak{B}_{	 K}\equiv\mathfrak{B}_{\mathbf{p}}/\Sigma_{\ell}$ thus we can compare this case with the analysis in Sec.\il(\ref{subsub:corot.mo}) it is then:
\begin{description}
\item[]
For
$\ell/M\in]0, 3 \sqrt{{3}/{2}}]$ solutions are for  $(K\in]0,1[, y_1)$, Fig.\il(\ref{Fig:SAnalogico}).
\item[]
For
$\ell/M\in]3 \sqrt{{3}/{2}},4[$: it is for \textbf{1.} $(K\in]0,K_{b}[,{y_1})$, and  \textbf{2.}
$({K_{b}},{y_{12}})$, or  \textbf{3.} $(K\in]K_{b},K_{a}[, y_{123})$ and \textbf{4.}
$(K_{a},y_{13}$. Finally   \textbf{5.}
 $(K\in]K_{a},1[, {y_1})
$.
\begin{figure}
\centering
\begin{tabular}{cc}
\includegraphics[scale=.13]{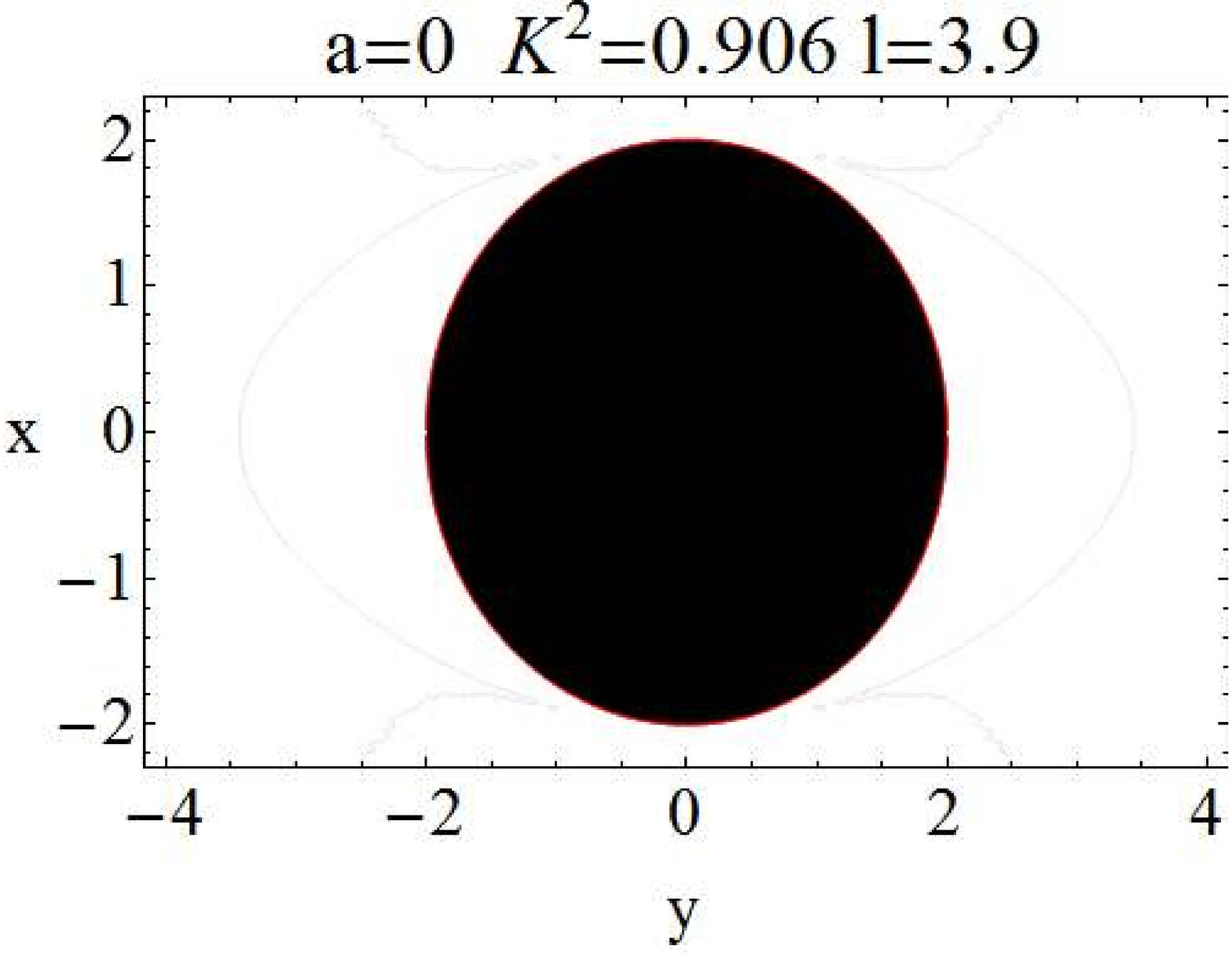}
\includegraphics[scale=.13]{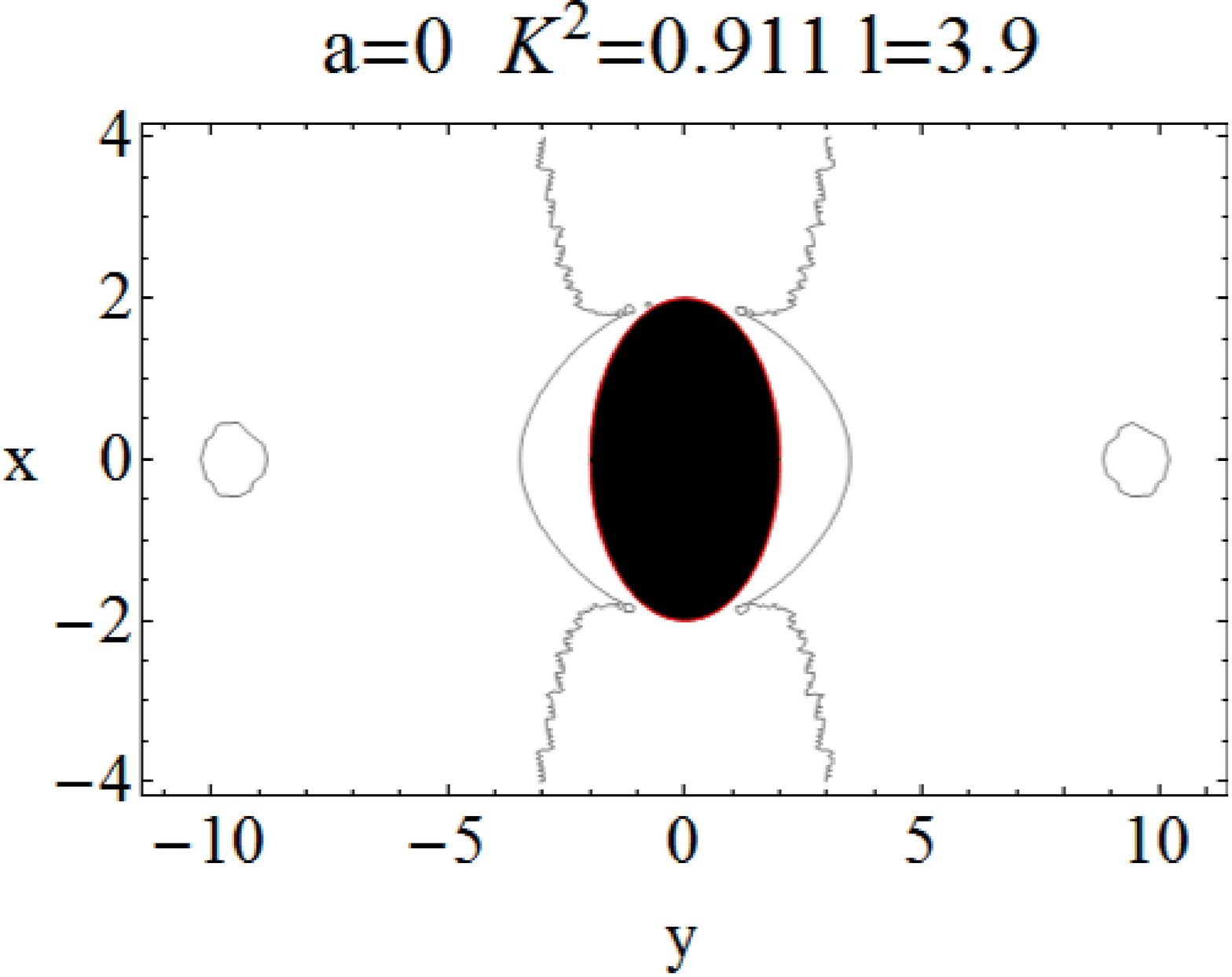}
\includegraphics[scale=.13]{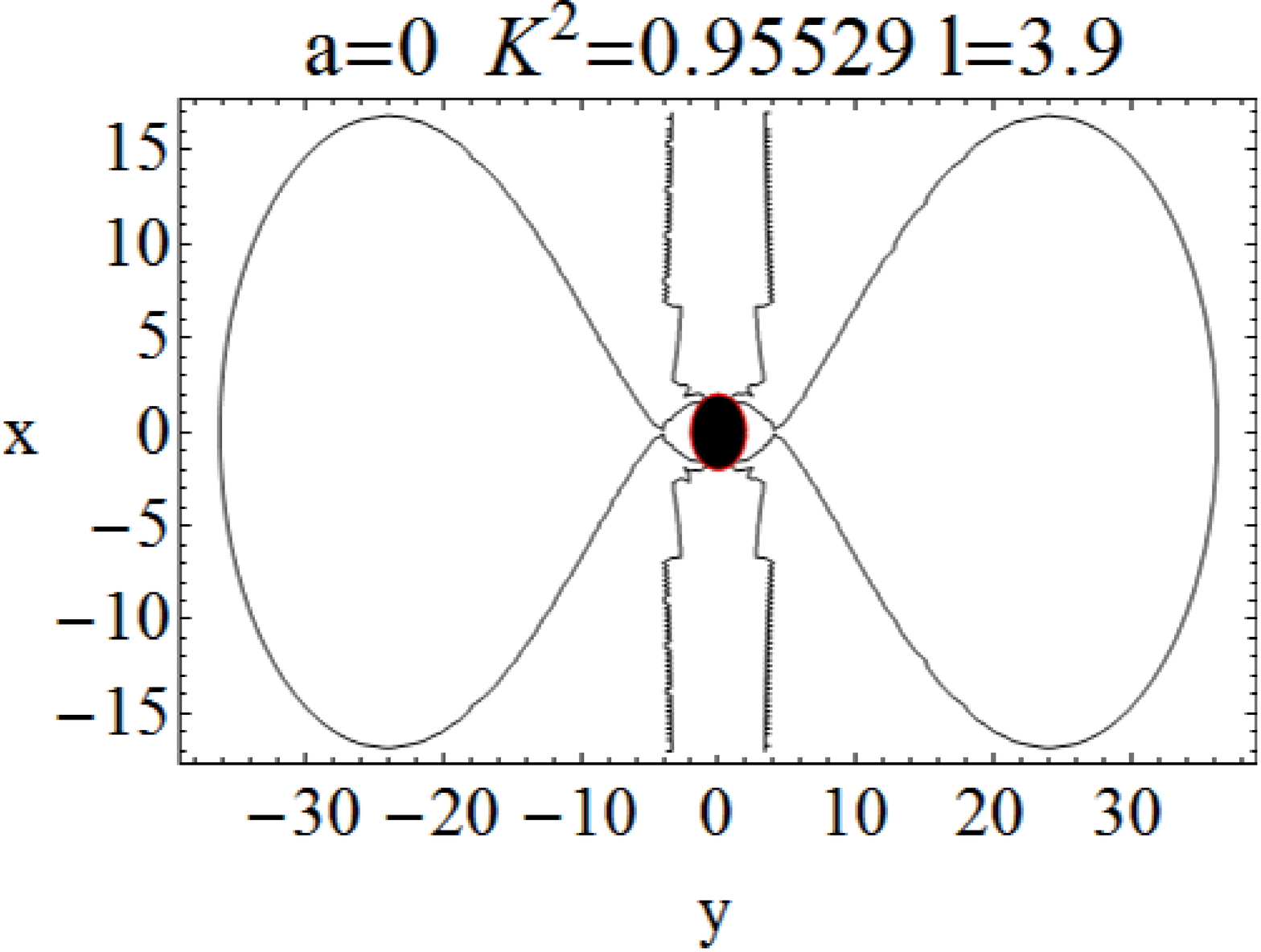}
\includegraphics[scale=.13]{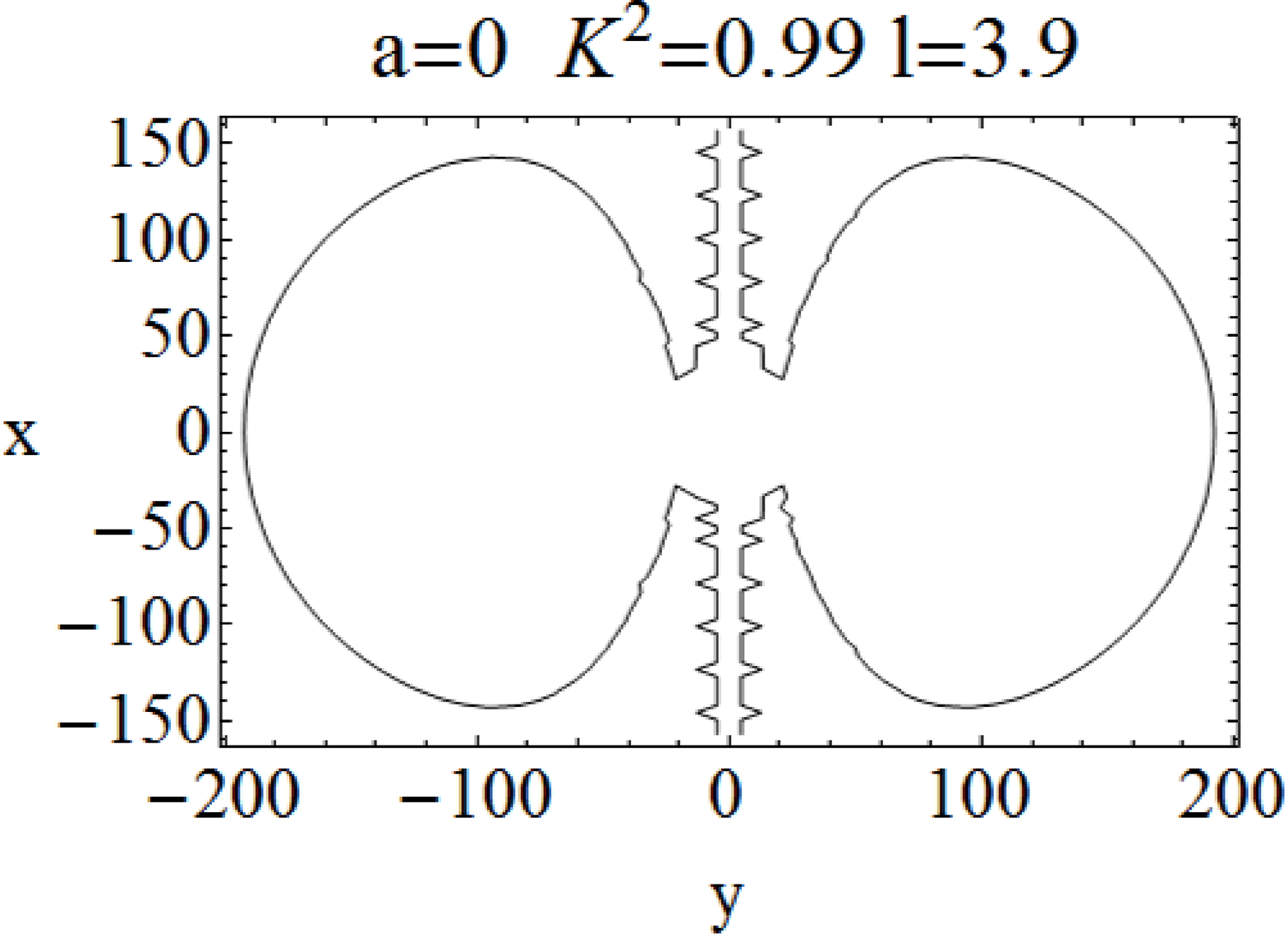}
\end{tabular}
\caption{{The Schwarzschild case: sequence $\mathcal{B}_{K}$ in  $\ell/M\in]3 \sqrt{{3}/{2}},4[$. It is  $\ell=3.9M$ in units of mass $M$ and $K^2_{i}=\{0.95529,  0.910634\}\quad i\in\{a,b\}$. The angular momentum is in units of mass $M$.}}\label{Fig:TSCrash}
\end{figure}
See Figs.\il(\ref{Fig:TSCrash}). 
At fixed $\ell$ as $K$ increases the
 ${B}$-configurations will form a nucleus of Boyer thin disk  quite far from the black hole increasing then  in thickness  and reconciling to the accretion configuration, to recreate again  the configuration ${B}$ i.e. one could consider the sequence
$\mathfrak{B}_{K}=[{B}, {C}, {C}_{x}, {B}]$.
\item[] For
$\ell\in[4, 3 \sqrt{3}]$: \textbf{1.} $(K\in]0,{K_{b}}[, {y_1})$, \textbf{2.} $(K_{b}, y_{12})$,  \textbf{3.} $(K\in]K_{b},1[, {y_{123}})$. See Figs.\il(\ref{Fig:SAnalogico}). In this case as in the previous region it is $\mathfrak{B}_K=[{O}, {B}, {C}]$
 and the surfaces along the axis  stretch on the equatorial plane.
\begin{figure}
\centering
\begin{tabular}{cc}
\includegraphics[scale=.2]{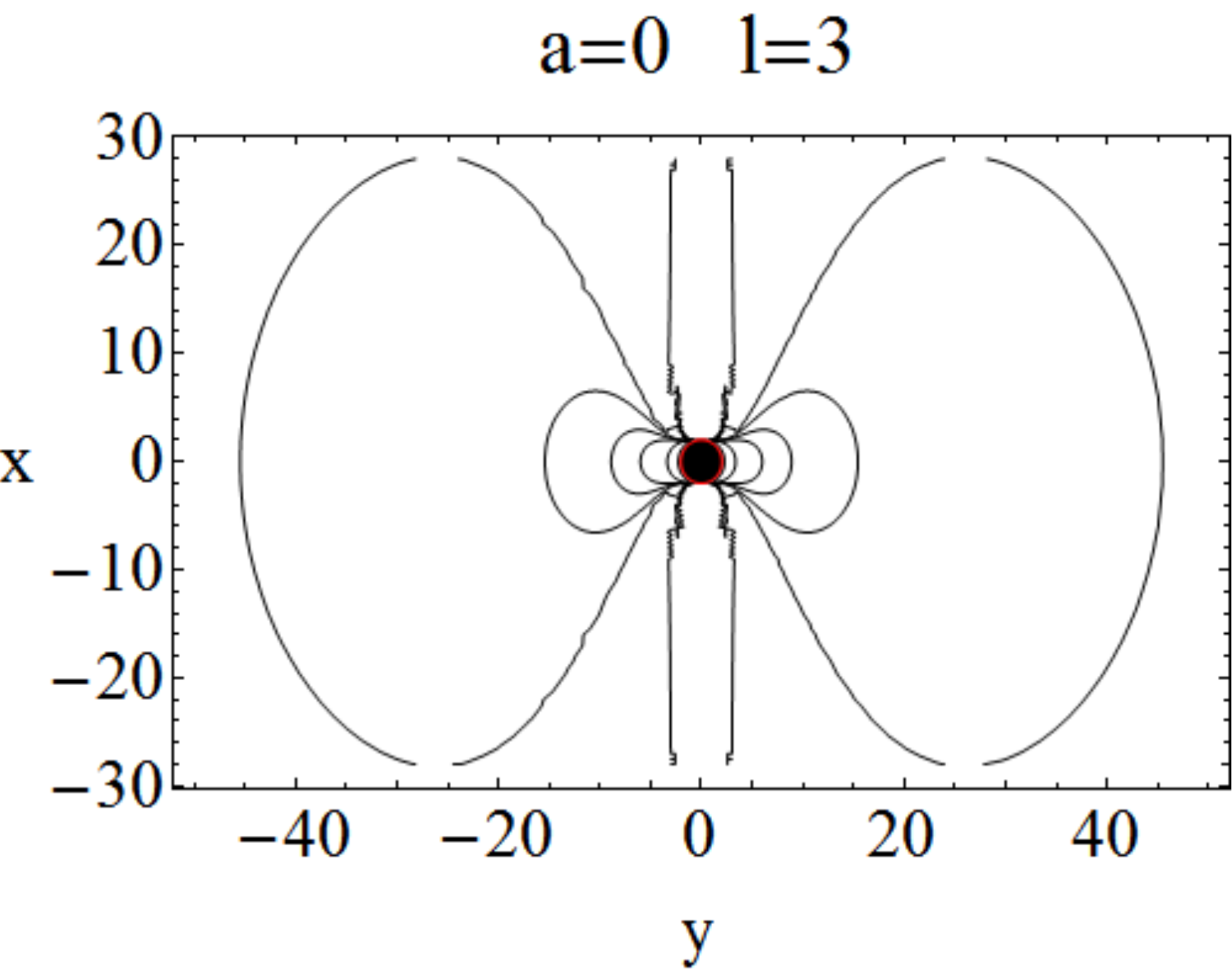}
\includegraphics[scale=.13]{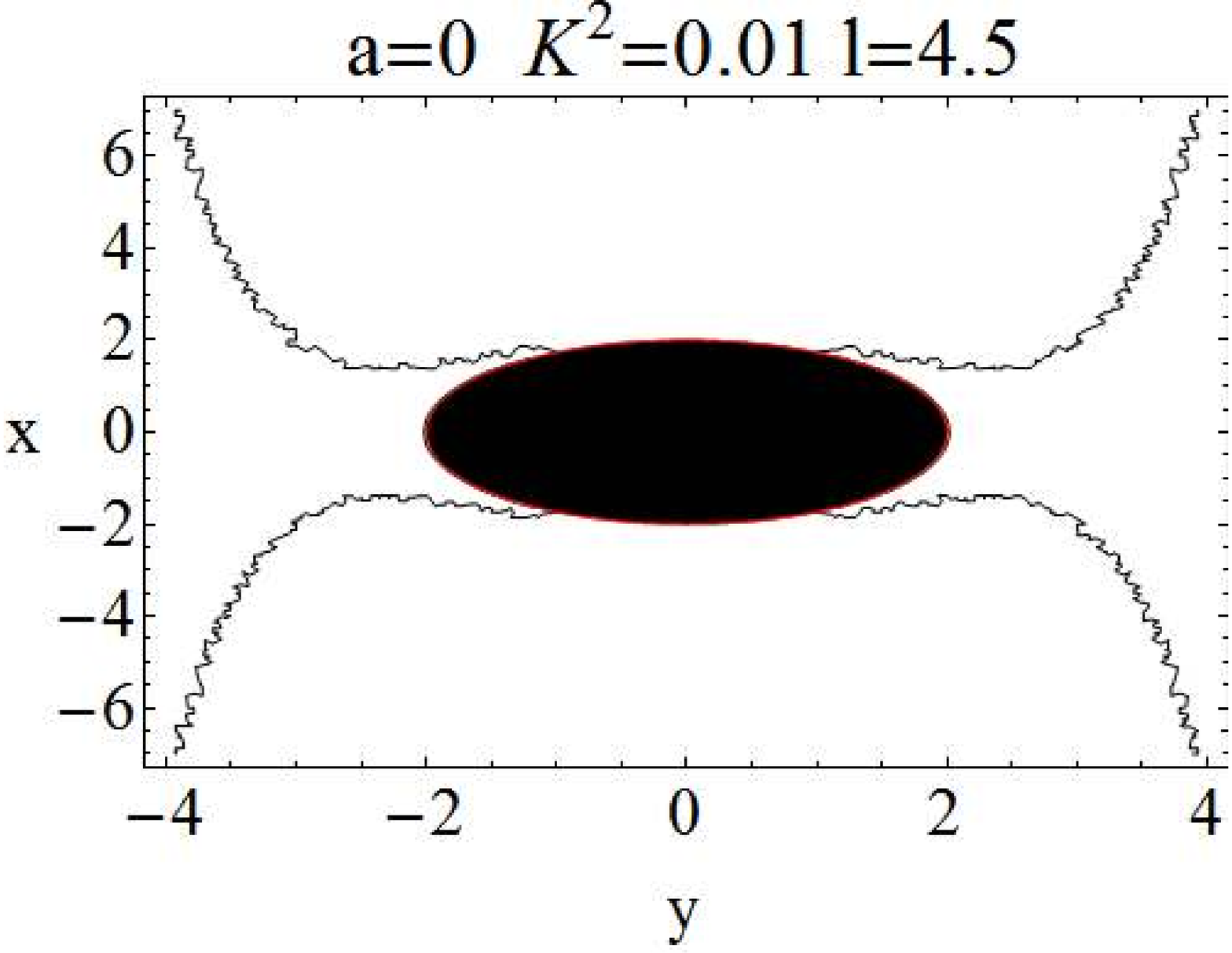}
\includegraphics[scale=.13]{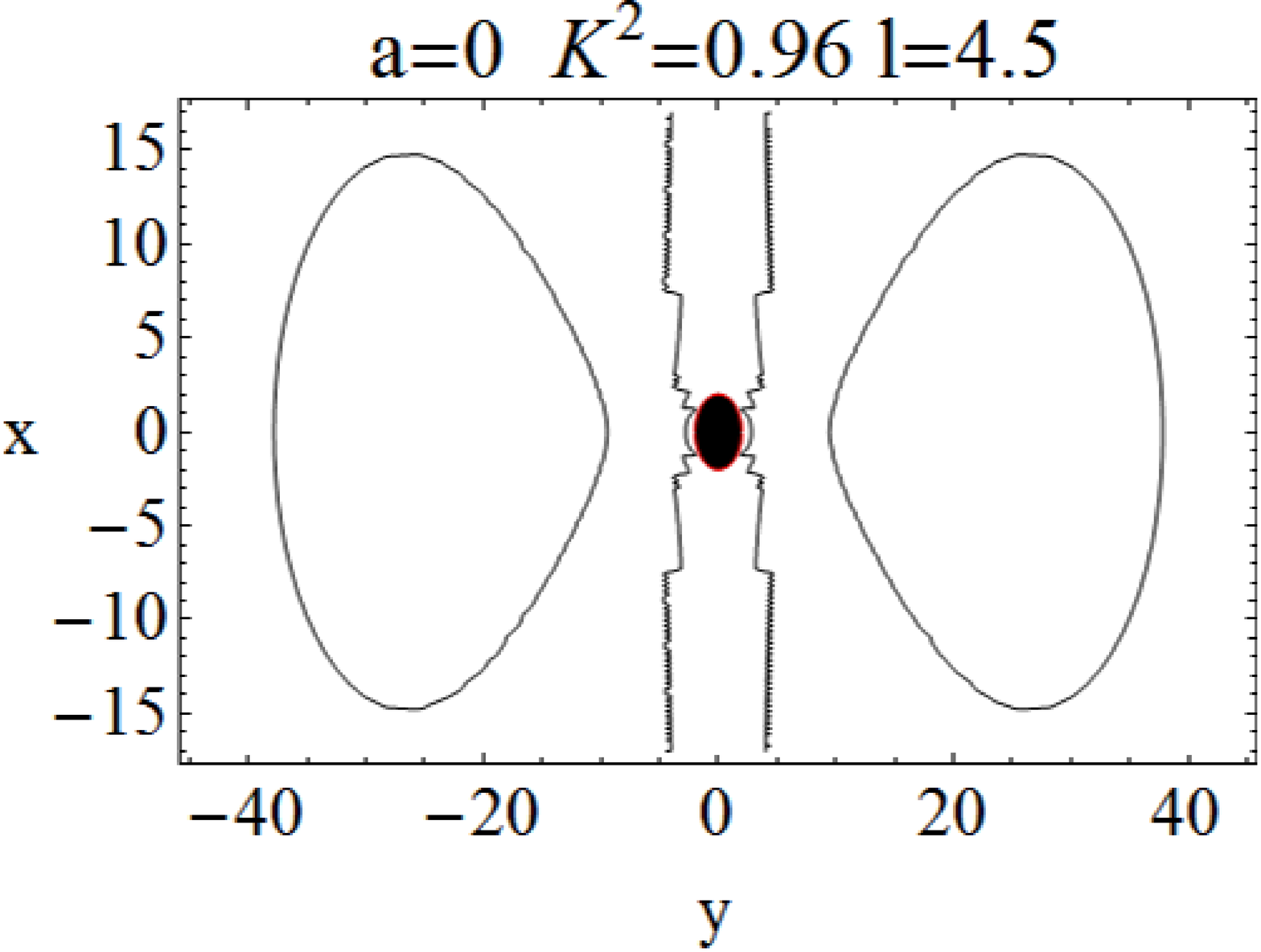}
\end{tabular}
\caption[font={footnotesize,it}]{\footnotesize{The Schwarzschild case sequence $\mathcal{B}_{K}$.  Left panel:$\ell/M\in]0, 3 \sqrt{{3}/{2}}]$ with  $(K\in]0,1[, y_1)$. Center and right panel: $\ell/M\in[4, 3 \sqrt{3}]$, it  is  $\ell=4.5M$ and $K^2_{i}=\{1.47765, 0.939964\}\quad i\in\{a,b\}$. The  angular momentum is in units of mass $M$.}}\label{Fig:SAnalogico}
\end{figure}
\item[] For
$\ell/M>3 \sqrt{3}$: \textbf{1.} $(K\in]0,K_{b}[, {y_1})$,  \textbf{2.}
$({K_{b}}, {y_{12}})$, \textbf{3. }
$(K\in]K_{b},1[, {y_{123}})$
\end{description}
\textbf{Fixed orbital parameter $K$}
We will consider the range $K>0$ and the evolutive sequences $\mathfrak{B}_{	 \ell}\equiv\mathfrak{B}_{\mathbf{p}}/\Sigma_{K}$.  We can compare this case with the analysis in Sec.\il(\ref{Subsun:barellM1}) and Sec.
\il(\ref{Subsubsec:lbarm1}) it is then:
\begin{description}
\item[{${K\in]0, \sqrt{{8}/{9}}]}$} ]
 there are the  solutions
$(\ell=0,y=3M)$ and  $(\ell>0,y=M)$, see Figs.\il(\ref{Fig:STGinocchio}). The limiting case $\ell=0$ is shown here, with increasing $\ell$  a ${B}$-configuration emerges.
%
\begin{figure}
\centering
\begin{tabular}{cc}
\includegraphics[scale=.13]{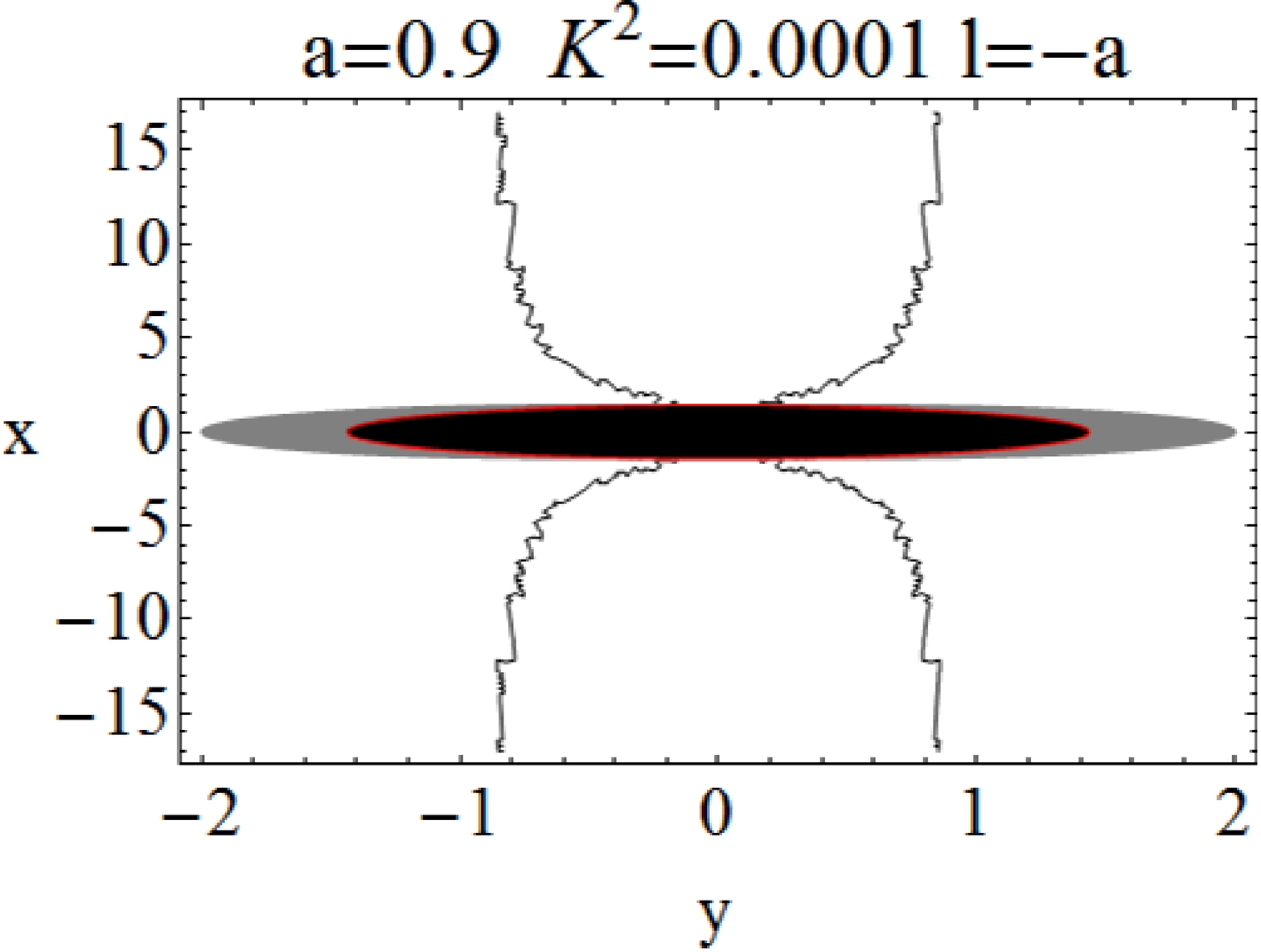}
\includegraphics[scale=.13]{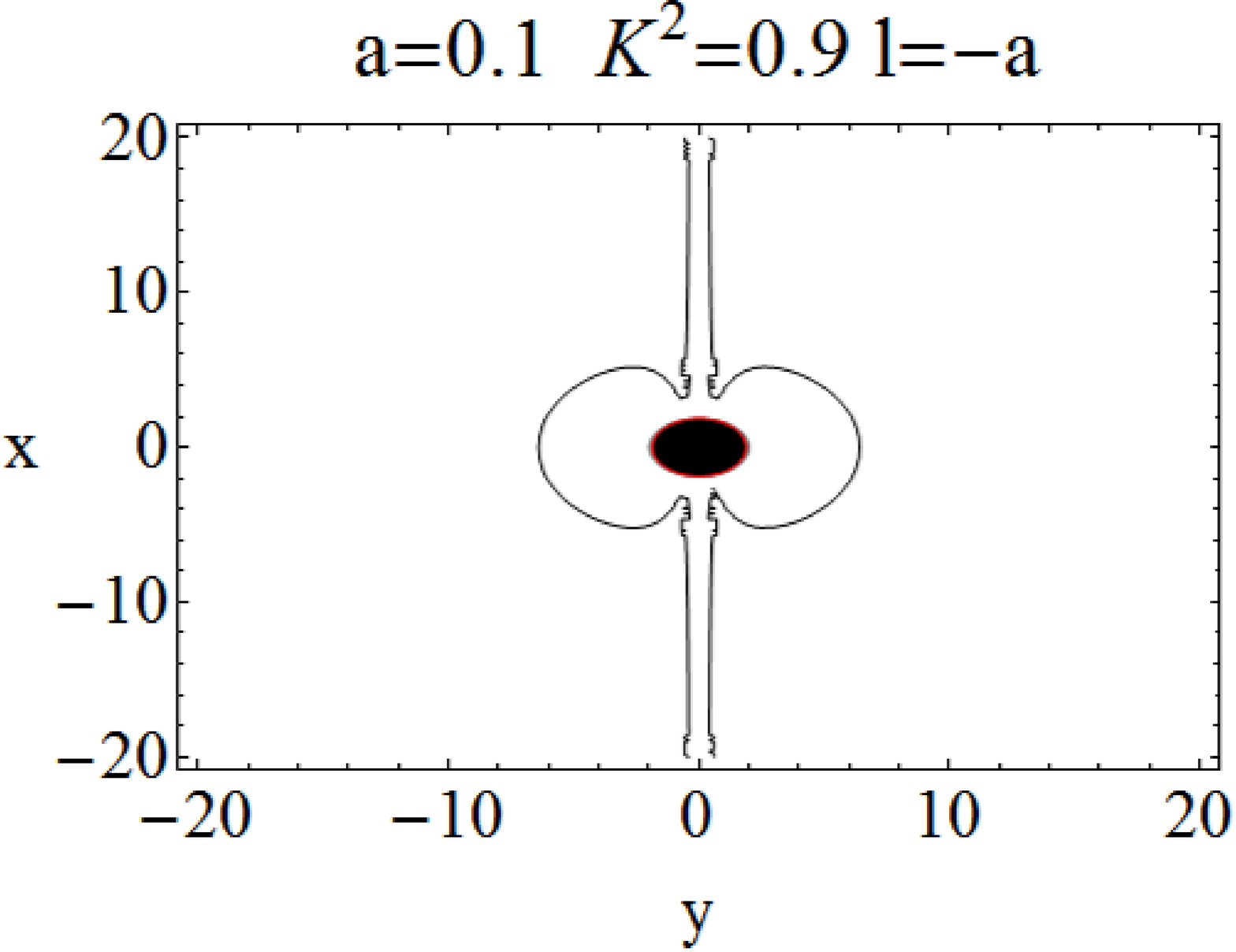}
\includegraphics[scale=.13]{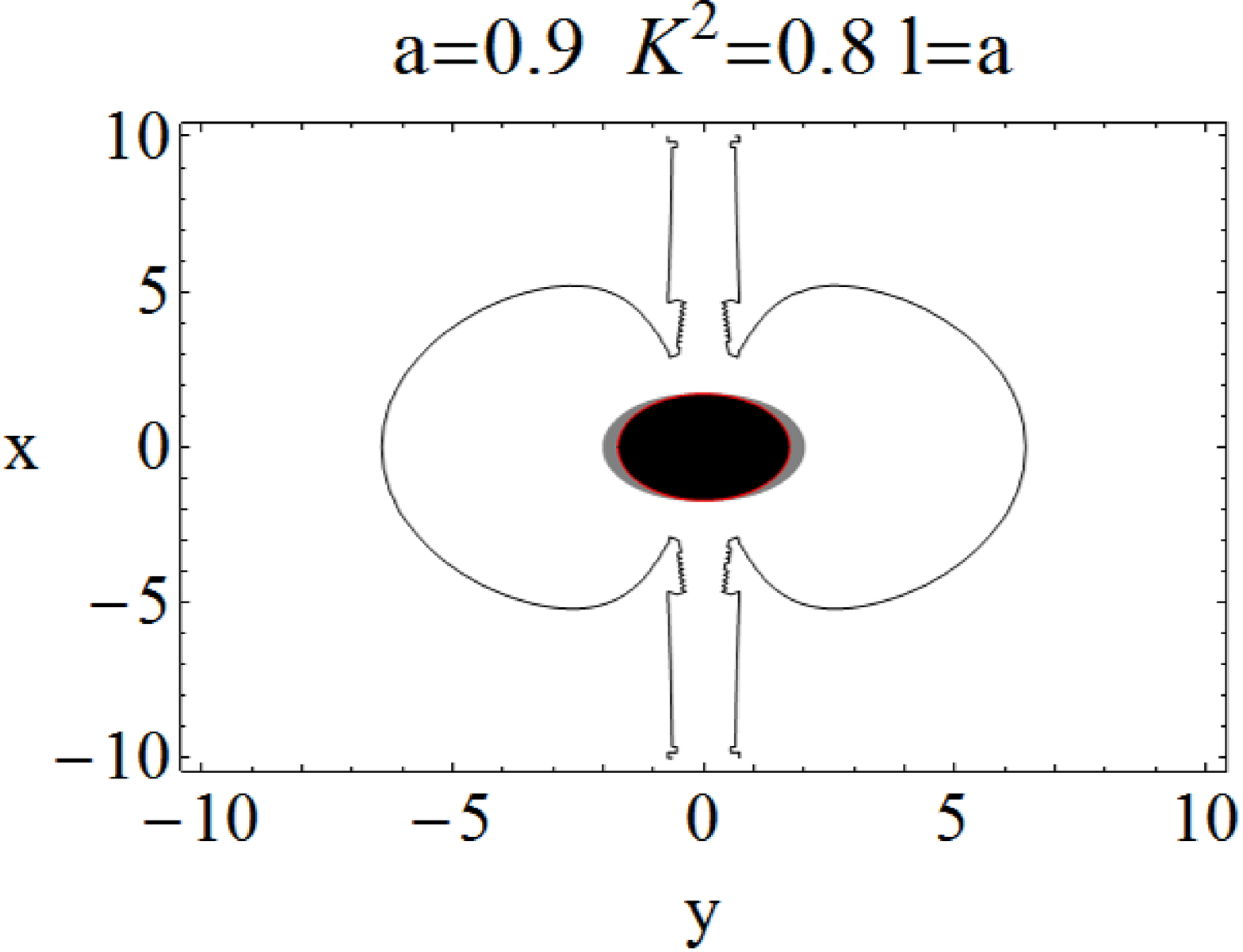}
\includegraphics[scale=.13]{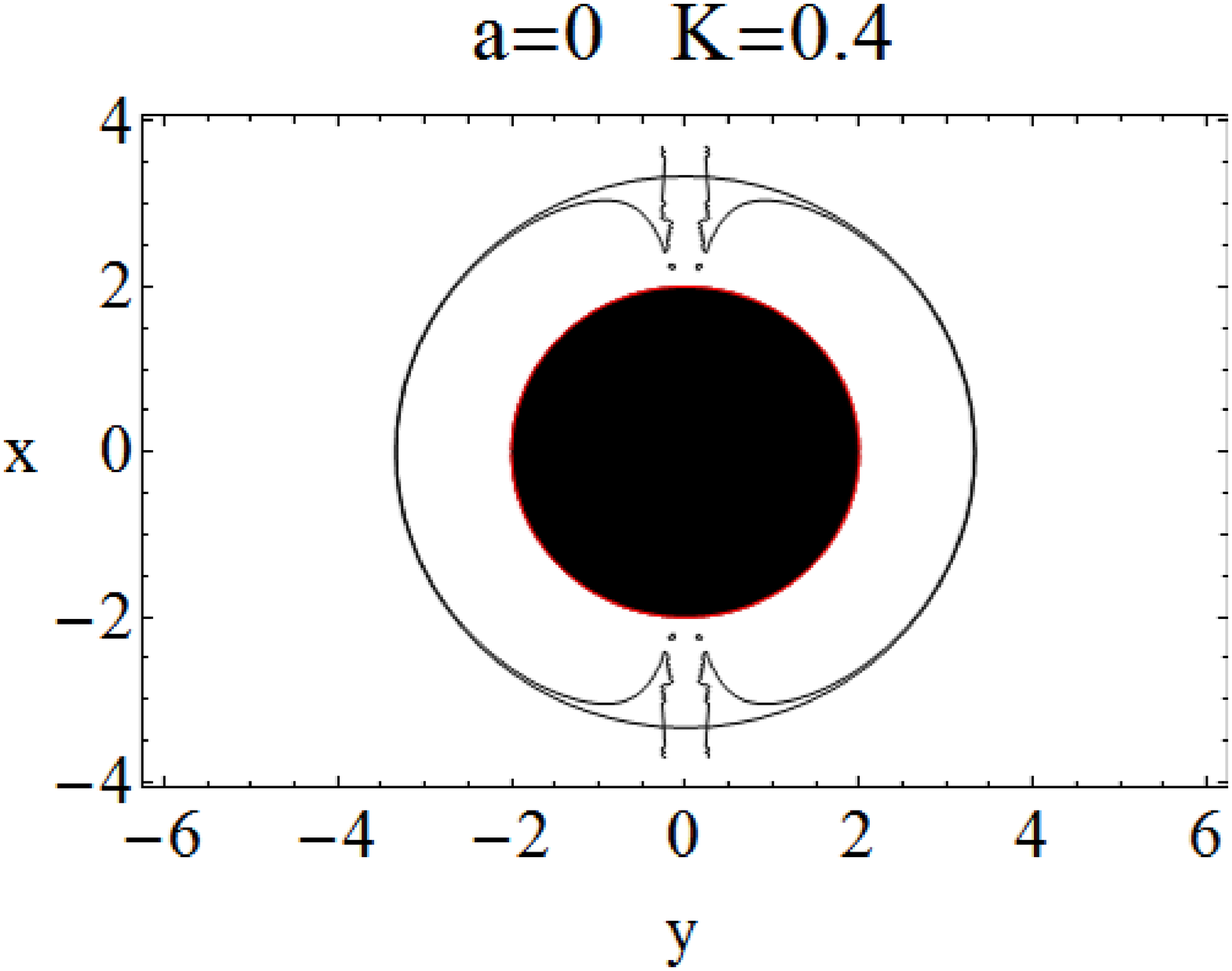}
\end{tabular}
\caption[font={footnotesize,it}]{\footnotesize{Configurations  $\ell=\pm a$   and different spin-mass ratio of the attractors.
The angular momentum is in units of mass $M$.
}}\label{Fig:STGinocchio}
\end{figure}
\item[{$K\in]\sqrt{{8}/{9}},1[ $} ]
  there are the following solutions for increasing angular momentum:
\textbf{1.}
$(\ell=0,y=3M)$,
\textbf{2.}
$(\ell\in]0,\ell_{\mathfrak{K}}^-[,y=M)$,
\textbf{3.}
$(\ell_{\mathfrak{K}}^-,y_{13})$,
\textbf{4.}
$(\ell\in]\ell_{\mathfrak{K}}^-,\ell_{\mathfrak{K}}^+[, y_{123})$,
\textbf{5.}
$(\ell_{\mathfrak{K}}^+,y_{12})$,
\textbf{6.}
$(\ell>\ell_{\mathfrak{K}}^+ ,y=M)$.
See Figs.\il(\ref{Fig:TSCavaliere}): increasing $\ell$ the configurations sequence becomes
$\mathfrak{B}_{\ell}=[{B},  {C}_{x}, {C}]$, closed surfaces finally disappear and only the interior ${B}$-surface, close to the black hole  with open funnels of matter aligned with the axis.
\begin{figure}
\centering
\begin{tabular}{cc}
\includegraphics[scale=.13]{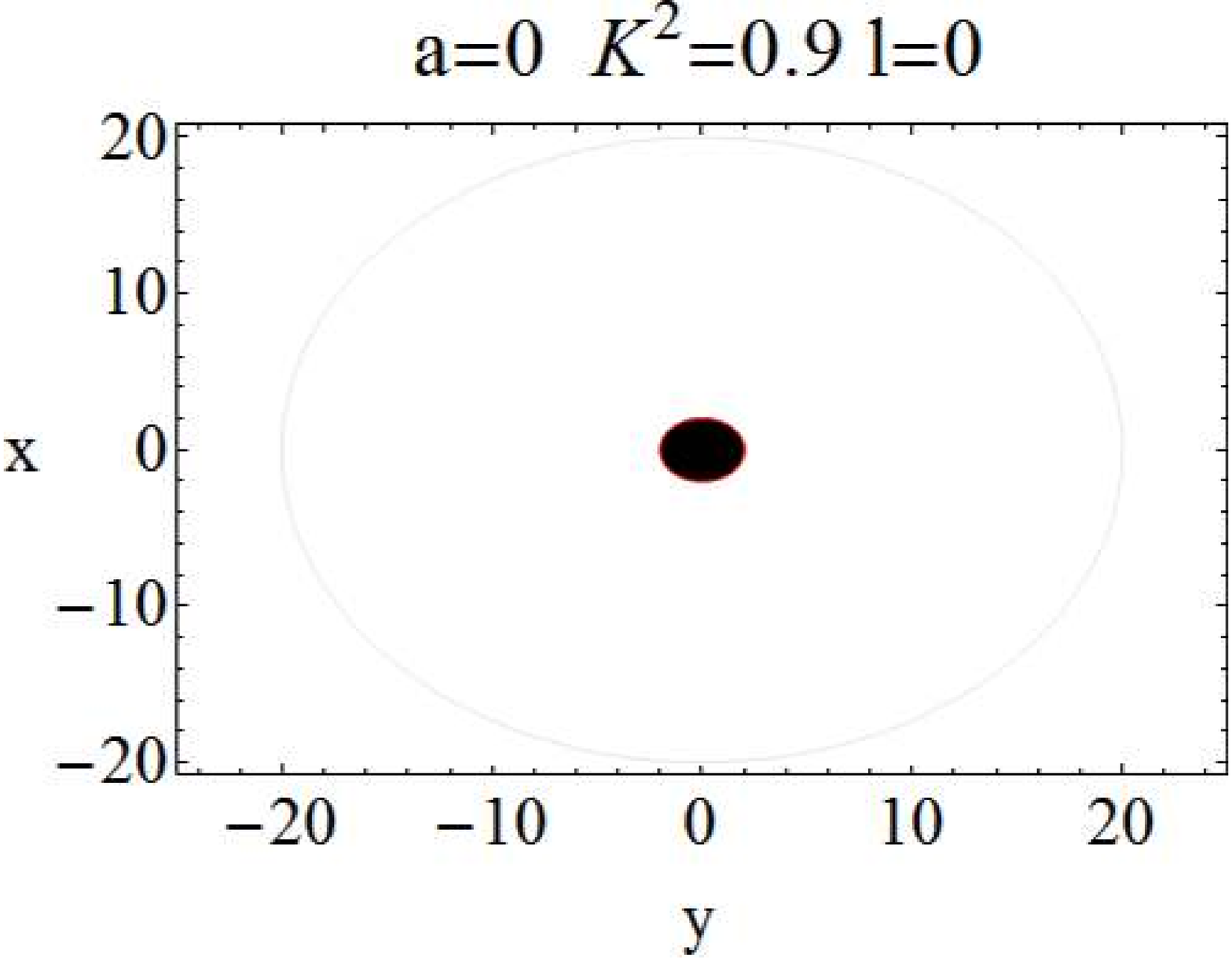}
\includegraphics[scale=.13]{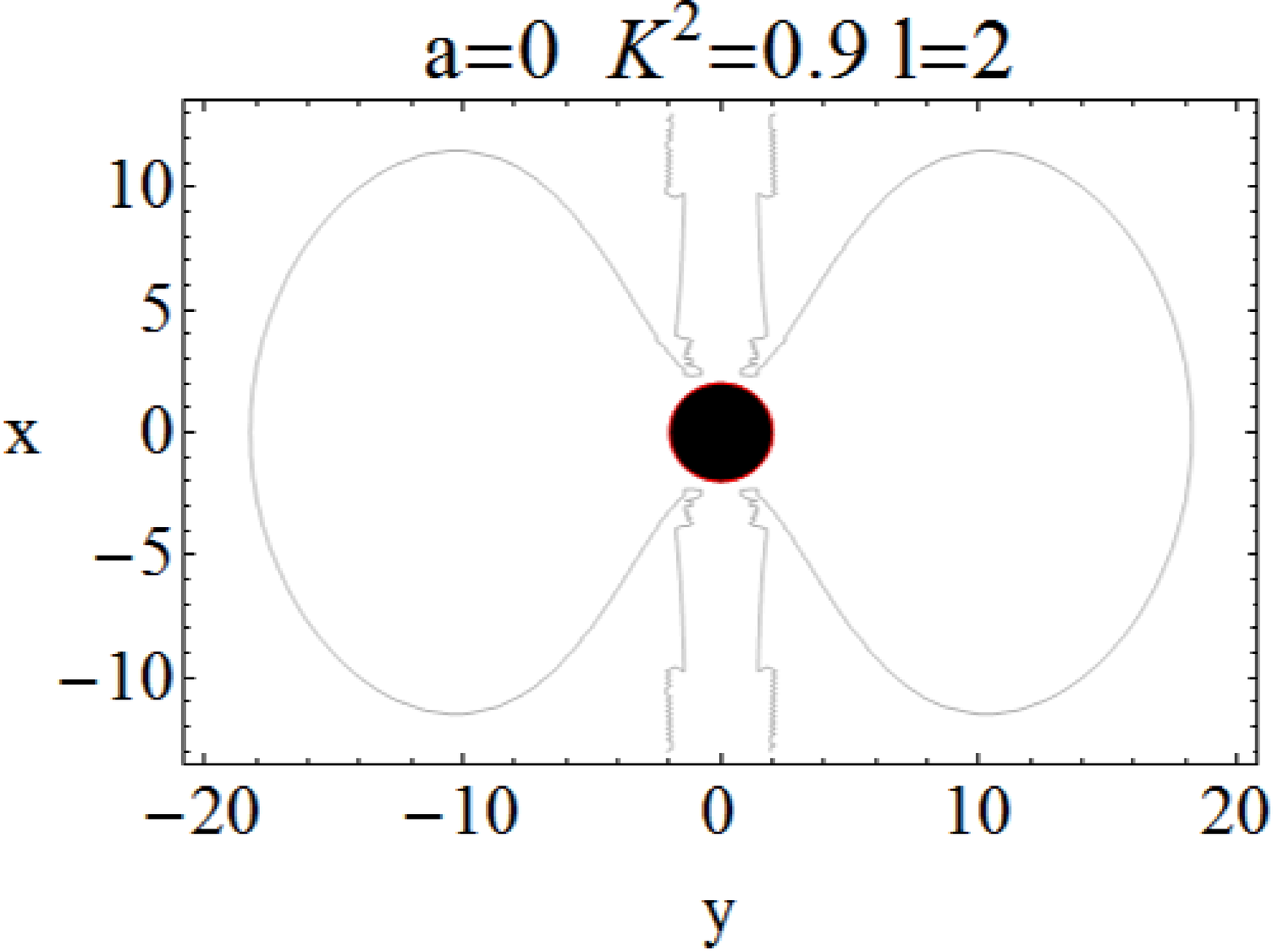}
\includegraphics[scale=.13]{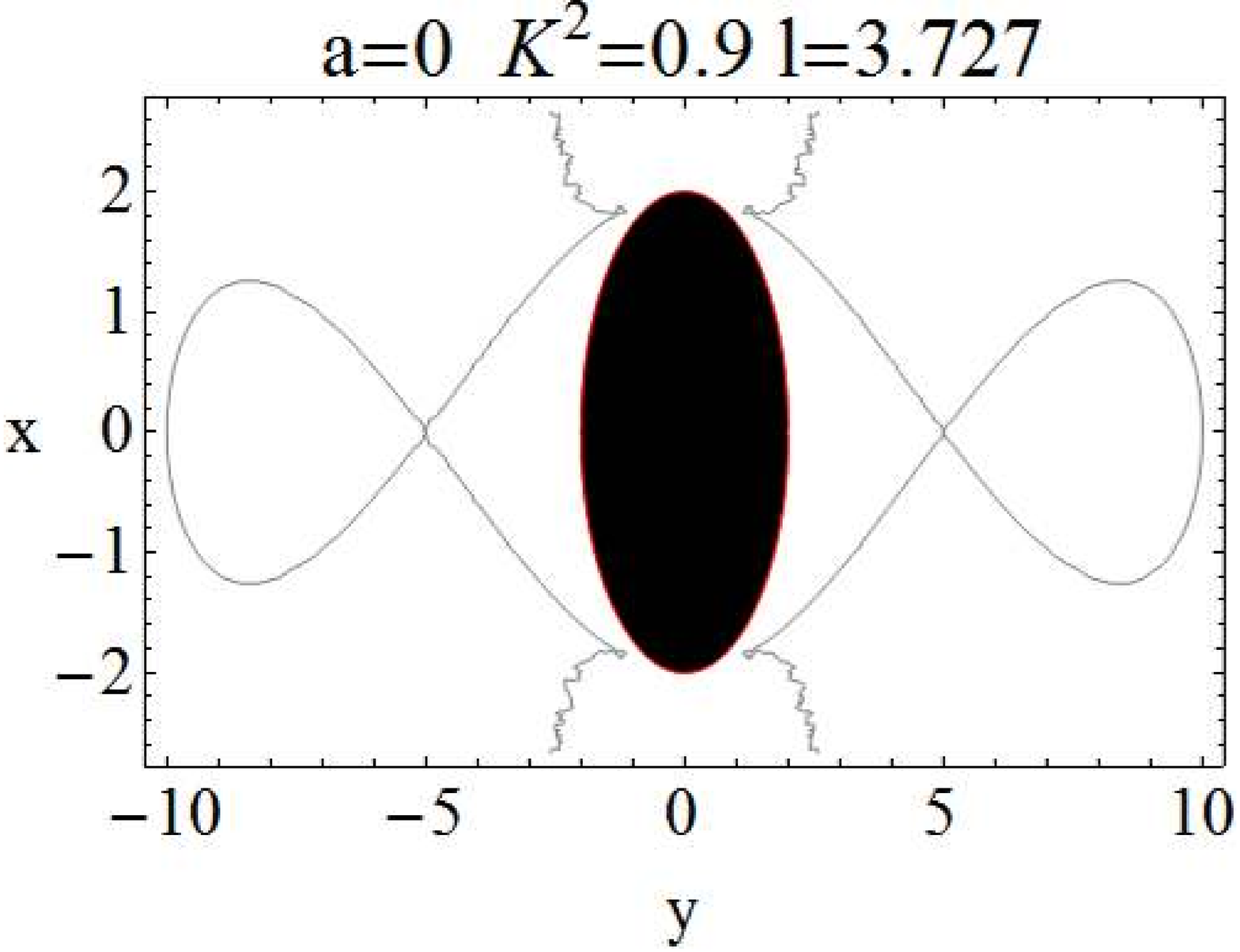}
\includegraphics[scale=.13]{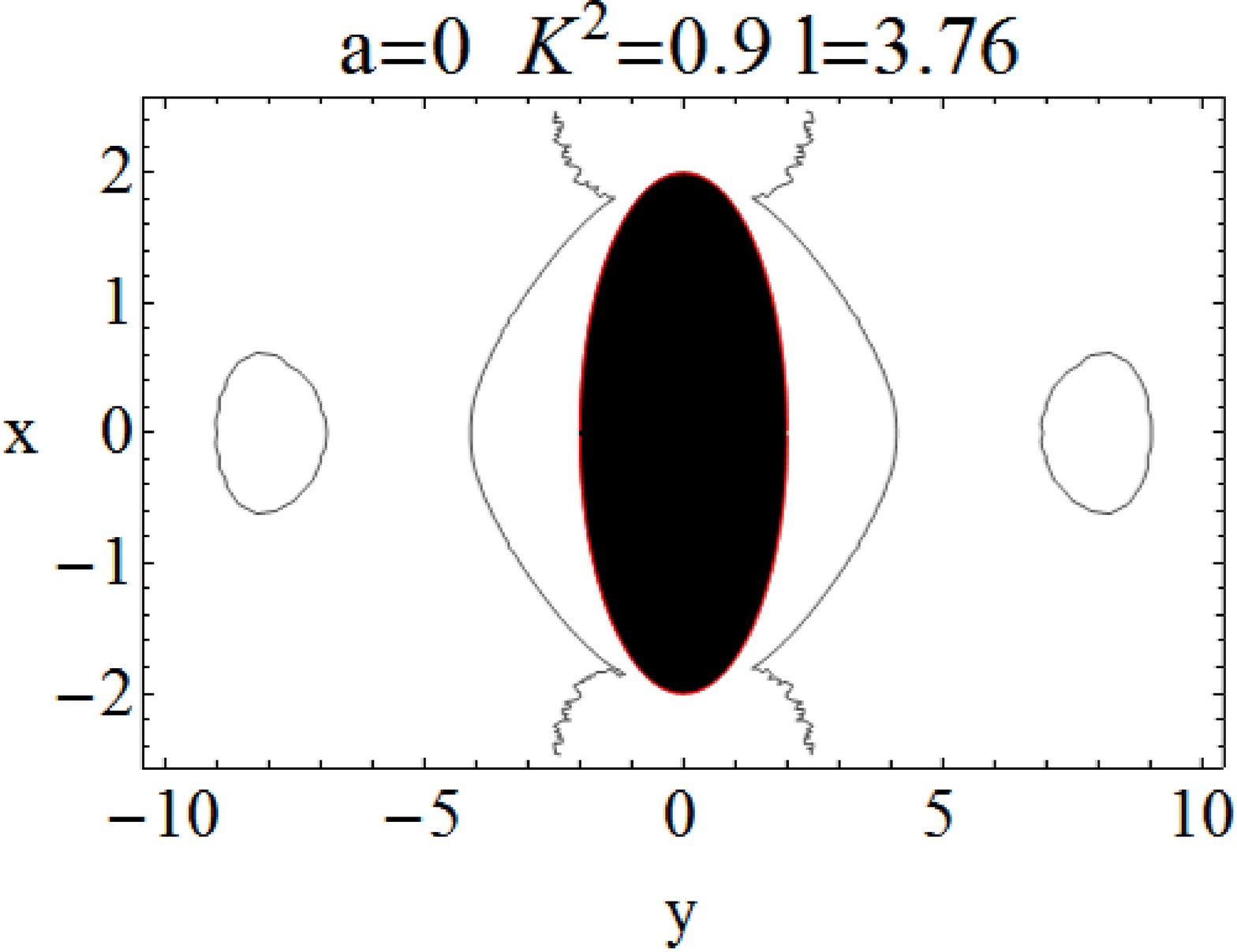}
\includegraphics[scale=.13]{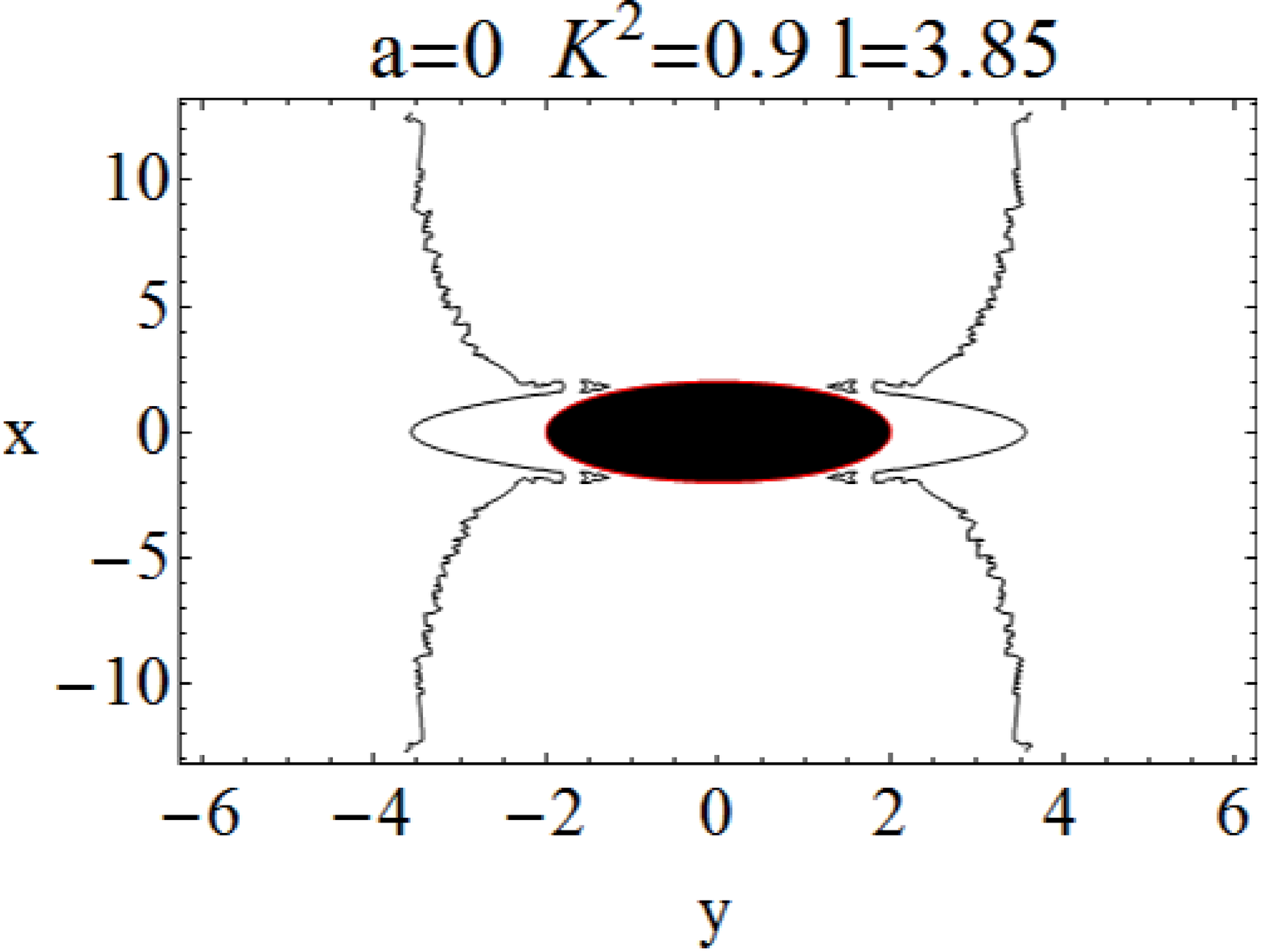}
\end{tabular}
\caption[font={footnotesize,it}]{\footnotesize{Non rotating attractor: sequences $\mathbf{B}_{\ell}$ with fixed $K\in]\sqrt{{8}/{9}},1[$. The angular momentum is in units of mass $M$. It is $K^2=0.9$ and $\ell_{\mathfrak{K}}^{\pm}\in\{3.72678, 3.77124\}$. }}\label{Fig:TSCavaliere}
\end{figure}
\item[$K=1$:]  \textbf{1.} $(\ell<-4M, y_{12})$, \textbf{2.}
$(\ell=-4M,, y=4M)$, \textbf{3.} $(\ell=4M,y=4M)$,
 \textbf{4.} $(\ell>4M, y_{12})$, see  Figs.\il(\ref{Fig:TSToti}-left). These are   open configurations,
these  solutions could simulate jets.
 \begin{figure}
\centering
\begin{tabular}{cc}
\includegraphics[scale=.13]{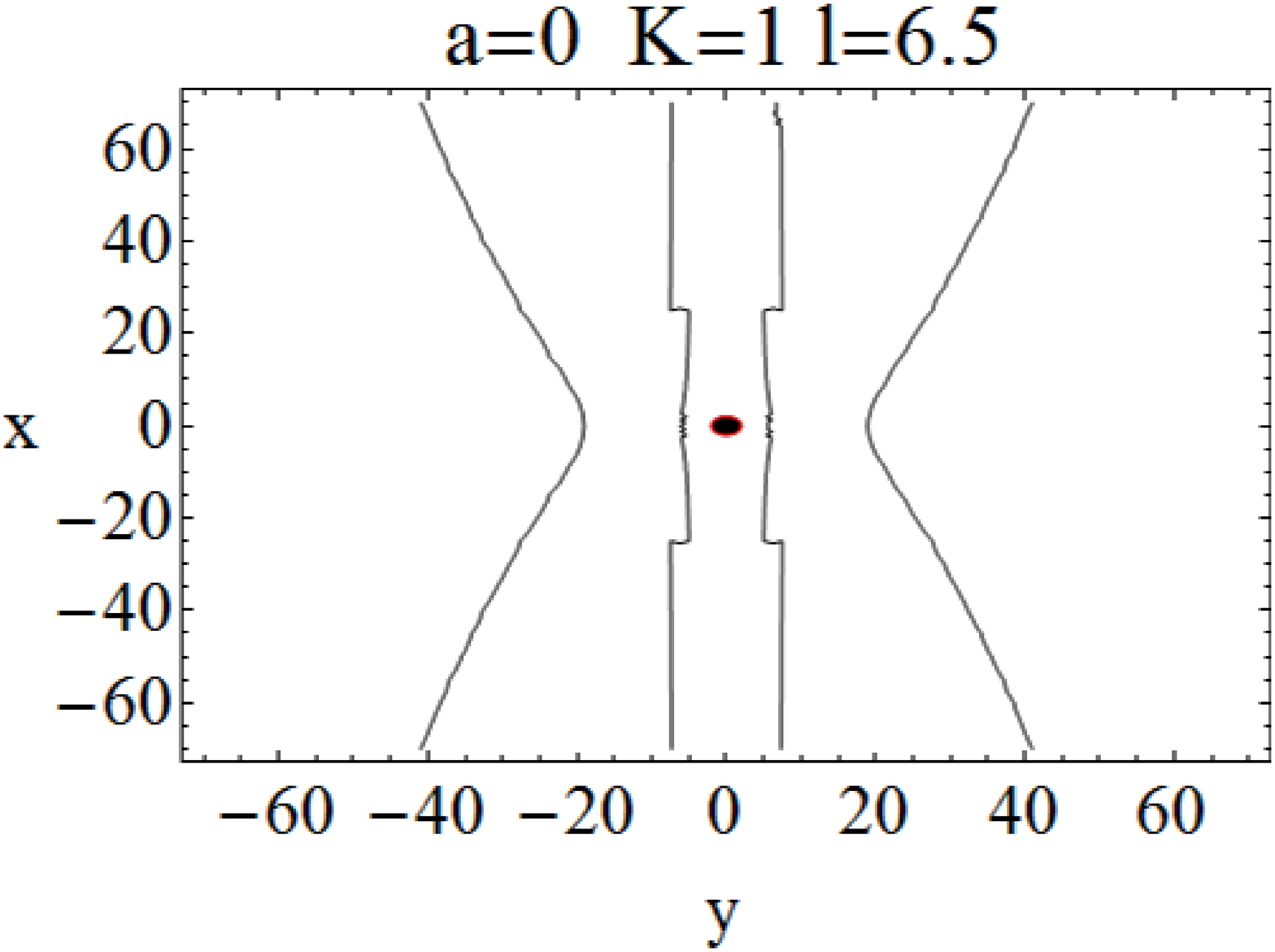}
\includegraphics[scale=.13]{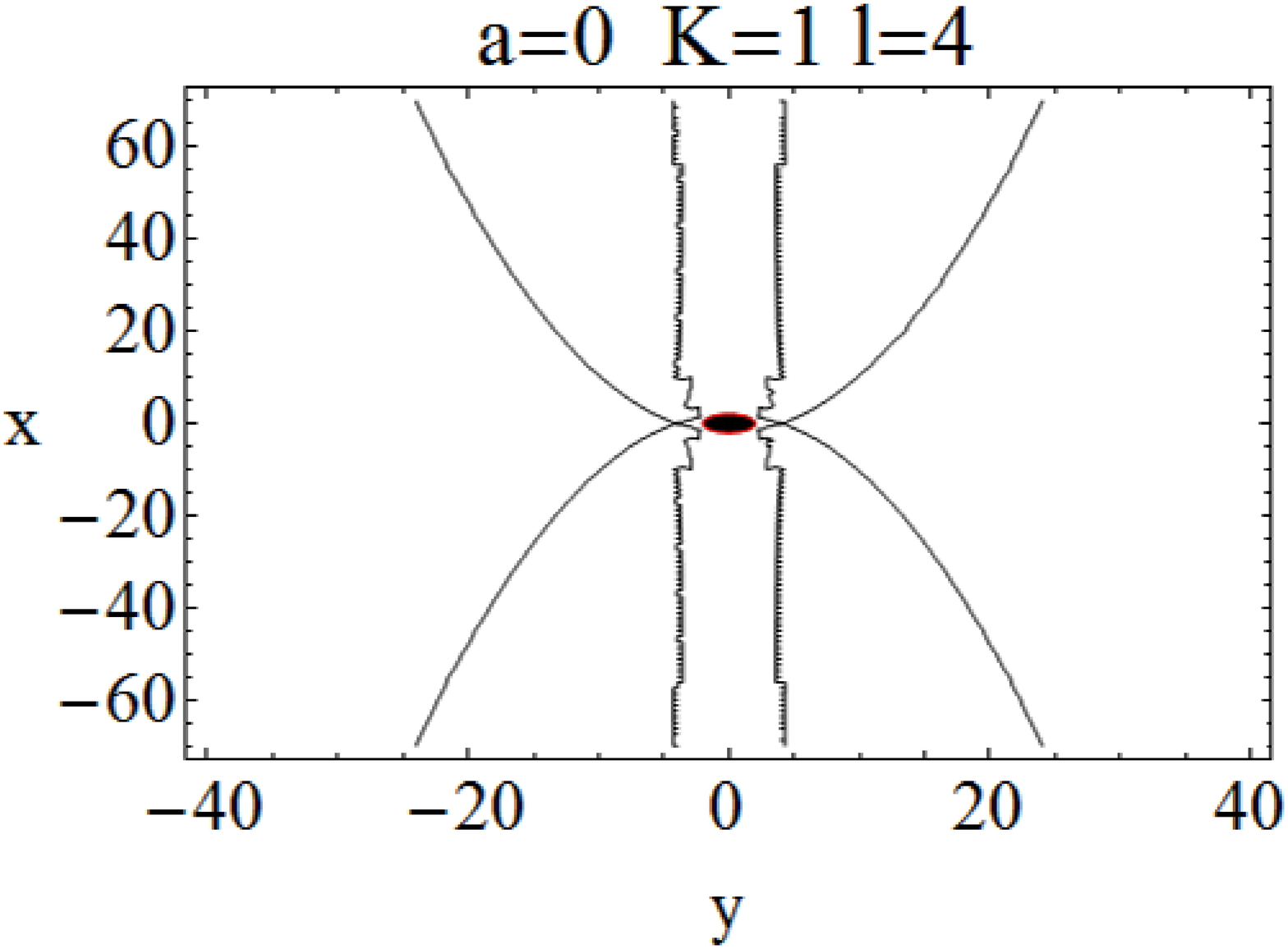}
\includegraphics[scale=.13]{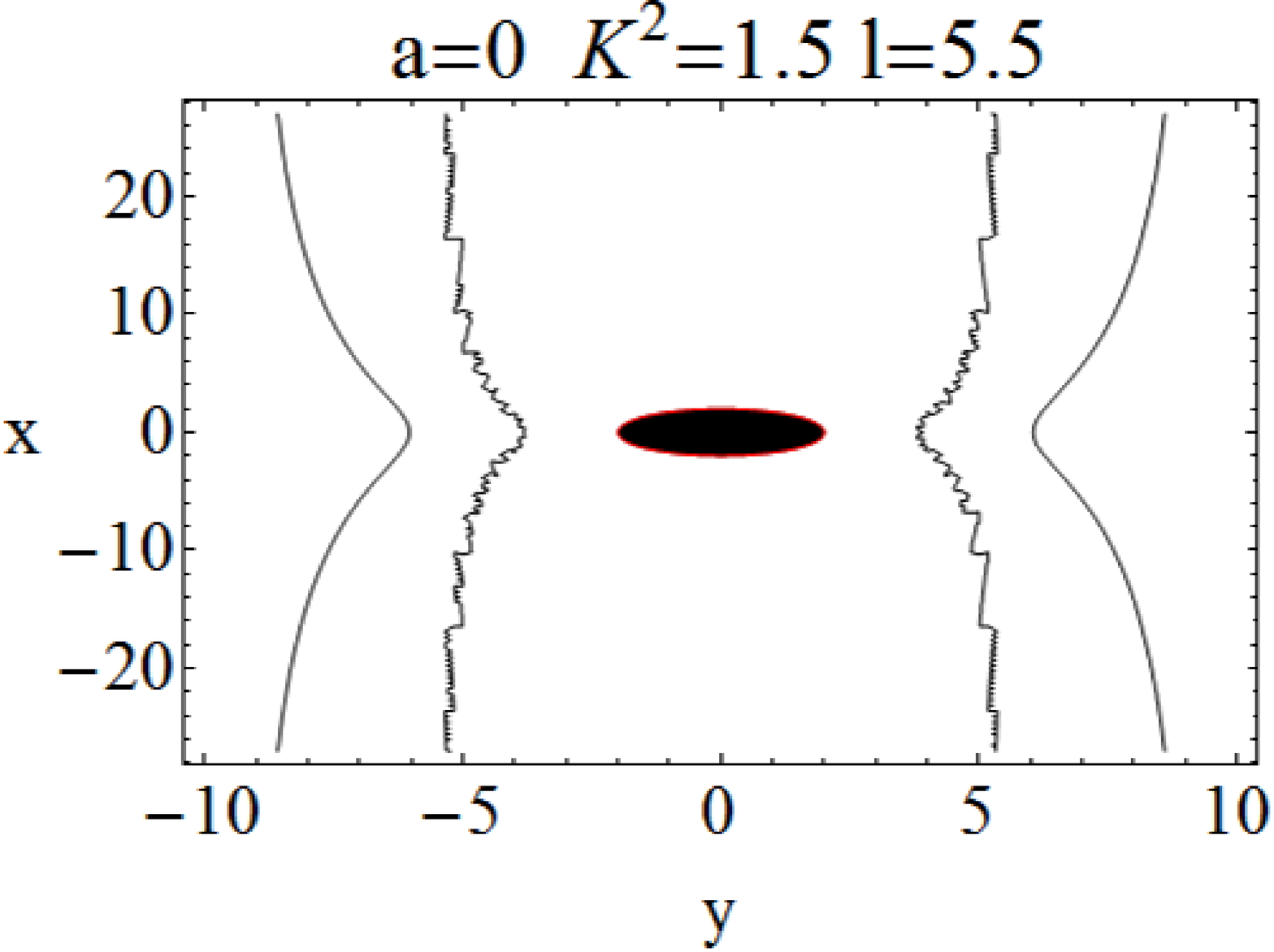}
\includegraphics[scale=.13]{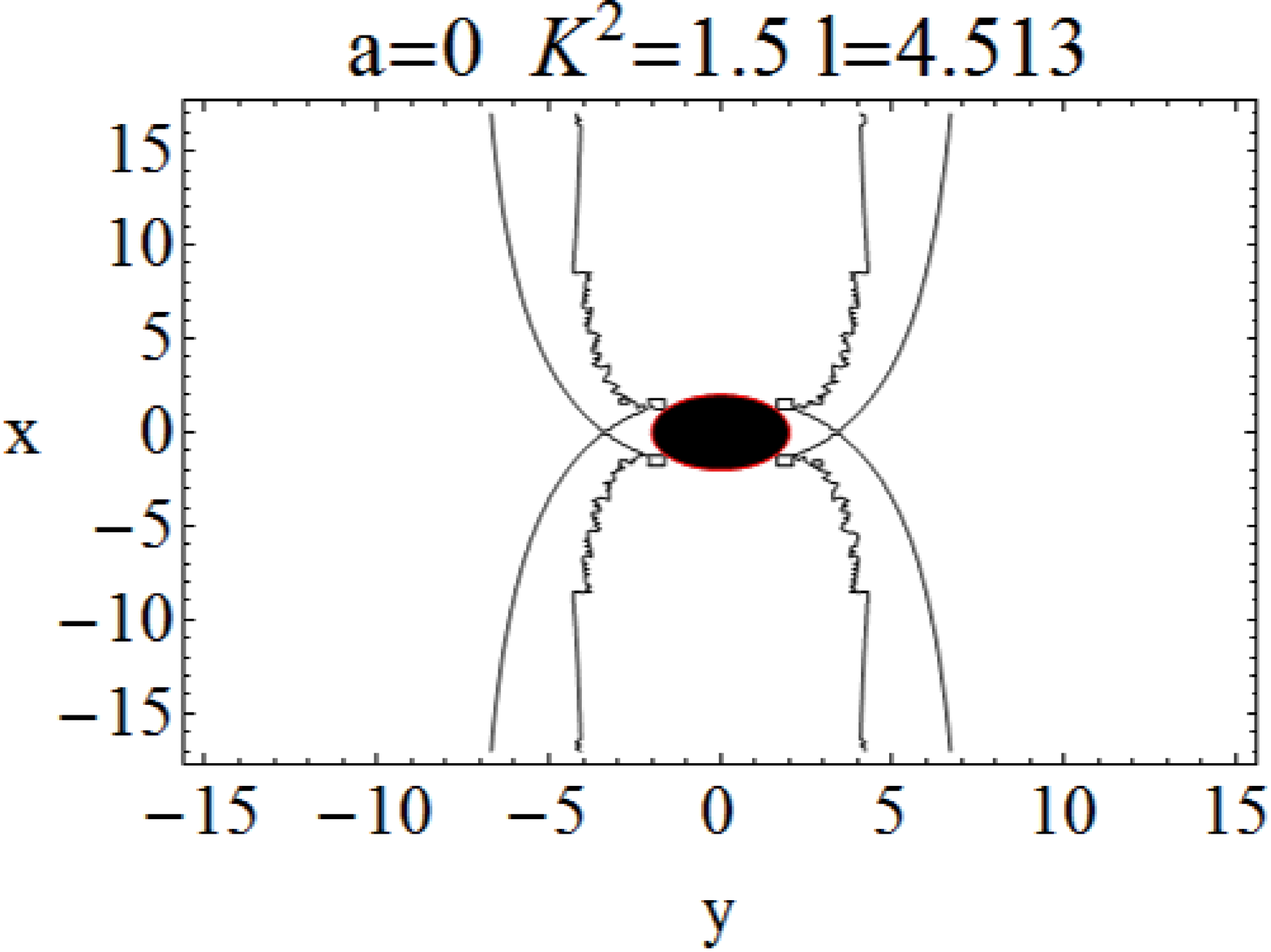}
\end{tabular}
\caption[font={footnotesize,it}]{\footnotesize{Non rotating attractor, sequences $\mathcal{B}_{\ell}$ for  $K=1$ and  $K>1$ where  $K^2=1.5$ and  $\ell_{\mathfrak{K}}^+=4.51276M$. }}\label{Fig:TSToti}
\end{figure}
\item[$K>1$:]  \textbf{1.} $(\ell_{\mathfrak{K}}^+, y=2M)$, \textbf{2.} $(\ell>\ell_{\mathfrak{K}}^+, y_{23})$. See Figs.\il(\ref{Fig:TSToti}) there are only open configurations.
\end{description}
In the limiting  case of the Schwarzschild geometry we  compare the two  evolutive sequences  $\mathfrak{B}_{\ell}$ and $\mathfrak{B}_{K}$,  tracing out some general considerations for the case $a\neq0$.
The first sequence $\mathfrak{B}_{\ell}$ has been addressed extensively for  the rotating case in Sec.\il(\ref{Subsec:sequence}) and $\mathfrak{B}_{K}$ has been considered in Sec.\il(\ref{Sec:K.L}), however we restrict our attention to the case the only sequences including the Boyer closed or closed crossed configurations.
Typical patterns are
$\mathfrak{B}^{>}_{\bar{\ell}}=[B, {C}_{x}, {C}, {O}]$
as seen in \underline{\textbf{Region II}}-Sec.\il(\ref{Subsun:barellM1}), and
$\mathfrak{B}^{<}_{\bar{\ell}}=[B, C, C_x, B]$
in  \underline{\textbf{Region III}} of
Sec.\il(\ref{Subsubsec:lbarm1}), finally it  is
$\mathfrak{B}_{K}=[{B}, {C}, {C}_{x}, {B}]$, analyzed for the case of a  Schwarzschild attractor.
We note a symmetry in the sequences $\mathfrak{B}^{>}_{\bar{\ell}}$ and $\mathfrak{B}^{<}_{\bar{\ell}}$  with respect to the corotating and counterrotating fluids and the configurations $(C, C_x)$ ,  clear also by the Tables\il(\ref{Fig:closed-Kmin-lneg}), actually the two sequences appear to be analogue once one considers the increasing values of the fluid angular momentum magnitude, irrespectively from  the cases $\ell a>0$ or $\ell a<0$: the evolutive sequence $(C, C_x)$ appears decreasing the fluid angular momentum magnitude or increasing, at fixed $\ell$, the $K$-parameter, see the $\mathfrak{B}_{K}$ sequence, this kind of symmetries will be also investigated in Sec.\il(\ref{Sec:multipleP-D}) and
 Sec.\il(\ref{SeC:poly}) addressing the analysis of the Boyer surfaces structures and the disk morphology.
\subsubsection{Configuration at $K=1$}\label{SeC:K1}
The case $K=1$ has been analyzed in  Sec.\il(\ref{Sec:KU1ana}), critical points are maximum of the effective potential and are located at $r_{b}^{\pm}$ with $\ell_f^{\pm}$ respectively.
In other words we consider the special sequences   $\mathfrak{B}_{	 \ell}\equiv\mathfrak{B}_{\mathbf{p}}/\left.\Sigma_{K}\right|_{K=1}$.  This case completes the analysis of  Sec.\il(\ref{Subsec:sequence}) and
Sec.\il(\ref{Sec:K.L})  and can be compared with the results in Sec.\il(\ref{Subsec:sequence}).  However we introduce  the  solutions
\be\label{Eq:ymp}
y_{\pm}\equiv\frac{1}{4 M} \left(\ell^2\pm\sqrt{[\ell^4-16M^2(-\Delta_{\ell}^{-})^2 ] }\right) ,
\ee
 limiting cases of the solutions $y_i$ in  Eqs.\il(\ref{Eq:5raidiff}) for $K=1$. Solutions  (\ref{Eq:ymp})   depend on the magnitude $|\ell-a|$  and  it is  $\ell_K^{\pm}(a; r_i, K_{b}^{\pm})=\ell_{b}^{\pm}\equiv \ell_f^{\pm}({r_b^{\pm}})$.
We summarize the results as follows:
\begin{description}
\item[The Schwarzschild case $a=0$:]
There are only two regions as follows:
 \textbf{Region I} where \textbf{1.} $(\ell<-4M,\; y_{\pm})$,\;\textbf{2.} $(\mp 4M,\;  y=4M)$, and
\textbf{Region II} with solution $(\ell>4M,\; y_{\pm})$.
\item[Kerr spacetime:]
$a\in]0,1[$:
\textbf{Region I}:
\textbf{1.}  $(\ell<\ell_b^+,y_{\pm})$,
 \textbf{2.}$((\ell_b^+,\ell_{\mu}^\Pi), y_-
)$, \textbf{3.} $(\ell\in]\ell_{\mu}^\Pi, a[,
 y_{\pm})$, \textbf{4.}
$(\ell=a\ y_+)$. This first region consider the values $\bar{\ell}\in[0,1]$, for counterrotating fluids ($\ell a<0$)  and corotating ones ($\ell a>0$) where no critical points are, and the values $|\bar{\ell}|>1$  at $\ell a<0$ where there is a maximum for the effective potential at $K=1$, see also Sec.\il(\ref{Subsubsec:lbarm1}).
\textbf{Region II}: In this region  it is $\bar{\ell}>1$ and it includes various subregions: \textbf{5.} $(\ell\in]a,\ell_f^{\pm}(r_-)[, y_{\pm})$,
\textbf{6.} $(\ell_f^{\pm}(r_-),y_-)$, \textbf{7.}
$(\ell\in]\ell_f^{\pm}(r_-),\ell^{\Pi}_{\mu}[, y_{\pm})$, \textbf{8.} $(-\ell^{\Pi}_{\mu}, y_-)$,
 \textbf{9.} $({\ell}_{b}^-, y_-)$, \textbf{10.} $(\ell\in({\ell}_{b}^-,\ell_f^{\pm}(r_+)), y_{\pm})$, \textbf{11.}
$(\ell_f^{\pm}(r_+), y_+)$,  \textbf{12.} $(\ell>\ell_f^{\pm}(r_+), y_{\pm})$. This region considerers $\bar{\ell}>1$ analysed for $K\neq0$ in Sec.\il(\ref{Subsun:barellM1}).
\item[Extreme Kerr Black hole:]
$a=M$
\textbf{1.} $(\ell<-2(1+ \sqrt{2})M, y_{\pm})$, \textbf{2.} $(\ell=-2(1+ \sqrt{2})M,y_-)$,
\textbf{3.}
  $(\ell=-2(1- \sqrt{2})M,y_-)$, \textbf{4.} $(\ell/M\in]-2(1- \sqrt{2}),1[, y_{\pm})$, \textbf{5.} $
(\ell=M,y={1}/{2}M)$, \textbf{6.} $(\ell/M\in]1,2[, y_{\pm})$, \textbf{7.} $(\ell>2M, y_{\pm}
)$
\end{description}
for $a=0$ some configurations are plotted in Fig.\il(\ref{Fig:TSToti}), however these are points of minimum pressure  and correspond to unstable fluid configurations in open funnels.
\subsubsection{Configurations with $\bar{\ell}=\pm1$}\label{subsubsec:lpm1}
We now focus on the cases $\bar{\ell}=\pm1$,
we know from the analysis in Sec.\il(\ref{Sec:barl}) that the limiting cases $\bar{\ell}=\pm 1$ do not admit any toroidal Boyer configurations, however the effective potential, given in Eqs.\il(\ref{Eq:effper1},\ref{Eq:effper2}), for these particular cases  is well defined in the region $r>r_+$.  In this section we  set the angular parameter $\ell=\bar{\ell}_{\theta=\pi/2}=\pm a$ as this is a relevant case for the    solutions of Eq.\il(\ref{Eq:scond-d},) therefore in the integration of the hydrodynamic equations on all  $\Sigma_{\theta}$ planes, we consider  the angular momentum  $\bar{\ell} =\pm1$. However, following the discussion in Sec.\il(\ref{Sec:barl}), for any plane $\Sigma_{\theta}$ the limit value for the angular momentum is $\bar{\ell}=\ell/a\sigma$ and  this more general definition should be  considered for example in the case  equatorial plane disk  not aligned with equatorial plane  of the rotating source.

\textbf{{Case: $\bar{\ell}=-1$}}
For a counterrotating fluid configuration at  $\ell=-a$, or $E=-L$
 we have from Eqs.\il(\ref{Eq:flo-adding}):
\bea
\Phi=\frac{L\Omega_z (\Omega_q-1) }{g_{tt} \Omega_q (1+\Omega_q \Omega_z)},\quad \Omega=\frac{ \Omega_z(\Omega_q-1)}{\Omega_q (\Omega_z+1)},\quad\Omega_q\equiv\frac{g_{t\phi}}{g_{tt}}, \quad\Omega_z\equiv-\frac{g_{t\phi}}{g_{\phi\phi}},\quad \left.V_{eff}^2(\bar{\ell})\right|_{\bar{\ell}=-1}=-\frac{g_{tt} \Omega_q (1+\Omega_q \Omega_z)}{\Omega_q(1+2 \Omega_z)-\Omega_z}
\eea
 $\Omega_z$ is the angular velocity of the observer at zero angular momentum (ZAMOs) $L=0$, the angular frequency is completely determined by the properties of the background only. This situation has been addressed along Sec.\il(\ref{Subsec:sequence}
and (\ref{Sec:K.L}) as limiting case.
 The first case we consider  is  the limit case $a=0$, where $\ell=0$.
\begin{description}
\item[Schwarzschild case:]
$
a=0 $: $(K\in]0,1[ ,y_3)$, see Fig.\il(\ref{Fig:TSCavaliere}, \ref{Fig:TSToti})
\item[Kerr spacetime:]
For rotating attractors,
$a\in]0, M]$, it is   for increasing values of the $K$ parameter \textbf{1.} $(K\in]0,K_2[, y_{123})$, \textbf{2.} $({K_2},y_{13})$, \textbf{3.} $(K\in]K_2,1[, y_1)$.
\end{description}
There are no configurations at $K\geq1$, see Fig.\il(\ref{Fig:STGinocchio}).

\textbf{Case: $\bar{\ell}=1$}
This is a critical configuration corotating  with the source i.e. $\ell=a$  where $E=L$ and
\be
\Phi=-\frac{L \Omega_z(\Omega_q+1)}{g_{tt} \Omega_q(1+ \Omega_q \Omega_z)},\quad\Omega=\frac{ \Omega_z(\Omega_q+1)}{\Omega_q(1-\Omega_z)},\quad \left.V_{eff}^2(\bar{\ell})\right|_{\bar{\ell}=1}=\frac{g_{tt} \Omega_q (1+\Omega_q \Omega_z)}{\Omega_q (2 \Omega_z-1)+\Omega_z}
\ee
\begin{description}
\item[Schwarzschild case:]
It is $a=0$: $(K\in]0,1[,\bar{\bar{y}}_{+})$, this case is analyzed above for $\bar{\ell}=-1$ where $\bar{\bar{y}}_{+}=y_3$.
\item[Kerr spacetime]
$a\in]0, M]$:
\textbf{1.} $(K\in]0,1[, \bar{\bar{y}}_{\pm})$,
\textbf{2.} $(1, y={a^2}/{2M})$,
\textbf{3.} $(K>1, \bar{\bar{y}}_{+})$.
See Fig.\il(\ref{Fig:STGinocchio}).
\end{description}
where:
\be
\bar{\bar{y}}_\pm \equiv-\frac{M}{K^2-1}\pm M\sqrt{\frac{M^2+a^2 (K^2-1)}{M^2(K^2-1)^2}}
\ee
are solutions $y_i$ in Eqs.\il(\ref{Eq:5raidiff}) on  $\ell=a$, however  $\bar{\bar{y}}_\pm $ are now functions of  $(K-1)$ only, not well defined at $K=1$.
At $r>r_+$  there are configurations   $K\in]0,1[$,
and $y=\bar{\bar{y}}_+$, nevertheless in $r\in]r_+,r_{\epsilon}^+]$ there are only solutions  for $K\in]0,a^2/4M^2]$ and on $r=r_{\epsilon}^+$  at $K=a^2/4M^2$,
In general,
the configurations have only one cross point on the equatorial plane, there are no closed surfaces.
\subsubsection{Configuration with $\ell=0$}\label{subsubsec:l0}
There are no  P-D configuration as $\ell=0$ (or $L=0$). We   characterized these configurations in particular for the Schwarzschild solution in Sec.\il(\ref{subsubsec:Schw}).
Thus it is :
\be
\Omega=\Omega_z,\quad E_z\equiv \left. V_{eff}(L)\right|_{L=0}=\left. V_{eff}(\ell)\right|_{\ell=0}=\sqrt{-g_{tt} (1+\Omega_q \Omega_z)},
\quad \Phi=-\frac{E \Omega_z}{g_{tt}(1+ \Omega_q \Omega_z)}.
\ee
Then for a rotating attractor, solutions are as follows:
for  \textbf{1.} $(K\in]0,K_2[, {y_{123}})$, for  \textbf{2.}
$(K_2, y_{13})$ and finally  \textbf{3.}
$(K\in]K_2,1[, {y_1})$.
Configurations with $\ell=0$ should be considered as  limiting cases  for very low constant angular momentum, as $\ell\gtrapprox0$ or $\ell\lessapprox0$, therefore  as  ``transition'' case from corotating to counterrotating fluids, as discussed also in Sec.\il(\ref{Sec:angularl}) and (\ref{Sec:limite.statico}). To be possible   a $\ell=0$ Boyer surface,  the background geometry  should have  a ``static limit'' for circular motion or a turning point of the radial acceleration, that is  an orbit $r_{stat}:\; \left. \partial_r\Omega\right|_{r_{stat}}=0$, such situations appear in geometries where  some repulsive  geometric or force effects compensate the gravitational attraction towards the source, for example in some cosmological solutions or in naked singularity sources \cite{2011,astro-ph/0605094,Stu-Sla-Hle:2000:ASTRA:,arXiv:0910.3184}.

\subsection{On the multiple thick configurations}\label{Sec:multipleP-D}
In section \il(\ref{Sec:s-pf}) we provided   a  characterization of the P-D tori in  nine classes of  spacetime backgrounds:  the  mutual spacing and arrangements of the P-D  disks and their  morphologic  characteristics  such as  the diameter, the extension on the equatorial plane and thickness  depending on the particular  black hole class where  their are located. Many  features   of the thick accretion disk model  are mostly determined by the  geometric   properties of the spacetime background  as given by  the set of radii $\mathfrak{{R}}$, in this sense one could equally say the Polish doughnut  be a  ``geometric''  model for  thick accretion disks. As a consequence of this in Secs.\il(\ref{Subsec:sequence}) and (\ref{Sec:K.L})
we  adopted  a disk ``evolutive'' interpretation for the  sequence  of configurations $\mathfrak{B}_{\mathbf{p}_j}$, analogue to the (time independent)  six-dimensional array $\mathfrak{B}_{\mathbf{p}}$,  introduced in Sec.\il(\ref{Sec:s-pf}) in accordance with
we provided a classification of Kerr attractors.
We
 assumed then one of the  parameter of the couple $\mathbf{p}$, as an evolutive or  chronological order parameter.
Here   we consider again  the set $\mathfrak{B}_{\mathbf{p}}$  as defined on a  $\Sigma_t$ surface and we focus on the multiple-configurations  possible  on the  fixed $\Sigma_t$:
we investigate  in particular   the existence of multiple ${C}$ or ${C}_{x}$ P-D  configurations  in the nine Kerr spacetime classes.
We question the existence of multi-structure toroidal accretion disks, i.e.  the aim is  to provide a  model for the internal structure of the multiple thick configuration, defined as $C\equiv \bigcup_i C_i$ with $C_i \bigcap C_j=\{\emptyset,y_{3}, y_1\}$, that is as the union of each closed, crossed or not torus, whose intersection is the null set or at last the inner or outer edge of the configurations (this analysis  does not take into account the inner disk associate to the solution $y_2$).
However  it is well  known that  the effective  potential (\ref{Eq:VltoVL}) for a Kerr-BH background  does  not have any double points of minimum or maximum:  the presence of two peaks or minimum of the hydrostatic pressure could  occur  in some special geometries  due to  the combining effect of the centrifugal repulsion and a repulsive component due for example to the presence of a cosmological constant or in spacetimes with  super spinning objects see for example \cite{2011,astro-ph/0605094,Stu-Sla-Hle:2000:ASTRA:,arXiv:0910.3184}. Thus  the sub-configurations $C_i\subset C$, can not be generated from the same effective potential $V_{eff}(\mathbf{p})$.
However, it is important to emphasize that, although not tied from the same  potential $V_{eff}(\mathbf{p})$,  their  presence as sub-configuration  in $C$ imposes some limits  constraining their angular momentum or thickness, and their dynamics   may even lead to  some instability phenomena for $C$, or feeding of  one disk from another. Each configuration  $C_i $ is a solution of an Euler equation for constant  $ \ell $, but any solutions  are linked each others  by the boundary conditions, in particular the $ K $-parameter that identifies the surface or the orbital range of integration, see also \cite{PuMonBe12}.

For a double configuration, with $(o)$ we refer here to  any quantity related  to  the external (outer) configurations  and $(i)$ to the more internal one  respect to the source, and for any quantity $\textbf{Q}$  we mean generally with $\textbf{Q}^{\pm}$ or $\textbf{Q}_{\pm}$ its value for  corotating or counterrotating matter respect to the source.

\begin{figure}
\centering
\begin{tabular}{cc}
\includegraphics[scale=.3]{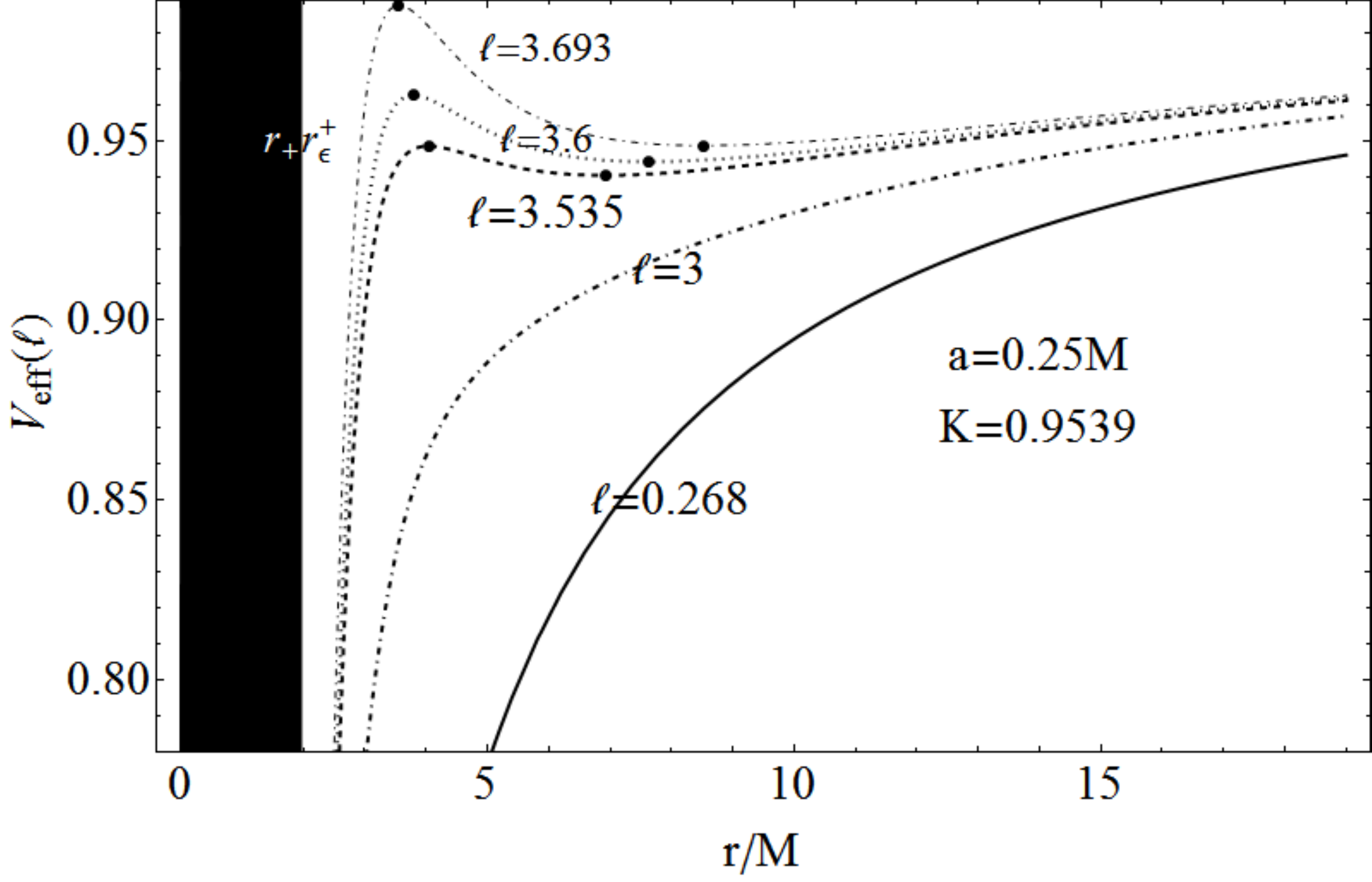}
\includegraphics[scale=.3]{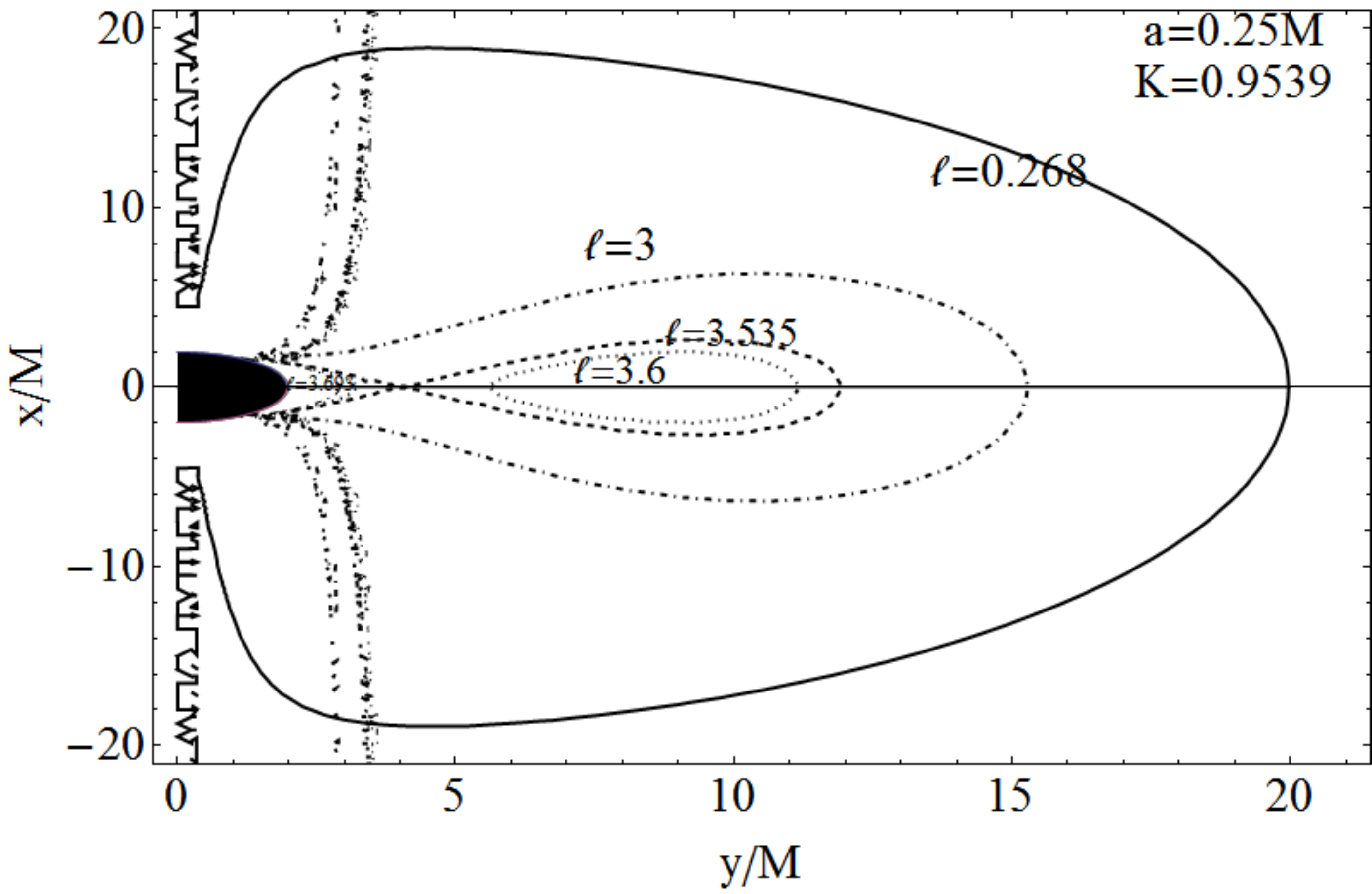}
\end{tabular}
\caption[font={footnotesize,it}]{\footnotesize{Effective potential $V_{eff}(\ell)$ (left-panel) and contour plot of the Boyer surfaces (sequences $\mathfrak{B}_{\ell}\equiv\mathfrak{B}_{\mathbf{p}}/\Sigma_{K}$  of ``loops'' of $C$, $C_x$ and  $B$-configurations) as function of $r/M$ at fixed spin $a=0.25M\in\mathbf{BHI}$ and  $K=0.9539$, and different  fluid angular momentum.
Each contour is at $\ell=$constant in units of mass $M$.
Black region is $r<r_+$ ($r_+$ is the outer horizon), gray region is $r\in]r_+,r_{\epsilon}^+]$  ($r_{\epsilon}^+$ is the static limit).}}\label{Fig:PlorrtC}
\end{figure}

We start our considerations by noting that
a double (closed) configuration  may exist if there is a couple of parameters $(\mathbf{{p}}_{(i)},\mathbf{{p}}_{(o)})$: $\mathbf{{p}}_{(i)}\neq \mathbf{{p}}_{(o)}$  and $y_1^{(i)}\leq y_3^{(o)}$.
In fact, two configurations can then be intertwined, ringed or separated at less than a double point: for
\emph{intertwined} configurations we mean    ${C}_{(o)}\cap {C}_{(i)}\neq\emptyset$ \emph{and} $\partial {C}_{(o)}\cap \partial {C}_{(i)}\neq\emptyset$  ($\partial {C}$ is the boundary of the closed configuration),  or \emph{ringed}, as ${C}_{(i)}\subset {C}_{(o)}$ on the equatorial plane that is no crossing loops of disks, should not be stable, and even if formed, (for example see Figs.\il(\ref{Fig:cPSAX01g},\ref{Fig:PlorrtC}))  they could possibly turn into  a single, energetically more favorable  configuration, for  the fluid fills the entire
contour  at equal $\ell$ $(\partial C)$ to ensure the existence and stability
of the  orbiting matter. These configurations  could have  however some relevance in  models with a non-constant  angular momentum along the disk\cite{Lei:2008ui}.
Configurations at   $\ell_{(i)}=\ell_{(o)}$  are \emph{centered} (i.e.  $ r^{(i)}_{min}=r^{(o)}_{min}$, $r_{min}$ being the disk center).
Thus we consider  here \emph{separated}   configurations i.e. ${C}_{(i)}\cap {C}_{(o)}=\emptyset$ or those with  $y_1^{(i)}=y_3^{(o)}$, that is the critical point, (maximum of potential at $K<1$) the inner point of the outer Boyer surface  or  of gravitational instability where the cross occurs in the external configuration (maximum at $K>1$)  coincides with the upper limit of the internal configuration, namely:  ${C}_{(i)}\cap {C}_{(o)}=\{y_1^{(i)}\equiv y_3^{(o)}\}$ or ${C}_{(i)}\cap {C}_{x}^{(o)}\equiv\{y_{crit}\}$  where  $y_1^{(i)}=y_3^{(o)}\equiv y_{crit}$. In this discussion we are neglecting the intersection with the inner surface that is the closest one embracing the black hole, or  $y_2$, assuming the presence of this surface, associated to the  closed configurations,  does not influence the more external ones in equilibrium or the configurations $C$. The basis of this   assumption  relies in the fact that for each closed toroidal P-D  $C$-configuration an inner surface (associated with the solution $y_2$), embracing the BH appears, however the two configurations are separated as far as the  parameter $\mathbf{p}$ of the  closed $C$-one remains far for the critical values where its morphology changes for $C$   to $C_x$ disk, the cross  can be interpreted then as a contact point with the inner source or, in other words $y_2=y_3$, so that as far the disk is regulated by Eq.\il(\ref{Eq:scond-d}), that is without considering any other possible interaction due to other ingredients of the accretion disk models as the magnetic field, these  two configurations  at the same $\mathbf{p}$ are, far from the critical phase, separated and  dynamically independent.

Thus, at $\mathbf{p}_{(i)}\neq \mathbf{p}_{(o)}$ three cases can occur: \textbf{1.} $K_{(i)}=K_{(o)}$ or \textbf{2.} $\ell_{(i)}=\ell_{(o)}$ or  finally \textbf{3. } $K_{(i)}\neq K_{(o)}$ and $\ell_{(i)}\neq\ell_{(o)}$.

\begin{figure}
\centering
\begin{tabular}{cc}
\includegraphics[scale=.3]{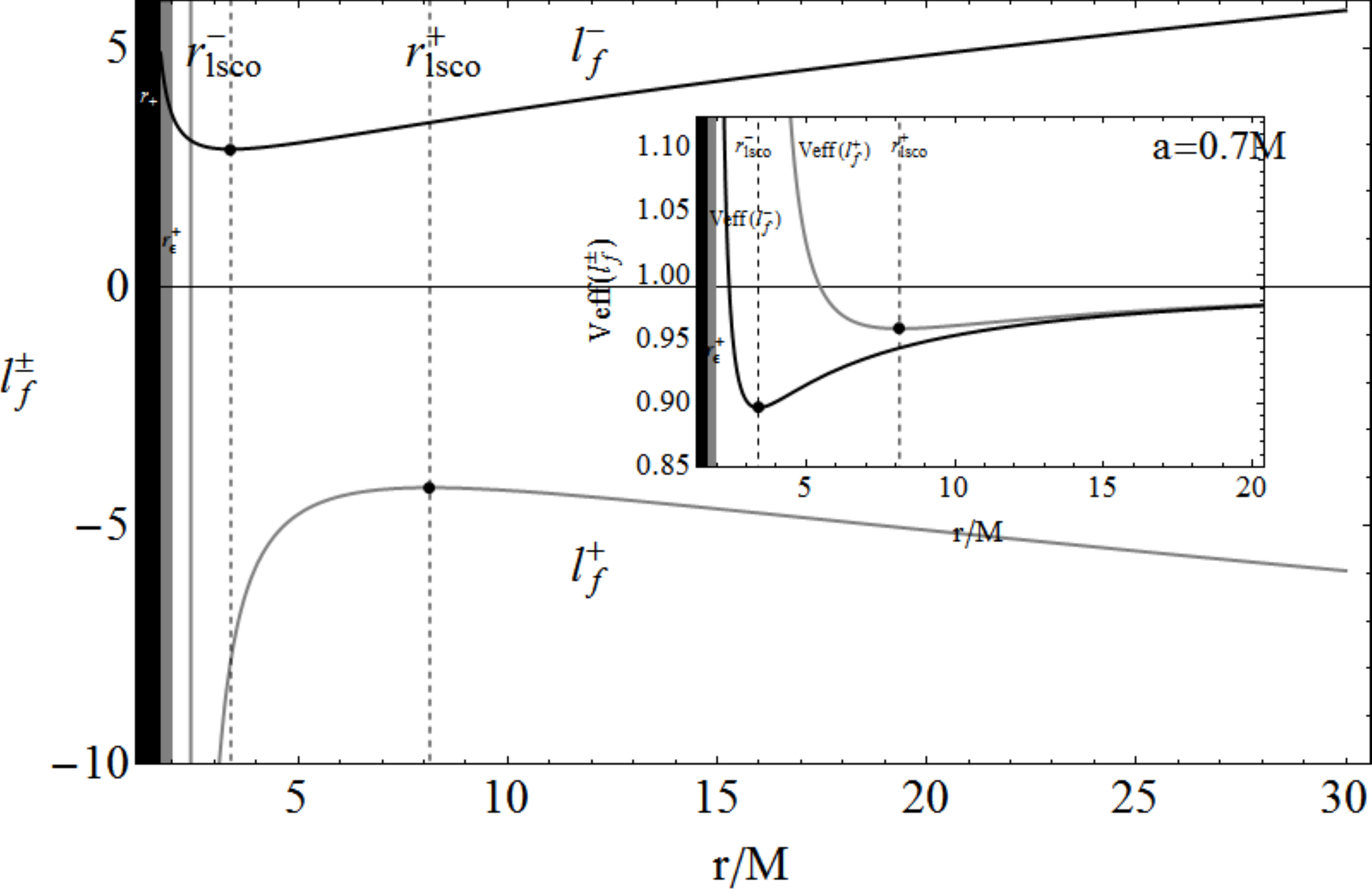}
\includegraphics[scale=.3]{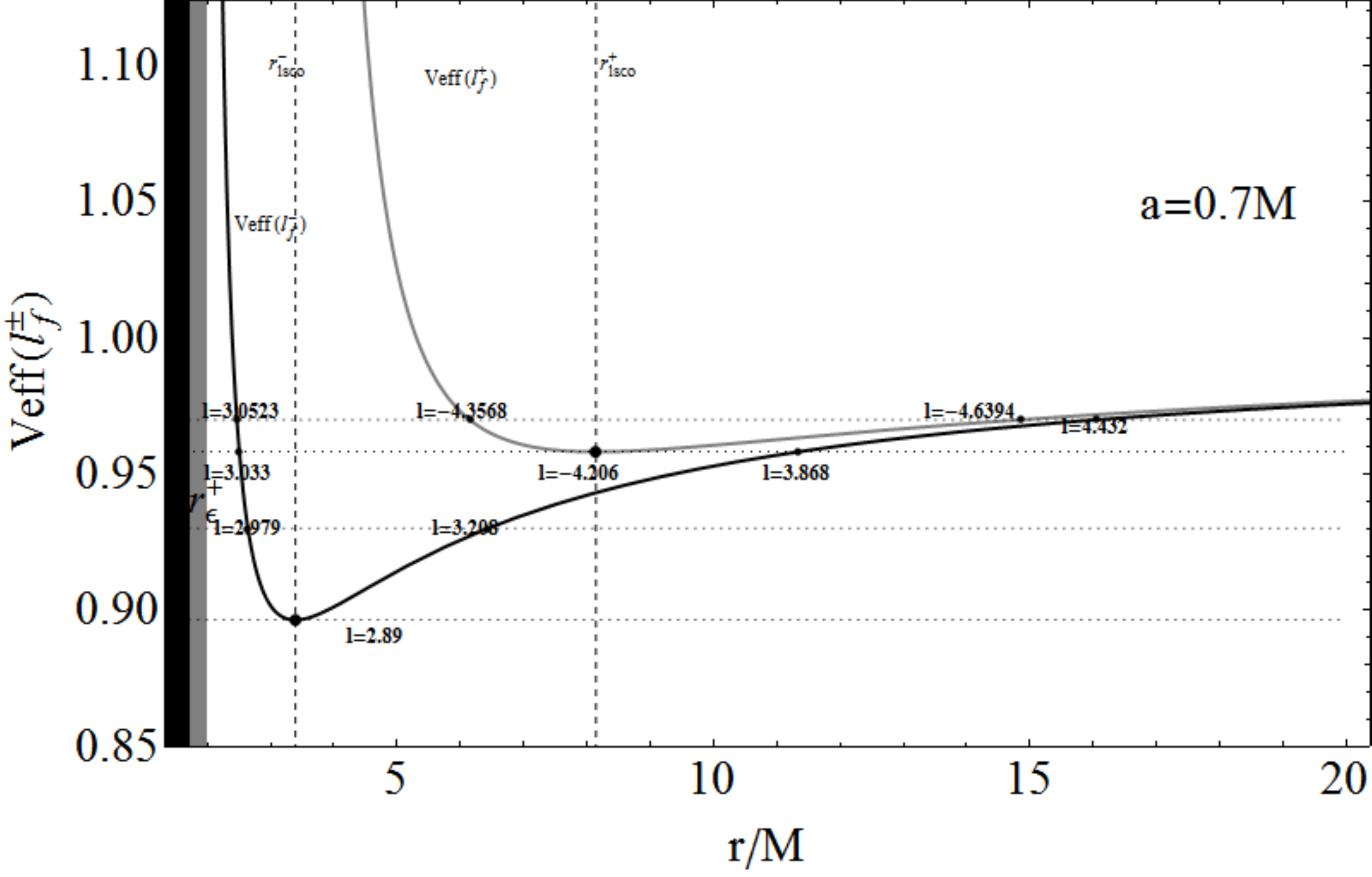}
\\
\includegraphics[scale=.3]{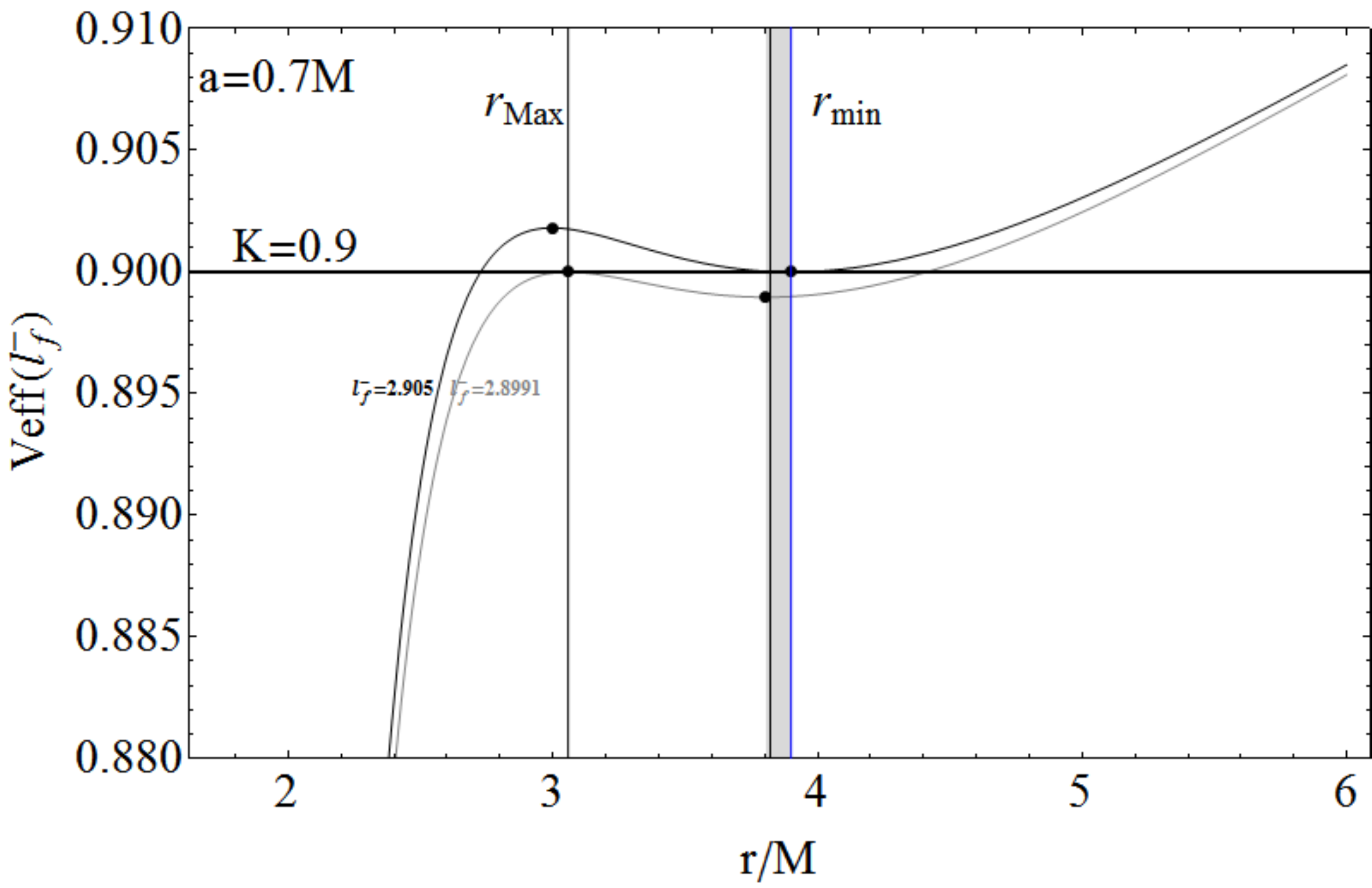}
\includegraphics[scale=.3]{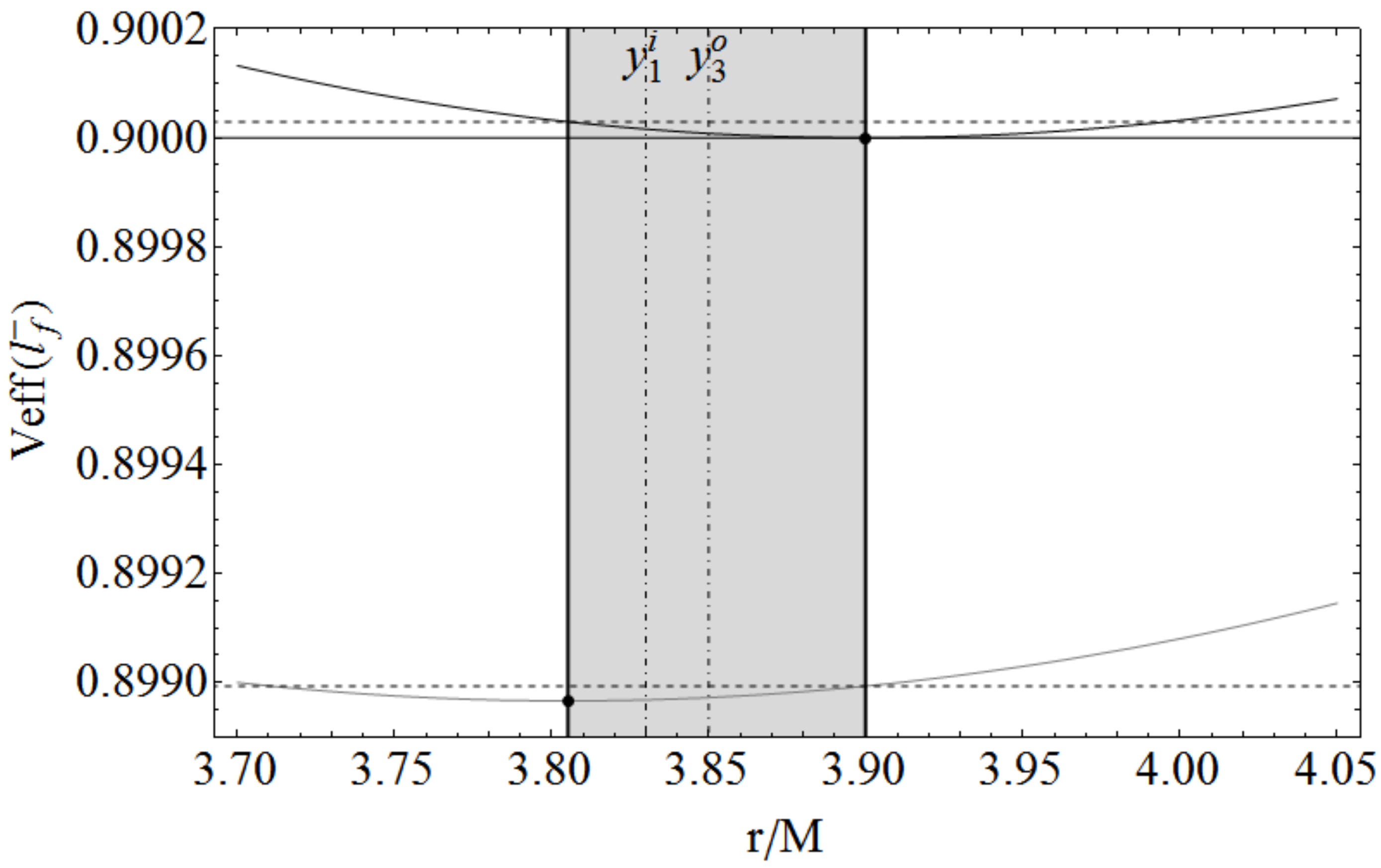}
\\
\includegraphics[scale=.3]{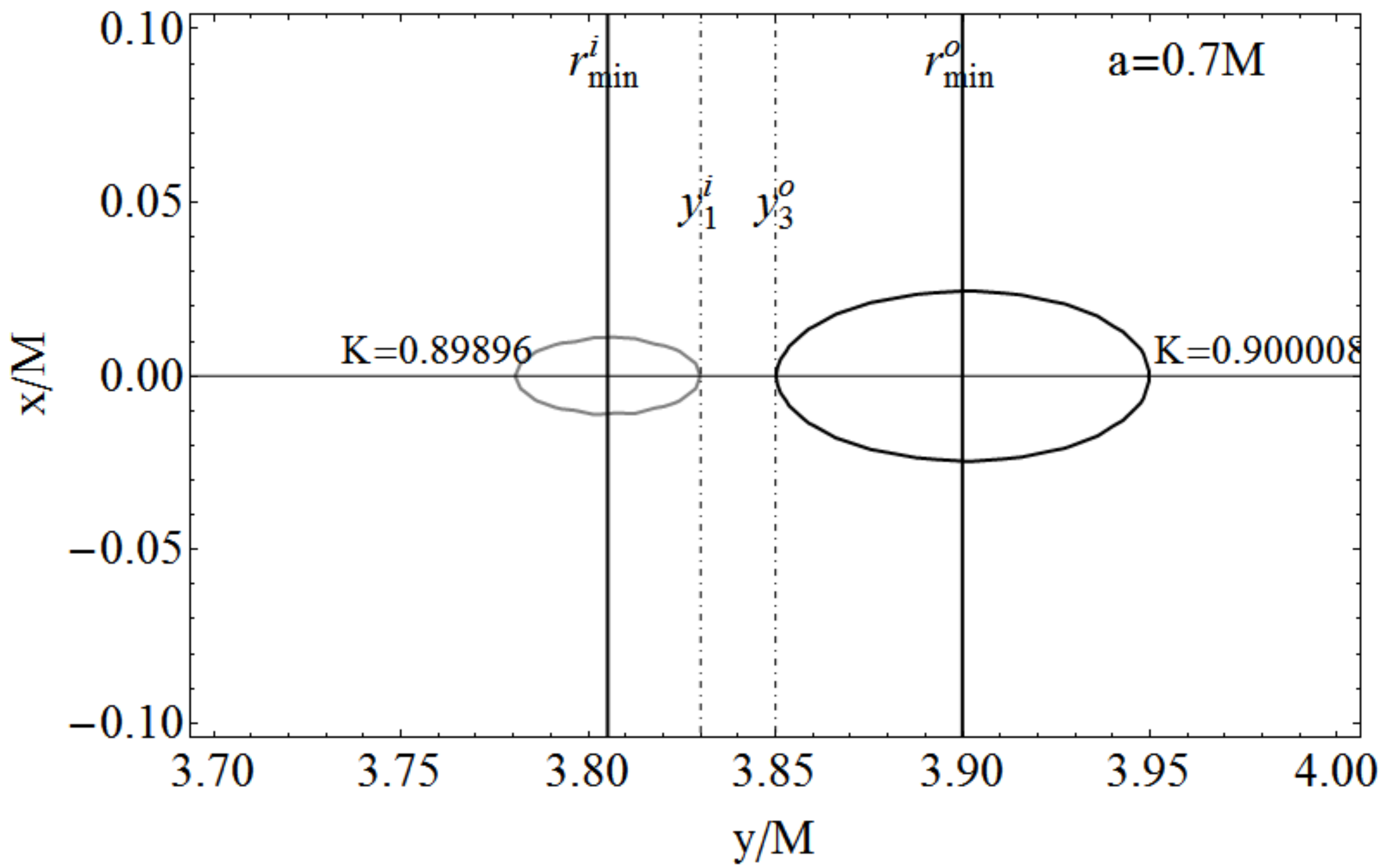}
\end{tabular}
\caption{Fluids orbiting in a  \textbf{BHIV}-spacetime  with $a=0.7M$. Left upper  panel:  the fluid angular momentum $\ell_{f}^{+}$
for counterrotating (gray curve) and corotanting $\ell_{f}^{-}$ (black curve)  matter as functions of the orbital radius $r/M$.
Black region is $r<r_+$, gray region is $r\in]r_+,r_{\epsilon}^+[$, the static limit is $r_{\epsilon}^+$, the outer horizon is the $r_+$.
Points on the curves set the critical points of the angular momentum, located at  on the last stable circular orbit  $r_{lsco}^{+}$ for counterotating matter and $r_{lsco}^{-}$ for corotating matter.  The marginally bounded orbits $r_{b}^{\pm}$ are also plotted. The lines at constant $K$-parameters are plotted with dotted lines.  Right panel: the effective potential  $V_{eff}(\ell_{f}^{+})$
for counterrotating (gray curve) and corotanting $V_{eff}(\ell_{f}^{-})$ (black curve)  matter as functions of the orbital radius $r/M$, the values  $V_{eff}(\ell_f^{\pm})$ are the $K$-parameter values of the Boyer surfaces, the values $\ell=$constant are signed with dotted lines.
Center panels: Plot of $V_{eff}(\ell_f^-)$ versus $r/M$.
the surfaces at $K=0.90$ are for $\ell_f^-=(2.899184,2.905348)$  gray and black curves respectively. At   $K=0.90$ are the maximum points  $r^{\pm}_{Max}$   and the minimum $r^{\pm}_{min}$  for the two curves respectively, double (separated and closed) Boyer configurations are allowed in the gray region where  $y_3^{(o)}$  and$ y_1^{(i)}$ must be located: the  boundaries of the region are
$r_{min}$ for  the outer configuration and the minimum point
of the inner one (signed by a point on the gray
curve)  i.e.
$[y_1^{(i)}, y_3^{(o)}]\subset[r^{(i)}_{min},r^{(o)}_{min}]$, left panel shows a zoom of the gray region
$[r^{(i)}_{min},r^{(o)}_{min}]$ as the location of $(y_1^{(i)}, y_3^{(o)})$
dotted-dashed lines. Bottom panel shows the couple of
corresponding  Boyer surfaces, an analogue couples
of configurations at $\ell=\ell_f^+$ are located at
$r>4M$. The angular momentum is in units of mass $M$. These are the maximum number $(n=4)$ of $C$
configuration in this spacetime.}\label{Fig:bCKGO}
\end{figure}

The second case,  \textbf{2.} $\ell_{(i)}=\ell_{(o)}$, is immediately ruled out, as these are evidently centered configurations (see also Fig.\il(\ref{Fig:bCKGO})), at constant $\ell$ the effective potential is uniquely defined  as function of $r/M$, and  there is only one maximum and one minimum.
For   \textbf{1.} $K_{(i)}=K_{(o)}$, configurations can be mutually corotating or
 {\textbf{\large{$ {\ell}$}}}corotating, meaning here $\ell_{(i)}\ell_{(o)}>0$ or {\textbf{\large{$ {\ell}$}}}counterrotating,   $\ell_{(i)}\ell_{(o)}<0$, see Fig.\il(\ref{Fig:bCKGO}), this case as well as the third \textbf{3.}  $K_{(i)}\neq K_{(o)}$ requires  further discussion.

\medskip

First, a  particular but interesting case  is for $ \Delta_{cri}^{(i)}\cap\Delta_{cri}^{(o)}=\emptyset$, ($\Delta_{crit}\equiv[r_{Max}, r_{min}]$ introduced in Sec.\il(\ref{Sec:limite.statico})) where  separated configurations can certainly exist as  indeed $r_{max}^{(o)} \in I^+_{\ti{$r_{min}^{(i)}$}}$,  where $I^+_{\ti{$r_{min}^{(i)}$}}$ is a  right neighborhood of $r_{min}^{(i)}$, i.e. $r_{max}^{(o)}>r_{min}^{(i)}$ for the continuity of the effective potential there exist a point $\bar{r}^{(o)}\in I^+_{\ti{$r_{min}^{(i)}$}}:\; \bar{r}^{(o)}<r_{Max}^{(o)}$ and it can be $\bar{r}^{(o)}=y_1^{(i)}$ for the configuration at $K_{(i)}=V_{eff}(p_{(i)},\bar{r})$. Thus, in conclusion a possible \emph{sufficient} but not necessary condition  for the formation of multiple surfaces is that  $ \Delta_{cri}^{(i)}\cap\Delta_{cri}^{(o)}=\emptyset$. It must be
$r_{lsco}^{(i)}<r_{min}^{(i)}<r_{Max}^{(o)}<r_{lsco}^{(o)}$, this consideration clearly excludes {\textbf{\large{$ {\ell}$}}}corotating matter configurations i.e. $\ell^{(o)} \ell^{(i)}>0$ (indeed  $r_b$ and $r_{lsco}$ are functions of the black hole spin only and  there are a couple of solutions $(r_b^{\pm},r_{lsco}^{\pm})$ for $\ell a\lessgtr0$ respectively).
For {\textbf{\large{$ {\ell}$}}}counterrotating disks, $\ell^{(o)} \ell^{(i)}<0$, it has to be
$r_{lsco}^{(i)}<r_{lsco}^{(o)}$. Then in conclusion multiple configurations of this kind, are necessarily {\textbf{\large{$ {\ell}$}}}counterrotating and with reference to  Fig.\il(\ref{Fig:radiilf1})-upper, it is $r_{lsco}^{(i)}=r_{lsco}^-$ and $r_{lsco}^{(o)}=r_{lsco}^+$, that  is the interior surfaces must be corotating respect to the black hole ($\ell_{(i)} a>0$), and  the region $[r_{min}^{(i)},r_{Max}^{(o)}]$ has to be included in $[r_{lsco}^-, r_{lsco}^+]$  or
$r_{lsco}^-<r_{min}^-<r_{Max}^+<r_{lsco}^+$. Furthermore  from  analogue consideration  it follows  that the maximum number of such multiple disks is  $n=2$ \footnote{Indeed, considering a  sequences of   $n=3$  closed configurations ${C}_{i}<{C}_j$ for  $i<j$, (with $<$ related to the matter configurations we mean the ${C}_i$ being the more internal one, closest to gravitational source, respect to ${C}_j$) it must be $r_{lsco}^{(1)}<r_{min}^{(1)}<r_{Max}^{(2)}<r_{lsco}^{(2)}<r_{min}^{(2)}<r_{Max}^{(3)}
<r_{lsco}^{(3)}$  which is contradictory for any kind  of rotating matter. }. The  range   $[r_{lsco}^-, r_{lsco}^+].$  increases with $a/M$ from the Schwarzschild case,  where we do not distinguish  multiple corotating or counterrotating configurations (meaning $\ell a\lessgtr0$ and \emph{therefore} $\ell_{(i)} \ell_{(o)}\lessgtr0$), to the extreme Kerr case, see also Fig.\il(\ref{Fig:radiilf1}).  The region $\Delta_{crit}$ increases with the spin and the configurations can be   more spaced,  the  innermost one approaching the black hole. However we need to distinguish  between the configurations at  spacetimes  $a\in]0,a_{\bullet}[$ where $r_b^+<r_{lsco}^-$ (\textbf{BHI} and \textbf{BHII}) and the region $a\in]a_{\bullet},M[$ where $r_b^+>r_{lsco}^-$: in  \textbf{BHI} and \textbf{BHII} there  are only  ${C}$-configurations.
Considerations on the last bounded orbital  $r_b^- $  make clear that the corotating internal configurations are necessarily closed. On the other hand, we have the double condition  $\Delta_{crit}\subset[r_b^+,r_{lsco}^+]\subset[r_{lsco}^-,r_{lsco}^+]$ at $a>a_{\bullet}$, otherwise for $\Delta_{crit}\subset[r_{lsco}^-,r_{b}^+]$  at $r_{Max}^{(o)}$ it is $K_{Max}^+>1$ and there is the possibility of outer (counterrotating) surface with a cross in $r>r_{Max}$ leading to an instability and funnel of matter in the region of the interior surface.
\medskip

More generally for the existence of the separated configurations it has to be  $r_{lsco}^{(i)}<r_{min}^{(i)}<r_{min}^{(o)}$, this relation also applies to the $\ell$-corotating disks  and it is  necessary for the existence of disjoint configurations for that  it can be considered a definition for  and a criterion to establish   the internal and external configurations. The case  $r_{Max}^{(i)}<r_{min}^{(i)}<r_{Max}^{(o)}<r_{min}^{(o)}$ has been explore above, now we focus on the case
$r_{Max}^{(o)}<r_{min}^{(i)}$ (it can be $r_{Max}^{(i)}<r_{Max}^{(o)}<r_{min}^{(i)}$ or $r_{Max}^{(o)}<r_{Max}^{(i)}<r_{min}^{(i)}$).
As it always exists, for definition of  critical point, a closed surface centered in the minimum one can always check for  two separated surfaces. But the location of the maximum $r_{Max}^{(o)}$ selects the kind  of rotating matter,  in fact if the disks are {\large{$\ell$}}corotating, $\ell_{(i)}\ell_{(o)}>0$,  then it is $r_{Max}^{(o)}<r_{min}^{(i)}$, if {\large{$\ell$}}counterrotating,  $\ell_{(i)}\ell_{(o)}<0$,  then it is $r_{lsco}^{(o)}>r_{lsco}^{(i)}$ and it follows that  $\ell_{(o)}>0$ (the corotating configuration respect to the black hole  is the exterior one) and $\ell_{(i)}<0$ the interior configuration is counterrotating $\ell_{(i)} a<0$ and viceversa from  $r_{lsco}^{(o)}<r_{lsco}^{(i)}$ it follows that  $\ell_{(o)}<0$ (counterrotating $\ell_{(o)} a<0$) and $\ell_{(i)}>0$ i.e. the internal configuration is corotating respect to the black hole.    For  $n>2$ the necessary condition is always verified for surfaces of all {\large{$\ell$}}corotating matter $\ell_{i}\ell_j>0$, however further discussion is required for the value of the $K$-parameter.

\medskip

\begin{figure}
\centering
\begin{tabular}{cc}
\includegraphics[scale=.3]{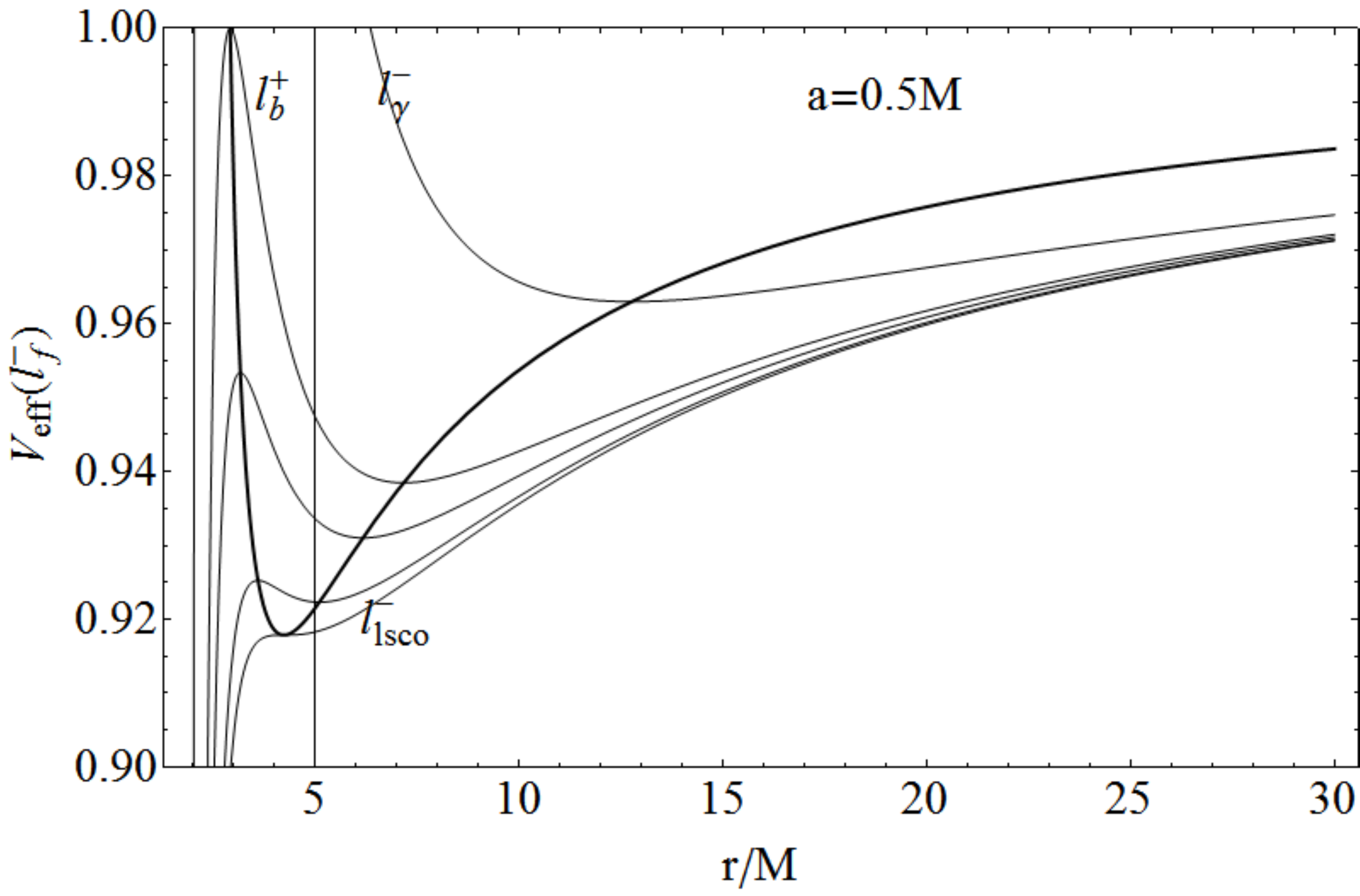}
\includegraphics[scale=.3]{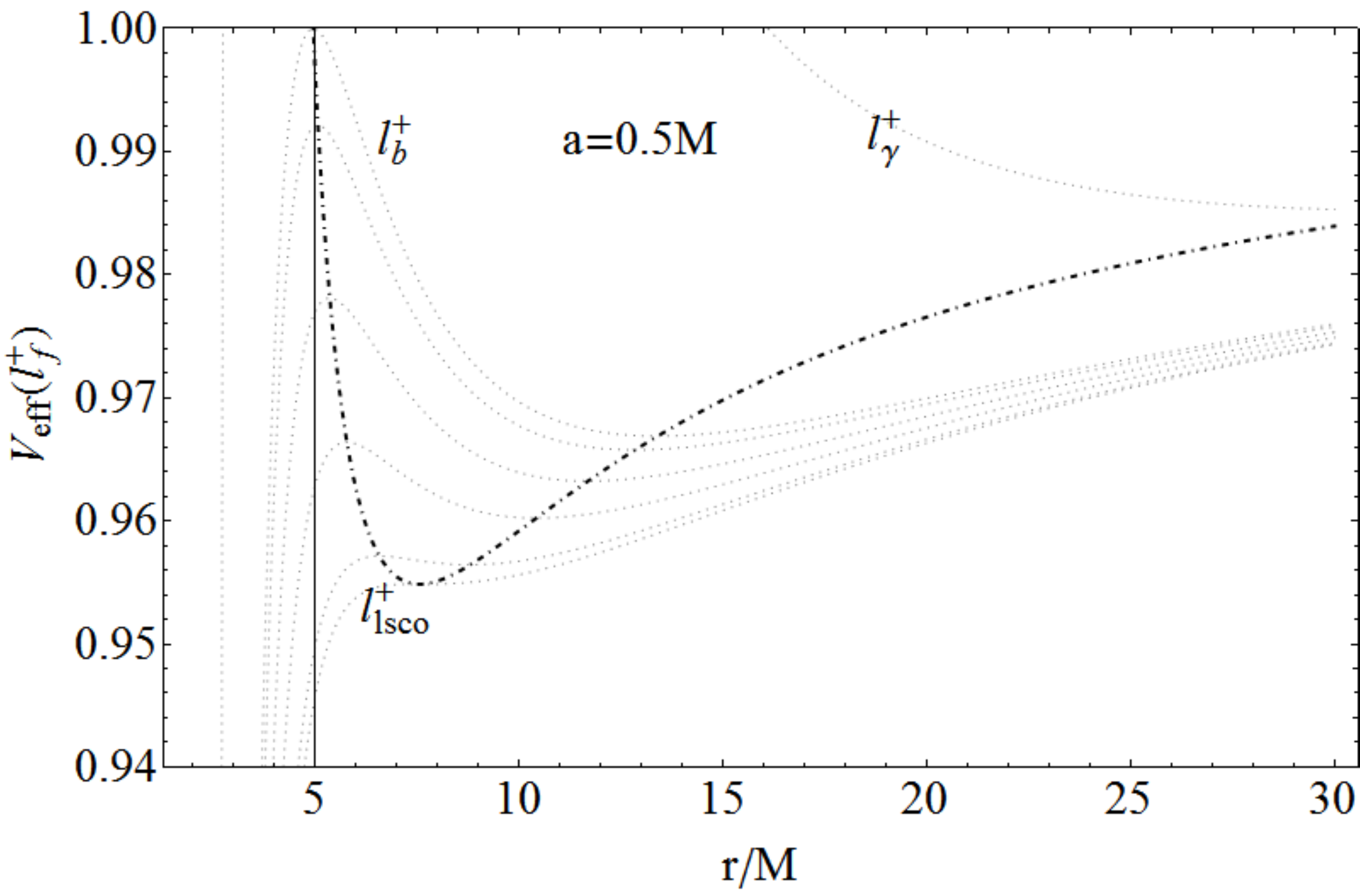}
\\
\includegraphics[scale=.3]{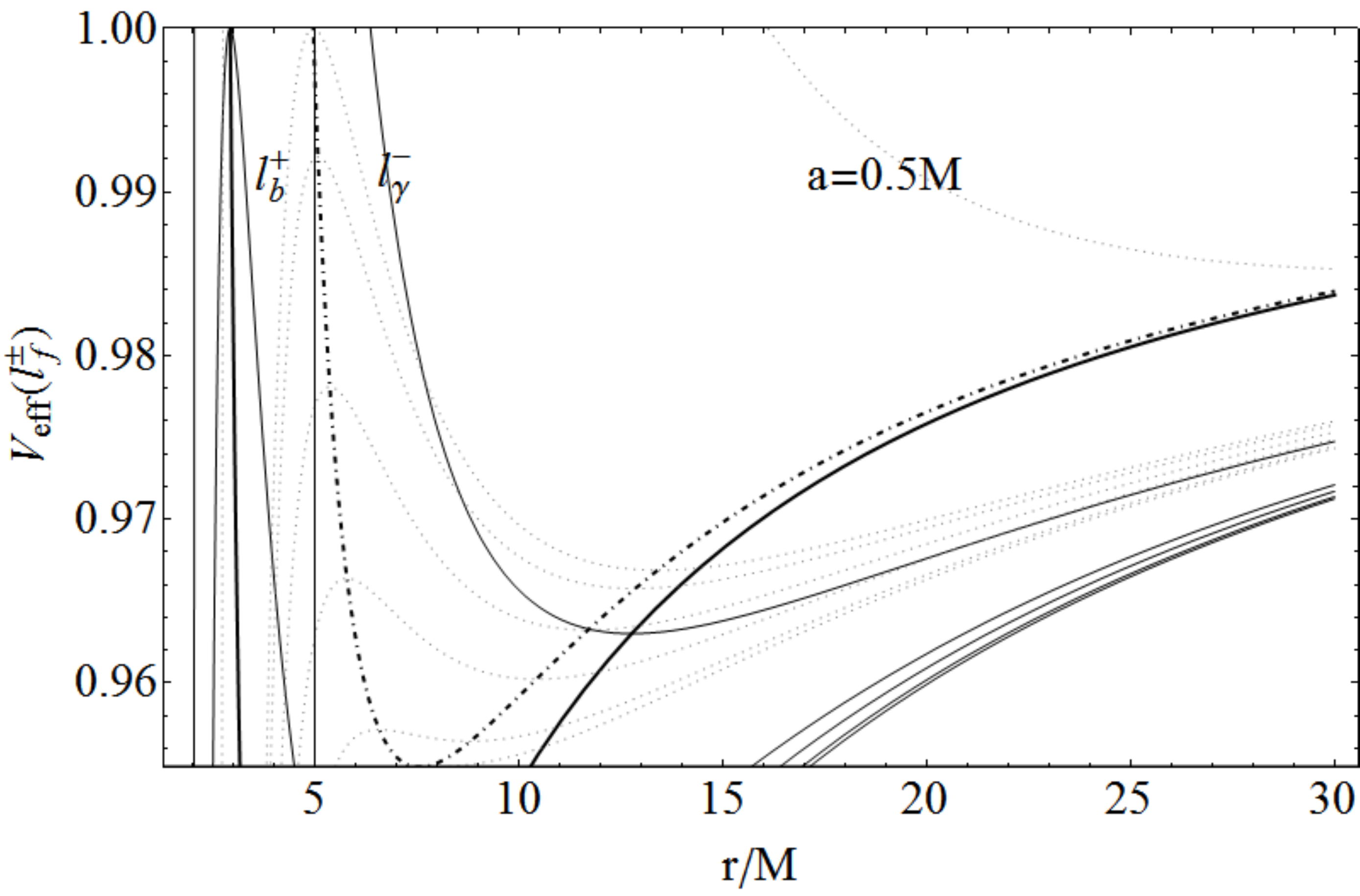}
\end{tabular}
\caption{{Effective potential $V_{eff}(\ell)$ as function of $r/M$ for a black hole spacetime at $a=0.5M\in\mathbf{BHIII}$, for different values of the fluid angular momentum $\ell_f^-$ (upper left  panel), $\ell_f^+$  (upper right panel), and $\ell_f^{\pm}$ bottom panel, in the range $\ell\in[\ell^{\pm}_{lsco},\ell^{\pm}_b]$ for corotating $(-)$ and counterrotating fluids  $(+)$, where $\ell^{\pm}_{lsco}\equiv \ell(r^{\pm}_{lsco})$ and  $\ell^{\pm}_{b}\equiv \ell^{\pm}(r_{b})$, where $r^{\pm}_{lsco}$ the radius of the last stable circular orbits, and $r^{\pm}_{b}$ the radius of the last circular bounded orbit, the potential at $\ell_{\gamma}^{\pm}\equiv \ell_(r_{\gamma}^{\pm})$ has been plotted as well where $r_{\gamma}^{\pm}$ is the last circular orbit. Thick lines mark the  critical points of the effective potentials.
The angular momentum is in units of mass $M$.}}\label{Fig:nesiferp}
\end{figure}

In  Figs.\il(\ref{Fig:bCKGO}) we show a procedure to select, at different $K$, multiple separated P-D  configuration, by requiring  $K_{Max}^{(i)}=K_{min}^{(o)}$, the maximum number of separated configurations is $n_{Max}=4$, two of then {\large{$\ell$}}counterrotating, $\ell_{(i)}\ell_{(o)}<0$, and the others {\large{$\ell$}}corotating, $\ell_{(i)}\ell_{(o)}>0$, as shown.
%
The case of $K_{(i)}=K_{(o)}$  is solved for any couple of fluid angular momenta  $\ell_{(i)}\neq\ell_{(o)}:\; \exists!K=K_{(i)}=K_{(o)}$ where $K_{(i)}$ ($K_{(o)}$) is the effective potential associated to the inner (outer) configuration.
This condition has to be completed by $r_{max}^{(i)}<r_{min}^{(i)}<y_1^{(i)}<y_3^{(o)}<y_{min}^{(o)}$ where $\left(\right)_{min}$ ($\left(\right)_{Max}$) refers as usually to the quantities at the minimum (maximum) of the effective potential. Thus it is $K_{min}^{(i)}<K<K_{Max}^{(o)}$. That is immediately verified for {\large{$\ell$}}corotating configurations $\ell_{(i)}\ell_{(o)}>0$.
We note that
in the  {\large{$\ell$}}corotating case, $\ell_{(i)}\ell_{(o)}>0$ at fixed $r$ the effective potential increases with the  magnitude of the angular momentum: $V_{eff}(\ell_b)>V_{eff}(\ell_a)$ as $|\ell_b|>|\ell_a|$ that is  $K_{min}(\ell_b)>K_{min}(\ell_a)$ where $r_{lsco}^a=r_{lsco}^b$ see Fig.\il(\ref{Fig:nesiferp}), therefore it is $r_{Max}^a<r_{Max}^b<y_3^b<y_3^a<y_{min}^b<y_{min}^a<y_1^a<y_1^b$ and they are centered configurations: $[y_3^a,y_1^a]\subset[y_3^b,y_1^b]$.
For the {\large{$\ell$}}counterrotating configurations it is  $r_{lsco}^-<r_{lsco}^+$, the inner configuration is corotating with respect to the source. It is necessary to distinguish the \textbf{BHI-II} sources at $a<a_{\bullet}:\;r_b^-<r_b^+<r_{lsco}^-<r_{lsco}^+$ and \textbf{BHIII-IX} $a>a_{\bullet}:\;r_b^-<r_{lsco}^-<r_b^+<r_{lsco}^+$. It is at fixed $r$ $V_{eff}^+>V_{eff}^-$, but it is required that $K_{Max}^-\geq K_{Max}^+$ and $K>K_{lsco}^+$ and it is always verified $r_b^{\pm}\leq r_{Max}^{\pm}<y_3^{\pm}<r_{lsco}^{\pm}<r_{min}^{\pm}<y_1^{\pm}$, but the assumption $y_1^-<y_3^+$ imposes the condition $r_{min}^-<r_{min}^+$ that is not satisfied as $r_{lsco}^-<r_{lsco}^+<r_{min}^+<r_{min}^-$ and $r_{max}^-<r_{Max}^+$ and
$r_{min}^+<r_{Max}^-$ with $V_{eff}^+>V_{eff}^-$, see Fig.\il(\ref{Fig:nesiferp})
However varying  $K\in]K_{min}, 1[$  there may be a continuous number of  many separated configurations, the pairs will denote configurations at different $K$.

We conclude this section with some final remarks: we  have investigated the possible existence of a structured accretion disks defined as the union configuration $C\equiv \bigcup_i C_i$ with $C_i \bigcap C_j=\{\emptyset,y_{3}, y_1\}$, obtained combining   sub-configurations of closed or closed crossed, P-D  configurations in such a way that they  can be separated and not intertwined or  in loops.
Then we discussed some of the properties of $C_i$ configurations,  showing that  they are deeply constrained  in number  and in the angular momentum, so that  we need to introduce  the concept of $\ell$corotating and $\ell$counterrotating rings. In some cases it was proved the existence of a maximum number of separated configurations. This first analysis shows  the possibility to  have a multi-structured disk of a limited number of rings.
Thus, the results traced here   should  be then completed including also the information related to the thickness and spatial distance of   each sub-configurations.
The dynamical equilibrium of the entire  structure should be  also investigated:
 there could arise   an instability mechanism for the $C$-configuration due to the presence of one or more  exterior $C_x$-configurations leading to a mutual destabilization  (of gravitational and hydrostatic equilibrium of each torii as foreseen  from the P-W model) among the sub-configurations,  this mechanism  imposes a   limit in the number of disks, location and fluid angular momentum for  the equilibrium of the multiple structures\cite{coop}.

\subsection{The polytropic equation of state and the  disk morphology}\label{SeC:poly}
We consider  a polytropic equation of state assuming the pressure $p$ be a function of the matter density $\varrho$: $p=k \varrho^{\gamma}$, where  $k>0$  and $\gamma$ is the polytropic index.
The case of non-rotating attractor  has been addressed in \cite{PuMonBe12},  mostly of the considerations traced out for $a=0$ are  valid for a more generic  P-W-effective potential, in particular  then for $a\neq0$.
It has been shown that for the Schwarzschild geometry there is a specific classification of eligible geometric polytropics (see  \cite{Raine}), and a specific class of polytropics is characterized by a discrete  range of values for the index $\gamma$.
A similar classification is also valid in the rotating case of  the  Kerr geometry.
Solving Eq.\il(\ref{Eq:scond-d}) for  $\varrho$  we find:
\bea\label{peter}
\bar{\varrho}_{\gamma}\equiv\left[\frac{1}{k}\left(V_{eff}^{-\frac{\gamma -1}{\gamma
}}-1\right)\right]^{\frac{1}{(\gamma-1)}}\quad\mbox{for}\quad\gamma\neq1,
\quad
\varrho_k\equiv V_{eff}^{-\frac{1+k}{ k}}\frac{1}{1+k},\quad\mbox{for}\quad \gamma=1.
\eea
We adopt the normalization: $\varrho_{\gamma}\equiv
k^{1/(\gamma-1)}\bar{\varrho}_{\gamma}$, which is independent from $k$. 

For $\gamma=1$ it is
\(
\varrho^{out}_{k}=\varrho^{in}_{k}\left({V^{out}_{eff}/V^{in}_{eff}}\right)^{-\frac{1+k}{k}},
\)
and for $\gamma\neq1$ it is
\(
k (\bar{\varrho})_{out}^{\gamma-1}=(k (\bar{\varrho})_{in}^{\gamma-1}+1)
\left({V^{out}_{eff}/V^{in}_{eff}}\right)^{\frac{1-\gamma}{\gamma}}-1
\),
where the integration range $[r_{in},
, r_{out}]\subset]r_+, \infty]$ is the range of existence for $V_{eff}(\ell)$.
It is $\varrho'=0$
when $p'=0$ and, being $\gamma>0$, the maxima (minima) of $p$ correspond to the maxima
(minima) of $\varrho$.
If the polytropic index is $\gamma\neq1$ then the density $\varrho=\varrho_{\gamma}$ is:
\(
{\varrho}_{\gamma}\equiv C^{1/(-1+\gamma )},
\)
with $C\equiv (V_{eff}^{-2})^{\frac{\gamma-1}{2 \gamma }}-1$.
We can distinguish  the following  two cases:
\textbf{1.} $C>0$ and the density $\varrho_{\gamma}$ is defined for all $\gamma$;
\textbf{2.} $C<0$ and the density $\varrho_{\gamma}$ is defined for
      $\gamma=\gamma_q\equiv1+\frac{1}{2q}$, where $|q|\geq 1$ are integers-
The condition $C>0$ is satisfied in two cases: where $V_{eff}^2<1$ and $\gamma >1$, in
a range  $\mathbf{R}_{\ti{I}}$,  and where $V_{eff}^2>1$ and $\gamma \in]0,1[$, in the ranges $\mathbf{R}_{\ti{II}}$.
When the polytropic index $\gamma=\gamma_{q}$, the fluid
density $\varrho$ is defined for the conditions $\mathbf{R}_{\ti{I}}\cup \mathbf{R}_{\ti{II}}$, when
$\gamma\neq\gamma_{q}$ it is defined only for the conditions $\mathbf{R}_{\ti{I}}$.
Then the only difference with the geometry $a=0$ lies in the identification of the regions $\mathbf{R}_{\ti{I}}$ and $ \mathbf{R}_{\ti{II}}$, these are easily identifiable in the equatorial plane and are studied   for the Boyer surfaces  in Sec.\il(\ref{Sec:KU1ana}).
Then for the more general classes including  the Boyer ${C}$-configurations, according to the  analysis in Sec.\il(\ref{Sec:Morfology}):
 $\mathbf{R}_{\ti{I}}$ is
$r\in]r_+,r_{\epsilon}^+[$ with ${\ell}<\hat{\ell}_\sigma^- \cup\ell> \hat{\ell}_{\sigma}^+$
and at $r>r_{\epsilon}^+$ $\ell\in]\hat{\ell}_\sigma^-,\hat{\ell}_\sigma^+[$, where
\be
\frac{\hat{\ell}_{\sigma}^{\pm}}{M}\equiv\pm\sqrt{2} \sqrt{\frac{r \Delta  \sigma ^2 \rho ^2}{M\left(\rho ^2-2 M r\right)^2 }}+\frac{2 a r \sigma ^2}{2M r-\rho^2},
\ee
(where $\sigma$ and  $\rho$ are introduced in Sec.\il(\ref{Sec:model}))
and  $\mathbf{R}_{\ti{II}}:$
$r\in]r_+,r_{\epsilon}^+[$ with ${\ell}\in]\hat{\ell}_\sigma^-, {\ell}_{\sigma}^-[\cup]{\ell}_{\sigma}^+,\hat{\ell}_\sigma^+[$
and at $r>r_{\epsilon}^+$  ${\ell}\in]{\ell}_{\sigma}^-,\hat{\ell}_\sigma^-,[
\cup]\hat{\ell}_\sigma^+,{\ell}_{\sigma}^+[$.
We can interpret these regions in terms of the orbits   $r_{\kappa}^{\pm}$  in Eq.\il(\ref{Eq:firsr-rver}) and  $r_{\ti{B}}^{\pm}$ in Eq.\il(\ref{Def:RBB}), projecting for simplicity to the plane $\theta=\pi/2$ thus:
$\mathbf{R}_{\ti{I}}$:
$r\in]r_+,r_{\kappa}^-[\cup r>r_{\kappa}^+$ where $\ell\leq\ell_b^+
\cup\ell>\ell_b^-$
and for $\ell\in]\ell_b^-,\ell_b^+[$ it is $r>r_+$.
and  $\mathbf{R}_{\ti{II}}$:
$r\in]r_{\kappa}^-,r_{\ti{B}}^-[\cup]r_{\ti{B}}^+, r_{\kappa}^+[$ for $\ell<\ell_{\gamma}^+\cup \ell>\ell_{\gamma}^-$
and $r\in]r_{\kappa}^-,r_{\kappa}^+[$ for $\ell\in]\ell_{\gamma}^+,\ell_b^+[\cup]\ell_b^-,\ell_{\gamma}^-[$.
\subsubsection{On the morphology of the Boyer surfaces}
\begin{figure}
\centering
\begin{tabular}{cc}
\includegraphics[scale=.31]{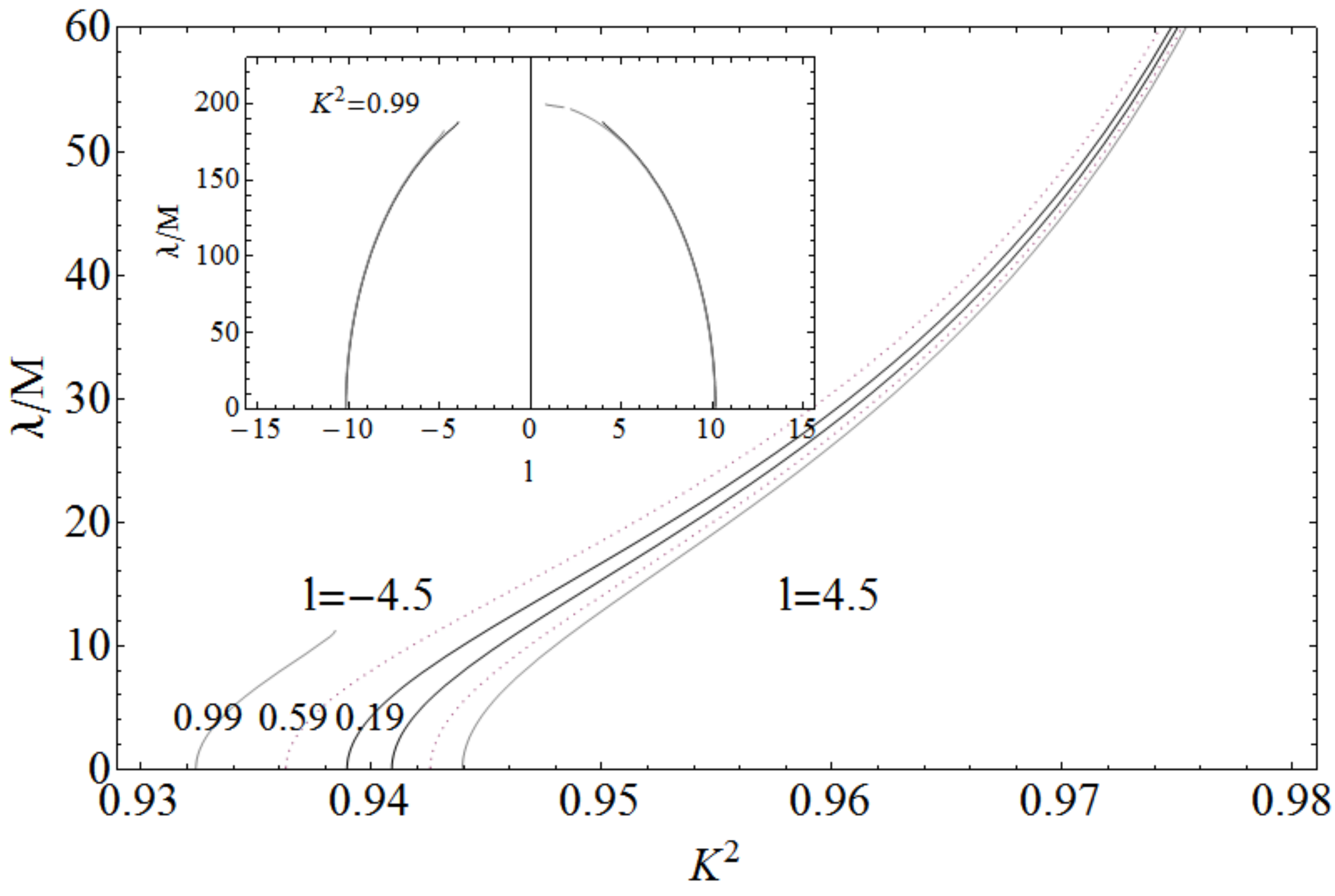}
\includegraphics[scale=.31]{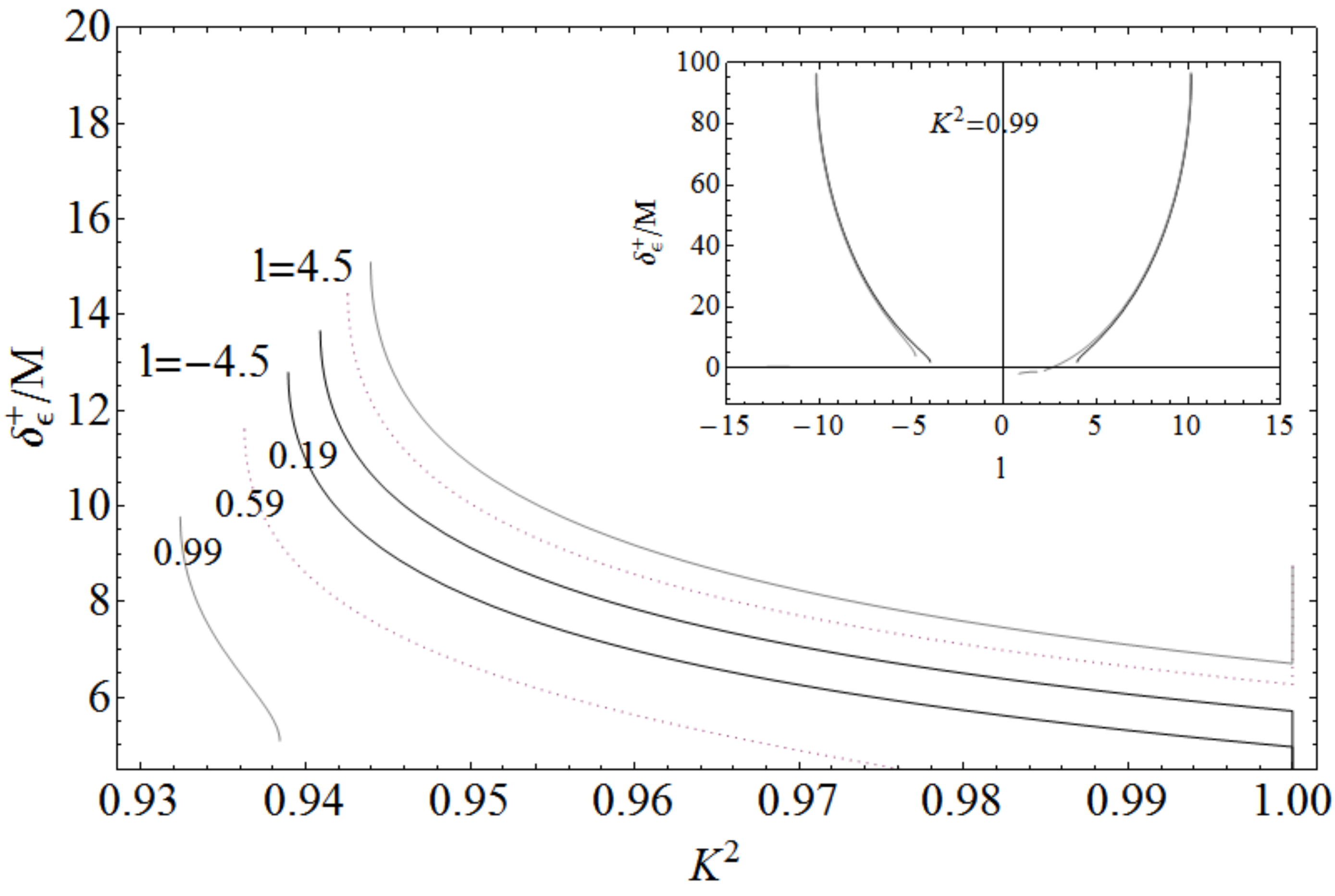}
\\
\includegraphics[scale=.31]{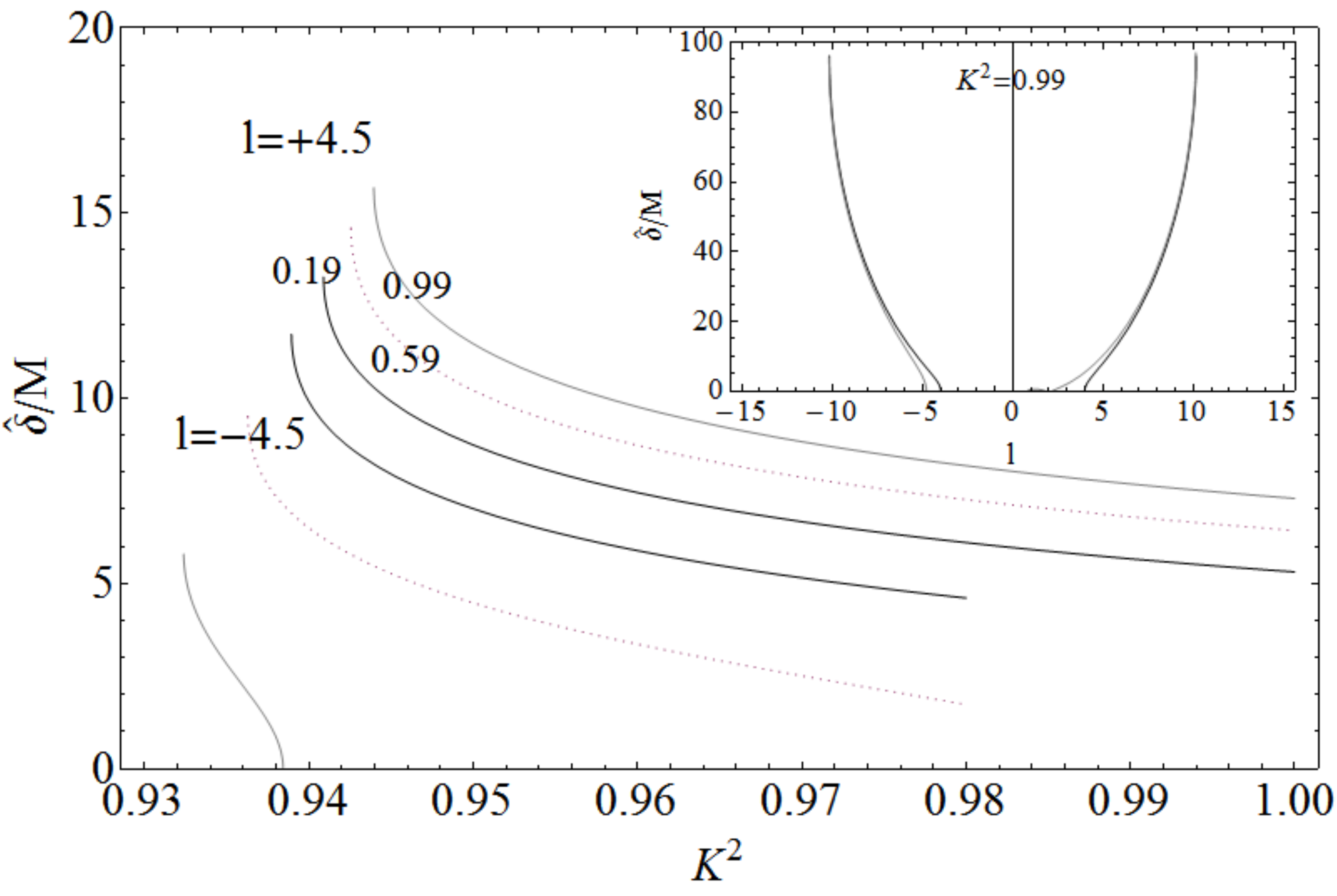}
\includegraphics[scale=.31]{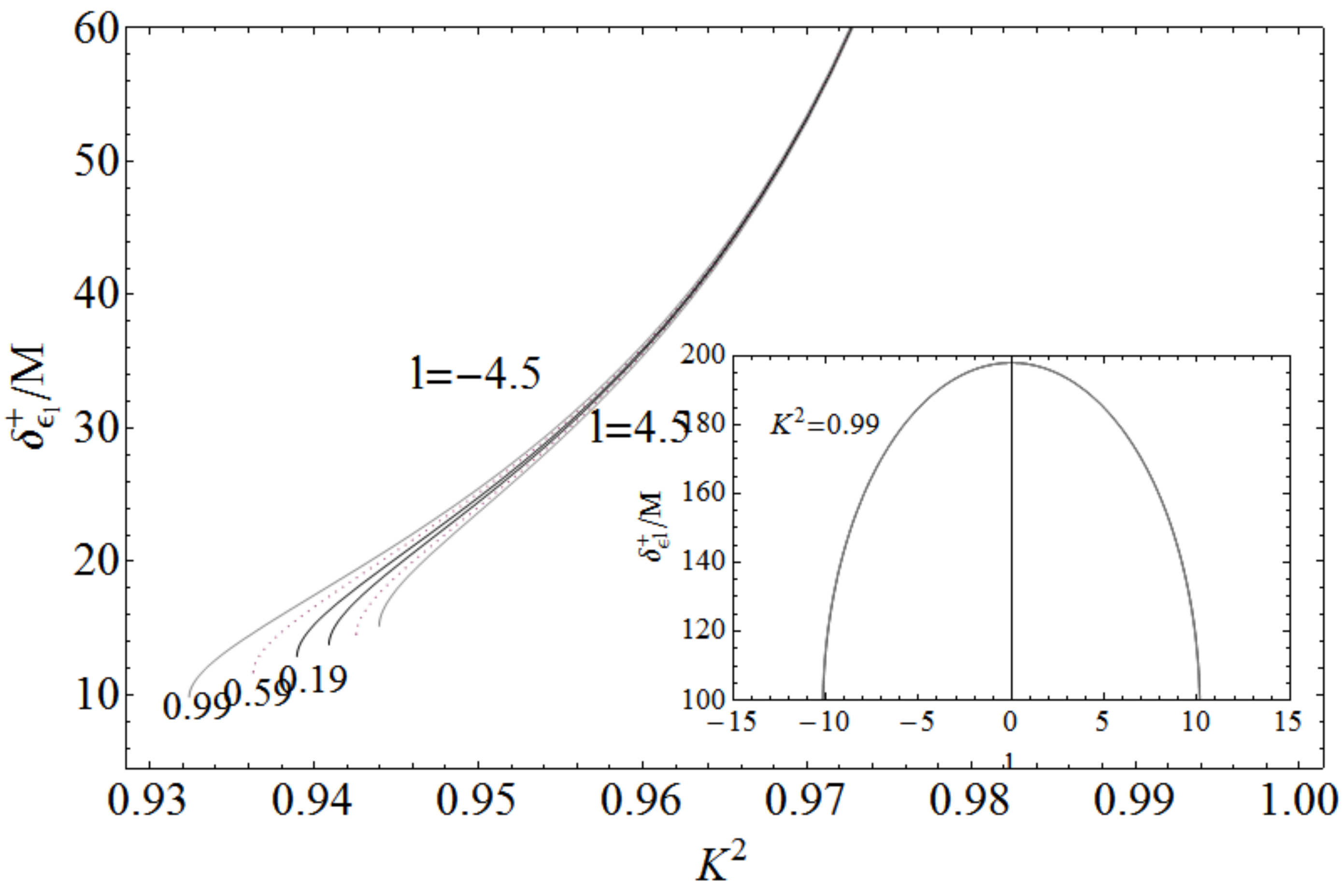}
\\
\includegraphics[scale=.31]{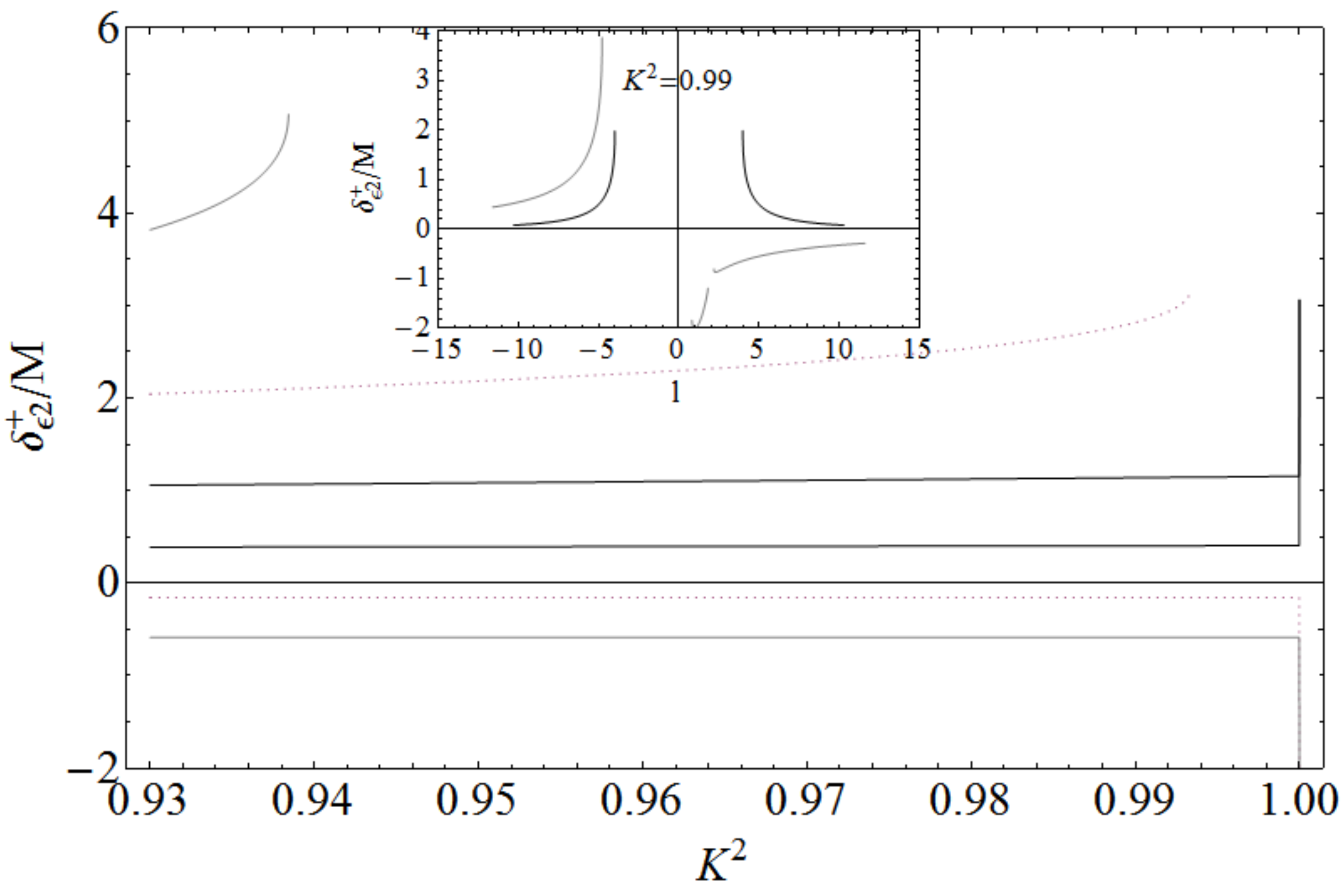}
\end{tabular}
\caption[font={footnotesize,it}]{\footnotesize{Plots of the function $\lambda\equiv y_1-y_3$ (left upper panel), surface of maximum torus diameter    and distance from the static limit surface $\delta_{\epsilon}^+\equiv y_3-r_{\epsilon}^+$ (right upper panel), the {distance of the torus inner surface from the structure inner surface}
$
\hat{\delta}\equiv y_3-y_2,
$ (bottom panel), the
{distance of the structure outer surface from the horizon}
    ${\delta}_{\epsilon_1}^+(a; \mathbf{p})\equiv y_1-2M$, and
 ${\delta}_{\epsilon_2}^+(a; \mathbf{p})\equiv y_2-2M$ where $y_2$ is the solution closest to the singularity, as functions of the parameter $K^2$, at different spacetime spin-mass ratios $a/M$: $a=0.99M\in\mathbf{BHIX}$ (gray curves), $a=0.59M\in\mathbf{BHIV}$ (dotted curves), $a=0.19M\in\mathbf{BHI}$ (black curves)  and  fluid angular momentum $\ell=-4.5$ (left set of curves) and  $\ell=4.5$ (right set of curves). Inset  plot shows each function of the parameter $\mathbf{p}\equiv(K,\ell)$ at fixed $K$, as function $\ell$, the black curve is for $a=10^{-5}M\in\mathbf{BHI}$, the gray curve is for $a=0.991M$. It is clear the symmetry of between the  $\ell$-counterrotating configurations respect to the transformation   $\ell\rightarrow-\ell$ and $\ell_f^-\rightarrow\ell_f^+$. The angular momentum is in units of mass $M$. }}\label{Fig:SLAMBA}
\end{figure}

The  morphology  of the thick accretion disks around a Kerr attractor changes with the parameter $\textbf{p}$ for, for example, the torus thickness, the disk elongation on the equatorial plane, the distance between the inner and outer surfaces of the configurations, and  many  properties are qualitatively poorly effected by the spin-mass ratio of the attractor \cite{PuMonBe12}.
However, for  $a\neq0$ there is  in general  a symmetry with the respect to the transformation $\ell\rightarrow-\ell$   for  different   accretion disk properties, see for example  Fig.\il(\ref{Fig:SLAMBA}.  and we can use this fact to analyze the disk morphology in terms of  the fluid angular momentum magnitude.
As it is   $y_2<y_3<y_1$ the {{surface maximum diameter}} can be defined as
\(
\label{E:ert}
\lambda(a; \mathbf{p})\equiv y_1-y_3
\). For $a=0$, it increases with the $K$-parameter, but
    decreases with the fluid angular momentum magnitude. The situation for $a\neq0$ is sketched in Fig.\il(\ref{Fig:SLAMBA}) where it  is clear  a symmetry  between the $\ell$counterrotating  configurations: $\lambda$ decreases with $a/M$ for corotating fluids and viceversa for the counterrotating ones.
Then we can  define the  {distance  on inner toroidal edge from the static limit surface $r_{\epsilon}^+$}   as:
\(
\delta_{\epsilon}^+(a; \mathbf{p})\equiv y_3-r_{\epsilon}^+,
\)
this is an increasing  function
 of the magnitude fluid angular momentum.
For corotating fluids, $\delta_{\epsilon}^+$
increases with $K$, and decreases with increasing $a$, at fixed $K$.
For counterrotating configurations,
$\delta_{\epsilon}^+$
increases with $K$ and $a$, see Fig.\il(\ref{Fig:SLAMBA})-right. For some configurations it can be $\delta_{\epsilon}^+<0$, according with the discussion in Sec.\il(\ref{Sec:limite.statico}).
 The {distance of the torus inner surface from the structure inner surface} can be defined as
\(
\hat{\delta}(a; \mathbf{p})\equiv y_3-y_2,
\)
this function
   increases with the angular momentum magnitude  and decreases with
    the energy. The function $\hat{\delta}(a; \mathbf{p})$  is clearly related to the unstable accretion configuration \cite{PuMonBe12}
and it is therefore essential for the determination of the accretion  (for $\hat{\delta}(a; \mathbf{p})=0$)  and the  multiple thick configurations as defined in Sec.\il(\ref{Sec:multipleP-D}).
Together with this, one can define
as well  the  {distance of the inner surface from the outer horizon} as:
\(
\breve{\delta}(a; \mathbf{p})\equiv y_2-r_+=\delta_{\epsilon}^+-\hat{\delta} +(r_{\epsilon}^+-r_+),
\)
for $a=0$ this quantity increases with increasing angular momentum of the fluid but decreases with
    increasing $K$.
Finally the {distance of the  outer edge from the static limit}
    ${\delta}_{\epsilon_1}^+(a; \mathbf{p})\equiv y_1-r_{\epsilon}^+$ increases with  $K$, increasing as the magnitude of the angular momentum decreases. This surface  is shown in Fig.\il(\ref{Fig:SLAMBA}),
    as function of  the parameter $\ell$, it is  clear the symmetry between the $\ell$counterrotating configurations. For  $ \ell<0$ at fixed  $K$, increasing the spacetime spin $a/M$, this quantity  ${\delta}_{\epsilon_1}^+$ increases on $\Sigma_{{\mathbf{p}}}$, viceversa  at $\ell>0$ the function ${\delta}_{\epsilon_1}^+$  is a decreasing function of $a/M$.

We introduce  the  function ${\delta}_{\epsilon_2}^+(a; \mathbf{p})\equiv y_2-2M$ where $y_2$ is the solution of the problem $\mathcal{V}_{eff}=0$ closest to the singularity, it in general  increases with the energy and decreases  with the
 magnitude of the    fluid angular momentum. This surface  is shown in Fig.\il(\ref{Fig:SLAMBA}), notably there  is  a partial loss of symmetry between $\ell$couterrotating configurations.
    Clearly, the symmetry is lost mostly with increasing the  \textbf{BH} spin,
 this fact should however not to be surprising as the properties of  ${\delta}_{\epsilon_2}^+(a; \mathbf{p})$ takes into account a general constant angular momentum $\ell$ (and therefore the general symmetry $\ell\rightarrow-\ell$), but no counterrotating Boyer solutions are allowed inside the region $]r_+,r_{\epsilon}^+[$ (see also discussion in Sec.\il(\ref{Sec:contro-l})
for some notes on the counterrotating configurations).
Therefore the results traced out from  Fig.\il(\ref{Fig:SLAMBA})  are valid for any general solutions of  the equation $\mathcal{V}_{eff}=0$, not  limited to the   Boyer surfaces, and they should be read in relation to the more general surface analysis presented in Secs.\il
(\ref{Subsec:sequence})  and (\ref{Sec:K.L}).
\subsubsection{On the relativistic angular velocity}
We focus now on the angular velocity $\Omega(\ell)$ on  $\theta=\pi/2$  as defined  in Eq.\il(\ref{Eq:flo-adding}).  $\Omega(\ell)$ increases always with the angular momentum $\ell$, i.e. $\partial_\ell\Omega(\ell)\nleq0$. For a counterrotating fluid it is  $
\Omega_f^+\equiv\Omega(\ell_f^+)<0$ the  minimum of the fluid relativistic angular velocity is located in $r_{\gamma}^+$, that is the function $\Omega_f^+(r)$ is always increasing with the radius $r>r_+$, or  $\partial_r\Omega_f^+>0$, and it is not well defined in $r_d$.
On the other hand for $\Omega_f^-\equiv \Omega(\ell_f^-)>0$ it is not well defined in $r_d$ and it is always decreasing with $r$, that is $\partial_r\Omega_f^-<0$.
In general, the function $\Omega(\ell)$  has a critical point i.e. $\partial_r\Omega(\ell)=0$  at $\ell<0$  (minimum  located at $r_{\xi}$)  and for
$\ell\in]a(M+r_+)/M,\ell_f^-(r_+)[$  (maximum  located at $r_{\xi}$).  For  $\ell\geq\ell_f^-(r_+)$ and $\ell\in]0, a(M+r_+)/M]$ the function $\Omega(\ell)$, where it is defined, decreases with $r/M$ where
\be
\frac{r_{\xi}}{M}\equiv
1-\frac{1}{\bar{\ell}}+2 \sqrt{\frac{(\bar{\ell}-1)^2}{\bar{\ell}^2}} \cos\left(\frac{1}{3} \mbox{arcsec}\left[-\frac{2 \sqrt{\frac{(\bar{\ell}-1)^2}{\bar{\ell}^2}} \bar{\ell}}{2+\bar{\ell} \left(a^2/M^2 \bar{\ell}-2\right)}\right]\right).
\ee

\section{Conclusions}\label{Sec:sum-con}
In  this work
we  investigated the structure and morphology of   thick  accretion disks orbiting around a Kerr attractor within    an hydrodynamic Polish doughnut (P-D) model.
The Boyer surfaces, associated with the critical points of the  effective potential, are solutions of the Euler equation. Keeping  the assumption of fluid angular momentum $ \ell $ constant we explored a  more general class of configurations including the Boyer tori and  constraining the role of the  hydrostatic pressure respect to the case of a Keplerian disk.
The model is regulated by the couple  $\mathbf{p}=(K,\ell)$ through the fluid effective potential  in the Euler equation.  Sequences  $\mathfrak{B}_{\mathbf{p}_j}\equiv\mathfrak{B}_{\mathbf{p}}/\Sigma_{\mathbf{p}_i}$   of corotating and  counterrotating configurations  are produced  fixing one  parameter $\mathbf{p_i}\in\mathbf{p}$  constant and  changing the remaining  $\mathbf{p}_j$.
 These collections of configurations, defined and characterized  in details in the  article, have been used here  many times in  different contexts   and  interpretations, providing  a useful tool for the characterization of both the properties of the  orbiting matter  both the background spacetime.
 We then investigated the orbital regions  where a P-D fluid configuration  exist,   proving   that these  regions depend on the gaps $\Delta_{\ell}^{\pm}(a)=\ell\pm a$, as a consequence of this it  was convenient the introduction of  a ``rationalized''  fluid momentum $ \bar{\ell}\equiv \ell/a $. Studying   the model in terms of the ratio $\bar{\ell}$ we identified and  discussed  the fluid properties   determined  by  the  dimensionless quantity   $\bar{\ell}$.
In particular we showed
the existence, in    the   P-D model, of a  maximum limit for   $\bar{\ell}$ and in  some cases  the  existence  of these configurations is mainly determined by the ratio $\bar{\ell}$ only.
According to the  analysis of the P-D disks,  we introduced  a  spacetime classification   in  nine sets  of attractors  named  accordingly   $\mathbf{BHI-BHIX}$. Defined by the spin-mass ratios, they are  characterized by the  sequences $\mathfrak{B}$: each class has a well defined and unique   arrangement of the Boyer surfaces in the   $\mathfrak{B}$ sequence, each slot filled with one or more configurations located in a defined orbital region, consequently   the toroidal configuration analysis reveals a  possible way to recognize a source from the investigation of  the toroidal accreting  surfaces. Finally, as the ergoregion  has an important role in the  of energy extraction  phenomena from the black hole, during the accretions and launching of jets, especially in the magnetohydrodynamic model \cite{Meier}, the cross of the static limit   is then analyzed in different black hole classes.
A more general class of configurations, including the P-D tori, have been also  considered,  these configurations  turn out in different topologies not arising from  the potential critical points. The array  of configurations, emerged from this analysis and including  the Boyer surfaces, is  thus   fitted   within  a dynamical interpretation adapted towards  a  possible comparison  with numerical simulations of   more extensive  dynamic models.
By varying $(\mathbf{p}_i,\mathbf{p}_j)\in \mathbf{p}$ we can construct   a set of nine matrices $\mathfrak{B}_{\mathbf{p}}$ on the surface $\Sigma_{\mathbf{p}}\equiv\Sigma_{K}\otimes\Sigma_{\ell}$ (or $\Sigma_{K}\otimes\Sigma_{\bar{\ell}}$), each for the nine \textbf{BH}-class of spacetimes.  The elements of   $\mathfrak{B}_{\mathbf{p_j}}$, matrix array or column,   figure different morphological phases related to  the history of a single $\mathbf{p}_i$-disk   pointing    $\mathbf{p}_j $ as an  evolutive or  ``chronological''  parameter, meaning that  we assume it  to follow  an   evolutionary model for the  configuration.
The  real disk  evolution  however can   occur along a diagonal or any other  sequence of  elements of $\mathfrak{B}_{\mathbf{p}}$, these     however could be easily fitted  according to a known dynamical law or  by comparison with  numerical simulations: considering  the matrix elements  following  a different order,
we should recognize the matrix elements  and identify then a proper exact chronological order.
As accretion is usually   modeled in terms of  angular momentum transport  inside the matter \cite{Balbus2011,Abramowicz:2011xu}, one   realistic choice for an evolutive parameter would be the fluid  momentum $\ell$.
The introduction of the   sequences $\mathfrak{B}$ was then considered
as defined on a  $\Sigma_t$
for  the analysis of
multi-structured toroidal accretion disks, that is     thick configurations structured in multiple  P-D sub-configurations $C_i$  (rings), defined as  the union of each closed, crossed or not torus  orbiting    the same attractor, whose mutual intersection is the null set or at last the inner or outer edge of the configurations.
To properly define these solutions we introduced a series of concepts as  the  \emph{separated} or  \emph{intertwined}  sub-configurations  and    \emph{loops} of  sub-configurations,   we  defined then  the multi-structured toroidal accretion disks as obtained combining  multiple  separated torus rings.
Our analysis has been performed considering  the situation in  the nine Kerr spacetime classes.
It is relevant here that,  by characterizing the properties of $C_i$ configurations,  emerged   the sub-configurations are deeply constrained  in number  and in the angular momentum, this first analysis indeed shows  the possibility in some cases of having a multi-structured disk made by a limited number of rings. Thus we introduced   the concept of $\ell$corotating and $\ell$counterrotating rings. As two rings, the inner one closest to the attractor   $(i)$  and the farther one $(o)$,  can be mutually corotating or
 {\textbf{\large{$ {\ell}$}}}corotating, meaning here $\ell_{(i)}\ell_{(o)}>0$ or {\textbf{\large{$ {\ell}$}}}counterrotating,   $\ell_{(i)}\ell_{(o)}<0$.
 The arrangement, in  the overall structure, of the {\textbf{\large{$ {\ell}$}}}corotating and {\textbf{\large{$ {\ell}$}}}counterrotating  rings of the  same disk,   is strongly constrained in the   fixed class of background geometries.
 Then we investigated these configurations  for  a polytropic equation of state and  analyzing the disk morphological properties in relation to the {\textbf{\large{$ {\ell}$}}}counterrotating rings, as the  torus thickness, the disk elongation on the equatorial plane, the distance between the inner and outer surfaces of each  configuration. For  $a\neq0$, the properties of the disks, as function of $\mathbf{p}$, show   a symmetry with the respect to the transformation $\ell\rightarrow-\ell$ or  a symmetry  between the $\ell$counterrotating  configurations. Some properties of the relativistic frequency are also analyzed.
 This analysis is in our view a major point in this  work.
The results traced here  we expect  will be     completed in a future work, investigating   the different  possible multi-structures not considered  in the present analysis, and  including also the information related to the thickness and spatial distance of   each sub-configurations\cite{coop}.
Thus the dynamical equilibrium of the entire  structure will be then  investigated:
the outer sub-configurations may affect the stability of the entire  structure,  generating a global  and internal instability phenomena, due to the presence of one or more  exterior $C_x$-configurations. The mutual destabilization  (as foreseen  by the Paczy\'nski mechanism due to
of violation of the hydrostatic equilibrium) among the sub-configurations  also occurs   with   exchange of
fluid elements among  the rings. This mechanism imposes indeed  a   limit in the number of disks, location,  fluid angular momentum  and thickness for  the equilibrium of the multiple structures.

\appendix
\begin{acknowledgments}
This work has been developed in the framework of the CGW Collaboration
(www.cgwcollaboration.it). DP wishes to thank the Blanceflor Boncompagni-Ludovisi, n\'ee Bildt Foundation, and would like to thank the institutional support
of   the Faculty of Philosophy and Science of the Silesian University of Opava.
\end{acknowledgments}
\section{Some further considerations on the effective potential}\label{Po}
\begin{figure}
\begin{tabular}{cc}
\includegraphics[scale=.23]{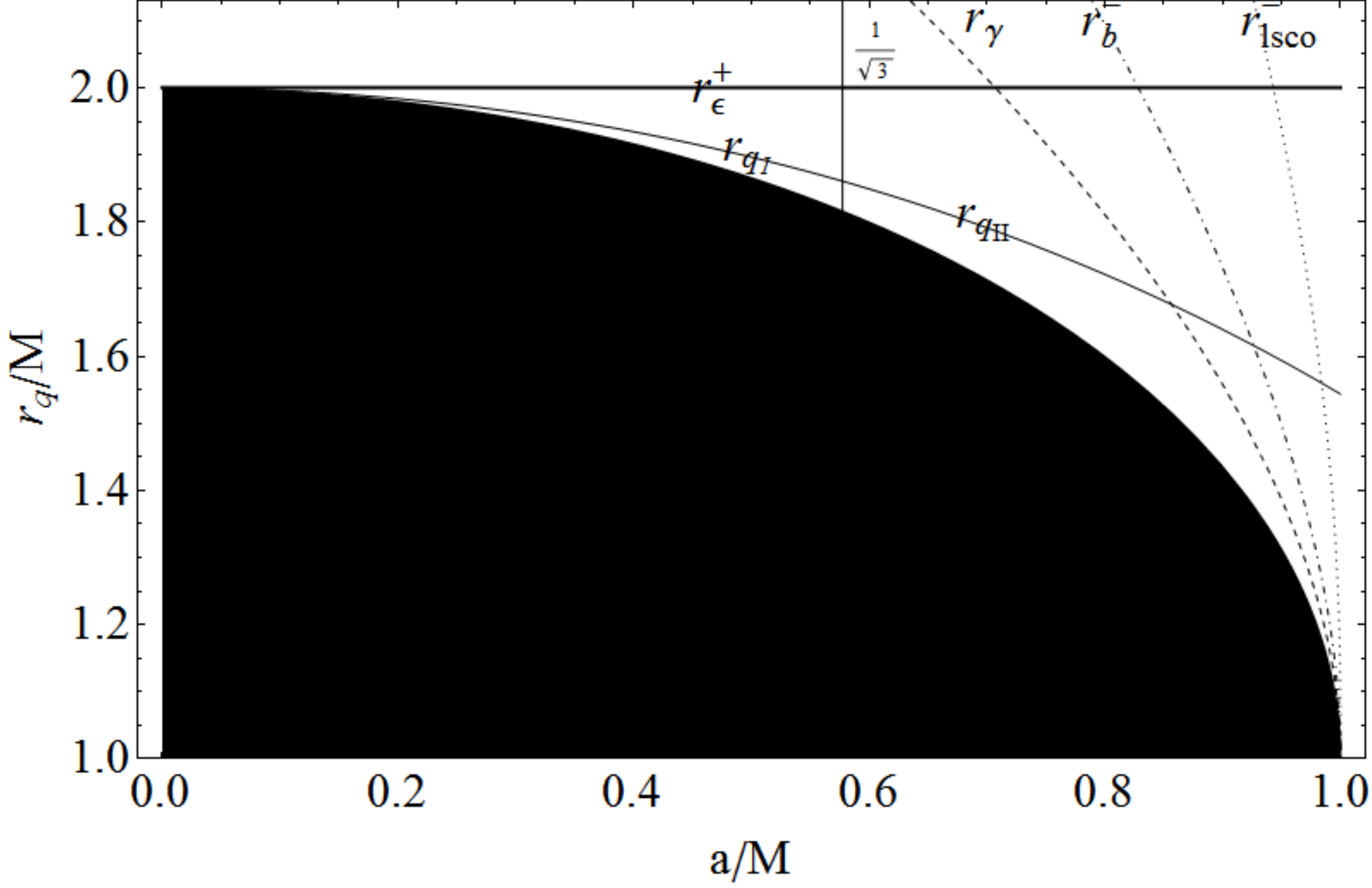}
\end{tabular}
\caption[font={footnotesize,it}]{\footnotesize{
Plot of $r_{\ti{qI}}/M$ and $r_{\ti{qII}}/M$ as function of $a/M$. Black region is $r<r_+$, the static limit $r_{\epsilon}^+$, the last circular orbit  $r_{\gamma}^-$, last bounded orbit $r_b^-$, last stable circular orbit radius  $r_{lsco}^-$ (test particle with $L=L_-$) are also plotted. }}
\label{Fig:Sollkinp}
\end{figure}
Angular momentum  and the energies for test particle circular orbits are
\be\label{LPM}
\frac{L_{\pm}}{\mu M}\equiv\frac{\left|\frac{a^2}{M^2}\pm
2\frac{a}{M}
\sqrt{\frac{r}{M}}+\frac{r^2}{M^2}\right|}{\sqrt{\frac{r^2}{M^2}\left(\frac{r}{M}-3\right)\mp
2\frac{a}{M}\sqrt{\frac{r^{3}}{M^3}}}}, {{E}^{(+)}_{\pm}}\equiv{{E}(L_{\pm})},\quad
{{E}^{(-)}_{\pm}}\equiv{{E}(-L_{\pm})} \ .\ee
for corotating $L_-$ ($(-)$) and counterrotating orbits $(-L_+)$ ($(+)$) orbits.
the photon orbits are and the last bounded orbits are respectively
\bea
{r_{\gamma}^{\mp}}\equiv 2 M\left(1+\cos \left[\frac{2 \arccos\left(\mp \frac{a}{M}\right)}{3}\right]\right),\quad
r_b^{\pm}\equiv2M\pm a+2 \sqrt{M}\sqrt{M\pm a},
\eea
the last stable circular orbits
\bea
r_{lsco}^{\mp}\equiv M\left(3+Z_2\mp\sqrt{(3-Z_1)(3+Z_1+2Z_2)}\right),
\eea
$Z_1\equiv1+\left[1-(a/M)^2\right]^{1/3}\left[(1+a/M)^{1/3}+(1-a/M)^{1/3}\right]$ and $Z_2\equiv\sqrt{3(a/M)^2+Z_1^2}$.

It is $V_{eff}(\ell)=M/\ell$ (i.e. $L=\mu M$), where $a\in]0,M]$: $ r\in]r_+,r_q[$ and   $\ell=\ell_s^{\pm}$, on
$r=r_q$ and  $\ell={r^3+a^2 (2M+r)}/{4 M a}$, finally it is   $r>r_q,\;\ell=\ell_s^+$,
where
\bea
&&
\ell_s^{\mp}\equiv-\frac{2 M^3a}{r \Delta-2M^3}\mp\sqrt{\frac{rM^2 \Delta \left(M^2 r+r^3+a^2 (2M+r)\right)}{\left(r \Delta-2M^3\right)^2}},\quad r_q=1.86102M \mbox{for}\quad a/M=1/\sqrt{3},
\\
&&
r_q=r_{\ti{qI}}\equiv\frac{2}{3} \left(M+\sqrt{M^2-3 a^2} \cos\left[\frac{1}{3} \arccos\left[\frac{26M^3-9 a^3}{\left(M^2-3 a^2\right)^{3/2}}\right]\right]\right)\quad\mbox{for}\quad a/M\in[0,1/\sqrt{3}[,
\\
&&
r_q=r_{\ti{qII}}\equiv
-\frac{2}{3} \left(-M+\sqrt{M^2-3 a^2} \sin\left[\frac{1}{3} \arcsin\left[\frac{26M^3-9 a^3}{\left(M^2-3 a^2\right)^{3/2}}\right]\right]\right)\quad\mbox{for}\quad a/M\in]1/\sqrt{3},1].
\eea
see Figs.\il(\ref{Fig:Sollkinp}).

\end{document}